\documentclass[a4paper, 12pt, twoside]{book}
\usepackage{a4}
\usepackage{amsfonts}
\usepackage{amssymb}
\usepackage{epsfig}

\newcommand{\be}{\begin{equation}}
\newcommand{\ee}{\end{equation}}
\newcommand{\bea}{\begin{eqnarray}}
\newcommand{\eea}{\end{eqnarray}}
\newcommand{\mbb}{\mathbb}
\newcommand{\ti}{\times}
\newcommand{\half}{\frac{1}{2}}
\newcommand{\third}{\frac{1}{3}}
\newcommand{\quarter}{\frac{1}{4}}
\newcommand{\twothirds}{\frac{2}{3}}
\newcommand{\mc}{\mathcal}
\newcommand{\K}{\mc{K}}
\newcommand{\vphi}{\varphi}

\newtheorem{Yau}{Theorem}

\begin{document}

\pagestyle{headings}

\thispagestyle{empty}
\newcommand{\HRule}{\rule{\linewidth}{1mm}}
\setlength{\parindent}{1cm}
\setlength{\parskip}{1mm}
\vspace*{\stretch{1}}
\noindent
\HRule
\begin{center}
{\Huge Moduli Stabilisation and Applications in IIB String Theory} \\[5mm]
\end{center}
\HRule
\vskip 2cm
\begin{center} \Large Joseph Patrick Conlon \end{center}
\begin{center} \Large
 Christ's College,
   Cambridge \\ and \\ Trinity College, Cambridge.

\vskip 4cm
\epsfig{file=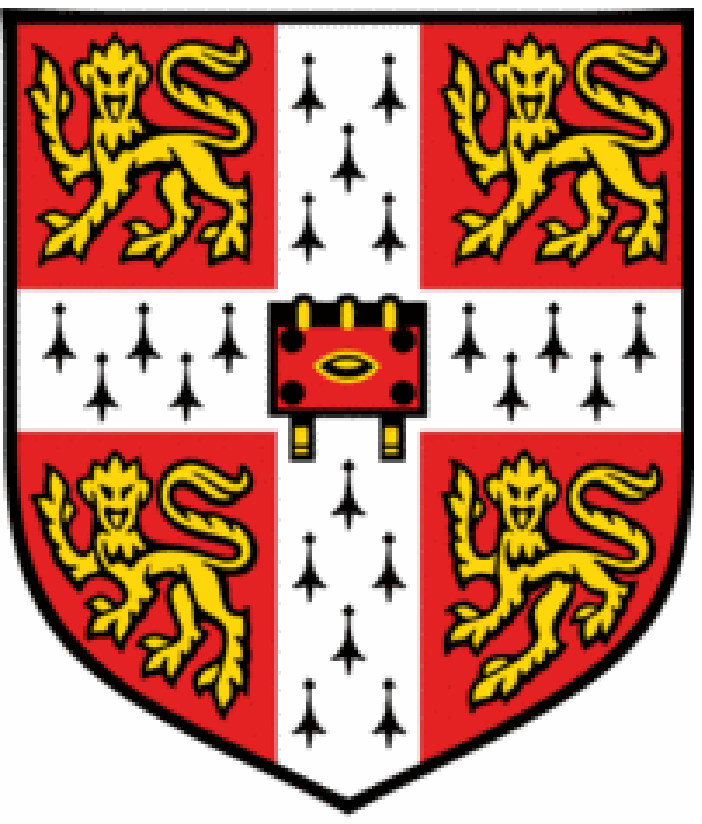, width=3.3cm, height=4cm}
\end{center}

\vspace*{\stretch{2}}
\begin{center}\Large \textsc {Dissertation submitted for the degree of
     Doctor of Philosophy \\ Cambridge University 2006}\end{center}

\newenvironment{dedication}   {\newpage \thispagestyle{empty}
\vspace*{\stretch{1}} \begin{center} \em}   {\end{center}
\vspace*{\stretch{3}} \clearpage}
\begin{dedication}\begin{flushright}{\it \Large{}}\end{flushright}\end{dedication}

\newenvironment{PreQuote}   {\newpage \thispagestyle{empty}
\vspace*{\stretch{1}} \begin{center} \em}   {\end{center}
\vspace*{\stretch{2}} \clearpage}
\begin{PreQuote} 
Per varios casus, per tot discrimina rerum \newline
tendimus in Latium; sedes ubi fata quietas \newline
ostendunt; illic fas regna resurgere Troiae. \newline
Durate, et vosmet rebus servate secundis. \phantom{g} \newline 
\newline
Aeneid I:204-7\end{PreQuote}

\begin{dedication}\begin{flushright}{\it \Large{}}\end{flushright}\end{dedication}

\newpage
\thispagestyle{empty}
\vspace*{\stretch{0.8}}
\noindent
\begin{center}
{\Large Acknowledgements}
\end{center}
I have many people to thank.
I am particularly grateful to my supervisor Fernando Quevedo for the
physics he has taught me, for his continual encouragement and for his
generosity of time and spirit in nurturing me in the world of science.
On the subjects contained in this thesis I have learned many things from my
collaborators Vijay Balasubramanian, Per Berglund, Fernando Quevedo
and Kerim Suruliz. I am also grateful to Ralph Blumenhagen and
Kerim Suruliz for collaboration on 
the unrelated paper \cite{hepth0404254} not described in this thesis.
Within Cambridge I have had stimulating discussions on various topics with many people
including Shehu Abdussalam, Ben Allanach, Heng-Yu Chen, Daniel
Cremades, Paul Davis, Maria Pilar Garcia del Moral, David Jennings, David Kagan, Chris Rayson,
Aninda Sinha, Julian Sonner,
Simone Speziale, Kerim Suruliz, Timo Weigand and Chris White.
Through conversations at conferences and email correspondence I have
also learnt many things from 
many physicists, and I am indebted to them for their time and careful
explanations. An incomplete list includes Cliff Burgess, Gabriele
Honecker, Shamit Kachru, Liam McAllister, Eran Palti and Michael Schulz.

For financial support, I am grateful to EPSRC for a studentship and to Trinity College, Cambridge
for a research fellowship. I would also like to thank both Christ's
College, Cambridge and Trinity College, Cambridge for contributions towards travel expenses.

Above all these I am grateful to my parents, for their love and
support throughout my life, and my fiance\'e Lucy, \emph{lux vitae} and whose smile is my
greatest joy.
\vspace*{\stretch{1}}

\newpage
\thispagestyle{empty}
\begin{center}
\Large{Declaration}
\end{center}
This dissertation is the result of my own work and includes nothing
which is the outcome of work done in collaboration except where
specifically indicated in the text. No part of this thesis has
previously been
submitted for a degree or other qualification at this or any other university.

This thesis
is based on the research presented in the following papers:
\begin{enumerate}

\item J. P. Conlon and F. Quevedo, {\it On the Explicit Construction and Statistics of Calabi-Yau
  Flux Vacua}, JHEP 0410:039 (2004), arXiv:hep-th/0409215.

\item  V. Balasubramanian, P. Berglund, J. P. Conlon and F. Quevedo, 
{\it Systematics of Moduli Stabilisation in Calabi-Yau Flux
Compactifications}, JHEP 0503:007 (2005), arXiv:hep-th/0502058.

\item J. P. Conlon, F. Quevedo and K. Suruliz, {\it Large Volume Flux Compactifications: Moduli Spectrum and
D3/D7 Soft Supersymmetry Breaking}, JHEP 0508:007 (2005), arXiv:hep-th/0505076.

\item  J. P. Conlon and F. Quevedo, {\it K\"ahler Moduli Inflation},
  JHEP 0601:146 (2006), arXiv:hep-th/0509012.

\item J. P. Conlon, {\it The QCD Axion and Moduli Stabilisation}, JHEP
  0605:078 (2006), arXiv:hep-th/0602233.

\item  J. P. Conlon and F. Quevedo, {\it Gaugino and Scalar Masses in
  the Landscape}, JHEP 0606:029 (2006), arXiv:hep-th/0605141.

\end{enumerate}

These papers are references \cite{hepth0409215,
  hepth0502058, hepth0505076, hepth0509012, hepth0602233, hepth0605141}
in the bibliography.

Sections 6.1, 6.2 and A.3 of this thesis are based on parts of 
the third paper above that were primarily
written by Kerim Suruliz. They are
included here to ensure completeness of the argument.

\vspace{2cm}

\begin{flushright}
JOSEPH CONLON 

Cambridge, Feast of Ss. Peter and Paul

29th June 2006
\end{flushright}

\newpage
\thispagestyle{empty}
\begin{center}
\Large{Summary} \\
\end{center}
 
String compactifications represent the most promising approach towards
unifying general relativity with particle physics. However, naive
compactifications give rise to massless particles (moduli) which would
mediate unobserved long-range forces, and it is therefore necessary to
generate a potential for the moduli. 

In the introductory chapters I review this problem and
recall how in IIB compactifications the dilaton and
complex structure moduli can be stabilised by 3-form fluxes. 
There exist very many possible discrete flux choices which motivates
the use of statistical techniques to analyse this discretuum of
choices. Such approaches generate formulae predicting the
distribution of vacua and I describe numerical tests of these formulae
on the Calabi-Yau $\mbb{P}^4_{[1,1,2,2,6]}$.
Stabilising the K\"ahler moduli requires nonperturbative
superpotential effects. I review the KKLT construction and explain why
this must in general be supplemented with perturbative K\"ahler corrections.
I show how the incorporation of such corrections generically leads
to non-supersymmetric minima at exponentially large volumes, giving
a detailed account of the $\alpha'$ expansion and its relation to
K\"ahler corrections. I illustrate this with explicit computations for
the Calabi-Yau $\mbb{P}^4_{[1,1,1,6,9]}$.

The next part of the thesis examines phenomenological applications of
this construction. I first describe how the magnitude of the soft supersymmetry parameters
may be computed. In the large-volume models the gravitino mass and
soft terms are volume-suppressed. As we naturally have $\mc{V} \gg 1$, this gives a
dynamical solution of the hierarchy problem.
I also demonstrate the existence of a fine structure in the soft
terms, with gaugino masses naturally lighter than the
gravitino mass by a factor $\ln \left( \frac{M_P}{m_{3/2}} \right)$. A
second chapter  
gives a detailed analysis of the relationship of moduli stabilisation to
the QCD axions relevant to the strong CP problem, proving a no-go theorem on the
compatibility of a QCD axion with supersymmetric moduli stabilisation.
I describe how QCD axions can coexist with nonsupersymmetric perturbative stabilisation
and how the large-volume models naturally
contain axions with decay constants that are phenomenologically
allowed and
satisfy the appealing relationship $f_a^2 \sim M_P M_{susy}$. 
A further chapter describe how a simple and predictive inflationary model can be
built in the context of the above large-volume construction, using the
no-scale K\"ahler potential to avoid the $\eta$ problem.

I finally conclude, summarising the phenomenological scenario and
outlining the prospects for future work.

\begin{dedication}\begin{flushright}{\it \Large{}}\end{flushright}\end{dedication}

\tableofcontents

\part{Introduction}

\chapter{Introduction and Motivation}
\label{Intro} \linespread{1.3}

`If string theory is the answer, what is the question?' 

The basic assumption
of this thesis is that the appropriate question is `What set of ideas
underlies, unifies and explains the physics of both general relativity
and the Standard Model?', and the basic aim of this thesis is to address how
one can connect the formal structure of string theory with the
particle physics and cosmology that can be tested experimentally.

General relativity and quantum mechanics are two of the best ideas of
all time and were the two major
achievements of twentieth century physics. The former originates in
classical electrodynamics and the special theory of relativity. This
was soon generalised by Einstein to its mature formulation as general
relativity. In this theory, space and time are themselves the dynamical
quantities. Particles follow geodesics in spacetime determined by the
spacetime metric, which is in turn determined by the distribution of matter.
The force of gravity is the manifestation of space-time curvature: the
attraction of particles corresponds to the approach of geodesics.
The action governing general relativity is the Einstein-Hilbert action
\be
\label{eqEHA}
S_{EH} =  \int d^4 x \sqrt{-g} \left( \frac{\mc{R}}{16 \pi G} + 2
\Lambda + \mc{L}_M \right).
\ee
This action describes the physics of black holes, the expansion of the
universe and the precession of the perihelion of
Mercury. It is tested everyday in
the use of GPS receivers, whose accuracy depends on the correct
inclusion of general
relativistic effects. It is a precise relativistic account of gravity,
the highpoint of classical physics, and is consistent with all
experimental tests to date \cite{grqc0103036}.

The other cornerstone of modern physics is quantum mechanics. While
general relativity is a theory of the big - tested on solar system and
cosmological scales - quantum mechanics is a theory of the small. It
originated in the study of physics at atomic length scales. The measurement of
atomic emission and absorption spectra led to the first rudimentary
quantisation laws of Bohr and Einstein. In the 1920's these were
developed into quantum theories of particle mechanics through the
Schr\"odinger and Dirac equations, 
\bea
\mc{H} | \psi \rangle & =  & i \hbar \frac{\partial}{\partial t} | \psi
\rangle \\
(i \Gamma^\mu \partial_\mu + m) \psi & = & 0.
\eea
The precision with which these equations worked - for example in the
spectrum of the Hydrogen atom -
confirmed quantum mechanics as a necessary part of the correct
description of the world. However, many years elapsed before quantum
theories of particles, such as electrons, were successfully extended to quantum
theories of fields, such as electromagnetism.

This advance was famously conceived at the 1947 Shelter Island conference. The
recent measurement of the Lamb shift in the hydrogen spectrum 
provoked the theorists present into developing the quantum field
theory of electromagentism, quantum electrodynamics (QED). Particles
were no longer understood as fundamental objects but rather as
elementary excitations of an underlying field: for example the photon
is the elementary excitation of the electromagnetic field. In modern
language, QED is a $U(1)$ gauge theory: a field theory with a local
$U(1)$ symmetry at every point in space time. In addition to
electromagentism, two other short distance forces are known: the
strong force, responsible for binding together atomic nuclei, and the
weak force, responsible for radioactive $\beta$ decay. The former is described
by an $SU(3)$ gauge theory and the latter by a (spontaneously broken)
$SU(2)$ gauge theory. It is not understood why we observe these forces
and these forces only.

In the first development of quantum field theories, a major problem
was encountered with infinities: naive attempts to calculate quantum
corrections to classical effects gave infinite answers. This problem
was addressed by the development of renormalisation: just because
a quantity was infinite did not mean that quantity was zero. Specifically, by
including a finite number of infinite counterterms, the infinities
could be cancelled and physical predictions in agreement with experiment
could be extracted. Originally this procedure was rather \emph{ad hoc},
and it was thought that the only sensible quantum field theories were
renormalisable ones. In the modern Wilsonian understanding,
the infinities requiring renormalisation come from integrating out high
energy degrees of freedom. Renormalisation is the procedure of doing
low-energy quantum physics without knowing the high-energy
completion. In non-renormalisable theories, the leading physics is
suppressed high-energy physics, and in order to do quantum physics
we must first understanding the underlying theory.

For example, we expect any quantum field theory to fail at the Planck
scale, $M_P = 2.4 \ti 10^{18} \hbox{GeV}$. Excepting the quadratic
Higgs divergence, all the divergences encountered in particle physics
are logarithmic. If a Planck-scale cutoff is introduced, the
`divergences' become finite,
\be
\int_{p_0}^\infty \frac{d^4 p}{p^4} \to \int_{p_0}^{M_P} \frac{d^4
  p}{p^4} \sim \ln\left(\frac{M_P}{p_0}\right).
\ee
The divergences are therefore not real divergences, but merely a
consequence of our lack of knowledge of the fundamental high energy
theory. The need to renormalise implies the existence of a more
fundamental theory in which the divergent quantities are actually finite.
The low-energy approximate theory
carries the imprint of the fundamental theory in the
texture of low-energy parameters such as masses and coupling
constants. 

The particular approximate theory that accurately describes the
physics seen at particle accelerators is known as the Standard
Model. It consists of an $SU(3) \ti SU(2) \ti U(1)$ gauge theory
coupled to three generations of chiral matter and has been rigorously
tested in many different settings \cite{PDG}. There is clearly 
structure sitting behind the Standard Model: for example, the matter
content is replicated three times - this is not explained. 
The masses of the particles in each generation show a
hierarchical structure - this is not explained. The strong
interactions do not violate CP, and the weak interactions do - this is
not explained. In the minimal Standard Model, the Higgs mass is
unstable against radiative corrections: this may be explained by
supersymmetry, but this is not known.
There is clearly a structure behind the Standard
Model, but it is not known what this structure is.

One further piece of physics definitely in nature and definitely not in the
Standard Model is the gravitational force. General relativity
does an excellent job of describing gravity, but it is an excellent
classical job. As quantum mechanics is integral to nature, there must
exist a quantum theory of the gravitational force not included in the Standard Model.
One may hope that the correct formulation of this theory will help
explain some of the structure visible and not understood in the Standard Model.
Another hope is that the quantum theory of gravity will allow a
description of
gravity common with the other known forces of nature. This is not because
of logical necessity, but rather an educated guess based on the historical trend
towards unification in physics.

The natural approach to finding a quantum theory of gravity is to 
use the same methods that worked so effectively for the other forces. 
Unfortunately this method fails irrevocably. The reason conventional
quantisation works for gauge theories is that in four dimensions gauge
theories are renormalisable. This tells us that at low energy we do
not need to know the high-energy theory in order to do quantum, as
opposed to classical, computations.
This is not true for general relativity, which is an irrevocably 
non-renormalisable theory. General relativity is an effective theory,
valid at energies $E \ll M_P$. In addition to the interactions 
of equation (\ref{eqEHA}), there will also be non-renormalisable
interactions suppressed by higher powers of $M_P$,
\be
\label{eqNRGR}
S = \frac{M_P^2}{2} \int \sqrt{g} \left( \mc{R} +
\frac{1}{M_P^2} \left( c_1 \mc{R}^2 + c_2 \mc{R}_{\alpha \beta}
\mc{R}^{\alpha \beta} + c_3 \mc{R}_{\alpha
  \beta \gamma \delta} \mc{R}^{\alpha \beta \gamma \delta} \right)  +
\ldots \right).
\ee
While unimportant at low energies, for scattering energies $E \gtrsim
M_P$ the higher-derivative interactions of equation (\ref{eqNRGR}) are
as equally important as the Einstein-Hilbert term. Finding a quantum
theory of gravity requires a way to calculate the constants $c_1,
c_2, \ldots $ in
order that scattering at energies $E \gtrsim M_P$ be
predictive. The quantum effects are encoded in the values of these
terms. Unfortunately, at high energies an infinite number of such
terms are required and so it is not possible to be predictive with any
finite number of measurements.
Clearly, we expect new degrees of freedom to appear at the Planck scale - 
however as the action (\ref{eqEHA}) describes an
effective theory it gives no clue as to what these degrees of
freedom should be.

String theory is a hypothetical account of physics at the Planck scale
and in particular of the new degrees of freedom that are present there.
The subject is well named - the fundamental perturbative
excitations are one-dimensional extended objects, strings. The new
degrees of freedom for energies
above the Planck scale 
are a tower of excited string states.\footnote{assuming the Planck and string scales to be
equivalent.} 
The strings contain
both gauge and gravitational degrees of freedom in the oscillations of
open and closed strings respectively. This explains the
principal appeal of string theory, in that it offers a framework in
which both gauge and gravitational physics can be simultaneously addressed.
In particular, as the higher-derivative corrections to the classical
gravitational action can be computed and are finite, string theory
represents a \emph{quantum} theory of gravity.

However, this is not as first advertised, for
string theory was originally formulated in the late 1960's as a theory
of the strong interactions. In this context it encountered three serious problems. First, the theory required
more than four dimensions. Secondly, there always existed a massless
spin two state present in the theory, while no such state was observed
in the
strong interactions. Finally, in string theory 
scattering amplitudes were exponentially soft at high energy while the
observed (QCD) amplitudes were power-law soft.
The interest in string theory therefore faded as
quantum chromodynamics (QCD) was realised to be
the correct description of the strong interactions. 
It was subsequently proposed by Scherk and Schwarz in 1974
\cite{ScherkSchwarz} to turn the vices into virtues by
re-interpreting the massless spin two particle as a graviton and treating
string theory as a theory of quantum gravity. As the critical
spacetime dimension for the superstring is ten, string theory gives
rise to a ten-dimensional theory of quantum gravity.

Despite its current pre-eminence, this approach was for a long time unpopular.
As a highly supersymmetric ten-dimensional theory, the low energy
limit of string theory is ten dimensional supergravity, and it was `known' that
supergravity theories with charged chiral fermions suffered from both gauge
and gravitational anomalies.
However in 1984 it was shown by Green and Schwarz in a celebrated
paper \cite{GreenSchwarz} that these anomalies exactly cancelled 
for gauge groups $SO(32)$ and $E_8 \ti E_8$ with the addition of extra terms arising from the low-energy limit of
string theory. 
This discovery provoked an intense study of the subject which has
continued to the present day.

For string theory - a ten-dimensional theory - to represent a
candidate description of nature, the extra dimensions have to be
compactified. The possibility that extra-dimensional geometry
underlies the observed forces and matter content is an old idea
dating back to Kaluza and Klein \cite{Kaluza, Klein}. The original Kaluza-Klein observation
was that five-dimensional general relativity compactified on a circle
gives a four-dimensional Einstein-Maxwell-scalar theory. String
compactifications involve six extra dimensions and the
compactification spaces - typically Calabi-Yau threefolds - are far
more complicated than a simple $S^1$.
However while the details differ considerably, the essence is nonetheless the same - the particle
content and forces in four dimensions are determined by the geometry of
the extra, compactified dimensions. 

In addition to the graviton, ten-dimensional supergravity theories
also contain gauge fields and p-form potentials. On dimensional
reduction, the particles observed in four dimensions come from the
zero modes of these fields (there being no experimental
evidence for Kaluza-Klein particle towers). Zero modes correspond
to zero eigenvalues of a particular differential operator: for the
reduction of scalar
fields, this is the ordinary Laplacian $\nabla^2$. It is a well-known
mathematical fact that there is a correspondence between 
zero modes of differential operators on a given space and topological
data of that space such as the dimension of appropriate cohomology
groups. For example, in the simplest heterotic string
compactifications the number of generations (fermions minus
antifermions) is determined by the Euler
number of the Calabi-Yau used for the compactification,
\be
N_{gen} = \half \vert \chi(M) \vert.
\ee
Yukawa couplings are another low-energy quantity that may be geometrically determined in
string compactifications. Recall that the masses of Standard Model
fermions are given by the fermion Yukawa coupling to the Higgs. In
heterotic compactifications, the structure of Yukawa couplings may be determined at tree-level by
the triple intersection form
$$
\kappa_{ABC} = \int_{X_6} e_A \wedge e_B \wedge e_C,
$$
where $e_A$ are a basis of 2-forms on the Calabi-Yau.

The above explains the reason why string theory has been studied so
intensely: it appears to represent a theory of quantum
gravity which also naturally includes the elements of particle physics.
Furthermore, in string theory there are no fundamental constants: 
all statements about
masses or couplings are statements about the vacuum state and thus
about the dynamics of the theory.
This intense study is further driven by the lack of any
serious rival: the problem of finding consistent quantum theories unifying gauge and
gravitational interactions is just hard, and may even have a unique
solution.

The study of string theory has
led to many profound results, some with deep connections to apparently
unrelated fields. Among these are the discovery of mirror symmetry
\cite{CLS, GreenePlesser}, an exact microscopic calculation of the
Bekenstein-Hawking black hole entropy \cite{hepth9601029}, a smooth
description of spacetime topology change \cite{hepth9301042,
  hepth9301043, hepth9504090, hepth9504145} and the AdS/CFT
correspondence \cite{hepth9711200}. 
There exist fully pedagogical accounts of the subject in textbooks 
such as \cite{GSW, PolchinskiBook}. Despite these successes, progress
in phenomenology has been much more limited than had been hoped in 1985: 
the origin of the structure of the Standard Model is no
better understood now than it was then. Advances in this area have
been mostly internal and a decisive low-energy test of string theory
does not seem possible.

There are two broad approaches to the study of the subject. The first primarily seeks a deeper understanding
of the fundamental theory. Such a physicist feels either that the most
interesting
aspects of string theory are the mathematical ones or that the
understanding of string theory is too poor to make contact with
experiment: the royal road to experiment lies squarely through the centre of
the M-theory duality web.
He hopes that apparent problems, such as the multiplicity of vacua,
may disappear once the theory is fully understood.
A physicist following the second approach does not disagree about the existence of important conceptual problems
in understanding and defining the theory. However he
feels that we should try and come as close as possible to observed
physics. This task sharpens our understanding of the underlying
physics,
and anyhow it is surely not the case that the deep structure and complexity of
the Standard Model will just pop out, however well we understand the theory.
These approaches are of course complementary; the possibilities for stringy model-building were greatly enriched
by the 1995 discovery of D-branes\cite{hepth9510017}, whereas the discovery of mirror symmetry followed scatter
plots of the Hodge numbers of known Calabi-Yaus\cite{CLS,
  GreenePlesser}. 
However, in this thesis I shall align myself more with
the second approach to the subject.

The ultimate objective of string model-building is to reproduce
and explain the entire structure of the Standard Model: scales, particle
content, masses, charges and interactions. As indicated above,
explaining even a very limited subset of these would constitute a
great success. This enterprise has one main conceptual problem and
several important technical problems. The conceptual problem is that
of vacuum selection. String theory seems to admit
a very large number of vacuum solutions and there does not seem to be a good way of choosing
between these. 
It is relatively easy to find vacua qualitatively
similar to the Standard Model, but extremely hard to find models
quantitatively the same. This conceptual problem is partly
philosophical and I shall not discuss it here: recent articles include
\cite{physics0604134, hepth0302219, hepth0303194}.

The particular technical problems encountered are to
\begin{enumerate}

\item Obtain the correct gauge group and chiral spectrum. 
\item Ensure the absence of exotic particles, either charged or
  uncharged.
\item Understand the physics of supersymmetry breaking.
\item Explain why the observed couplings take the values they do and
  predict new effects.
\item Describe the cosmology of the universe from the Planck epoch to nucleosynthesis.
\end{enumerate}
These problems are separate but inter-related. In this thesis I
shall discuss all problems in the above list except the first.
Part II of this thesis will focus on the second problem,
 and in particular on a class of uncharged scalar particles known as
 moduli. The failure to observe fifth forces implies such particles
 should be given masses and hence potentials. This is the problem of
 moduli stabilisation. 
Solutions to the problem of moduli stabilisation are generally
prerequisites for solutions of the third and
 fourth problems, and in part III of this thesis we shall apply the
 results of part II to study these problems, discussing
 soft supersymmetry breaking, inflation and
 the strong CP problem.

The work contained in this thesis is based on the papers
\cite{hepth0409215} (chapter \ref{StatisticsReview}),
\cite{hepth0502058} (chapter \ref{chapterLargeVol}),
\cite{hepth0505076} (chapters \ref{ChapterKahlerModuli},
\ref{chapterLargeVol} and \ref{chapterSoftSusy}), \cite{hepth0509012}
(chapter \ref{InflationModel}),
  \cite{hepth0602233} (chapter \ref{ChapterAxions}) and
  \cite{hepth0605141} (chapter \ref{chapterSoftSusy}),  
As indicated in the preface I am grateful to my collaborators Vijay
Balasubramanian, Per Berglund, Fernando Quevedo and Kerim Suruliz. 

I have a final note on references. I have tried to cite relevant
work where appropriate, but it is inevitable that there are lapses. 
As this is a thesis rather than a review article I have focused
primarily on my own work and the results most directly relevant to it, with 
the consequence that I have failed to cite many interesting and important
articles. I apologise in advance to the authors of
these papers.

\chapter{Moduli and Fluxes: A Brief Review}
\label{ChapterModuliAndFluxes} \linespread{1.3}

The purpose of this chapter is to review background material on moduli
and flux compactifications. An inexhaustive list of useful
references extending this discussion is
\cite{CHSW, CandelasDeLaOssa, hepth9702155, hepth0105097,
  hepth0405068}, and in particular the review of flux compactifications
\cite{hepth0509003}.

\section{String Compactifications}

The most straightforward formulation of string theory is in flat
ten-dimensional Minkowksi space. There are then five string theories:
the IIA and IIB closed string theories, the type I (type IIB
orientifold) $SO(32)$ theory, and the heterotic $E_8 \ti E_8$ and
$SO(32)$ theories. These are all unsatisfactory in (at least) two ways: the
gauge group is much larger than that of the Standard Model and the
number of dimensions is six too many.

The standard resolution of these problems is
compactification, in which the ten dimensional space
becomes a (possibly warped) product of our 4-dimensional
world with an internal 6-dimensional space $X$.
The most general
metric reproducing 4-dimensional Poincare invariance is 
\be
ds_{10}^2 = e^{-2 A(y)} ds_4^2 + e^{2 A(y)} g_{mn} dy^m dy^n.
\ee
The $y^m$ parametrise the internal space $X$ and
the factor of $e^{-2A(y)}$ allows for the possibility of warping in the
ansatz, 
Large variations in the warp factor $A(y)$ may generate
interesting physical effects as in the Randall-Sundrum scenario
\cite{hepph9905221, hepth9906064}. However it will not be
important for our purposes, and so unless explicitly stated
we shall not regard warping as significant. 
The characteristic signals of extra dimensions are either the production of
Kaluza-Klein (KK) copies of the Standard Model or the modification of
gravity at short distances. Neither have been observed, and so
consistency requires 
$X$ to be compact and sufficiently small that the resulting physics below
the Kaluza-Klein scale is effectively four-dimensional. 

Compactification clearly reduces the number of large dimensions, but it also
permits a reduction in the rank of the gauge group. In heterotic compactifications,
this is achieved by giving vevs to extra-dimensional components of
the gauge potential or field strength. To be unbroken, a gauge symmetry must
leave such vevs unaffected, and so the surviving gauge group is the
commutant of the turned-on fields in the ten-dimensional gauge group.
In the type II orientifold compactifications we will be most
interested in, the orientifold tadpole must be cancelled by the
D-branes present. The loci and intersections of these branes determine
the low-energy gauge group. The rank of this is
model-dependent but may naturally be small.

This philosophy of compactification says very little about the internal space
$X$. Fundamentally, $X$ should be determined by the
cosmology of the very early (Planckian)
universe. As spacetime topology change has a smooth description
in string theory \cite{hepth9301042, hepth9301043, hepth9504090, hepth9504145}, such transitions ought to occur in the
early universe. The choice between compactification manifolds $X_1$
and $X_2$ is then determined by cosmological dynamics.
However, as Planck-scale cosmology is extremely
speculative we instead look to particle physics for a guiding principle.

\begin{figure}[ht]
\linespread{0.2}
\begin{center}
\makebox[8cm]{ \epsfxsize=8cm \epsfysize=4cm \epsfbox{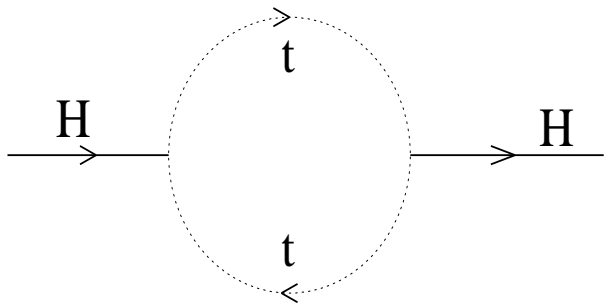}}
\end{center}
\caption{The Higgs mass is radiatively unstable in the Standard Model
  due to quadratic divergences from top loops.}
\label{HiggsMass}
\end{figure}

This guiding principle is taken to be the existence of $\mc{N} = 1$
supersymmetry at low energy. 
There are a variety of reasons why this is phenomenologically
desirable.
The most well known is the 
quadratic sensitivity
of the Standard Model Higgs mass to the cutoff $\Lambda$ used in loop
integrals such as in figure \ref{HiggsMass}. In the Standard Model
unitarity bounds require the Higgs mass to be $m_H \lesssim 1 \hbox{TeV}$.
The diagram in figure \ref{HiggsMass} gives
\be
m_H^2 \sim \int^\Lambda d^4 p \frac{1}{p^2} \sim m_{H,bare}^2 + \Lambda^2.
\ee
Physically, the cutoff represents the scale of new physics. The
Standard Model is renormalisable and thus in principle valid up to the
Planck scale. However, if we take $\Lambda = M_P$ we must fine-tune
$m_{H,bare}^2$ to an unacceptable degree of precision (around one part
in $10^{30}$) in order to have
$m_H$ around the electroweak scale - this is the electroweak hierarchy problem.
This suggests that the cutoff $\Lambda$ should instead appear at the TeV scale.
A good candidate for the cutoff physics is supersymmetry - a symmetry
relating bosons and fermions.
This tends to cancel divergences in
loop integrals, as particles and their superpartners give exactly
opposite contributions when running around the loop. 
Of course, supersymmetry must be a broken symmetry, as no superpartners have
currently been observed. 

There are other attractions of low energy supersymmetry. Through the
tachyonic running of the up-type Higgs mass term, it can
explain why the scale of electroweak symmetry breaking
and the scale of the physics stabilising the
Higgs mass should be comparable.
TeV-scale supersymmetry with the Minimal Supersymmetric Standard Model
(MSSM) matter content also leads
to the unification of the MSSM gauge coupling constants at the GUT
scale $10^{16}$ GeV. 
Finally, supersymmetry also provides a natural dark matter candidate 
in the lightest neutralino, whose stability is protected by R-parity.
We shall not dwell further on low energy $\mc{N} = 1$
supersymmetry as an attractive phenomenological scenario. We shall
return to supersymmetric phenomenology in chapter
\ref{chapterSoftSusy}, but for now mention
\cite{hepph0312378} as an extensive and detailed account
with thorough references.

We add one cautionary note. The justification for Calabi-Yau compactification, on which most
string model building (including this thesis) has been built, is closely tied to low energy
supersymmetry. If the Large Hadron Collider (LHC) 
finds that physics other than supersymmetry is responsible for
stabilising the Higgs mass, this would remove the prime motivation
for low-energy supersymmetry. There would then be little theoretical reason to 
find supersymmetry below the Planck scale and the relevance of Calabi-Yau
compactification to nature would be questionable.

The central topic of this thesis is moduli stabilisation. To frame this
as a problem, we first review the \emph{fons et origo} of
string compactifications \cite{CHSW}. Although involving the
heterotic string rather than the IIB case we are primarily interested
in, both results and techniques are general.

The heterotic string, either with $SO(32)$ or $E_8 \ti E_8$ gauge
group, is one of the five consistent string theories in ten flat
dimensions.
At low energy, string theory reduces to supergravity and we only focus
on the (massless) supergravity degrees of freedom.
The massless fields are the graviton $g_{MN}$,
dilaton $\Phi$, antisymmetric tensor field $B_{MN}$ and a gauge field
$A_{M}$. These last have 3- and 2-form field strengths, $H_3 = dB_2$
and $F = dA$.

We ask when it is possible to preserve a global $\mc{N}=1$ supersymmetry
in compactifications of the heterotic string.
If we do not turn on $H_3$,  
it was shown in \cite{CHSW} that in order to preserve
$\mc{N} = 1$ supersymmetry the fields must satisfy
\be
ds_{10}^2 =  \eta_{\mu \nu} + g_{mn} dy^m dy^n, 
\ee
\be
H_{mnp} = 0, 
\ee
\be
\Phi = \Phi_0, 
\ee
\be
\label{eqDUY}
F_{ij} = F_{\bar{i} \bar{j}} = G^{i \bar{j}} F_{i
  \bar{j}} = 0, 
\ee
\be
\label{eqRF}
\hbox{Tr}(\mc{R}_2 \wedge \mc{R}_2) - \hbox{Tr}_v (F_2 \wedge F_2) = 0, 
\ee
where the trace is taken in the vector representation of the gauge group.
$g_{mn}$ is the metric on the internal space,
which must be a Ricci-flat K\"ahler manifold. This does not seem a
strenuous requirement - surely \emph{some} Ricci-flat metric must exist on any given K\"ahler manifold. However,
in general this is true only locally, and there is a topological obstruction to the global existence of
such a metric. It was conjectured by Calabi and proved by Yau that
\begin{Yau}[Yau]
Any K\"ahler manifold $X$ with vanishing first Chern class admits a Ricci-flat metric, unique up to the complex
structure of $X$ and the cohomology class of the K\"ahler form.
\end{Yau}
Such manifolds are called Calabi-Yau manifolds. 
Compactification of either heterotic (as above) or type I strings on a Calabi-Yau
gives rise to $\mc{N} = 1$ supersymmetry in four
dimensions (compactifying type II strings preserves $\mc{N} =2 $
supersymmetry).
There are consistency conditions (\ref{eqDUY}) and (\ref{eqRF}) on the
gauge bundle, which can most simply be satisfied by identifying $F$ and $\mc{R}$.
The resulting gauge group is the commutant of $F$ in $E_8 \ti E_8$ (or
$SO(32)$).
For the standard embedding $F = \mc{R}$, the tangent and gauge bundles
are $SU(3)$ bundles. The commutant of $SU(3)$ in $E_8$ is $E_6$ and so
this gives the low-energy gauge group $E_6
\ti E_8$.

It is also possible to look for compactifications with non-vanishing 3-form flux, with $\hbox{d}
H_{mnp} = 0$ but $H_{mnp} \neq 0$. In this case supersymmetry requires
that $X$ be an $SU(3)$ structure manifold with nontrivial torsion
\cite{Strominger1986, hepth0509003}. However, despite recent
interest in this area we shall not consider
heterotic flux compactifications further in this thesis.

\section{Moduli}

Given a particular compactification, the principal desiderata are the spectrum
and interactions of the light particles. 
`Light' implies the dropping of stringy states with
characteristic masses of $\mc{O}(\frac{1}{\sqrt{\alpha'}})$, and
Kaluza-Klein states with characteristic masses $\mc{O}\left(\frac{1}{R
  \sqrt{\alpha'}}\right)$ for $R$ the compactification radius in
string units. This leaves us with the states that are present in the
supergravity spectrum and survive dimensional reduction to four dimensions.

Any compactification leads to a particular particle spectrum and gauge group.
 In one sense, the most interesting feature of this will be the gauge sector of
 particles, as we can compare this to that present in the Standard
Model or its various extensions. We may define this as
consisting of all particles charged under the various gauge groups present.
In heterotic compactifications these are determined by the gauge bundle over the Calabi-Yau,
whereas in orientifold compactifications these come from
open strings ending on D-branes. However, we shall not
discuss this interesting and important issue further (see
\cite{GSW, hepth0204089, hepth0405014, hepth0502005} for
reviews), and instead focus on the moduli sector.

Moduli are \emph{senso stricto} massless scalars $\phi_i$ that parametrise continuous families of
vacua.
They are characteristic of supersymmetric compactifications which
typically come with large moduli spaces. They may or may not be
charged. For example, the process of Higgsing a supersymmetric D-brane
stack may be described as giving a vev to a charged modulus. However,
we shall use the word mainly to refer to uncharged scalars whose only
interactions are gravitational, and shall also broaden the use to include
scalar fields after a potential has been generated for them.

We would like to enumerate the different kinds of moduli.
Moduli parametrise continuous families of nearby vacua, and so for
Calabi-Yau compactifications the most obvious moduli are those
parametrising the space of Calabi-Yau manifolds. 
As a continuous deformation cannot alter toplogical properties, we are
restricted to the space of topologically equivalent
Calabi-Yaus.\footnote{We shall not discuss extremal transitions and
  paths in moduli space connecting topologically \emph{distinct} Calabi-Yaus.} The mathematical question asked is, given
that $g_{mn}$ is the metric for a Calabi-Yau space $Y$, under what circumstances is $g_{mn} + \delta g_{mn}$ a
metric for a neighbouring Calabi-Yau space? Ricci-flatness implies
$$
R_{mn}(g) = R_{mn}(g + \delta g) = 0.
$$
This implies that $\delta g$ satisfy the Lichnerowicz equation, 
\be 
\nabla^l \nabla_l  \delta g_{mn} +
2R_{\phantom{l}m\phantom{r}n}^{l\phantom{m}r\phantom{n}} \delta g_{lr}
= 0. 
\ee 
For K\"ahler manifolds the solutions to this equation
are independent and associated with either pure ($\delta g = \delta g_{\bar{\mu} \bar{\nu}}$) or mixed ($\delta
g = \delta g_{\mu \bar{\nu}}$) deformations. These are in one-to-one correspondence with harmonic $(2,1)$- and
$(1,1)$-forms respectively. For mixed solutions this correspondence is
given by 
\be 
\delta g_{\mu \bar{\nu}}
\leftrightarrow \delta g_{\mu \bar{\nu}} d x^\mu \wedge d
x^{\bar{\nu}} \in H^{1,1}(M), 
\ee 
whereas for pure
solutions $\delta g_{\bar{\mu}
  \bar{\nu}}$ we have
\be \delta g_{\bar{\mu} \bar{\nu}} \leftrightarrow
 \Omega_{\phantom{\nu} \kappa \lambda}^{\bar{\nu}} \delta g_{\bar{\mu} \bar{\nu}} dx^\kappa \wedge d x^\lambda
\wedge d x^{\bar{\mu}} \in H^{2,1}(M), \ee and likewise for $\delta g_{\mu \nu}$ and $(1,2)$-forms. As the
structure of harmonic differential forms is well known to be
isomorphic to that of tangent bundle cohomology classes, the
number of 
geometric moduli in compactifications on $Y$ is determined by the cohomology of $Y$.

We note this is a general feature of string compactifications - the
light particle spectrum is determined by topological considerations
and the number of particles of given type is equivalent to the
dimension of appropriate (sheaf) cohomologies.

The moduli associated with (1,1)-forms are called K\"ahler moduli and the moduli associated with (2,1)-forms are
called complex structure moduli. This nomenclature is because the former modify the K\"ahler class
of the manifold whereas the latter alter the complex structure. The first claim is manifest: under $g \to g +
\delta g$, the K\"ahler form behaves as
$$
J = i g_{\mu \bar{\nu}} dx^\mu \wedge dx^{\bar{\nu}} \to i (g_{\mu
  \bar{\nu}} + \delta g_{\mu \bar{\nu}}) dx^\mu \wedge
dx^{\bar{\nu}}.
$$
To verify the latter, we first note that $g + \delta g$ is a K\"ahler
metric.
This follows because the K\"ahler form is unaltered so long as $\delta g$ is a pure deformation. Thus 
 there exists a coordinate system in
which all pure components of the metric vanish. However, under general coordinate transformations $x^m \to x^m +
f^m(x^n)$, $\delta g$ transforms as \be \delta g_{mn} \to \delta g_{mn} - \frac{\partial f^r}{\partial x^m}
g_{rn} - \frac{\partial f^r}{\partial x^n} g_{mr}. \ee Consequently holomorphic transformations $f = f(x^\mu)$
cannot alter the pure components $g_{\mu \nu}$ of the metric. A coordinate system in which the metric is
K\"ahler can only be obtained by a non-holomorphic coordinate change, which by definition will change the
complex structure. This can also be understood more succinctly from Yau's theorem: as the Ricci-flat metric is
unique up to the complex structure and the cohomology class of the K\"ahler form, the physical moduli should
parametrise the possible modifications of these.

The above are geometric moduli. In addition to the geometric fields string
theory contains antisymmetric form fields, which also give 
rise to moduli. For example, heterotic compactifications contain a
2-form $B_2$ field, with a field strength 
$$
H_3 = dB_2 + CS(F) - CS(\mc{R}),
$$
where $CS$ refers to the Chern-Simons 3-form. This field strength is
unaltered by a transformation
$$
B_2 \to B_2 + b^i e_i,
$$
where $e_i \in H^2(Y, \mbb{R})$. The fields $b_i(x)$ associated with
internal 2-cycles thus do not affect the
field strength and are also moduli. This feature is general across
string compactifications - for example, in type IIB compactifications, the
$B_2$, $C_2$ and $C_4$ potentials all also give rise to moduli. 

There are in addition other sources of moduli such as brane and bundle moduli. Brane
moduli are moduli associated with the positions of branes in the
compactification manifold. These may be charged and can
act as Higgs fields for the gauge group on the branes. Bundle moduli
occur when nontrivial gauge backgrounds exist in the theory and
describe the possible deformations of the gauge bundle that are consistent with
the equations of motion. 

We shall mostly focus on the bulk moduli in this thesis. 
These are more generic: the gauge sector in string
compactifications is heavily model-dependent, but the bulk sector is
very similar across different string theories and compactifications.
As mentioned above, in the simplest compactifications such as
\cite{CHSW} the moduli represent massless uncharged scalar particles.
The existence of such massless scalars is inconsistent with
experiment. Moduli couple gravitationally to ordinary matter and so
can generate forces due to particle exchange. For a modulus of mass
$m_{\phi}$, the characteristic range of such forces 
is $R \sim \mc{O}(\frac{1}{m_{\phi}})$. As
fifth force experiments have probed gravity to submillimetre distances,
this requires that $m_{\phi} > \mc{O}(10^{-3})$ eV \cite{hepph0405262}. Consequently
experiment requires the 
existence of potentials 
giving masses to the moduli.

We note as an aside that the above does not apply for one particular class of
particle, axions. Axions have topological couplings to gauge groups
and derivative couplings to matter. Derivative couplings vanish for
long-distance interactions (i.e. small momentum transfer) and so do not give rise to fifth
forces. We shall discuss axions further in chapter \ref{ChapterAxions}.

As creating and controlling moduli potentials is a hard task, we may
prefer to
look for compactifications without moduli.
However while massless scalars are inconsistent with experiment, the dynamics
of massive scalars is very important. The physics of inflation - the dominant theory
of structure formation - is determined by a scalar field potential.
Scalar potentials also play an important role in theories of
supersymmetry breaking, a topic we will return to in chapter \ref{chapterSoftSusy}.

Our aim is now clear. Given that massless moduli are both a generic feature of
string compactifications and experimentally disallowed, we need
techniques that will create a potential for these moduli and give them
masses. Fluxes are a powerful example of such a technique: it is to
these we now turn.

\section{Fluxes and Moduli Stabilisation}
\label{sec22}

The above establishes the need to give potentials to moduli. A
useful way of doing this is through flux compactifications. These are
distinguished from ordinary compactifications by having
antisymmetric tensor fields that are nonvanishing in the
compact space. They have a 
long history in string
theory\cite{Strominger1986, Hull1986,  
hepth9606122, hepth9510227, hepth9908088, hepth0105097, hepth0201028}. 
Our main interest will be 3-form
fluxes in IIB orientifold compactifications. 
However, we
first want to develop a general intuition for why fluxes can stabilise
moduli. As a simple example, we take the geometry of the principal avatar of the AdS/CFT
correspondence\cite{hepth9711200}, namely type IIB string theory on
$AdS_5 \ti S^5$. This has 5-form flux present on the $S^5$. The
ten-dimensional supergravity can be solved exactly to give
\bea 
ds^2 & = & h(r)^{-\half}\left(-dt^2 + dx^2 +dy^2 + dz^2\right) +
h(r)^\half \left(dr^2 + r^2 d\Omega_5^2 \right), \nonumber \\
\Phi & = & \Phi_0, \nonumber \\
F_5 & = & \mc{F}_5 + \mc{F}_5^{*}, \quad \textrm{ with } \mc{F}_5 = 16
\pi \alpha'^2 N \textbf{vol}(S^5). 
\eea
Here $h(r) = 1 + \frac{L^4}{r^4}$, where $L = (4 \pi g_s N \alpha'^2)^{\quarter}$ is the radius of both $AdS_5$
and $S^5$ in the $AdS_5 \ti S^5$ near-horizon limit of this
geometry. However, our interest here is less in the full
ten-dimensional solution than
in understanding why the $S^5$ radius $L$ obtains the value it
does. To this end,
we reduce this geometry to a
5d effective theory and evaluate the potential energy as a function of $L$.

The 10-dimensional type IIB supergravity action is given in Einstein frame by\cite{GSW, PolchinskiBook}
 \be \label{IIBaction} S_{IIB} =
\frac{1}{2 \kappa_{10}^2} \int d^{10} x \sqrt{-g} \left[
  \mc{R} - \frac{\partial_M \tau \partial^M
    \bar{\tau}}{2(\textrm{Im}\tau)^2} - \frac{G_3 \cdot \bar{G}_3}{12
    \textrm{Im} \tau} - \frac{\tilde{F}_5^2}{4 \cdot 5!} \right] + S_{CS},
\ee where $G_3 = F_3 - \tau H_3$ and $S_{CS}$ is the Chern-Simons
 term. Here $\kappa_{10}^2 \sim l_s^8$, with $l_s = 2 \pi
 \sqrt{\alpha'}$. $\tau$ is the dilaton-axion and $G_3$ and $F_5$ are
 3-form and 5-form fluxes.

We now suppress the dilaton-axion, 3-form fluxes and numerical coefficients as not relevant for
our purposes. Compactifying on an $S^5$ of radius $L = R l_s$ with $N$ units of $F_5$ flux on the $S^5$, we
obtain schematically \bea S_{compactified} & = & \frac{1}{l_s^8} \int d^{10}x \sqrt{-g} \left(
\mc{R}_{5D} + \mc{R}_{S^5} - F_5^2 \right) \nonumber \\
& = & \frac{1}{l_s^8} \int d^{5}x \sqrt{-g} \left( \mc{R}_{5D}(R^5 l_s^5) + \frac{(R^5 l_s^5)}{R^2 l_s^2} -
\frac{N^2 l_s^5}{R^5 l_s^2} \right). \eea We have used the
quantisation condition for $F_5$, 
\be
\int_{S^5} F_5 = (2 \pi)^4 \alpha'^2 N \sim N l_s^4. 
\ee 
We now convert this to 5-dimensional Einstein
frame. Under a metric rescaling, 
$$
g =  \tilde{g} \Omega \Rightarrow \sqrt{-g} \mc{R} = \sqrt{-\tilde{g}} \tilde{\mc{R}} \Omega^{\frac{3}{2}}.
$$
Therefore if we take \be \Omega = \frac{l_s^2}{l_5^2} R^{-\frac{10}{3}}, \ee where $l_5$ is the 5D Planck mass,
the resulting action for $\tilde{g}$ is canonical and given by 
\be 
S_{5D} = \frac{1}{l_5^3} \int d^5 x
\sqrt{-\tilde{g}} \left( \tilde{\mc{R}} + \frac{1}{l_5^2} R^{-\frac{16}{3}} - \frac{1}{l_5^2} N^2
R^{-\frac{40}{3}} \right). 
\ee 
The effective potential for the radial modulus is then \be V(R) =
-R^{-\frac{16}{3}} + N^2 R^{-\frac{40}{3}}. \ee The fluxes contribute positively to the potential energy and the
curvature negatively. This potential has an AdS minimum at $R \sim
N^\quarter$, with scaling behaviour consistent with the full
solution.

Physically, the intuition we derive is that fluxes wrapped on a cycle
carry a potential energy which depends on the volume and geometry of the
cycle. As moduli parametrise the geometry, this automatically generates a potential for the moduli describing
these cycles. This potential can be analysed from an effective field theory
viewpoint. There may be other contributions to the potential - here
the $S^5$ curvature is important - and these must also be
included. Putting everything together gives a potential for the $S^5$ radius.
For completeness, we note that in the above
example the $S^5$ radius 
is not properly a modulus, because without the 5-form flux the solution is
unstable rather than flat.  However, this will not be the case in the more realistic flux
compactifications we now study.

\section{Flux Compactifications of IIB Orientifolds}

Let us move from the above motivational example to the setup we
shall use for the rest of this thesis. This is a class of flux
compactifications of type IIB orientifolds 
with D3/D7 branes and O3/O7 planes
\cite{hepth0105097}
developed by Giddings, Kachru and Polchinski. Discrete NSNS and RR 3-form fluxes
are present which generically give
masses to dilaton, complex structure and D7 brane moduli. At this level
the K\"ahler moduli are unstabilised and for these moduli the
potential is in fact no-scale. 

One useful feature of these compactifications is that they can be
described in both ten- and four-dimensional language. While
Kaluza-Klein reduction to deduce low energy actions is to some extent
a necessity, given our lack of knowledge of, \emph{inter alia}, Calabi-Yau
metrics, it is not a completely ideal approach. 
This is because it entails a truncation, rather than a Wilsonian 
integration out, of the heavy Kaluza-Klein modes. One can argue that 
the resulting low energy action is nonetheless correct, (see
e.g. chapter 16 of \cite{PolchinskiBook}), but one
would still prefer a ten-dimensional picture. It is an attractive
feature of the compactifications of \cite{hepth0105097} that 
we can adopt either a four- or ten-dimensional viewpoint.

As these compactifications will play a fundamental role in the rest of the
thesis we now review them in first ten- and then
four-dimensional language.

\subsection{Ten-dimensional Perspective}
\label{tendpic}

The models of \cite{hepth0105097} are orientifold compactifications. That is, in
addition to the purely geometric background there are non-dynamical
negative tension orientifold planes. The cancellation of RR tadpoles
then necessitates the inclusion of D-branes. As both O-planes and
D-branes are localised sources,   
the appropriate action is ten dimensional IIB supergravity 
coupled to localised sources.

In Einstein frame this is
\be
S = \frac{1}{(2 \pi)^7 \alpha'^4} \int d^{10}x \sqrt{-g} \left[ \mc{R}
  - \frac{\partial_M \tau \partial^M \bar{\tau}}{2 (\textrm{Im}
    \tau)^2} - \frac{G_3 \cdot \bar{G}_3}{12 \textrm{Im} \tau} -
  \frac{\tilde{F}_5^2}{4 \cdot 5!} \right] + S_{cs} + S_{loc},
\ee
where $G_3 = F_3 - \tau H_3$ with $\tau = C_0 + i
e^{-\phi}$ the axio-dilaton. The 5-form $\tilde{F_5}$ is defined by
\be
\tilde{F}_5 = F_5 - \half C_2 \wedge H_3 + \half B_2 \wedge F_3.
\ee
The Chern-Simons term is
\be
S_{cs} = \frac{1}{4 i (2 \pi)^7 \alpha'^4} \int C_4 \wedge G_3 \wedge
\bar{G}_3.
\ee
Note that the presence of nontrivial 3-form flux $G_3$ sources a tadpole for
the $C_4$ field, i.e. for D3-brane charge. 

The localised sources are D3/7 branes and O3/7 planes. Dropping $B$
and $F$ fields, the action for
a D$p$-brane wrapping a cycle $\Sigma$ is
\be
S_{loc} = - T_p \int_{R^4 \times \Sigma} d^{p+1} \xi \sqrt{-g} + \mu_p
\int_{R^4 \times \Sigma} C_{p+1}.
\ee
In comparison to a D-brane, an orientifold plane has opposite tensions and
RR charges.

In the above setup
there are several  discrete choices to be
made, the most obvious of which is the compactification manifold. It
is also necessary to specify the cohomology of the 3-form
fluxes. These are 
integrally quantised with
\be
\frac{1}{(2 \pi)^2 \alpha'} \int_{\Sigma} F \in \mbb{Z}, \quad
\textrm{ and } \quad \frac{1}{(2 \pi)^2 \alpha'} \int_{\Sigma} H \in
\mbb{Z},
\ee
where $F$ and $H$ are the RR and NSNS 3-form fluxes
respectively. Given these choices, the equations of motion for the
fields may be solved and the resulting solutions described as follows.

The metric takes the warped form
\be
\label{gkpmetricsolution}
ds^2 = e^{2A(y)} \eta_{\mu \nu} dx^\mu dx^\nu + e^{-2A(y)}
\tilde{g}_{mn} dy^m dy^n,
\ee
giving a warped product of Minkowski space with a conformally
Calabi-Yau internal space. This latter property is key, as it allows
much of the well-established technology of Calabi-Yau
compactifications to be reused. The warp factor is related to the
self-dual 5-form flux by
\be
\tilde{F}_5 = (1 + *) \left[d\alpha \wedge d x^0 \wedge d x^1 \wedge d
  x^2 \wedge d x^3 \right],
\ee
with $\alpha = e^{4 A(y)}$. The 3-form flux is imaginary self-dual (ISD),
satisfying
\be
\label{ISDcondition}
G_3 = i *_6 G_3,
\ee
where $*_6$ is the Hodge dual on the internal space.

The resulting solution has $\mc{N} = 1$ supersymmetry, which may be
spontaneously broken by the fluxes.  The condition that supersymmetry
be preserved is that the 3-form flux $G_3$ is (2,1) and primitive.
The primitivity condition $G_3 \wedge J = 0$ holds automatically for  a Calabi-Yau. The
ISD condition requires that the (1,2) and (3,0) parts of $G_3$ vanish,
and so the residual requirement for the existence of supercharges leaving
the vacuum invariant is that the (0,3) part of $G_3$ vanish.

The ISD condition fixes the dilaton and complex structure
moduli. Although naively a condition on $G_3$, we
should instead regard it as a condition on the complex structure
moduli, as the definition of $*_6$ depends on the complex
structure of the internal space. The complex structure adjusts to
satisfy condition (\ref{ISDcondition}), thereby fixing all complex structure
moduli. As the dilaton appears in the definition of $G_3$, it is also
fixed by the fluxes.

The K\"ahler moduli are unfixed as there exists a continuous family
of solutions (\ref{gkpmetricsolution}) related by rescalings of the internal volume. The
warp factor transforms nontrivially under these but vanishes at large volume as
\be
\alpha = e^{4 A} \sim 1 + \frac{1}{\mc{V}^{2/3}}.
\ee
In the large-volume limit the warping therefore becomes negligible.

As mentioned above, the fluxes generate a tadpole for $C_4$. There are
other sources for this tadpole; in particular, curvature couplings
on wrapped D7-branes also generate negative D3 charge. In the language
of F-theory, this charge is determined by the Euler number of the
F-theory fourfold $Z$:
\be
Q_3^{D7} = - \frac{\chi(Z)}{24}.
\ee
The resulting tadpole equation is
\be
\label{eqTCE}
\frac{1}{(2 \pi)^4 \alpha'^2} \int H_3 \wedge F_3 + N_{D3} - N_{\bar{D}3} =
\frac{\chi(Z)}{24}.
\ee
(There may also be extra sources of D3-charge; for example,
from gauge fields on D7 branes. We shall not concern ourselves with
these.)
The number of D3 branes that must be introduced varies with the discrete flux
choices. 
If we wish to avoid the need to include D3 branes, we can always take
the fluxes to saturate the tadpole condition.

The above is a consistent solution of 10D supergravity coupled to
localised sources at leading order in the $g_s$ and $\alpha'$
expansions.
We now analyse it from a four-dimensional perspective.

\subsection{Four-dimensional Perspective}

On general grounds, the above compactifications ought to be described
at low energy by a 4-dimensional $\mc{N} = 1 $ supergravity
theory. Here `low-energy' refers to energies substantially below the
Kaluza-Klein scale.
Physically, this simply says that the ten-dimensional
solution carries far more information than is needed. For probe energies much
below the Kaluza-Klein scale, the
precise geometry blurs and we can only resolve bulk properties such as
the overall volumes, but not fine-grained structure.

A four-dimensional $\mc{N} = 1 $ supergravity theory is specified at low energies
(i.e. at two derivatives) by its matter content and by the K\"ahler
potential $\mc{K}$,
superpotential $W$ and gauge kinetic functions $f_i$. The latter run
over gauge groups and give the
complexified gauge coupling for each gauge group. One
advantage of working in 4-dimensional language is that
nonrenormalisation results are available that would be very difficult
to see directly in ten dimensions. Let us state these, using
$\phi_i$ to denote the chiral superfields present.
\begin{enumerate}

\item The K\"ahler potential $\mc{K}(\phi_i, \bar{\phi}_i)$ is 
a real non-holomorphic function of the chiral
  superfields.
It is renormalised at all orders in perturbation
  theory and nonperturbatively.

\item The superpotential $W(\phi)$ is a complex holomorphic function of the chiral
  superfields. It is not renormalised in perturbation theory, but may
  receive nonperturbative corrections.

\item The gauge kinetic functions $f_i(\phi)$ are complex holomorphic
  functions of the chiral superfields. They are perturbatively exact
  at one loop, but will receive nonperturbative corrections.

\end{enumerate}

To specify the low energy theory, it is first necessary to specify the low
energy fields. We are focusing on moduli, and so will neglect any
charged matter that may be present. The reason why this is sensible is
that moduli obtain large classical vevs, 
while the vevs of Standard Model matter are vanishing or
hierarchically small (as with the Higgs). It is then consistent to
neglect the dynamics of charged
matter while studying the moduli potential.

We shall primarily be concerned with the closed string moduli which 
are more generic as they occur in all string
compactifications. It is also possible to be more explicit when
treating them. However, in the next section we shall also briefly 
discuss the open string moduli and how they mix with the closed string fields.

In the above IIB flux compactifications the dilaton superfield is defined by
\be
S = e^{-\phi} + i C_0,
\ee
where $e^\phi$ is the string coupling and $C_0$ the RR
0-form. Strictly this is only the scalar component of the full
superfield, but we will generally overload notation and use the same
symbol to denote both the superfield and its scalar component.
Sometimes the alternative definition $\tau = -C_0 + i e^{-\phi} \equiv
iS$ is used.

 There are various equivalent ways to define the 
complex structure moduli. For a Calabi-Yau, they are uniquely
specified by the periods
$$
\Pi^i = \int_{\Sigma_i} \Omega,
$$
where $\Sigma_i$ is a 3-cycle, $\Omega$ the holomorphic (3,0) form
and $\Pi^i$ the period vector. $\Omega$ is only defined up to overall
rescalings, and so one can use the period vector as projective
coordinates for the complex structure moduli. Many
Calabi-Yaus are hypersurfaces in weighted projective
spaces and specified by a defining equation $f(z_i) = 0$. In this case we
may instead define the complex structure moduli by the coefficients of
the defining polynomial - this will be the approach adopted in section
\ref{Tests}.

The K\"ahler moduli are defined by
\be
T_i = \sigma_i + i \rho_i,
\ee
where $\sigma_i$ is the Einstein-frame volume of a 4-cycle $X_i$ and $\rho_i =
\int_{X_i} C_4$. $\sigma_i$ can be expressed in terms of 2-cycle
volumes $t^i$ by
\be
\label{eqKMD}
\sigma_i = \frac{\partial \mc{V}}{\partial t^i} = \half k_{ijk} t^j t^k.
\ee

We can now specify the K\"ahler and superpotentials for the
compactifications of \cite{hepth0105097}.
The superpotential is that of Gukov, Vafa and
Witten and takes the form \cite{hepth9906070} 
\be 
\label{GKVsuperpotential} 
W = \frac{1}{(2 \pi)^2 \alpha'} \int_M G_3 \wedge \Omega, 
\ee
where $G_3 = F_3 + iS H_3$.
As the periods specify the complex structure, it
follows that the superpotential depends on all the complex structure
moduli.
However, note that (\ref{GKVsuperpotential}) is independent of the K\"ahler
moduli.  In F-theory, 
(\ref{GKVsuperpotential})
generalises to
\be
W = \int_M G_4 \wedge \Omega.
\ee
In F-theory, the D7 position moduli correspond to complex structure
moduli of the fourfold, and thus these are also fixed by the fluxes.

The K\"ahler potential is to leading order given by 
\be 
\label{noscaleK}
\mc{K}_{no-scale} = -2 \log \left[\mc{V} \right] - \log\left[-i \int_M \Omega \wedge \bar{\Omega}\right] -
\log\left[ S + \bar{S} \right]. 
\ee 
$\mc{V} \equiv \frac{1}{6} k_{ijk} t^i t^j t^k$ should be regarded as an implicit function of the
K\"ahler moduli, as in general the relations (\ref{eqKMD}) cannot be inverted.
(\ref{noscaleK}) possesses no-scale structure. 
That is, in the $\mc{N}= 1$ F-term scalar potential, 
\be 
\label{Vn=1sugra} V =
e^{\mc{K}} \left[\mc{K}^{i \bar{j}} D_i W \bar{D}_j \bar{W} - 3 \vert W \vert^2
  \right], 
\ee 
with $i,j$ running over all moduli, the sum over K\"ahler moduli
cancels the $3\vert W \vert^2$ term and the resulting potential is given by \be V_{no-scale} = e^{\mc{K}} \mc{K}^{ab}D_a W
\bar{D}_b \bar{W}, \ee where $a$ and $b$ run over dilaton and
complex-structure moduli only. 

It follows that  we can stabilise the dilaton and complex structure moduli at a minimum of the potential by
solving 
\be 
\label{eqSTS}
D_S W \equiv \partial_S W + (\partial_S \mc{K}) W = 0,
\nonumber 
\ee 
\be
\label{csstabilisation} D_a W \equiv \partial_a W + (\partial_a \mc{K})
W = 0. 
\ee 
We denote the resulting value of $W$ as $W_0 = \left\langle \int G_3
\wedge \Omega \right\rangle$. 
It can be shown that 
for solutions of equations ($\ref{csstabilisation})$, the flux tadpole $\int G_3 \wedge \bar{G}_3$ becomes
positive definite. In this approximation, the K\"ahler moduli are
unfixed due to the no-scale structure. As 
\be
D_T W = (\partial_T \mc{K}) W \propto \int G_3 \wedge \Omega,
\ee
the F-terms corresponding to the K\"ahler moduli are nonvanishing
unless $W = 0$. This is the same condition to preserve supersymmetry 
as arose in the 10-dimensional picture: $W=0$ is
equivalent to $G_3$ having no (0,3) component. It can likewise be seen
that requiring $D_S W = D_\phi W = 0$ eliminates the (3,0) and
(1,2) components of $W$. Thus - as expected - we recover the same
results in both four-dimensional and ten-dimensional approaches.

This construction will stabilise the dilaton and all complex structure moduli for a generic choice of fluxes. However, its use
would be limited unless there were practical methods to compute the K\"ahler and superpotentials as a function
of the moduli. In particular, in order to discuss complex structure
moduli stabilisation in as explicit as fashion as possible, we need to
be able to compute the periods of Calabi-Yaus. 
To this end, we now briefly review some standard material on complex geometry
which will be needed in the next chapter.

For any Calabi-Yau 3-fold, the middle homology and cohomology are naturally expressed in terms of a symplectic
basis. That is, there exists a basis of 3-cycles $A^a$, $B_b$ and a
basis of 3-forms $\alpha_a$, $\beta^b$
(where $a,b=1,2 \ldots, h^{2,1}+1$), such that in homology \bea
A^a \cap B_b = -B_b \cap A^a & = & \delta^a_b, \nonumber \\
A^a \cap A^b = B_a \cap B_b & = & 0, \eea and \bea
\int_{A^b} \alpha_a = - \int_{B_a} \beta^b & = & \delta^b_a ,\\
\int_{\mathcal{M}} \alpha_a \wedge \beta^b = -\int_{\mathcal{M}} \beta^b \wedge \alpha_a & = & \delta^b_a . \eea
Such a symplectic basis is only defined up to $Sp(2n,\mathbb{Z})$ transformations, as these preserve the
symplectic intersection form. The periods are defined as the integral of the holomorphic 3-form $\Omega$ over
this basis of cycles, \be \int_{A_a} \Omega = z^a,\qquad \qquad
 \int_{B^b} \Omega = \mathcal{G}_a.
\ee The periods are encapsulated in the period vector, $\Pi = \left( \mathcal{G}_1, \ldots
,\mathcal{G}_n,z_1,\ldots, z_n \right)$, where $n = h^{2,1} + 1$. This inherits the holomorphic freedom of
$\Omega$ and is defined up to holomorphic rescalings $\Omega \to
f(\phi_i)\Omega$. Then $\Pi =
\Pi(\phi_i)$, where $\phi_i$ are the complex structure moduli of the Calabi-Yau.

In terms of the periods, the Gukov-Vafa-Witten superpotential is \be
\label{gvwpot} W = \int G_3 \wedge \Omega =
(2 \pi)^2 \alpha' (f + i S h)\cdot \Pi(\phi_i), \ee where $f = (f_1, \ldots, f_6)$ and $h = (h_1, \ldots, h_6)$
are integral vectors of fluxes along the cycles.
 The quantity of $D3$-brane
charge carried by the fluxes is 
\be 
\label{Nflux} 
N_{flux} = \frac{1}{(2 \pi)^4 \alpha'^2} \int H_3 \wedge
F_3 = f^T \cdot \Sigma \cdot h. 
\ee

Given the vector of periods $\Pi(\phi_i)$, the K\"ahler potential on complex structure moduli space is given by
\bea \mc{K}_{cs}(S, \phi_i) & = &  -\ln(S + \bar{S}) - \ln \left( -i \int
\Omega \wedge \bar{\Omega} \right) \nonumber \\
& = & -\ln(S + \bar{S}) - \ln(-i\Pi^\dagger
\cdot \Sigma \cdot \Pi) \\
& \equiv & \mc{K}_S + \mc{K}_\phi. \nonumber \eea where
$$
\Sigma = \left( \begin{array}{cc} 0 & \mathbf{1}_n \\
-\mathbf{1}_n & 0 \end{array} \right).
$$
We can then compute the metric on moduli space, 
\be 
\label{eqKM}
g_{\alpha \bar{\beta}} =
\partial_\alpha
\partial_{\bar{\beta}} \mc{K}, \ee and the Riemann and Ricci curvatures
\bea 
\label{eqKMC}
R^{\lambda}_{\mu \bar{\nu} \rho} & = & -\partial_{\bar{\nu}} (g^{\lambda \bar{\alpha}} \partial_\mu g_{\rho
\bar{\alpha}} ),
\nonumber \\
R_{\mu \bar{\nu}} = R^{\lambda}_{\mu \bar{\nu} \lambda} & = & -
\partial_\mu \partial_{\bar{\nu}} \log (\det g). 
\eea
The K\"ahler metric (\ref{eqKM}) determines the moduli kinetic terms. The
expressions (\ref{eqKMC}) will be useful in chapter \ref{StatisticsReview}.

Thus in order to solve equations (\ref{eqSTS}) and (\ref{csstabilisation}), the essential technical datum
is an explicit expression for the period vector $\Pi(\phi_i)$ in a symplectic basis. It is in general a highly
non-trivial task to find this, but we shall see in section \ref{Tests} that there are examples for which it is
known. 

However, before we do this we first review for completeness the definition of the
chiral coordinates in the presence of open string fields.

\section{Chiral Coordinates for $\mc{N} = 1$ Supergravity} 
\label{seCCS}           

In any $\mc{N} =1$ supersymmetric theory, we can always package the scalar
fields into chiral superfields. It is only in terms of these that the
action takes the canonical form used above and given in full detail in \cite{WessBagger}.
Thus if we wish to use the formulae of
4-dimensional supergravity with the conventions of \cite{WessBagger}, it is an important and necessary task to
identify the chiral superfields. 

For example, in heterotic 
compactifications 
the K\"ahler moduli are defined in terms of the 2-cycles of the
internal space, and are given by
\be
T_i = t_i + i b_i.
\ee
$t_i$ is the string-frame volume of a 2-cycle $\Sigma_i$ and $b_i$ is
the component of the NSNS 2-form $B_2$ along that cycle. 
The K\"ahler form $J$ and NSNS 2-form $B$ can be expressed in terms of a basis 
$e_i$ of $H^{1,1}(M, \mbb{R})$
$$
J = \sum t^i e_i, \quad \textrm{ and } \quad B = \sum b^i e_i.
$$
In IIB orientifold compactifications, this definition is no longer appropriate.
The reason for this is twofold. First, the orientifold action projects out 
fluctuations of $B$ from the low energy spectrum and so these can no longer appear in chiral multiplets.
Secondly, the correct definition of the K\"ahler moduli now 
involves 4-cycle rather than 2-cycle volumes. 

As well as geometric moduli, there are also moduli corresponding 
to brane positions - for both D3 and D7 branes - and to Wilson lines on branes, and in general these
mix. To determine the correct K\"ahler coordinates a careful
dimensional reduction must be performed of the orientifold model with branes
and fluxes. This problem has been studied in great detail by Louis and
collaborators \cite{hepth0312232, hepth0403067, hepth0409098,
  hepth0412277, hepth0502059} and we shall simply state their results.

In D3/D7 compactifications the orientifold action $\sigma$ must leave the K\"ahler form
invariant. The
2-cycles can be written in a basis in which they are either even or
odd under $\sigma$. We denote the $(h^{1,1})^+$ even elements by
  $e_\alpha$ and the $(h^{1,1})^-$ elements by $e_a$. As the volume form
$J \wedge J \wedge J$ is even, it follows that the triple intersectons
$k_{\alpha \beta c}$ and $k_{abc}$ vanish.

The general expressions for the K\"ahler coordinates are rather
complicated, and as we do not need the fully general form we will
leave these to the references. In particular, we will make the
simplifying assumption that
$(h^{1,1})^- = 0$ - this will be the case for the explicit model 
considered in chapter \ref{chapterLargeVol}. In this case, the K\"ahler coordinates in the presence
of D3 and D7 branes are \cite{hepth0409098}
\bea
\label{eqBM}
S & = & S_0 + \kappa_4^2 \mu_7 \mc{L}_{A \bar{B}} \zeta^A
\bar{\zeta}^{\bar{B}} \\
T_\alpha & = & \sigma_\alpha + i \rho_\alpha + 2 i \kappa_4^2 \mu_7 l^2
\mc{C}_\alpha^{I \bar{J}} a_I \bar{a}_{\bar{J}} + i \mu_3 l^2
(\omega_\alpha)_{i \bar{j}} \hbox{tr} \Phi^i (\bar{\Phi}^{\bar{j}} -
\frac{i}{2} \bar{z}^{\tilde{a}} (\bar{\chi}_{\tilde{a}})^{\bar{j}}_l
\Phi^l ). \nonumber
\eea
Here $S_0 = e^{-\phi} + iC_0$ is the dilaton-axion, $\sigma_i$ is the
Einstein frame volume of a 4-cycle, $\rho_i$ is the component of $C_4$ along this cycle,
$\zeta$ is a D7-brane
position modulus, $\Phi$ is a D3-brane position modulus and $a_I$ are
Wilson line moduli. The complex structure moduli surviving the
orientifold projection are unchanged in definition. 

To simplify these expressions, we assume first that 
no wandering D3 branes are present; this we
may always achieve by saturating the tadpole with fluxes. We
also assume that there are no Wilson line moduli on the
D7-branes. A Calabi-Yau never has bulk one-cycles: what this assumption
says is that there are also no one-cycles inside four-cycles.
In this case the bulk moduli simplify to
\bea
S  & = & S_0 + \kappa_4^2 \mu_7 \mc{L}_{A \bar{B}} \zeta^A
\bar{\zeta}^{\bar{B}} \nonumber \\
T_\alpha & = & \sigma_\alpha + i \rho_\alpha.
\eea
The K\"ahler potential is unaltered from (\ref{noscaleK}). However,
now $S_0 + \bar{S}_0$ must be regarded as an implicit function of
the dilaton-axion and D7-brane moduli. The K\"ahler moduli part of
the K\"ahler potential, $- 2 \ln (\mc{V})$, is unaffected by the brane
moduli. 
This will be important later, as it implies the D7-brane moduli are fixed at
tree-level by the fluxes and do not interfere with the stabilisation
of K\"ahler moduli.\footnote{If wandering D3s are present, dealing
  with the  K\"ahler moduli becomes more complicated as these branes mix
  nontrivially in the definition of the K\"ahler moduli.}

The above expressions for the chiral coordinates have been derived by
dimensional reduction of 10-dimensional actions. This process is
rather laborious and we note there is a more intuitive way to understand some
of the results, through noting that instantonic effects should be holomorphic
in the chiral coordinates. For example, in heterotic theory the
relevant instantons are worldsheet instantons. The worldsheet action
is
\be
S_{worldsheet} = \frac{1}{2 \pi \alpha'} \int (\sqrt{g} + i B).
\ee
As worldsheet instantons correspond to calibrated embeddings of the
worldsheet on spacetime 2-cycles, it follows that the natural chiral
coordinates are
$t_i + i b_i$. In the IIB case, it is
D3-instantons that are present, with action
\be
\label{D3InstAction}
S_{D3} = \frac{1}{(2 \pi)^3 \alpha'^2} \int e^{-\phi} \sqrt{g} + i
C_4.
\ee
The action (\ref{D3InstAction}) shows
why chiral coordinates now involve combinations of Einstein frame
4-cycle volumes with $C_4$ axions.
This explanation
become more cumbersome at one-loop level; for example the Wilson line
contribution 
in (\ref{eqBM}) is loop-suppressed and the holomorphicity derivation
requires a string loop calculation \cite{hepth0404087}.

Having established conventions, notation and background we now
move in part II to a more detailed study of moduli stabilisation. 

\part{Moduli Stabilisation}

\chapter{Statistics of Moduli Stabilisation}
\label{StatisticsReview} \linespread{1.3}

The second half of this chapter is based on the paper \cite{hepth0409215}.

In the compactifications of \cite{hepth0105097} reviewed in chapter
\ref{ChapterModuliAndFluxes}, 
the dilaton and complex
structure moduli were stabilised by the fluxes, but the K\"ahler moduli
still remain unfixed. It may seem natural to now discuss K\"ahler moduli
stabilisation.
However,
we shall defer this to chapters \ref{ChapterKahlerModuli} and
\ref{chapterLargeVol} and shall instead give
a detailed account of aspects of the
flux-stabilisation. In particular, we discuss the question of where
the complex structure moduli are stabilised and the extent to which
there are preferred loci in moduli space.

At leading order in $g_s$ and $\alpha'$, the 
calculational procedure is 
well-defined. We need to know the Calabi-Yau periods, which is a clean
mathematical problem, and specify flux choices $F_3$, $H_3 \in H^3(M,
\mbb{Z})$.
These determine the F-term equations (\ref{eqSTS}) and (\ref{csstabilisation}),
which we solve to
stabilise the dilaton and complex structure moduli.
Throughout this chapter I shall abuse terminology and
refer to such solutions as vacua. Clearly a genuine vacuum must also,
at least, stabilise the K\"ahler moduli.

The total number of such flux vacua appears extremely large. There are
$h^3 = h^{3,0} + h^{2,1} + h^{1,2} + h^{0,3} = 2(h^{2,1} + 1) \equiv
K$ cycles on which to wrap fluxes. For
a typical Calabi-Yau, $K \sim \mc{O}(100)$. We specify $F_3$ and $H_3$ 
by choosing integers $f_i$ and $h_i$ for each 3-cycle, giving $2K$
integers in total. For $\omega^i$ a symplectic basis of 3-forms,
\be
F_3 = \sum_{i=1}^K f_i \omega^i, \qquad H_3 = \sum_{i=1}^K h_i
\omega^i.
\ee
We must also satisfy the
 tadpole cancellation condition (\ref{eqTCE}). Although
 this appears to be of indefinite sign, it can be shown that for
 solutions of (\ref{eqSTS}) and (\ref{csstabilisation}) we
 have\footnote{Intuitively, this is beacause supersymmetric solutions require
   the fluxes to be pseudo-BPS, carrying D3 tension and D3 (rather
   than anti-D3) charge.}
\be
\label{Fluxgecon}
\int H_3 \wedge F_3 \ge 0.
\ee

The conditions (\ref{eqTCE}) and (\ref{Fluxgecon}) heuristically restrict the
 fluxes to a ball of radius $\sqrt{L}$ in flux space, where $L = \frac{\chi}{24}$.
An estimate for the number of consistent flux choices is then $\sqrt{L}^{2K} = L^K$.
For F-theory compactifications and their IIB orientifold limits
$L \sim \mc{O}(1000)$, and so $L^K$ may easily take values of order $\sim 1000^{200} \sim 10^{600}$. The main area of
physics in which such numbers are successfully encountered is that of
 statistical mechanics.
This motivates the
use of statistical techniques to tame and understand the multiplicity of vacua.

\section{Derivation of Vacuum Distributions}

In a series of papers\cite{hepth0303194, hepth0307049, hepph0401004,mathcv0402326,
  hepth0404116, hepth0405279, mathcv0406089, hepth0411183, mathph0506015} Douglas and collaborators have developed
techniques to count solutions to equations (\ref{eqSTS}) and
  (\ref{csstabilisation}) 
and to make rigorous statements about the ensemble of flux vacua. The
  purpose of this section is to
give a technical account of the methodology and to derive the simplest formula
thereby attainable.
We shall also summarise other results attained in this fashion. In the
  subsequent section \ref{Tests} we describe our tests of these formulae.

We aim to explain the derivation of the Ashok-Douglas formula
\cite{hepth0307049} for the index density of vacua. This computes
\be 
\label{indexdensity} 
d \mu_I(z) = \sum_{\textrm{vacua}}
\textrm{sign}(\det D^2 W) 
\ee 
as a function of complex structure moduli space $z$,
and evaluates it to be
\be
\label{indexdenre}
\int_{\mc{F}} d \mu_I(z) = \frac{(2
\pi L_{max}^K)(-1)^{\frac{K}{2}}}{\pi^{n+1} K!} \int_{\mc{F}}
\det(-\mc{R} - \omega).
\ee
Here $L_{max}$ is the total
available D3-brane charge from the orientifold, $K$ the number of 3-cycles, and $n$ the number of moduli.
$\mc{F}$ is a fundamental region in the combined dilaton-axion and complex structure moduli
space, while $\omega$ and $\mc{R}$ represent the K\"ahler and curvature
2-forms on this space. These are both fully determined by the K\"ahler
potential of the moduli space,
\be
\mc{K} = - \ln(S + \bar{S}) - \ln \left( i \int \Omega \wedge
\bar{\Omega} \right).
\ee
We reemphasise that here `vacuum' simply refers
to a choice of fluxes satisfying the tadpole constraint and
generating a superpotential $W = \int G_3 \wedge \Omega$ such that, at the point in moduli
space denoted by $z$, 
\be 
D_S W(z) = D_{\phi_i}W(z) = 0. 
\ee 
The phrase `index density' refers to the fact that the vacua are
counted with signs.
As the index density is much easier to calculate, it is this quantity
rather than the absolute density that is computed.

Nonetheless, the derivation of (\ref{indexdenre}) is by no means
obvious. The calculation is structured as follows.
We first compute $\langle d\mu_I(z) \rangle$ within a Gaussian
ensemble of superpotentials, counting vacua with weights.
Once this has been computed, it is relatively straightforward to convert this
into a formula for the actual index density $d \mu_I(z)$, in which vacua are only weighted
by a sign. This conversion is essentially an inverse Laplace transform
on $\langle d \mu_I (z) \rangle$.

The ensemble considered is the ensemble of flux superpotentials. The
flux superpotential is a linear combination of periods $\int_{\Sigma_i} \Omega$
and so the ensemble 
consists of all superpotentials $W(z)$ expressible as 
\be
\label{SuperpotentialEnsemble} 
W(z) = \sum_{\alpha = 1}^K N^\alpha \Pi_\alpha(z), 
\ee 
where $N^\alpha$ are complex numbers
and $\Pi_\alpha(z)$ are the Calabi-Yau periods. The
main approximation made is to treat the $N^\alpha$ as continuous,
whereas flux quantisation renders them integral.
In writing (\ref{SuperpotentialEnsemble}) we also assume a fixed dilaton; we will
remedy this assumption later.

The expectation value of a quantity $\mc{A}(W)$ in the ensemble is 
\be 
\langle \mathcal{A}(W)
\rangle = \frac{1}{\mathcal{N}} \int DW f(W) A(W), 
\ee 
with the measure $f(W)$ and normalisation $\mc{N}$ defined to be
\bea
\mathcal{N} & = & \langle 1 \rangle = \int DW f(W), \nonumber \\
f(W) & = & \int d^2 N^\alpha \exp (-Q_{\alpha \beta} N^\alpha \bar{N}^\beta) \delta (W - \sum N^\alpha
\Pi_\alpha). \eea In the case at hand,
\be 
Q_{\alpha \beta} = \eta_{\alpha \beta} = \left( \begin{array}{cc} 0 & \mathbf{1}_n \\
-\mathbf{1}_n & 0 \end{array} \right). 
\ee 
However it is useful to leave $Q_{\alpha \beta}$ general.

The measure $f(W)$ has no intrinsic significance: it is
merely an efficacious stepping stone on the way to computing
(\ref{indexdensity}). 
The purpose of these definitions is that computations in this ensemble reduce to
integrals computable using the techniques of quantum field theory. Our
aim is to compute 
\be
\label{TargetComputation} 
\langle d\mu_I (z_0) \rangle = \langle \delta^n(DW(z_0))
\delta^n(\bar{D}W^{*}(\bar{z}_0)) \det D^2 W(z_0) \rangle. 
\ee 
Once (\ref{TargetComputation}) is computed, an inverse transform
allows the derivation of $d\mu_I(z)$.
The factor of $\det D^2 W(z_0)$ accounts for the change of variables
from $z$ to $DW(z)$ in defining the $\delta$-functions. We also recall
that $D^2 W(z_0) = \partial DW(z_0)$ so long as $DW(z_0) = 0$.
However, in order to compute (\ref{TargetComputation}),
there are several necessary preliminary calculations. First,
$$
\mathcal{N} =\int DW f(W) =  \frac{\pi^K}{\det Q}.
$$
This follows straightforwardly, as the integral reduces to
$$
\int d^2 N^\alpha \exp\left(-Q_{\alpha \beta} N^{\alpha} \bar{N}^\beta \right) = \frac{\pi^K}{\det Q}.
$$
The cornerstone of the computation is the 2-point function 
$\langle W\left(z_1\right) W^{*}\left(\bar{z}_2\right) \rangle$,
which we denote by $G(z_1, \bar{z}_2)$. Through standard quantum field theory
arguments this evaluates easily to 
\be
\langle W\left(z_1\right) W^{*}\left(\bar{z_2}\right) \rangle = \left(Q^{-1}\right)^{\delta \gamma}
\Pi_\gamma\left(z_1\right) \Pi^{*}_\delta \left(\bar{z_2}\right). 
\ee 
For the specific ensemble of flux superpotentials,
\be 
\langle W\left(z_1\right) W^{*}\left(\bar{z_2}\right) \rangle = -\eta^{\delta \gamma}
\Pi_\gamma\left(z_1\right) \Pi^{*}_\delta \left(\bar{z_2}\right) =
\exp(-\mc{K}(z_1, \bar{z}_2)), 
\ee 
where $\mc{K}$ is the K\"ahler potential for the complex structure moduli. The importance of $G(z_1, \bar{z}_2)$ is
that, by Wick's theorem, all other quantities in the ensemble are expressible in terms of it.

A second important building block is
$$
\mathcal{X} \equiv \left\langle \delta^{2n}\left(DW\left(z_0\right)\right) \right\rangle.
$$
We compute this as follows 
\bea \langle \delta^{2n}\left(DW\left(z_0\right)\right) \rangle & = &
\frac{1}{\mathcal{N}}\int DW f\left(W\right) \delta^n \left(D_i W\left(z_0\right)\right) \delta^n
\left(\bar{D}_j
W^{*}\left(\bar{z_0}\right)\right) \nonumber \\
& = & \frac{1}{\mathcal{N}} \int dN^{\alpha} d\bar{N}^\beta \exp\left(-Q_{\alpha \beta} N^\alpha \bar{N}^\beta
\right) \delta^n \left(N^\gamma D_\mu \Pi_\gamma \left(z_0\right)\right) \delta^n \left(\bar{N}^\delta
\bar{D}_\nu \Pi^{*}_\delta \left(z_0\right)\right) \nonumber 
\eea
$$
=  \frac{1}{\mathcal{N} \times \pi^{2n}}\int d^2 N d\lambda_\mu d\bar{\lambda}_\nu \exp\left(-Q_{\alpha \beta}
N^\alpha \bar{N}^\beta + i \lambda^\mu N^\gamma D_\mu \Pi_\gamma \left(z_0\right) + i \bar{\lambda}^\nu
\bar{N}^\delta \bar{D}_\nu \Pi^{*}_\delta \left(z_0\right)\right),
$$
where we have implemented the $\delta$-function constraints by introducing new integration variables
$\lambda_\mu$. This integral may be performed by completing the square in $N^\alpha$ and shifting \bea
N^\alpha & \to & N^\alpha + i\left(Q^{-1}\right)^{\gamma \alpha} \bar{\lambda}^\nu \bar{D}_\nu
\Pi^{*}_\gamma \left(z_0\right), \nonumber \\
\bar{N}^\beta & \to &  \bar{N}^\beta  + i\left(Q^{-1}\right)^{\beta \delta} \lambda^\mu D_\mu \Pi_\delta
\left(z_0\right). \nonumber \eea The integral becomes \be \int dN^\alpha d \bar{N}^\beta d\lambda^\mu
d\bar{\lambda}^\nu \exp\left(-Q_{\alpha \beta} N^\alpha \bar{N}^\beta\right)
\exp\left(-\left(Q^{-1}\right)^{\gamma \delta} \lambda^\mu \bar{\lambda}^\nu D_\mu \Pi_\delta\left(z_0\right)
\bar{D}_\nu \Pi^{*}_\gamma \left(z_0\right)\right). 
\ee 
The $N$ and $\lambda$ integrals have separated, and
this evaluates to
$$
\frac{1}{\pi^n} \times \frac{1}{\det_{\mu, \nu} \left( \left( Q^{-1} \right)^{\delta \gamma} D_\mu \Pi_\gamma
\left( z_0 \right) \bar{D}_\nu \Pi^{*}_\delta \left(z_0 \right) \right)} = \mathcal{X}.
$$
If $G \left(z_1, \bar{z_2} \right) = \exp \left(- \mc{K}\left(z_1, \bar{z_2} \right) \right)$, as is
appropriate for the flux ensemble, this becomes
$$
\frac{1}{\pi^n} \times \frac{e^{-n \mc{K}(z_0, \bar{z}_0)}}{\left(\det g\right) },
$$
where $g$ is the moduli space metric and $n$ the number of $\lambda
\bar{\lambda}$ integrations performed, 
which equals the number of complex structure moduli. Therefore
$$
\mathcal{X} = \frac{1}{\pi^n} \times \frac{e^{-n \mc{K}(z_0, \bar{z}_0)}}{\left(\det g\right) }.
$$
The final prerequisite is $G_{z_0}(z_1, \bar{z}_2) = \langle \delta^{2n}\left(DW\left(z_0\right)\right)
W\left(z_1\right) W^{*}\left(\bar{z}_2\right) \rangle$. This can be computed by identical techniques: we complete
the square in $N$ and perform the $N$ and $\lambda$ integrals separately. The calculation is messy but
not difficult, and we obtain \be G_{z_0}(z_1, \bar{z}_2) = \mathcal{X} \times \left(G\left(z_1, \bar{z}_2\right)
- D_{z_0^\mu} G\left(z_0, \bar{z_2}\right) \left(D_\mu \bar{D}_\nu G\left(z_0, \bar{z}_0\right)\right)^{-1}
\bar{D}_{\bar{z}_0^{\bar{\nu}}} G\left(z_1,
\bar{z}_0\right)\right). 
\ee 
The reason for this computation 
is that $\langle d \mu_I (z) \rangle$ is most simply expressed in terms of $G_{z_0}(z_1, \bar{z}_2)$.

We now have sufficient ingredients to compute the ensemble index
density 
\be \langle d \mu_I(z_0) \rangle
= \langle \delta^{2n}\left(DW\left(z_0\right)\right) \det D^2 W
\rangle. 
\ee 
The calculation is unenlightening and we shall
not perform all steps explicitly, but shall instead merely show the structure. We first of all write
$$
\det D^2 W = \int d^2 \psi^i \, d^2 \theta^i \, \exp \left[ \theta^a \psi^b \partial_a D_b W + \bar{\theta}^a
\psi^b \bar{\partial}_a D_b W + \bar{\theta}^a \bar{\psi}^b \bar{\partial}_{\bar{a}} \bar{D}_{\bar{b}} W^{*} +
\theta^a \bar{\psi}^b
\partial_a \bar{D}_b W^{*} \right],
$$
where $\theta$ and $\psi$ are Grassmann variables. We seek
to compute 
\bea 
\frac{1}{\mathcal{N}}
\times \frac{1}{\pi^{2n}} \int d^2 N \, d\lambda_\mu d\bar{\lambda}_\nu \, d^2 \psi^i \, d^2 \theta^i \,
\exp\left(-Q_{\alpha \beta} N^\alpha \bar{N}^\beta + i \lambda^\mu N^\gamma D_\mu \Pi_\gamma \left(z_0\right) +
i \bar{\lambda}^\nu \bar{N}^\delta \bar{D}_\nu
\Pi^{*}_\delta \left(z_0\right)\right) & & \nonumber \\
\exp \left[ \theta^a \psi^b \partial_a D_b W + \bar{\theta}^a \psi^b \bar{\partial}_a D_b W + \bar{\theta}^a
\bar{\psi}^b \bar{\partial}_{\bar{a}} \bar{D}_{\bar{b}} W^{*} + \theta^a \bar{\psi}^b
\partial_a \bar{D}_b W^{*} \right]. & & \nonumber
\eea
We complete the square in $N$ and integrate to obtain \bea \frac{1}{\pi^{2n}} \int d \lambda^\mu
d\bar{\lambda}^\nu d^2 \psi^i \, d^2 \theta^i \, \exp \Big( \left(Q^{-1}\right)^{\gamma \delta}\left(i
\bar{\lambda}^\nu \bar{D}_\nu \Pi^{*}_\gamma\left(z_0\right) + \bar{\theta}^a \bar{\psi}^b \bar{\partial}_a
\bar{D}_b \Pi^{*}_\gamma\left(z_0\right) +
\theta^a \bar{\psi}^b \partial_a \bar{D}_b \Pi^{*}_\gamma \left(z_0\right)\right)  & & \nonumber \\
\times \left(i \lambda^\mu D_\mu \Pi_\delta\left(z_0\right) + \theta^a \psi^b \partial_a D_b
\Pi_\delta\left(z_0\right) + \bar{\theta}^a \psi^b \bar{\partial}_a D_b \Pi_\delta \left(z_0\right)\right)
\Big). & & \nonumber \eea 
The integral is now quadratic in $\lambda$. We integrate over $\lambda$
to obtain 
\bea
& & \mathcal{X} \times \int d^2 \theta^i d^2 \psi^i \nonumber \\
& &
 \exp \Big[\left( \theta^c \psi^d \partial_{1c} D_{1d} +  \bar{\theta}^c \psi^d \bar{\partial}_{1c} D_{1d}\right)
 \left(\bar{\theta}^a \bar{\psi}^b \, \bar{\partial}_{2a} \bar{D}_{2b} \, +
\theta^a \bar{\psi}^b \partial_{2a} \bar{D}_{2b}\right) \, G_{z_0}\left(z_1, \bar{z}_2\right)\Big] |_{z_1 = z_2
= z_0}. \nonumber \eea We may write the terms inside the exponential as
$$
 \theta^c \psi^d \bar{\theta}^a \bar{\psi}^b \partial_{1c} D_{1d} \, \bar{\partial}_{2a} \bar{D}_{2b}
G_{z_0}\left(z_1, \bar{z}_2\right) +  \bar{\theta}^c \psi^d \theta^a \bar{\psi}^b \bar{\partial}_{1c} D_{1d}
\partial_{2a} \bar{D}_{2b} G_{z_0}\left(z_1, \bar{z}_2 \right),
$$
with other terms vanishing. We recall that $\mathcal{X} = \frac{1}{\pi^n} \times \frac{1}{\left(\det g\right)
e^{n \mc{K}\left(z_0, \bar{z}_0\right)}}$ and bring the $ e^{n \mc{K}\left(z_0, \bar{z}_0\right)}$ into the Grassmann
exponent. We then need to compute
$$
F_{cd\vert \bar{a} \bar{b}} = e^{-K\left(z_0, \bar{z}_0\right)} D_{1c} D_{1d} \bar{D}_{2a} \bar{D}_{2b}
G_{z_0}\left(z_1, \bar{z}_2\right)
$$
and
$$
F_{\bar{c}d\vert a \bar{b}} = e^{-K\left(z_0, \bar{z}_0\right)} \bar{D}_{1c} D_{1d} D_{2a} \bar{D}_{2b}
G_{z_0}\left(z_1, \bar{z}_2\right).
$$
These can be evaluated by brute force - as $G_{z_0}\left(z_1, \bar{z}_2\right)$ is a combination of the Kahler
potential and its derivatives, there is no difficulty of
principle in the calculation. We find 
\bea F_{cd| \bar{a}
\bar{b}} & = & R_{c \bar{a} d \bar{b}} + \left(g_{d \bar{a}}
g_{c \bar{b}} + g_{c \bar{a}} g_{d \bar{b}}\right), \nonumber \\
F_{\bar{c} d \vert a \bar{b}} & = & g_{a \bar{b}} g_{d \bar{c}}. 
\eea 
We then have \be \langle d
\mu_I(z) \rangle = \frac{1}{\pi^n \times \det g} \int d^2 \psi^i d^2 \theta^i \exp \left(-\theta^a
\theta^{\bar{c}} \psi^b  \psi^{\bar{d}} R_{a \bar{c} b \bar{d}} - g_{a \bar{c}} \theta^a \theta^{\bar{c}} \,
g_{b \bar{d}} \psi^b \bar{\psi}^{\bar{d}}\right). \ee 
Performing the $\theta$ integral, we
get \bea & & \frac{1}{\pi^n \times \det g} \det_{a, \bar{c}} \left(  -\psi^b  \psi^{\bar{d}} R_{a \bar{c} b
\bar{d}}-
 g_{a \bar{c}} \, g_{b \bar{d}} \psi^b \bar{\psi}^{\bar{d}} \right) \nonumber \\
& = & \frac{1}{\pi^n} \det \left( (g^{-1}) \times  \left(  - \psi^b  \psi^{\bar{d}} R_{a \bar{c} b \bar{d}}-
 g_{a \bar{c}} \, g_{b \bar{d}} \psi^b \bar{\psi}^{\bar{d}} \right) \right). \nonumber
\eea We now promote $\psi^i$ and $\bar{\psi}^j$ to differential forms - this is possible because forms have the
same Grassmann properties as fermionic variables. We finally obtain \be \langle d \mu_I(z) \rangle =
\frac{1}{\pi^n} \det(-\mc{R} - \omega), \ee where $\mc{R}$ is the curvature two-form and $\omega$ the volume
form on moduli space.

This computes the index density within the Gaussian ensemble. However, we would like to count vacua without
including a Gaussian weight. Let us state how to do this and then justify it. The claim is that if
we can compute \be \label{vacuasum} N_{vac} \left(\alpha\right)  =
\sum_{vacua} e^{-2 \alpha \left(\rm{Im }
\tau\right) \eta_{\alpha
    \beta} N_{RR}^\alpha N_{NS}^\beta},
\ee then the total number of vacua, weighted by the index alone, is
given by
\be
\label{InverseTransform}
N_{vac}\left(L \le
L_{max}\right) = \frac{1}{2 \pi i} \int_{C} \frac{d
  \alpha}{\alpha} e^{2 \alpha \left({\rm{Im \,}} \tau\right) L_{max}} N_{vac}\left(\alpha\right).
\ee
Why is this true? (\ref{vacuasum}) contains a measure $\exp \left(-2
\alpha \left({\rm{Im \,}}
\tau\right) \eta_{\alpha \beta} N_{RR}^\alpha N_{NS}^\beta\right)$.
This is analogous to
computing \be \tilde{f}(\lambda) = \int f(x) e^{-\lambda x^2} dx \ee
for some value of $\lambda$. However, given
$\tilde{f}(\lambda)$, we may find $f(x)$ by an appropriate inverse
transform. For the flux ensemble, this is given by
(\ref{InverseTransform}). We then evaluate
$$
N_{vac}(\alpha) = ({\rm{Im \,}} \tau)^{-K} \int DW \int d^2N^\alpha \exp (-i \alpha \eta_{\alpha \beta} N^{\alpha}
\bar{N}^\beta) \delta^{2n}(DW(z_0)) \det D^2 W
$$
$$= \frac{\pi^{K-n} (-1)^{K/2}}{(\alpha {\rm{Im \,}} \tau)^K} \det(-\mc{R}
- \omega).
$$
The $({\rm{Im \,}} \tau)^{-K}$ prefactor arises from converting $\sum_{N_{RR},
  N_{NS}}$ to $\int dN d\bar{N}$.
We can then take the inverse transform of this to get the total number
of vacua.

As mentioned above, we must actually be a little more subtle.
We have regarded the dilaton as fixed and have focused on the complex
structure moduli, not requiring
$D_\tau W = 0.$ We ought to repeat the entirety of the
above analysis including the dilaton. If this is done,
we finally obtain
\be
\label{ADdensity} d \mu_I(z) = \frac{(2
\pi L_{max}^K)(-1)^{\frac{K}{2}}}{\pi^{n+1} K!}
\det(-\mc{R} - \omega),
\ee
as above.
This gives the index density of vacua stabilised at a point $z$ in
moduli space.
$\omega$ and $\mc{R}$ represent the K\"ahler and curvature
2-forms on the combined dilaton-axion/complex structure moduli space.

(\ref{ADdensity}) describes the distribution of flux vacua in complex
structure moduli space. Although the final form is simple, it contains
a lot of information about where the complex structure moduli tend to
be stabilised. To a first approximation, the distribution is
uniform in volume, being given by $\det(-\omega)$. However, the form of
(\ref{ADdensity}) also shows that vacua tend to cluster in regions of
high moduli space curvature. An example of such a region is the vicinity of a
conifold singularity, where the moduli space curvature diverges
(although the integrated curvature is finite).

The statistical techniques above can be extended and applied to compute
other quantities of interest in the ensemble of flux vacua.
I leave full details to
the references \cite{hepth0307049, hepth0404116, hepth0411183} and
simply state the results that will be most relevant for the tests that are
described in the next section.
The first concerns the number of vacua with small tree-level
superpotential. This is of interest as it is related to the gravitino
mass and supersymmetry breaking scale. This is also the condition
required for control of the approximations in the KKLT scenario.
\be
N(\textrm{vacua s.t. } e^{\mc{K}} \vert W \vert^2 < \epsilon) \sim \epsilon.
\ee 
The second concerns the distribution of the stabilised string coupling
constant. This is of interest if we wish to remain within weakly
coupled string theory.
\be
\label{gsADdistribution}
N(\textrm{vacua s.t. } \frac{1}{g_s} > \epsilon) \sim \epsilon.
\ee 
(The distribution (\ref{gsADdistribution}) is actually implicit in
(\ref{ADdensity})). 

Let us expand on the relationship of $e^{\mc{K}} \vert W
\vert^2$ to the supersymmetry breaking scale. We work here in a
truncation for which the K\"ahler moduli are excluded.
If \emph{all} moduli are
included, $m_{3/2}^2 = e^{\mc{K}} \vert W \vert^2$ and the gravitino mass
$m_{3/2}$ does give a true
measure of the scale of supersymmetry breaking, either directly or as required
to uplift a supersymmetric AdS minimum to Minkowski space.
While the size of $e^{\mc{K}} \vert W \vert^2$ with the truncation may allow us
to say whether the gravitino mass is (relatively) `larger' or
`smaller', in the absence of the vacuum values of the K\"ahler moduli
we cannot make statements about the absolute magnitude of $m_{3/2}$.

Having described the derivation of the statistical formulae concerning the number and distribution of
solutions to the dilaton and complex structure moduli F-term
equations, 
let us now turn to our tests of these results.

\section{Tests of the Statistical Predictions}
\label{Tests}

This section is based on the paper \cite{hepth0409215}.

For an idealised flux ensemble, (\ref{ADdensity}) gives the distribution of vacua
over moduli space. There are two ways in which (\ref{ADdensity}) deviates from the actual density of vacua.
First, in its derivation the quantisation of the 3-form fluxes has been neglected and a
continuum approximation used. Secondly, (\ref{ADdensity}) does not count the absolute number
of vacua but instead counts vacua with signs. It is therefore
interesting to test whether the predictions of (\ref{ADdensity})
actually work in real examples.
The most direct way to test this is through Monte
Carlo simulation: we randomly generate fluxes, study the distribution
of stabilised moduli, and test for 
agreement with the predictions of equation (\ref{ADdensity}).

This procedure was carried out for a 1-modulus
Calabi-Yau in \cite{hepth0404243}. 
In this section we examine
the distribution of the complex structure moduli, the dilaton-axion, and the `susy breaking scale' $e^{\mc{K}}
\vert W_0 \vert^2$ for a particular 2-modulus Calabi-Yau, the
hypersurface in $\mbb{P}^4_{[1,1,2,2,6]}$.

\subsection{The Model Used: $\mbb{P}^4_{[1,1,2,2,6]}$}

There exist many Calabi-Yau manifolds that can be realised as hypersurfaces in weighted projective space. The
weighted projective space $\mbb{P}^4_{[k_0, k_1, k_2, k_3, k_4]}$ is
defined by the complex coordinates $w_0$, $w_1$, $w_2$, $w_3$, $w_4$
subject to 
$$
(w_0,w_1,w_2,w_3,w_4) \equiv (\lambda^{k_0} w_0, \lambda^{k_1} w_1, \lambda^{k_2} w_2, \lambda^{k_3} w_3,
\lambda^{k_4} w_4),
$$
and has four complex dimensions. To obtain a space of three complex
dimensions we must restrict to a hypersurface
$P(w_i) = 0$, where $P$ is polynomial in the $w_i$. 
We require the hypersurface to be Calabi-Yau, i.e. to have vanishing
first Chern class. Using standard results from algebraic geometry
(e.g. see chapter 5 of H\"ubsch \cite{HubschBook}), this occurs provided 
$$
\textrm{deg}(P) = \sum_{i=0}^4 k_i.
$$
We will make use of the Calabi-Yau hypersurface in
$\mbb{P}^4_{[1,1,2,2,6]}$, which requires a polynomial homogeneous of
degree 12. The simplest form of this polynomial is
\be 
\label{polyeqn}
 w_0^{12} + w_1^{12} + w_2^{6} + w_3^{6} + w_4^{2} = 0.
\ee 
We shall denote this manifold by $\mathcal{M}$. $\mathcal{M}$ has $h^{1,1} = 2$ and $h^{2,1} = 128$. The
number of complex structure moduli is determined by the number of
monomial deformations of degree 12, modulo coordinate redefinitions, that can be
added to (\ref{polyeqn}). There are two complex structure moduli of
particular relevance, which we shall
denote by $\psi$ and $\phi$. These perturb (\ref{polyeqn}) to 
\be 
\label{manifoldeq} 
P(w_i) \ =\ w_0^{12} +
w_1^{12} + w_2^{6} + w_3^{6} + w_4^{2} -12\psi\left( w_0 w_1 w_2 w_3 w_4\right) - 2\phi\left( w_0^6
w_1^6\right) = 0. 
\ee 
Their significance is due to their survival in the mirror manifold $\mathcal{W}$, obtained
by identifying points in $\mathcal{M}$ related by the action of $G$, the maximal group of scaling symmetries of
(\ref{polyeqn}). Here $G$ is $\mbb{Z}_2 \times \mbb{Z}_6^2$, and its action is represented by
$$
\mbb{Z}_2:(w_1,w_4)  \to   (\alpha w_1, \alpha w_4)\textrm{ with } \alpha^2=1,
$$
$$\mbb{Z}_6:
(w_1,w_3)  \to  (\beta^5 w_1,  \beta w_3) \textrm{ with } \beta^6=1,$$
$$\mbb{Z}'_6:
(w_1,w_2)  \to  (\gamma^5 w_1, \gamma w_2) \textrm{ with } \gamma^6=1.$$ These manifestly leave equation
(\ref{manifoldeq}) invariant, whereas all other degree 12 deformations of equation (\ref{manifoldeq}) are not well
defined on the mirror. As $\mathcal{W}$ has $h^{1,1} = 128$ and $h^{2,1} = 2$, it is an example of a Calabi-Yau
with two complex structure moduli. Both $\mathcal{M}$ and $\mathcal{W}$ develop a conifold singularity when $864
\psi^6 + \phi = 1$. This condition follows from requiring $P(w_i) = \underline{\nabla} P(w_i) = 0$ with
$\underline{\nabla} \, \underline{\nabla} \, P$ non-singular.

To test the distribution (\ref{ADdensity}) on the manifold $\mc{M}$ we need its form for the case of two complex
structure moduli. This is evaluated to be\cite{kgprivate}
\footnote{In this section we use $\tau = C_0 + i e^{-\phi} = iS$ to
  denote the dilaton.} \be \label{indexDensity} d \mu = g_{\tau \bar{\tau}}
d\tau \wedge d \bar{\tau} \wedge \left( 4 \pi^2 c_2 - \det(g_{a \bar{a}}) d \psi^1 \wedge d \psi^{\bar{1}}
\wedge d \psi^2 \wedge d \psi^{\bar{2}} \right), \ee where $g_{a\bar{b}} = \partial_a
\partial_{\bar{b}} \mc{K}$ and $c_2$ is the second Chern class for the complex structure moduli space of
$\mathcal{M}$, given by \be c_2 = \frac{1}{8 \pi^2} \left( \mathrm{tr}(\mathcal{R} \wedge \mathcal{R}) -
\mathrm{tr} \mathcal{R} \wedge \mathrm{tr} \mathcal{R} \right). \ee (\ref{indexDensity}) may be rewritten as \be
d \mu =  g_{\tau \bar{\tau}} d\tau \wedge d \bar{\tau} \wedge d \psi^1 \wedge d \psi^{\bar{1}} \wedge d \psi^2
\wedge d \psi^{\bar{2}}
 \left[ \epsilon^{ab} \epsilon^{\bar{a}\bar{b}} \left(R^1_{\phantom{1}
     a \bar{a}  1} R^2_{\phantom{2} b \bar{b} 2} - R^1_{\phantom{1} a \bar{a} 2}
   R^2_{\phantom{2} b \bar{b} 1}\right) -
\det{g_{a \bar{a}}}\right]. \ee

To make further progress we need a knowledge of the periods. This is in general a highly non-trivial task.
However, the manifold defined by (\ref{manifoldeq}) is of a class that has been extensively studied. The
relevant periods have been computed in the mirror symmetry literature 
\cite{hepth9308005,hepth9308083}, following the classic treatment of the
quintic \cite{quintic}, and we shall borrow these results. For the
Calabi-Yau described by the hypersurface 
\be P=
\sum_{j=0}^4 x_j^{d/k_j} - d \psi x_0 x_1 x_2 x_3 x_4 - \frac{d}{q_0} \phi
x_0^{q_0}x_1^{q_1}x_2^{q_2}x_3^{q_3}x_4^{q_4} = 0, 
\ee 
the fundamental period in the large $\psi$ region is
given by 
\be 
\label{largepsifund} \varpi_f(\psi,\phi) = \sum_{l=0}^{\infty} \frac{(q_0 l!) (d \psi)^{-q_0 l}
(-1)^l} {l! \Pi_{i=1}^4 (\frac{k_i}{d}(q_0 - q_i)l)!} u_l(\phi), 
\ee 
where \be u_l(\phi) = (D\phi)^l
\sum_{n=0}^{[\frac{l}{D}]} \frac{    l! \Pi_{i=1}^4 (\frac{k_i}{d}(q_0 - q_i)l)!(-D\phi)^{-Dn} } { (l-Dn)!n!
\Pi_{i=1}^4 (\frac{k_i q_i}{q_0}n + \frac{k_i}{d}(q_0 -
  q_i)l)!} .
\ee This is obtained by direct integration of $\int \Omega$ and satisfies the Picard-Fuchs equation. There are other
independent solutions to the Picard-Fuchs equation having a logarithimic dependence on $\psi$. In total there
are six independent solutions, one for each 3-cycle, and the actual periods are a linear combination of these.

The two regions of moduli space that will most interest us are the vicinities of the Landau-Ginzburg point $\psi
=\phi =0$ and the conifold locus $864 \psi^6 + \phi = 1$. To obtain a basis of periods in the small $\psi$
region, (\ref{largepsifund}) may be analytically continued to obtain \be \label{smallpsi}
\varpi_f\left(\psi,\phi\right) = -\frac{2}{d} \sum_{n=1}^{\infty} \frac{\Gamma\left(\frac{2n}{d}\right) \left(-d
\psi\right)^n u_{-\frac{2n}{d}}\left(\phi\right)} {\Gamma\left(n\right) \Gamma\left(1-\frac{n}{d}\left(k_1
-1\right)\right) \Gamma\left(1 - \frac{k_2 n}{d}\right)
  \Gamma\left(1-\frac{k_3 n}{d}\right) \Gamma\left(1-\frac{k_4 n}{d}\right)} .
\ee Here $u_\nu(\phi)$ is related to the hypergeometric functions and is defined through the contour integral
\be \label{smallpsiu} u_\nu (\phi) = \frac{2^\nu}{\pi} \int_{-1}^{1} \frac{d \zeta}{\sqrt{1
    - \zeta^2}} (\phi - \zeta)^\nu .
\ee The contour integral is initially defined for $\textrm{Im}(\phi) > 0$ and then defined over the rest of the
plane by deforming the integral contour. The branch cuts, which are unavoidable when $\nu$ is non-integral,
start at $\pm 1$ and run to $\pm \infty$.

We may derive a basis of periods from the fundamental period in a simple manner. If we define \be
\label{periods} \varpi_j(\psi,\phi) = \frac{-(2 \pi i)^3}{\psi} \varpi_f(\alpha^j \psi, \alpha^{j q_0} \phi) \ee
then $\varpi(\psi, \phi) = (\varpi_0(\psi,\phi), \varpi_1(\psi,\phi), \varpi_2(\psi,\phi), \varpi_3(\psi,\phi),
\varpi_4(\psi,\phi), \varpi_5(\psi,\phi))$ gives a basis of periods known as the Picard-Fuchs basis. Naively
(\ref{periods}) would seem to give 12 independent periods, but there are interrelations discussed in
\cite{hepth9308005}. The net result is that, as expected, there are
six independent periods. 
However these periods do not correspond to a symplectic basis of 3-cycles.
A symplectic basis is given by \be \label{pidef} \Pi(\psi,\phi)  = m \cdot \varpi(\psi,\phi).
\ee 
For $\mbb{P}^4_{[1,1,2,2,6]}$,
$m$ was computed in \cite{hepth9912147, hepth0312104} and is given by
$$
m = \left( \begin{array}{cccccc} -1 & 1 & 0 & 0 & 0 & 0 \\
\frac{3}{2}
  &  \frac{3}{2} &  \frac{1}{2}
&  \frac{1}{2} & - \frac{1}{2} & - \frac{1}{2} \\
1 & 0 & 1 & 0 & 0 & 0 \\ 1 & 0 & 0 & 0 & 0 & 0 \\ - \frac{1}{2} & 0 &
\frac{1}{2} &  \frac{1}{2} & 0 & 0\\
 \frac{1}{2} &  \frac{1}{2} & - \frac{1}{2} &  \frac{1}{2} & - \frac{1}{2} &  \frac{1}{2} \end{array} \right).
$$

In principle, equation (\ref{pidef}) completely determines the periods near $\psi = 0$. However, it involves the
integral expression (\ref{smallpsiu}) for $u_\nu (\phi)$ which is
unsuitable for use in a computational treatment.
This is facilitated by the power series expansion of $u_\nu(\phi)$ in the small $\phi$ region given
in \cite{hepth9308005},
\be 
\label{usmallphi} u_\nu\left(\phi\right) = \frac{e^{\frac{i\pi \nu}{2}}
\Gamma \left(1 + \frac{\nu\left(k_1 - 1\right)}{2}\right)}{2\Gamma\left(-\nu\right)} \sum_{m=0}^\infty
\frac{e^{i \pi m/2} \Gamma\left(\frac{m-\nu}{2}\right) \left(2\phi\right)^m}{m! \Gamma\left(1 - \frac{m - \nu
k_1}{2}\right)} .
\ee

Let us briefly discuss the
region of convergence of equations (\ref{smallpsi}),
(\ref{smallpsiu}), and (\ref{usmallphi}), being specific to the 
particular Calabi-Yau $\mbb{P}^4_{[1,1,2,2,6]}$. The Calabi-Yau develops a
conifold singularity when $\phi + 864\psi^6 = \pm 1$ and there is a
further singularity at $\phi
= \pm 1$. The singularities determine the regions of convergence; all three equations are only valid for
$|\frac{864 \psi^6}{\phi \pm 1} | < 1$ and equation (\ref{usmallphi}) has the additional restriction $|\phi| <
1$.

We will also be interested in the periods near the conifold locus. As will be further discussed in section
\ref{coniStatistics}, the periods here take a certain standard form. However, for their exact numerical
determination, we will use a neolithic approach, parametrising them
through direct evaluation of the (implicit) power series (\ref{pidef}) near the
conifold locus.

Finally, the Calabi-Yau we use is the original manifold $\mathcal{M}$ defined by the locus of
(\ref{manifoldeq}), and not its mirror $\mathcal{W}$. $\mathcal{M}$ has a total of 128 complex structure moduli.
Some will be removed by the orientifold projection, but there are
still many not considered directly.
The validity of this was explained in \cite{hepth0312104}. The group $G$ is a symmetry of (\ref{manifoldeq}) and
if we only turn on fluxes invariant under this symmetry then the superpotential can only have a higher-order
dependence on the other moduli. It is thus consistent to set all other moduli equal to zero and focus only on
the moduli in equation (\ref{manifoldeq}) and their associated fluxes. We will comment briefly on the general
situation in section \ref{Discussion}.

We will now test the Ashok-Douglas formula in the vicinities of the Landau-Ginzburg point
$\psi = \phi = 0$ and the conifold locus $864 \psi^6 + \phi = 1$. We solve \be \label{P11226equations} D_\tau W
= D_\phi W = D_\psi W =0, \ee and study the distribution of vacua.

\subsection{The Landau-Ginzburg Point $\psi = \phi = 0$}
\label{LGpoint}

In equation
(\ref{pidef})
a symplectic basis for the periods was given.
Let us untangle this in the vicinity of
$\psi = 0$. We can expand $\Pi(\psi, \phi)$ as \be \Pi = \underline{a}(\phi) + \underline{b}(\phi) \psi^2 +
\underline{c}(\phi) \psi^4 + \mathcal{O} \left(\psi^6 \right). \ee Here $\underline{a}, \underline{b}$ and
$\underline{c}$ are vector functions of $\phi$ whose explicit form arises from the combination of equations
(\ref{smallpsi}), (\ref{periods}), (\ref{pidef})
and (\ref{usmallphi}). It can be checked that
$\underline{a}^\dagger \cdot \Sigma \cdot \underline{b} = \underline{a}^\dagger \cdot \Sigma \cdot \underline{c}
= 0$, implying
$$
\Pi^\dagger \cdot \Sigma \cdot \Pi = (\underline{a}^\dagger \cdot \Sigma \cdot \underline{a}) +
(\underline{b}^\dagger \cdot \Sigma \cdot \underline{b}) \psi^2 \bar{\psi}^2 + \mathcal{O}\left(|\psi|^6
\right),
$$
and consequently \bea \mc{K}_\phi(\psi, \phi) & = & - \ln \left(-i \Pi^\dagger \cdot \Sigma
\cdot \Pi \right) \nonumber\\
& = & - \ln \left(-i \underline{a}^\dagger \cdot \Sigma \cdot \underline{a} \right) - \ln \left(1 +
\frac{(\underline{b}^\dagger \cdot \Sigma \cdot \underline{b})}{(\underline{a}^\dagger \cdot \Sigma \cdot
  \underline{a})} \psi^2 \bar{\psi}^2  + \mathcal{O}\left(|\psi|^6\right) \right) \nonumber\\
& = & -\ln\left(-i \underline{a}^\dagger \cdot \Sigma \cdot \underline{a}\right) - \frac{(\underline{b}^\dagger
\cdot \Sigma \cdot \underline{b})}{(\underline{a}^\dagger \cdot \Sigma \cdot
  \underline{a})} \psi^2 \bar{\psi}^2 + \mathcal{O}\left(|\psi|^6\right).
\eea Equations (\ref{P11226equations}) then have the form \bea \label{smallphi1} \frac{1}{\psi} D_\psi W = 0
\Rightarrow & (f-\tau h) \cdot \left( \underline{\alpha}_1 (\phi) + \underline{\alpha}_2 (\phi) \psi^2 +
\underline{\alpha}_3
(\phi) \bar{\psi}^2 \right) & = 0, \\
\label{smallphi2} D_\phi W = 0 \Rightarrow & (f - \tau h) \cdot \left( \underline{\beta}_1 (\phi) +
\underline{\beta}_2 (\phi) \psi^2 \right)
& = 0, \\
\label{smallphi3} D_\tau W = 0 \Rightarrow & (f - \bar{\tau} h) \cdot \left( \underline{a}(\phi) + \underline{b}
(\phi) \psi^2 \right) & = 0, \eea where we have dropped terms of $\mathcal{O}\left(|\psi|^4\right)$. Here
$\underline{\alpha}(\phi)$ and $\underline{\beta}(\phi)$ are complicated functions of $\phi$ depending on the
integral expressions for $u_\nu(\phi)$. However, when $\vert \phi \vert < 1$ the use of the power series
expansion in equation (\ref{usmallphi}) converts $\underline{\alpha}(\phi)$ and $\underline{\beta}(\phi)$ to a
more tractable form. The leading behaviour of the metric $g_{\alpha \bar{\beta}} =
\partial_\alpha \partial_{\bar{\beta}} \mc{K}$ is given by
$$
g_{\psi \bar{\psi}} \sim \psi \bar{\psi}, \qquad g_{\psi \bar{\phi}} \sim \psi \bar{\psi}^2 \bar{\phi}, \qquad
g_{\phi \bar{\psi}} \sim \psi^2 \bar{\psi} \phi, \qquad g_{\phi \bar{\phi}} \sim 1.
$$
This is consistent with expectation, as at the Landau-Ginzburg point
$\psi = \phi = 0$ the moduli space metric becomes
singular. The curvature 2-form $\mathcal{R}$ and the Chern class $c_2$ may be calculated straightforwardly from
the full expressions for the metric using a symbolic algebra program. Evaluating the Ashok-Douglas density, we
find it has leading behaviour
$$
d \mu \sim g_{\tau \bar{\tau}}\, d \tau \wedge d \bar{\tau} \wedge \left( \psi \bar{\psi} d \psi \wedge d
\bar{\psi} \wedge d \phi \wedge d \bar{\phi} \right).
$$
The regions on which we compare the Ashok-Douglas formula to our empirical results are balls in $\psi$ and
$\phi$ space. We then expect as leading behaviour \bea N(\textrm{vacua s.t.}  \quad \vert \psi \vert <
r_1) & \sim & r_1^4, \nonumber\\
N(\textrm{vacua s.t.}  \quad \vert \phi \vert < r_2) & \sim &
r_2^2. \nonumber \eea
However, in the actual plots we evaluated the Ashok-Douglas density numerically rather than
simply using the leading behaviour.

To test this expectation, we generated random fluxes and sought solutions of equations (\ref{smallphi1}) to
(\ref{smallphi3}) using a numerical root finder. The range of fluxes used was (-20, 20). This is not as large as
one might prefer. However, a larger range of fluxes resulted in solutions being produced insufficiently rapidly
for our purposes. In order to be able to trust our truncation of the power series, we only kept solutions
satisfying \be \label{solnconds} \left\vert \frac{864 \psi^6}{\phi \pm 1} \right \vert < 0.5 \textrm{
  and  } \vert \phi \vert < 0.75 .
\ee 
While processing the numerical results, there is an important subtlety
that must be accounted for\footnote{We thank S. Kachru for
  bringing this to our attention.}. It is well known that there is an exact
  $SL(2,\mathbb{Z})$ symmetry of type IIB string theory,
$$
\tau \to \frac{a \tau + b}{c \tau + d}, \qquad \left(
\begin{array}{c} F_3 \\ H_3 \end{array} \right) \to \left(
\begin{array}{cc} a & b \\ c & d \end{array} \right) \left(
\begin{array}{c} F_3 \\ H_3 \end{array} \right).
$$
where $a,b,c,d \in \mathbb{Z}$ and $ad - bc = 1$. Thus each vacuum found has many physically equivalent $SL(2,
\mathbb{Z})$ copies that we should not double-count. One way to deal with this would be to fix the gauge
explicitly and then perform the Monte-Carlo analysis. Our approach is instead to weight each vacuum by the
inverse of the number of copies it has within the sampled flux range. The purpose of this is to ensure that
vacua with many $SL(2, \mbb{Z})$ copies are not unduly preferred.

As well as the $SL(2, \mbb{Z})$ symmetry, there is a monodromy near the Landau-Ginzburg point that needs similar
treatment. It follows from the definition of the periods
(\ref{periods}) that they have a monodromy under
$(\psi, \phi) \to (\alpha \psi, - \phi)$, where $\alpha^{12} = 1$, of
$$
\varpi(\psi, \phi) \to a \cdot \varpi(\psi, \phi),
$$
where
$$
a = \left( \begin{array}{cccccc} 0 & 1 & 0 & 0 & 0 & 0 \\ 0 & 0 & 1 &
  0 & 0 & 0 \\ 0 & 0 & 0 & 1 & 0 & 0 \\ 0 & 0 & 0 & 0 & 1 & 0 \\
0 & 0 & 0 & 0 & 0 & 1 \\ -1 & 0 & 0 & 0 & 0 & 0 \end{array} \right).
$$ In the symplectic basis, the monodromy matrix $A$ is given by $m
\cdot a \cdot m^{-1}$. The effect of this monodromy is to generate a family of physically equivalent solutions
related by \bea
(\psi, \phi) & \to & (\alpha^{-1} \psi, - \phi), \nonumber \\
f & \to & f \cdot A, \nonumber \\
h & \to & h \cdot A. \eea When weighting vacua we need to find the
total number of copies due to symmetries and monodromies
lying within the
sampled flux range. 
This has important systematic effects as vacua with
smaller values of $f_i$ and $h_i$, and thus smaller values of $N_{flux}$, have more copies. A naive counting
that neglects the symmetries or monodromies that are present places undue emphasis on vacua with smaller
values of $N_{flux}$.

\begin{figure}[ht]
\linespread{0.2}
\begin{center}
\makebox[10cm]{ \epsfxsize=10cm \epsfysize=7cm \epsfbox{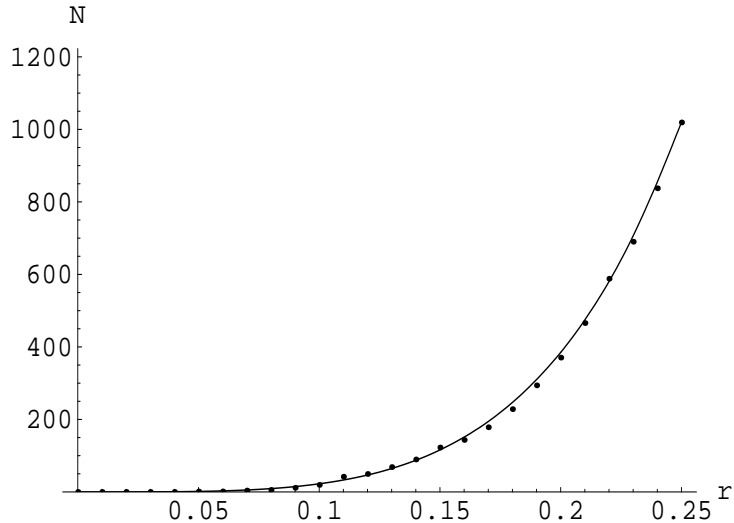}}
\end{center}
\caption{ Number of vacua with $\vert \psi \vert < r$. The value of $N$ plotted includes a weighting due to the
$SL(2,\mathbb{Z})$ copies of
  each vacuum within the range of fluxes sampled. The dots represent
  the numerical results, and the continuous line the
  (rescaled) numerical integration of $\int_{\vert \psi \vert
  < r} d \mu$, where $d \mu$ is the index density. The flux range used
  was (-20,20).}
\label{psismallresults}
\end{figure}

\begin{figure}[ht]
\linespread{0.2}
\begin{center}
\makebox[10cm]{ \epsfxsize=10cm \epsfysize=7cm \epsfbox{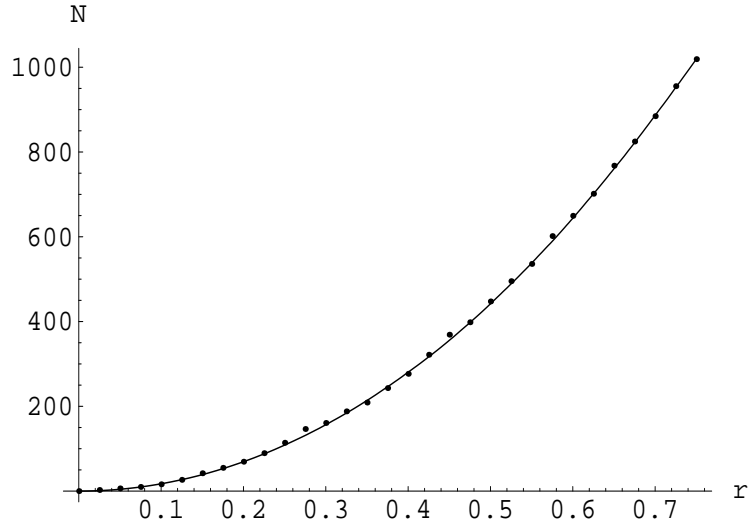}}
\end{center}
\caption{The weighted number of vacua, $N$, with $\vert \phi \vert <  r$. The dots represent the numerical
results and the continuous line the
  numerical integration of $\int_{\vert \phi \vert < r} d
  \mu$, rescaled by the same factor as in diagram \ref{psismallresults}.}
\label{phismallresults}
\end{figure}

\begin{figure}[ht]
\linespread{0.2}
\begin{center}
\makebox[10cm]{ \epsfxsize=10cm \epsfysize=7cm \epsfbox{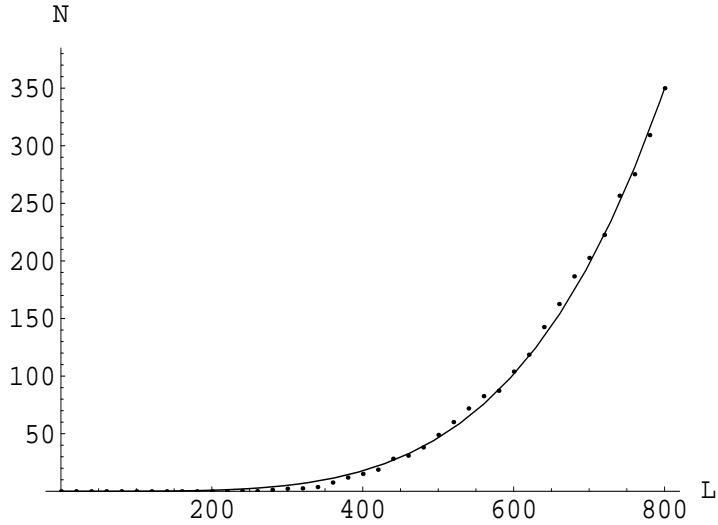}}
\end{center}
\caption{The weighted number of vacua $N$ with
  $N_{flux} < L$. $4.5 \times 10^6$ sets of
  flxues were generated, with values of $L$ equally distributed
  between $0$ and $800$ and the range of fluxes being $(-20,20)$. The
  results are fit by $N \sim L^{4.3}$.}
\label{Ldependence}
\end{figure}

We examined the distribution of vacua within fixed balls in $\psi$ and $\phi$ space. In figure
\ref{psismallresults} we plot the number of vacua satisfying (\ref{solnconds}) and having $| \psi | < r$. The
results are seen to agree well with the theoretical prediction.
Likewise, figure \ref{phismallresults} shows the distribution of vacua for a ball $\vert \phi \vert < r$ in
$\phi$ space. The continuous line again represents the cumulative number of vacua and the dots the rescaled
numerical integration of  $\int_{\vert \phi \vert
  < r} d \mu$. The empirical results are
again well captured by the theoretical prediction.
Finally, in figure \ref{Ldependence} we examine the dependence of the number of vacua on the distance in flux
space $N_{\mathrm{flux}} = f^T \cdot \Sigma \cdot h$. The graph is fit by $N \sim L^{4.3}$. This is surprising,
as the expected scaling is $L^6$. Furthermore, in the vicinity of the Landau-Ginzburg point for an analogous
one-modulus example, the correct $L^4$ scaling is found \cite{hepth0411061}. As discussed further in section
\ref{Discussion}, we believe our results are an artifact of the small flux range used, and that were a larger
flux range used we would obtain the correct scaling. As we will shortly describe, in the vicinity of the
conifold locus we do obtain the expected scaling with a flux range of $(-40,40)$.

\begin{figure}[ht]
\linespread{0.2}
\begin{center}
\makebox[8cm]{ \epsfxsize=12cm \epsfysize=7cm \epsfbox{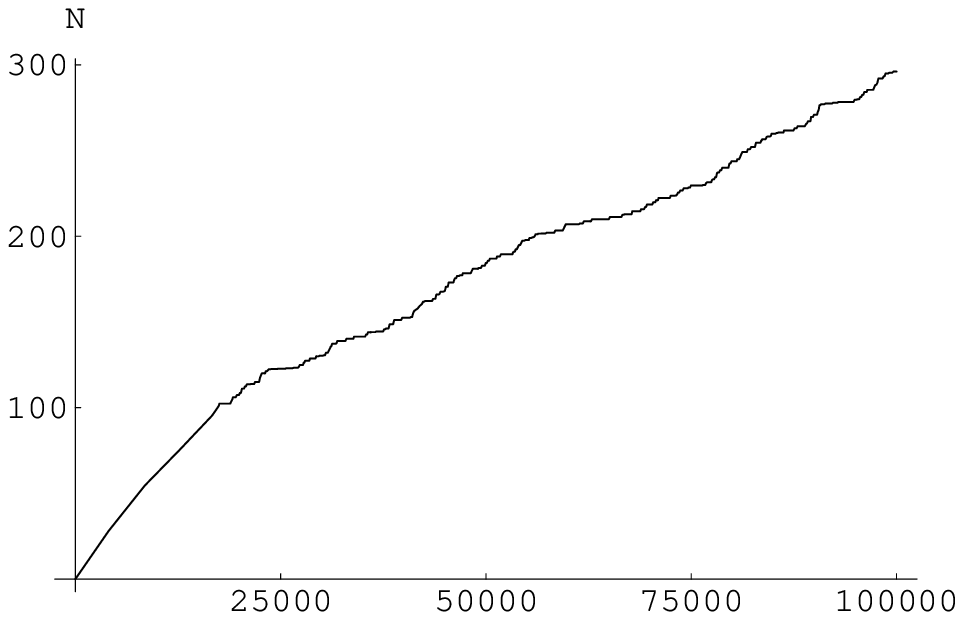}}
\end{center}
\caption{The value of $(2 \pi \sqrt{\alpha'})^4 e^{\mc{K}} |W|^2$ in units of $(\alpha')^2$
  for vacua in the vicinity of $\psi = \phi = 0$, for
  $(2 \pi)^4 e^{\mc{K}} |W|^2 < 100000$. The flux range was (-20, 20), and the
  vacua satisfied the conditions (\ref{solnconds}).}
\label{mainSusyScale}
\end{figure}

It is also of interest to study the supersymmetry breaking scale, as
measured by $(2 \pi \sqrt{\alpha'})^4 e^{\mc{K}} \vert
W \vert^2$, for vacua in the vicinity of $\psi = \phi = 0$. This is shown in figure \ref{mainSusyScale}. The
most noticeable thing about this graph is that this distribution is uniform near the
origin.

\subsection{The Conifold Locus $864 \psi^6 + \phi =1 $}
\label{coniStatistics}

The Calabi-Yau $\mathcal{M}$ has a codimension one conifold degeneration along the moduli space locus $864\psi^6
+ \phi = 1$. As the moduli space curvature diverges near a conifold
point, the expectation from (\ref{ADdensity}) is that vacua should
cluster near the conifold locus. A conifold singularity is locally a
cone over $S^2 \ti S^3$. At the singularity, both the $S^2$ and $S^3$
cycles collapse and the (spatial) curvature diverges.\footnote{We
  note that two entirely separate curvatures are
  diverging. At the conifold locus in moduli space, the moduli space
  curvature diverges. At this same locus, a point-like conifold
  singularity appears in the Calabi-Yau, at which the spatial
  curvature of the Calabi-Yau also diverges.} The periods measure the
cycle of a size, and so as we approach the locus in moduli space
at which the cycle collapses, the period of the collapsing cycle must
vanish. Indeed the Calabi-Yau periods take a certain standard form in
the vicinity of a conifold singularity in moduli space. We denote the
period of the collapsing $S^3$ cycle by $\mc{G}_1$ and the period of its dual cycle 
by $z_1$. If $\epsilon$ measures the moduli space distance from the
singular locus, we have
\bea
\mc{G}_1 & = & \epsilon, \nonumber \\
z_1 & = & - \frac{1}{2 \pi i} \mc{G}_1 \ln \left( \mc{G}_1 \right) +
\textrm{ analytic},
\eea
with all periods entirely regular in the vicinity of the singularity.

To study these vacua numerically, we must restrict attention to a small region near the conifold locus where we
can compute the periods explicitly. We take this region to be the neighbourhood of the point $\phi = 0,  \psi =
\psi_0 = 864^{-\frac{1}{6}}$. If we write $\psi = \psi_0 + \xi$ and truncate the periods at first order in $\xi$
and $\phi$, then $\Pi(\xi, \phi) = (\mathcal{G}_1, \mathcal{G}_2, \mathcal{G}_3, z_1, z_2, z_3)$ is approximated
by \bea
\mathcal{G}_1 & = & 3202\xi + 171.8 \phi + \mathcal{O}(\xi^2, \phi^2, \xi \phi) \nonumber \\
\mathcal{G}_2 & = & 4323 - i(1553\xi + 107.4 \phi) + \mathcal{O}(\xi^2, \phi^2, \xi \phi)\nonumber \\
\mathcal{G}_3 & = &  (-492.7 + 1976.8i) + (371.0 - 300.2i)\xi + (-259.0
-59.0i)\phi  + \mathcal{O}(\xi^2, \phi^2, \xi \phi) \nonumber \\
z_1 & = &  \frac{-1}{2\pi i}\mathcal{G}_1 \ln \left(\mathcal{G}_1 \right) + 784.8i -
  2306i\xi -44.35i\phi  + \mathcal{O}(\xi^2, \phi^2, \xi \phi) \nonumber \\
z_2 & = & (-994.6 - 184.8i)  + (861.9 + 476.5i)\xi  + (9.91 -
      112.7i)\phi  + \mathcal{O}(\xi^2, \phi^2, \xi \phi) \nonumber \\
z_3 & = &  i(369.5 - 953.0\xi + 225.4\phi) + \mathcal{O}(\xi^2, \phi^2, \xi \phi) \eea The numerical values were
found by evaluating the series (\ref{smallpsi}) up to one hundred terms in $\psi$ and twenty-five terms in
$\phi$. The values for the coefficients of the $\mathcal{O}(\xi, \phi)$ terms were sensitive to the number of
terms used in the power series at the level of a couple of percentage points. We did not keep terms quadratic in
$\xi$ and $\phi$ - inclusion of these would lead to $\mathcal{O}(\xi)$ corrections to the results below.

The general form of the periods is as anticipated above. 
The cycles corresponding to $\mathcal{G}_2, \mathcal{G}_3, z_2$ and
$z_3$ are all remote from the conifold singularity and the associated
periods are both regular and
non-vanishing near the conifold degeneration. 
The cycle corresponding to $\mathcal{G}_1$ is the $S^3$
 that shrinks to zero size along the conifold locus.
 The period along this cycle is regular near the conifold locus and
 vanishing along it. Finally, the cycle corresponding to $z_1$ is that
 dual to the $S^3$. It is not uniquely defined and in
 fact has a
monodromy under a loop in moduli space around the conifold locus. As
above, its period takes the form $\left( -\frac{1}{2 \pi
i}\mathcal{G}_1 \ln (\mathcal{G}_1) + \textrm{analytic terms} \right)$.

It is convenient to define $Z = \left( \frac{3202}{171.8} \right) \xi + \phi$ and rewrite the periods as \bea
\mathcal{G}_1 & = & 171.8 \, Z, \nonumber \\
\mathcal{G}_2 & = & 4323 + 107.4 i \, Z -3554 i \, \xi, \nonumber \\
\mathcal{G}_3 & = & 784.8 - (259 + 59i)\, Z + (5198 + 799 i) \, \xi, \nonumber \\
z_1 & = &  \frac{-1}{2\pi i}171.8 \, Z \ln \left( Z \right) + 784.8
-44.4i \, Z -1479 i \, \xi, \nonumber \\
z_2 & = & (-994.6 - 184.8i) + (9.91 - 112.7i) \, Z + (677 + 2577 i)\, \xi, \nonumber \\
z_3 & = & i(369.5 + 225.4 \, Z -5154 \, \xi). \label{zPeriods} \eea

Z is a measure of the distance from the conifold locus. We are interested in vacua extremely close to the
conifold locus - typically $\ln{|Z|} < -5$ - and thus we will regard $ |Z| \ll |\xi| \ll 1$. Having set up the
periods (\ref{zPeriods}), we can now compute the Ashok-Douglas expectation for the index density and compare
this with numerical results.

Let us first solve equations (\ref{P11226equations}). We have, 
\be 
\label{taustabilisation} 
D_{\tau} W = 0
\Rightarrow  (f - \bar{\tau}h)\cdot \Pi  =  0
\Rightarrow \tau = \frac{f \cdot \Pi^\dagger}{h \cdot \Pi^\dagger} .\\
\ee This can be written as \be \label{tauvalue} \tau = \frac{a_0 + a_1 \bar{\xi}}{b_0 + b_1 \bar{\xi}} +
\mathcal{O}(Z \ln Z), \ee where $a_i$ and $b_i$ are flux-dependent quantities. Next, \be \label{xivalue} D_\xi W
= 0 \Rightarrow (f - \tau h)\cdot (c_0 + c_1 \xi + c_2 \bar{\xi} + \mathcal{O}(\xi^2)) = 0. \ee Using
(\ref{tauvalue}) this becomes a linear equation for $\xi$, easily solved to determine $\xi$ and $\tau$. We
finally need the value of $\ln Z$. This is obtained by considering \be \label{zstabilisation} D_Z W  = 0
\Rightarrow (f_4 - \tau h_4)\ln Z = (d_0 +d_1 \tau)+ (d_2 + d_3 \tau) \xi + (d_4 + d_5 \tau) \bar{\xi} \ee
Substituting in for $\tau$ and $\xi$ from (\ref{tauvalue}) and (\ref{xivalue}) gives the value of $\ln Z$.

When analysing the results we must account for the $SL(2, \mathbb{Z})$ copies discussed in section
\ref{LGpoint}. There is also a monodromy near the conifold. When solving for $\ln Z$, we impose no restriction
on the imaginary part of $\ln Z$. There is a monodromy \bea
\ln Z & \to & \ln Z + 2 \pi i, \nonumber \\
(f_1, f_2, f_3, f_4, f_5, f_6) & \to & (f_1 + f_4, f_2, f_3, f_4, f_5,
f_6), \nonumber \\
(h_1, h_2, h_3, h_4, h_5, h_6) & \to & (h_1 + h_4, h_2, h_3, h_4, h_5, h_6). \eea corresponding to a loop in
moduli space around the conifold locus. This gives a further source of physically equivalent solutions that
should not be double-counted. For each vacuum, we count the total number of copies within the specified
flux range and weight by the inverse of this number.

\begin{figure}[ht]
\linespread{0.2}
\begin{center}
\epsfxsize=0.85\hsize \epsfbox{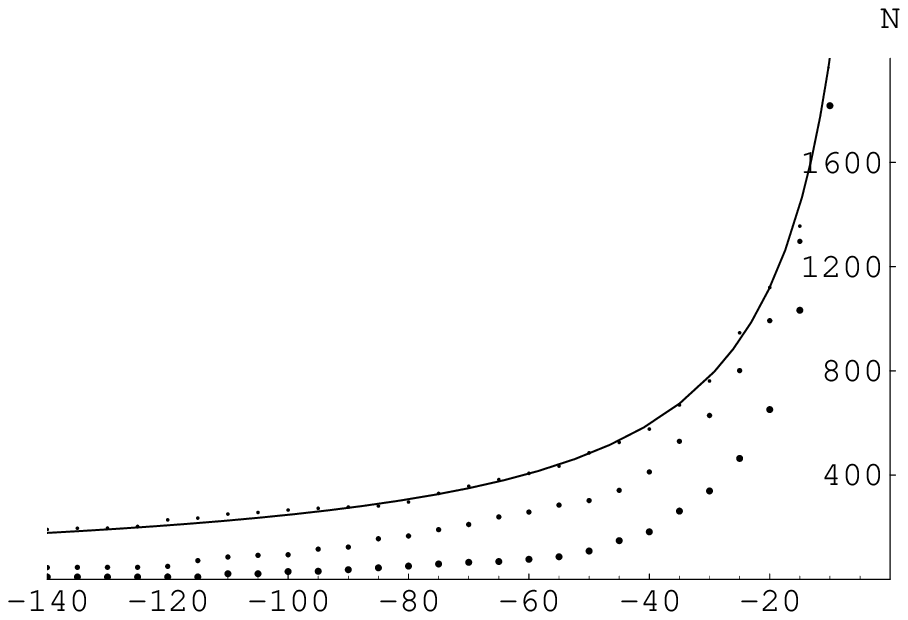}
\end{center}
\caption{The distribution of vacua transverse to the conifold. We plot
  the number of vacua having $\ln Z < D$ for $D \in (-120, -5)$
  against $D$, restricting to $\vert \xi \vert < 0.05$. The dots refer to the
  vacua found numerically and the smooth line to the Ashok-Douglas
  prediction. We include results for three flux ranges - (-20, 20),
  (-30, 30) and (-40, 40). The fit of the results to the expected
  distribution improves markedly as the flux range is increased.}
\label{ConiLogVacuaPlot}
\end{figure}

\begin{figure}[ht]
\begin{center}
\epsfxsize=0.85\hsize \epsfbox{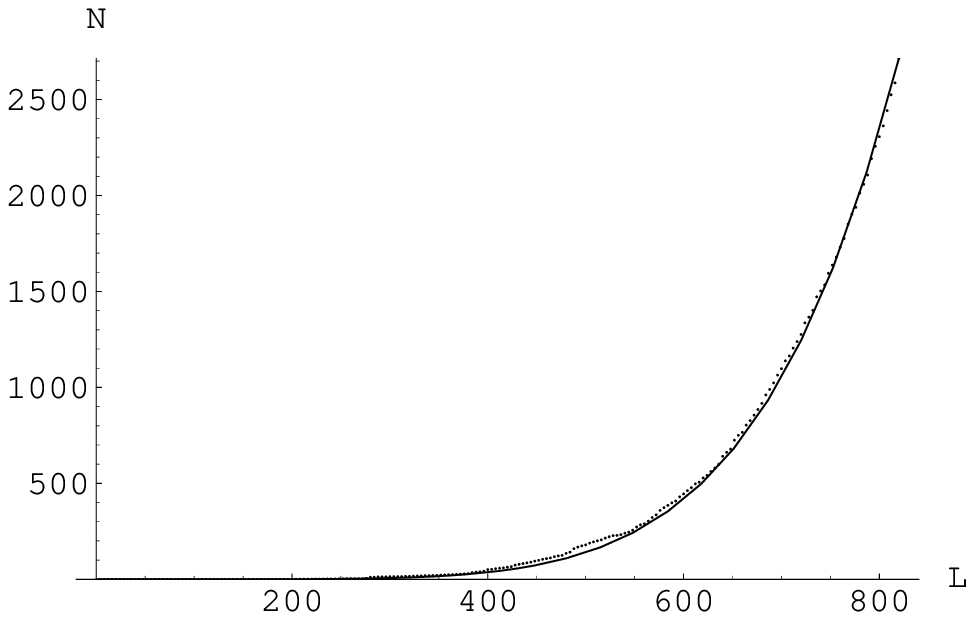}
\end{center}
\caption{The weighted number of vacua with $N_{\mathrm{flux}} < L$. The curve
  is fit well by $N \propto L^{6}$. The range of fluxes used was (-40,40).}
\label{ConiFluxVacuaNumbers}
\end{figure}

\begin{figure}[ht]
\linespread{0.2}
\begin{center}
\epsfxsize=0.8\hsize \epsfbox{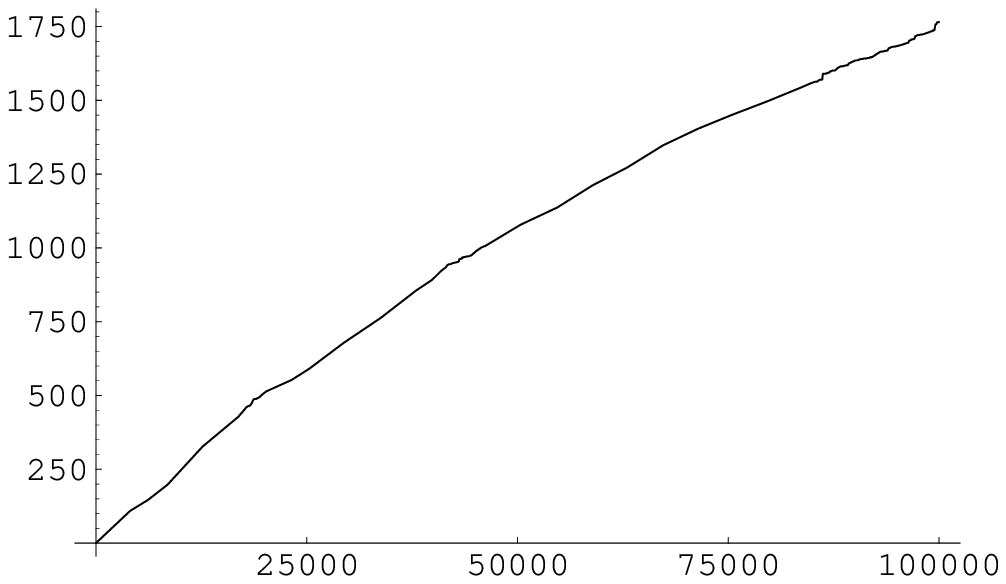}
\end{center}
\caption{The value of $(2 \pi )^4 e^{\mc{K}} |W|^2$ in units of $(\alpha')^2$ for vacua near the conifold, for
  $(2 \pi)^4 e^{\mc{K}} |W|^2 < 100000$. The flux range was (-40, 40) and we
  restricted to vacua satisfying $\vert \xi \vert < 0.05$.}
\label{coniSusyScale}
\end{figure}

Results are shown below. In figure \ref{coniClustering} we show the clustering of vacua by plotting the
distribution of vacua transverse to the conifold locus.
\begin{figure}[ht]
\linespread{0.2}
\begin{center}
\epsfxsize=0.85\hsize \epsfbox{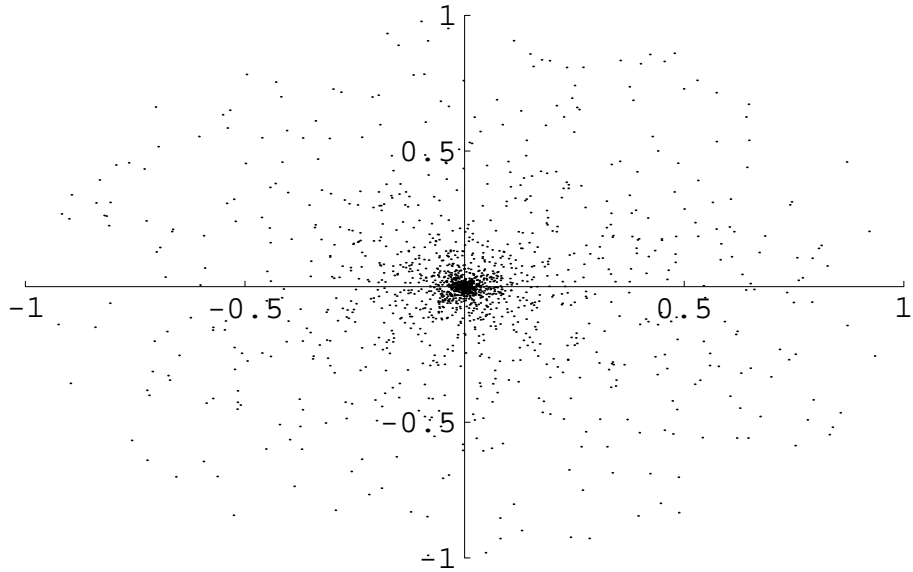}
\end{center}
\caption{The value of $Z$ for vacua near the conifold. We have restricted to
  $|Z| < 0.0001$ and have rescaled the above plot by $10^4$. The flux range used was (-40,40).}
\label{coniClustering}
\end{figure}
A similar plot of vacua parallel to the conifold locus shows no such clustering, indicating that there is no
preferred position along the conifold locus. In figure \ref{ConiLogVacuaPlot} we perform a quantitative
comparison with the expected vacuum density, finding very good agreement over a large range of values of $\ln
Z$.
In figure \ref{ConiFluxVacuaNumbers} we examine the scaling with $L$ of the number of vacua having $N_{flux} <
L$. This reproduces the expected $L^6$ scaling of (\ref{ADdensity}).
We also computed the `susy breaking scale', $(2 \pi \sqrt{\alpha '})^4
e^{\mc{K}} \vert W \vert ^2$. 
We reemphasise that the actual susy breaking scale can only be
computed once the K\"ahler moduli have been fixed.
This distribution 
is plotted in figure \ref{coniSusyScale}, and is again uniform near zero.

\subsection{Results and Limitations}
\label{Discussion}

Let us briefly summarise the results of these tests.
\begin{enumerate}
\item We constructed a large class of flux vacua for the two-moduli
  Calabi-Yau threefold $\mbb{P}^4_{[1,1,2,2,6]}$. We independently
  computed the Ashok-Douglas density and compared with our results.
We find good agreement which improves as the range of fluxes is increased. The number of vacua was limited
mostly by working in two patches in
  moduli space: the region near $\psi=\phi=0$ and the region close to
  the conifold singularity $\phi+864 \psi^6=\pm 1$.
\item We found a large concentration of vacua close to the conifold
  singularity as predicted, with the detailed distribution of the vacua being
  in close accordance with expectation.
\item In both regions, the values of $e^{\mc{K}} \vert W_0 \vert^2$
  are uniformly distributed near zero. As a consequence, large values of the
  superpotential, corresponding to a large `no-scale' supersymmetry breaking scale, are
  more abundant than small ones.
\item In both regions the number of models scaled as a power of the maximum permitted value of $N_{flux}$,
$L_{max}$. In the vicinity of the conifold we reproduced the expected $L^6$ scaling. In the region close to the
Landau-Ginzburg point we only found a scaling of $L^{4.3}$. We attribute the failure to achieve the expected
scaling in this region  to the smaller range of fluxes used there.
\end{enumerate}

An important feature of our analysis is that the flux range used has a significant effect on the results. If the
flux range is insufficiently large, then the distribution of vacua found numerically will not fit with the
theoretical density. This is most strikingly illustrated for the case of the conifold in figure
\ref{ConiLogVacuaPlot}. Similar behaviour was seen for the scaling of the number of vacua with $L$ - in the
vicinity of the conifold locus, a reduction of the flux range to (-20,20) caused a reduction in the power of the
scaling from $\approx 6$ to $\approx 5$. Given this, we believe that the scaling of $N(\textrm{vacua } \vert
N_{flux} < L) \sim L^{4.3}$ found in the vicinity of the Landau-Ginzburg point is simply an artefact of the flux
range used.

\begin{figure}[ht]
\label{dilatonPlot} \linespread{0.2}
\begin{center}
\epsfxsize=0.8  \hsize \epsfbox{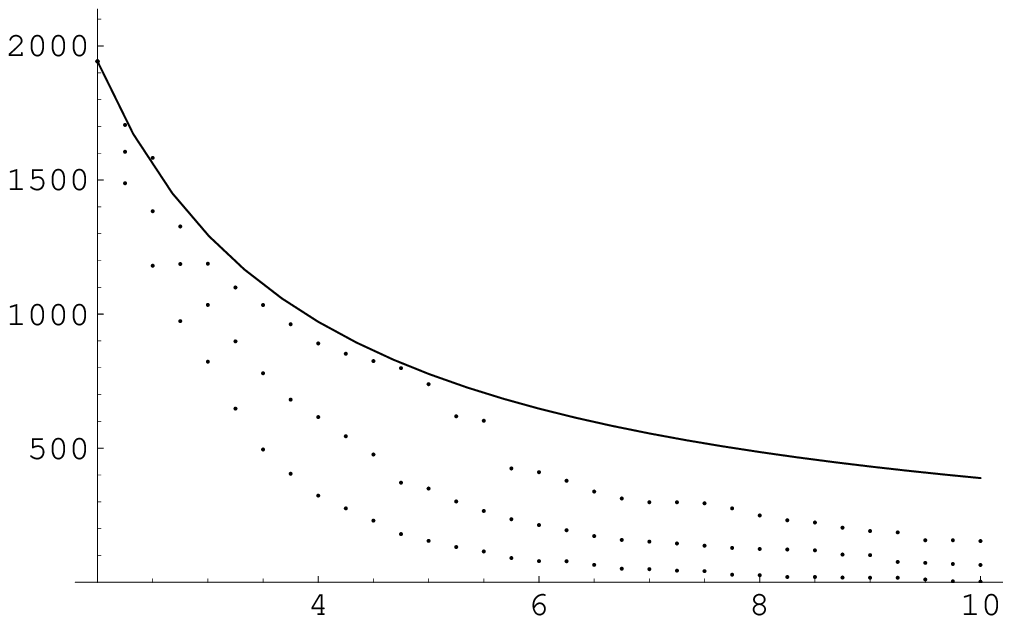}
\end{center}
\caption{Distribution of vacua near the conifold locus with $\rm{Im}(\tau) > y$ for $y \in
  (2,10)$. Results have been brought to the same scale and
plotted for three separate flux ranges:  (-20,20), (-40,40) and
  (-60,60). We see that as the flux range increases the empirical plot
  moves closer to the expected result (represented by a smooth line).}
\end{figure}

We find a similar dependence on the range of fluxes used in the distribution of the dilaton. After being
transformed to the $SL(2, \mbb{Z})$ fundamental region, the expectation from equation (\ref{indexDensity}) is
that the number of vacua with $\textrm{Im}(\tau) > \tau_0$ should
scale like $\frac{1}{\tau_0}$. In figure \ref{dilatonPlot} 
we compare this with numerical results arising from using flux ranges $(-20,20)$, $(-40,40)$ and $(-60,60)$.

We see that as the flux range increases the empirical distribution tends towards the theoretical one, and also
that even with a flux range of $(-60,60)$ the two distributions do not yet fully match. In general terms these
results are reassuring in the sense that arbitrarily large values of the dilaton can be obtained, consistent
with the weak coupling aproximation, even though they are not
statistically preferred. 
We have only explored a small set of the flux vacua on the given
Calabi-Yau. Nonetheless, the results are consistent with expectations and
support the notion that relatively simple formulae can capture the
distribution of flux vacua.

Let us finally make a general observation unrelated to the particular
model above. 
The Calabi-Yau considered above has $h^{1,1} = 2$ and $h^{2,1} = 128$.
We have turned on fluxes only along cycles corresponding to two of the 128 complex structure moduli. We
expect that some of the remaining moduli should be frozen out by the orientifold symmetry, but it would
still be obviously impractical to attempt either to write down or to solve the moduli stabilisation equations
with all fluxes turned on. However, if we assume that the Ashok-Douglas density remains valid then we can say
something about the generic situation. Suppose we have K cycles supporting flux and that - as holds for this and
many other F-theory models - \be N_{D3} + N_{flux} = L_{max} \sim 1000. \ee The Ashok-Douglas density
(\ref{ADdensity}) tells us that \be N(\textrm{vacua } \vert  N_{flux} < L_*) \sim L_*^K. \ee The fraction of
vacua having $N_{D3} \ge n$ is then estimated by \be \frac{N(\textrm{vacua } \vert  N_{flux} \le L_{max}
-n)}{N(\textrm{vacua
  } \vert  N_{flux} \le L_{max})} =
\frac{(L_{max} - n)^K}{L_{max}^K}. \ee For $K \sim 200$, then for $n=5$ this is approximately $\frac{1}{e}
\approx 0.36$. We then see that despite the large amount of $D3$-brane charge we have to play with, generic
choices of flux come close to saturating the D3-brane tadpole (a
similar observation is made in \cite{hepth0409218}).
  This is appealing
if the standard model were to live on D3 branes as in the
models of \cite{hepth0005067, hepth9909172, hepph0001083, hepth0312051}.
Clearly however actual numbers depend on the particular Calabi-Yau,
 the number of 3-cycles surviving the orientifold projection and
the modifications to the Ashok-Douglas density when $L \approx (\textrm{a
  few})K$. This also requires the Standard Model to be realised on D3
branes: we will argue in chapter \ref{ChapterAxions}
that to solve the strong CP problem it is instead preferable to realise the
Standard Model on D7-branes.

\chapter{K\"ahler Moduli Stabilisation}
\label{ChapterKahlerModuli}

This chapter is based on aspects of the paper \cite{hepth0505076}.

\section{Recall of Definitions}

We have discussed above techniques to
stabilise complex structure moduli and described how statistical methods may be
used to characterise the
resulting solutions.
However, these results are incomplete as it is necessary to stabilise all the
compactification moduli and in particular the K\"ahler moduli.
This problem is arguably more important, as these
determine the overall size of the internal space, which in turn
determines the ratio of the string and Planck scales.
The value of the string scale is the most basic question in any
attempt to do string phenomenology.
In addition, the K\"ahler moduli also determine the coupling strength for
gauge groups living on D7 branes.

As the K\"ahler moduli determine the size of the extra dimensions,
there do exist limited direct experimental bounds. 
In the braneworld scenarios relevant here, the extra-dimensional radii
must satisfy 
$r \lesssim 10^{-5}$m \cite{hepph0405262}. Expressed as a mass, this also bounds 
the lightest moduli to have masses $m \gtrsim 10^{-4}$ eV. 
While rather weak,
these bounds should be kept in mind - in particular, the bound on moduli masses
will be relevant later.

We want to discuss the moduli in an $\mc{N}=1$ supergravity context.
Therefore let us recall from section \ref{seCCS} the chiral
coordinates for the K\"ahler moduli in $D3/D7$
IIB orientifold compactifications. We shall for simplicity assume
the absence of 2-form fluxes on D7-branes and that no
wandering D3 branes are present in the compactification; this latter can be
achieved by taking the fluxes to saturate the $C_4$ tadpole. We do not
expect these assumptions to affect the physics substantially.

The K\"ahler moduli are defined by
\be
T_i = \tau_i + i b_i,
\ee
where $\tau_i$ is the Einstein frame 
volume of a 4-cycle $\Sigma_i$, measured in units of
$l_s = (2 \pi) \sqrt{\alpha'}$, and $b_i$ is the component of 
the RR 4-form $C_4$ along this cycle: $\int_{\Sigma_i} C_4 = b_i$. The
4-cycle volumes $\tau_i$ may be related to 2-cycle volumes
$t_i$. If the overall volume is
\be
\mc{V} = \frac{1}{6} \int_{CY} J \wedge J \wedge J = \frac{1}{6} k_{ijk}t^i t^j t^k,
\ee 
the 4-cycle moduli $\tau_i$ are defined by
\be
\label{2to4}
\sigma_i = \partial_i \mc{V} = \half k_{ijk} t^j t^k.
\ee
The K\"ahler moduli do not appear in the tree-level
superpotential. This follows because a tree-level presence of the
K\"ahler moduli would violate the exact axionic shift symmetry $b_i \to b_i
+ 2 \pi$.
The K\"ahler potential is given by \cite{hepth0403067}
\be
\mc{K} = - 2 \ln (\mc{V}) - \ln \left( -i \int \Omega \wedge
\bar{\Omega} \right) -
\ln (S + \bar{S}).
\ee
This can be derived by dimensional reduction.
$\mc{V}$ should be understood as an implicit function of the K\"ahler
moduli, as the relations (\ref{2to4}) cannot generally be inverted.
This K\"ahler potential is no-scale and satisfies
\be
\sum \mc{K}^{i \bar{j}} \partial_{i} \mc{K} \partial_{\bar{j}} \mc{K} = 3,
\ee
where the sum is over K\"ahler moduli.

There are two important results concerning the generation of
potentials for the K\"ahler moduli. Although simple in themselves, these
have profound consequences. 
\begin{enumerate}

\item At tree level, the
  K\"ahler moduli are unfixed and do not appear in the superpotential.

\item In supersymmetric field theories, the superpotential is not
  renormalised at any order in perturbation theory.

\end{enumerate}
The immediate consequence is that the K\"ahler moduli can only 
appear nonperturbatively in the superpotential. 

In general, such
nonperturbative effects both can and do appear. Terms of the form $e^{-aT}$ are
manifestly 
compatible with the shift symmetry given above. There are two
known ways for the K\"ahler moduli to appear nonperturbatively in the
superpotential.
The first is through Euclidean D3-brane instantons\cite{hepth9604030}. In type IIB language, this
corresponds to a Euclidean D3 brane wrapped on a 4-cycle in the
Calabi-Yau. The instanton action is 
\be
S_{D3} = - \frac{1}{(2 \pi)^3 \alpha'^2} \int \sqrt{g} + \frac{i}{(2
  \pi^3) \alpha'^2}  \int C_4.
\ee
A BPS brane is calibrated with respect to $e^J$.
This implies that for a BPS instanton on a cycle $\Sigma_i$,
\be
\frac{1}{l_s^4} \int_{\Sigma_i} \sqrt{g} = \frac{1}{2} \int_{\Sigma_i}
J \wedge J,
\ee
and the resulting instanton action is holomorphic in the 
K\"ahler moduli.  An expansion about an instanton background gives a contribution to the
path integral weighted as $e^{-S_{D3}} = e^{-2 \pi T}$. To contribute to
the superpotential, the instanton must have exactly two fermionic zero
modes, in order that the integral over collective coordinates
generates a term
$$
\int d^4 x d^2 \theta ( \ldots ).
$$
For the brane instantons relevant here, there are
topological restrictions on when this can occur. In the absence of
flux, a necessary condition is that \cite{hepth9604030}
\be
\chi_g(D) = \sum_{i=0}^3 (-1)^i h^{i,0}(D) = 1,
\ee
where $D$ is the (effective) divisor wrapped by the brane.
In the presence of 3-form flux, this topological constraint can be 
 relaxed \cite{hepth0407130, hepth0501081, hepth0503072, hepth0503125, hepth0503138, hepth0507069}.

The other source of nonperturbative contributions is gaugino
condensation on wrapped D7-branes. Expanding the DBI and Chern-Simons terms, 
the D7-brane worldvolume action is
\bea
-\frac{1}{(2 \pi)^7 \alpha'^4} \int_{\mbb{R}^4 \times \Sigma}
e^{-\phi} \sqrt{g}
- \frac{1}{4 (2 \pi)^5 \alpha'^2} \int_{\mbb{R}^4 \times \Sigma}
\sqrt{g} e^{-\phi} F_{\mu \nu} F^{\mu \nu} 
& +  & \frac{i}{(2 \pi)^7 \alpha'^4} \int_{\mbb{R}^4 \times \Sigma} C_8 + \nonumber \\
 + \frac{i}{2 (2 \pi)^5 \alpha'^2} \int_{\mbb{R}^4 \times \Sigma} F
\wedge F \wedge C_4. & & 
\eea
By examining the 4-dimensional $F^2$ terms we see that the holomorphic gauge
kinetic function is 
$$
f_i = \frac{T_i}{2 \pi},
$$
with $\frac{1}{g^2} = \hbox{Re}(f)$. 
Gaugino condensation on such branes can produce effects
described by a nonperturbative superpotential, which appears at order
$e^{-\frac{4 \pi^2}{N g_{YM}^2}} \sim e^{-\frac{2 \pi T}{N}}$, where $N$ is the
rank of the gauge group.

\section{The KKLT Scenario}

The gist of the above discussion is that the K\"ahler moduli are not
stabilised at tree level and only appear nonperturbatively in the superpotential.
This is an apposite place to introduce an important scenario that is in
many ways a reference point for work on moduli stabilisation. This is
the now-famous paper
\cite{hepth0301240} by Kachru, Kallosh, Linde and Trivedi (KKLT), 
outlining a procedure to construct de Sitter vacua by first
stabilising K\"ahler moduli to create a supersymmetric AdS vacuum, and subsequently  
introducing effects to uplift this to a de Sitter vacuum. 

This approach starts by observing that in the flux compactifications of
\cite{hepth0105097} reviewed in chapter \ref{ChapterModuliAndFluxes}, 
the complex structure moduli receive tree-level masses of $\mc{O}(m_s/\sqrt{\mc{V}})$,
whereas the K\"ahler moduli are massless. It should therefore be possible to integrate out
the complex structure moduli by setting them equal to their vevs, and
subsequently to consider a low energy effective
field theory for the K\"ahler moduli.

If we assume there to be only one K\"ahler modulus $T$, 
the superpotential for the K\"ahler moduli effective theory is
\be
W = W_0 + A e^{-aT}.
\ee
Here $W_0 = \langle \int G_3 \wedge \Omega \rangle$ 
is the (now constant) tree-level flux superpotential, $a = \frac{2 \pi}{N}$ is the coefficient of
the exponent, and $A$ represents threshold effects. As $A$ depends
only on complex structure moduli it is here regarded as a constant.

The K\"ahler potential remains no-scale,
\be
\label{Kintegrateout}
\mc{K} = \mc{K}_{cs} -3 \log (T + \bar{T}).
\ee
$\mc{K}_{cs}$ is a constant from integrating out the dilaton and complex structure moduli. With these potentials,
the scalar potential can be extremised by solving
\be
D_T W \equiv \partial_T W + (\partial_T \mc{K}) W = 0.
\ee
This equation gives
\be
-a A e^{-a T} - \frac{3}{T + \bar{T}} \left(W_0 + A e^{-a T} \right) =
0.
\ee
Solving this stabilises the K\"ahler modulus at $\hbox{Re}(T) \sim \frac{1}{a} \log W_0$.
As this is a supersymmetric AdS minimum, the vacuum energy is given by
\be
V_{AdS} = -3 e^{\mc{K}} \vert W \vert^2.
\ee

The K\"ahler potential (\ref{Kintegrateout}) will receive perturbative
$\alpha'$ and $g_s$ corrections.
In order to suppress these, it is necessary that the $T$ modulus be stabilised at large
values, which in turn requires that $W_0$ be sufficiently small. However,
as $T$ depends logarithmically on $W_0$, rather small values of $W_0$ are necessary to obtain large values of
$T$. The value of $W_0$ arises from the possible choices of integral fluxes. It is argued that, because
of the vast range of possible flux choices, some will generate the
necessary small 
values of $W_0$. The large discretuum of flux choices is thus essential to
the KKLT construction, as it is the source of the claim that the
necessary small values of $W_0$ are possible.

As discussed in chapter \ref{StatisticsReview}, this last statement was quantified in
\cite{hepth0404116, hepth0411183}. Specifically, if we adopt a measure
in which each flux choice receives equal weight, then 
\be
N(e^{\mc{K}_{cs}} \vert W_0 \vert^2 < \epsilon) \sim \epsilon
\ee
for small $W_0$. $W_0$ is typically $\mc{O}(1)$ and thus values for $W_0$ of $\mc{O}(10^{-5})$ require a
tuning of discrete fluxes of $\mc{O}(10^{-10})$. Given the large
numbers of discrete flux choices, there will be however, in absolute terms, many such choices
that realise these values. 

To uplift to a dS solution and break supersymmetry, \cite{hepth0301240} propose to
add an anti-D3 brane living at the bottom of a warped 
throat. The warping red-shifts the energy allowing it to be tuned to
sufficient precision
to uplift to de Sitter space. There also exist other proposals
for the uplift \cite{hepth0309187, hepth0402135}.
However it is generally agreed that this step is the least tractable
part of the construction and I shall not discuss it significantly.

\section{Perturbative  Effects in $\mc{N}=1$ Supergravity }
\label{Alpha'Section} \linespread{1.3}

Despite the simplicity of the above scenario, some caveats naturally
arise. The most obvious of these is that only nonperturbative
corrections are included in the potential - the perturbative
corrections appearing in the K\"ahler potential are absent. In order
to justify this, it is necessary that the tree level superpotential $W_0$
be extremely small, which necessitates a considerable degree of fine-tuning.

We shall now make this more precise and address the question of when
$\alpha'$ corrections may be neglected in IIB flux
compactifications, obtaining the answer `almost never'. However, it is
helpful to view this as a particular example of more general behaviour.

\subsection*{The General Case}

Let us start with an $\mc{N}=1$ supergravity theory with tree-level K\"ahler potential $\mc{K}$ and tree-level
superpotential $W$. In general $\mc{K}$ receives corrections at every order in perturbation theory, $\mc{K}_p$,
and non-perturbative corrections $\mc{K}_{np}$, whereas $W$ is not renormalised in perturbation theory and only
receives non-perturbative corrections $W_{np}$. Therefore we can write\footnote{The quantities $J$ and $\Omega$
introduced here should
  not be confused with the Calabi-Yau forms.}
\bea
{\K} & = & \K_0\ + \K_p + \K_{np}\approx \K_0+J, \\
W &  = &  W_0 + W_{np} \approx W_0+\Omega, \eea where $J$ represents the leading (perturbative) correction to
$\mc{K}$ and $\Omega$ the leading non-perturbative correction to $W$. We ask when it is
safe to neglect the corrections $J$ or $\Omega$.

The F-term scalar potential is 
\be 
\label{FScalarPot} 
V_F = e^{\K}\left[ \mc{K}^{i\bar{k}} D_i W D_{\bar k}{\bar W} \
 - 3 |W|^2 \right]. \ee This can be expanded in powers of $J$ and $\Omega$: \be
V_F = V_0 + V_J + V_\Omega + \cdots , \ee where
$$
V_0 \sim W_0^2, \quad V_J \sim J W_0^2, \quad V_\Omega \sim \Omega^2 + W_0 \Omega,
$$
and the ellipses refer to higher-order terms combining $J$ and $\Omega$. The exact expressions for these
quantities may be explicitly computed but are not relevant for our
immediate purposes.

Normally, the structure of $V$ is primarily determined by $V_0$, while the other terms provide small
corrections in a weak coupling expansion. However if the tree-level potential has a flat direction along which
$V_0$ is constant, then the structure of the potential, and in particular its critical points, are determined by
the corrections. Such behaviour is characteristic of the no-scale
models common in string theory, with the
K\"ahler moduli of IIB flux compactifications being just one example. 
No-scale models have a K\"ahler potential satisfying $\mc{K}^{i\bar{k}} \K_i
  \K_{\bar k} =3$ and a constant superpotential $W_0$, implying 
$V_0=0$. The corrections then determine the structure of the potential. Although both $V_J$ and
$V_\Omega$ will play a role, typically we expect $V_J$ to dominate over $V_\Omega$, as the former is
perturbative in the coupling and the latter non-perturbative. However, since $\Omega$ is the only correction to
$W_0$ it cannot be totally neglected.

There are some special cases where the
 $\mc{O}(J)$ corrections may be neglected.
 First, if $W_0=0$ then $V_J=0$ automatically and the
leading correction comes from $V_\Omega$. For example, this occurs in the heterotic string racetrack
scenario. Likewise, if $W_0 \ll 1$ in suitable units, then $\Omega$
 may be comparable to $W_0$. In
this case, we have
$$
\Omega \sim W_0 \Rightarrow V_\Omega \sim \Omega^2 \textrm{ and } V_J \sim J \Omega^2,
$$
and therefore \be \frac{V_J}{V_\Omega} \ll 1. \ee This is the relevant behaviour for the KKLT scenario. However,
once $W \gtrsim \frac{\Omega}{J}$ (which is $ \ll 1$ as $\Omega \ll J$), the perturbative effects dominate and must be
included.

It is worth noting that the limit $W \sim \Omega$, in which the tree level superpotential is comparable to its
non-perturbative corrections, is very unnatural. There is furthermore no need to restrict to this limit if we
have information on the perturbative corrections to $\mc{K}$, which is true in both type IIB and heterotic
cases.\footnote{Note that the original proposal for
  gaugino condensaton in the heterotic string
  \cite{Derendinger:1985kk, DRSW, hepth9505171} included a
  constant term in the superpotential from the antisymmetric
  tensor $H_{mnp}$ as well as the nonperturbative, gaugino
  condensation superpotential for the dilaton. This was abandoned
  because the constant was found to be quantised in string units and
  could not be of order the nonperturbative correction. It may be
  interesting to reinvestigate this problem including perturbative
  corrections to $\mc{K}$.}

\subsection*{Application to Type IIB Flux Compactifications}
Let us illustrate these issues in the concrete setting of type IIB flux
compactifications. As discussed above, the K\"ahler and
superpotentials take the forms, 
\bea
W & = & \int G_3 \wedge \Omega + \sum A_i e^{- a_i T_i}, \nonumber \\
\mc{K} & = & - 2\ln (\mc{V}_E) - \ln \left(-i \int \Omega \wedge
\bar{\Omega}\right) - \ln (S + \bar{S}). 
\eea
There is also a normalisation factor in front of $W$ that is
unimportant here but shall be made more precise later. 
The volume $\mc{V}_E$ and moduli $T_i$ are measured in Einstein frame ($g_{\mu \nu, E} =
e^{-\frac{\phi}{2}}g_{\mu \nu, s}$) and in units of $l_s = 2 \pi \sqrt{\alpha'}$. The non-perturbative superpotential is generated by either D3-brane
instantons ($a_i = 2\pi$) or gaugino condensation ($a_i = \frac{2 \pi}{N}$). If the dilaton and complex
structure moduli have been fixed by the fluxes, these potentials
reduce to 
\bea 
\label{FluxPotentials}
W & = & W_0 + \sum A_i e^{- a_i T_i}, \nonumber \\
\mc{K} & = & \mc{K}_{cs} - 2\ln(\mc{V}_E). 
\eea 
In the language of the previous section, the scalar
potential derived from (\ref{FluxPotentials}) includes $V_\Omega$ but not $V_J$. We now show that for almost all
flux choices and moduli values, the use of (\ref{FluxPotentials}) without $\alpha'$ corrections is inconsistent.
Although the argument extends easily to any number of K\"ahler moduli,
for illustration we shall consider one K\"ahler
modulus and the geometry appropriate to the
quintic. (\ref{FluxPotentials}) then reduces to 
\bea
\label{OneModulusPotentials}
W & = & W_0 + A e^{-a T}, \nonumber \\
\mc{K} & = & \mc{K}_{cs} - 3 \ln (T+\bar{T}).
\eea 
For the quintic, $\mc{V}_E = \frac{5}{6}t^3 =
\frac{1}{6\sqrt{5}}(T+\bar{T})^{\frac{3}{2}} = \frac{\sqrt{2}}{3\sqrt{5}} \sigma^{\frac{3}{2}}$, where $T =
\sigma+ib$. The leading $\alpha'$ correction to the K\"ahler potential is \cite{hepth0204254} \be
\label{alpha'correction} \mc{K}_{\alpha'} = \mc{K}_{cs} - 2 \ln \left(\mc{V}_E + \frac{\xi}{2
g_s^{\frac{3}{2}}}\right), \ee where $\xi = \frac{-\chi(M) \zeta(3)}{2(2 \pi)^3} = 0.48$. The factor of
$g_s^{-\frac{3}{2}}$ arises from our working in Einstein frame; it would be absent in string frame, in which
$\mc{V}_s = \mc{V}_E g_s^{\frac{3}{2}}$. The resulting scalar potential is \be \label{StringCorrectedPotential}
V = e^{\mc{K}} \Bigg( \overbrace{\frac{4 \sigma^2 (A a)^2 e^{-2 a
      \sigma}}{3} - 4 \sigma W_0 (A a) e^{-a \sigma}}^{V_\Omega} +
      \overbrace{\frac{9 \sqrt{5} \xi W_0^2}{4 \sqrt{2} g_s^{\frac{3}{2}} \sigma^{\frac{3}{2}}}}^{V_J} \Bigg). \ee The $\alpha'$
correction $V_J$ dominates at both small and large volume. Although perhaps counter-intuitive, this is to be
expected - the $\alpha'$ correction is perturbative in volume whereas the competing terms are non-perturbative.

We may quantify when the neglect of $\alpha'$ corrections is permissible. The allowed range of $\sigma$ is \be
\textrm{Max}\left(\mc{O}\left(\frac{1}{g_s}\right), \sigma_{min}\right) < \sigma < \sigma_{max}, \ee where $\sigma_{min}$
and $\sigma_{max}$ are taken to be the solutions of $\vert V_J (\sigma) \vert = \vert V_\Omega (\sigma) \vert$,
that is \be \label{SigmaEquations} \frac{4 \sigma^2 \vert A^2 \, a^2 \vert e^{-2 a \sigma}}{3} - 4 \sigma \vert
  W_0 A \, a\vert e^{-a \sigma}
= \frac{9 \sqrt{5} \xi \vert W_0 \vert ^2}{4 \sqrt{2}
  g_s^{\frac{3}{2}} \sigma^{\frac{3}{2}}}. \ee The $\mc{O}(\frac{1}{g_s})$
bound on $\sigma$ comes from requiring  $\mc{V}_s > 1$ in order to control the $\alpha'$ expansion. In general
this regime is rather limited. For very moderate values of $W_0$, equation (\ref{SigmaEquations}) has no
solutions and there is \emph{no} region of moduli space in which it is permissible to neglect $\alpha'$
corrections to (\ref{OneModulusPotentials}).

For concreteness let us consider superpotentials generated by D3-brane instantons and let us
take\footnote{Properly this should be
  $e^{\mc{K}_{cs}/2}A$ to be K\"ahler covariant,
but we drop the $e^{\mc{K}_{cs}/2}$.} $A=1$ and $g_s = \frac{1}{10}$.  Then we require $W_0 <
\mc{O}(10^{-75})$ to have any solutions at all of (\ref{SigmaEquations}) with $\mc{V}_s > 1$. This can be
improved somewhat by using gaugino condensation with very large gauge groups. Thus using $W_0 = 10^{-5}$, $g_s =
\frac{1}{10}$ and $N = 50$, $\sigma_{max} \sim 150$, which corresponds to $\mc{V}_s \sim 12$. However, as
generic values of $W_0$ are $\mc{O}(\sqrt{\frac{\chi}{24}}) \sim \mc{O}(10)$ and $W_0^2$ is uniformly
distributed \cite{hepth0404257}, $W_0 = 10^{-5}$ represents a tuning of one part in $10^{12}$, and even then the
range of validity is rather limited.

We therefore conclude that for generic values of $W_0$ there is no regime in which the perturbative corrections
to the K\"ahler potential can be neglected; the inclusion solely of non-perturbative corrections is
inconsistent. Furthermore, even for the small values of $W_0$ for which there does exist a regime in which
non-perturbative terms are the leading corrections, this is still only true for a small range of moduli values
and in particular never holds at large volume.\footnote{We do not consider the special case $\chi(M) = 0$ , when
the $\alpha'^3$ corrections considered above vanish. We do not know the
  status of higher $\alpha'$ corrections in such models, but we would
  expect these to likewise dominate at large volume.}

In the next chapter we shall study the effect of systematically
including the perturbative K\"ahler corrections.

\chapter{Large Volume Flux Compactifications}
\label{chapterLargeVol}

This chapter is based on the papers \cite{hepth0502058} and \cite{hepth0505076}.

The previous chapter has argued that, if we wish to study the scalar
potential across the full range of flux choices and moduli values, it is
essential to include perturbative corrections. The aim of this chapter
is to describe how the inclusion of such corrections leads very
generally to a non-supersymmetric minimum of the scalar potential at
exponentially large volumes. This minimum was first identified in
\cite{hepth0502058}. The rest of this thesis will be devoted to
describing the construction, properties and phenomenological
applications of this minimum.

The structure of this minimum depends crucially on the inclusion of
$\alpha'$ corrections.
A natural worry is that it is not possible to only include a subset of the perturbative
$\alpha'$ corrections: if some are important, all must be. 
We shall address this point below; the fundamental point is that
the $\alpha'$ expansion is an expansion in inverse volume and thus at
large volume can be controlled.

The leading corrections to the K\"ahler potential were computed in \cite{hepth0204254} and
 arise from the ten-dimensional $\mc{O}(\alpha'^3)$ $\mc{R}^4$ term.
Measuring dimensionful quantities in units of $l_s = 2\pi
 \sqrt{\alpha'}$, the resulting K\"ahler potential takes
the form
\be
\label{eqKP}
\frac{\mc{K}}{M_P^2} = - 2 \ln \left( \mc{V}_s + \frac{\xi
  g_s^{3/2}}{2 e^{3 \phi/2}} \right) - \ln( S + \bar{S}) -
\ln \left( -i \int \Omega \wedge \bar{\Omega} \right).
\ee
where $\xi = - \frac{ \chi(M) \zeta(3)}{2 (2 \pi)^3}$. $\mc{V}_s$ is
the internal volume, measured with an Einstein frame metric defined by
$g_{\mu \nu,E} = e^{(\phi_0 - \phi)/2} g_{\mu \nu,s}$. This is defined
  so
that at the minimum the Einstein and string frame metrics
  coincide.\footnote{This is why the subscript $_s$ is used.} $S =
e^{-\phi} + iC_0$ is the dilaton-axion.
We
will require $\xi > 0$, which is equivalent to $h^{2,1} > h^{1,1}$:
i.e., more complex structure than K\"ahler moduli. Here $g_s = \langle
e^\phi \rangle = e^{\phi_0}$ is the value of the stabilised dilaton.
The superpotential receives non-perturbative corrections causing it to depend on
the K\"ahler moduli. These arise either from D3-brane instantons
\cite{hepth9604030} or gaugino condensation from wrapped D7-branes
\cite{Derendinger:1985kk, DRSW, hepth9505171}. It takes the form
\be
\label{eqFS}
\hat{W} = \frac{g_s^{3/2} M_P^3}{\sqrt{4 \pi} l_s^2} \left( \int G_3
\wedge \Omega + \sum
A_i e^{-\frac{a_i T_i}{g_s}} \right).
\ee
The factor of $1/g_s$ in the exponent comes from the definition of
Einstein frame used.
The prefactor is derived in section \ref{appendixsec1} of the Appendix
by a careful dimensional reduction. The $A_i$
represent threshold effects and depend on
the positions of complex structure moduli and wandering D3-branes (if present).
Here  $a_i = \frac{2 \pi}{K}$ with $K \in \mbb{Z}_+$ and $K=1$ for D3-instantons.

Although the internal volume is measured in units of $l_s$, the
 $\mc{N} = 1$ supergravity potential is in units of $M_{pl}$. This
 arises from the string theoretic starting point by dimensional reduction
and  Weyl rescaling to 4-d Einstein frame.
The scalar potential is given by the $\mc{N} =1$ supergravity
formula
\be
\label{scalarpotential}
V = \int d^4 x \sqrt{-g_E} e^{\mc{K}/M_P^2} \left[ \mc{K}^{a \bar{b}}
D_a \hat{W} D_{\bar{b}} \bar{\hat{W}} - \frac{3}{M_P^2} \hat{W} \bar{\hat{W}}  \right].
\ee

Thus equations (\ref{eqKP}) and (\ref{eqFS}) completely specify the potential.
However, trying to directly visualise this is not illuminating.
We therefore follow KKLT \cite{hepth0301240} and first integrate out
the complex structure moduli.
Technically, we stabilise the dilaton and complex structure moduli
through solving (\ref{csstabilisation}), and then regard their values
as fixed.
This leaves a theory only depending on
K\"ahler moduli, which we then stabilise separately. It is important
to ask whether
the resulting critical point of the full potential, including the
moduli that have been integrated out,
is genuinely a minimum or merely a saddle point. For the simplest implementation of  the KKLT scenario with one
K\"ahler modulus and a rigid Calabi-Yau, the resulting potential has no minima \cite{hepth0411066}.
However, in multi-modulus models true minima can indeed occur \cite{hepth0506090}.
We shall show below that the vacua we find, whether AdS or uplifted dS,
are automatically tachyon-free.

After integrating out the dilaton and complex structure moduli the K\"ahler and superpotential become
\bea
\label{Kalpha1}
\frac{\mc{K}}{M_P^2} & = & \mc{K}_{cs}
-2 \log \left[\mc{V} +
\frac{\xi}{2} \right], \\
\label{Wcsfixed}
\hat{W} & = & \frac{g_s^2 M_P^3}{\sqrt{4 \pi}} \left( W_0 + \sum_n A_n e^{-
    \frac{a_n T_n}{g_s}} \right).
\eea
The extra factor of $g_s^{\half}$ comes from absorbing the
    dilaton-dependent part of $e^{\mc{K}_{cs}/2}$ into $\hat{W}$.
Although they are not shown explicitly above, the dilaton-dependent parts in
(\ref{Kalpha1}) should be used when determining the precise form of
    the inverse metric, and in particular the cross terms between
    dilaton and K\"ahler moduli.
This will be relevant when computing D3 soft terms in chapter \ref{chapterSoftSusy}.
We note for subsequent use that as $\mc{V} \to \infty$ the K\"ahler potential behaves as
\be
\label{Klargevol}
e^{\mc{K}} \sim \frac{e^{\mc{K}_{cs}}}{\mc{V}^2} + \mc{O}\left( \frac{1}{\mc{V}^3} \right).
\ee
If we substitute (\ref{Kalpha1}) and (\ref{Wcsfixed}) into equation (\ref{scalarpotential}), we obtain the
following potential \cite{hepth0408054}:
\bea
\label{potential}
V & = & e^{\mc{K}} \left[ \mc{K}^{\rho_j \bar{\rho}_k} \left(a_j A_j a_k \bar{A}_k
  e^{- \left(a_j T_j + a_k \bar{T}_k \right)}
- \left( a_j A_j e^{- a_j T_j} \hat{\bar{W}} \partial_{\bar{T}_k} \mc{K} + a_k \bar{A}_k e^{- a_k \bar{T}_k}
\hat{W} \partial_{T_j} \mc{K} \right) \right) \right. \nonumber \\
& + & \left. 3 \xi \frac{\left(\xi^2 + 7\xi \mc{V} +
\mc{V}^2\right)}{\left(\mc{V} - \xi\right)\left(2\mc{V} +
\xi\right)^2}
|\hat{W}|^2\right] \\
& \equiv & V_{np1} + V_{np2} + V_{\alpha'}. \nonumber
\eea
To simplify notation we shall now absorb factors of $\frac{1}{g_s}$ into
the $a_i$: the possible values for $a_i$ are now $\frac{2 \pi}{N g_s}$. We
shall also drop factors of $M_P$ and the prefactors in $\hat{W}$, returning
to these later when we discuss scales.

\section{Large Volume Limit}
\label{sseLVL}

We now study the large-volume limit of
the potential (\ref{potential})
for a general Calabi-Yau manifold.
We then illustrate these ideas through explicit computations on a
particular orientifold model,
$\mbb{P}^4_{[1,1,1,6,9]}$. The aim is to establish the existence of a
non-supersymmetric mininum at exponentially large volume.

\subsection{General Analysis}
\label{GeneralAnalysis}

The argument for a large-volume AdS minimum of the potential
(\ref{potential}) has two stages. We first
show that there will  in general be
a decompactification direction in moduli space along which
\begin{enumerate}
\item The divisor volumes $\tau_i \equiv \textrm{Im}(\rho_i) \to \infty$.
\item $V < 0$ for large $\mc{V}$, and thus the potential approaches zero from below.
\end{enumerate}
This leads to an argument that there must exist a large volume AdS
 vacuum.  Naively we would expect that the positive
 $(\alpha')^3$ term, scaling as $+\frac{1}{\mc{V}^3}$, will
dominate at large volume
over the non-perturbative terms which are exponentially suppressed.
However, care is needed: the $(\alpha')^3$ term is perturbative in the volume of the entire Calabi-Yau,
whereas the naively suppressed terms are exponential in the
divisor volumes separately.  Hence, in a large volume limit in which
 some of the
divisors are relatively small the non-perturbative terms can compete with the peturbative ones.

The $\left(\alpha'\right)^3$ term in (\ref{potential}) denoted by $V_{\alpha'}$ is easiest to analyse
in the large $\mc{V}$ limit. Owing to the large volume behaviour (\ref{Klargevol})
of the K\"ahler potential, this scales as
\be
\label{Valpha'}
V_{\alpha'} \sim + \frac{3\xi}{4 \mc{V}^3} e^{\mc{K}_{cs}}|\hat{W}|^2 + \mc{O}\left(\frac{1}{\mc{V}^4}\right).
\ee
We observe that $V_{\alpha'}$ is always positive and
depends purely on the overall volume $\mc{V}$.

However, $V_{np1}$ and $V_{np2}$ both depend explicitly on the K\"ahler moduli and we must be more precise in
specifying the decompactification limit.
We first consider the $\mc{V} \to \infty$ limit in moduli space where $\tau_i \to \infty$ for all moduli
except one, which we denote by $\tau_s$. There are two conditions on $\tau_s$. The first is that this limit
be well-defined; for example, the volume should not become formally negative in this limit. The second
is that $\tau_s$ must appear non-perturbatively in $W$.
It is however not essential for this purpose
that all K\"ahler moduli appear non-perturbatively in the superpotential.

Let us study $V_{np1}$ in this limit.
From (\ref{potential}),
\be
\label{Vnp1General}
V_{np1} = e^{\mc{K}} \mc{K}^{T_j \bar{T}_k} \left(a_j A_j a_k \bar{A}_k e^{-
  \left(a_j T_j + a_k \bar{T}_k \right)}\right).
\ee
This is seen to be positive definite.
As we have taken $\tau_i$ large for $i \ne s$, the only term not exponentially
suppressed in (\ref{Vnp1General}) is that involving $T_s$ alone. $V_{np1}$ then reduces to
\be
V_{np1} =  e^{\mc{K}} \mc{K}^{T_s \bar{T}_s} a_s^2 |A_s|^2 e^{-2 a_s \tau_s} .
\ee
We need to determine the inverse metric $\mc{K}^{T_i \bar{T}_j}$. With $\alpha'$ corrections included, this
is given by (e.g. see \cite{hepth0412239})\footnote{The conventions
  used for the K\"ahler moduli in \cite{hepth0412239} differ
  slightly from ours; as we are in this section only interested in the overall
  sign of the potential these are not important.}
\be
\label{inversemetric}
\mc{K}^{T_i \bar{T}_j} = -\frac{2}{9} \left(2\mc{V} + \xi\right) k_{ijk}t^k + \frac{4 \mc{V} - \xi}{\mc{V} - \xi} \tau_i \tau_j.
\ee
In the large $\mc{V}$ limit this becomes
\be
\mc{K}^{T_i \bar{T}_j} = -\frac{4}{9} \mc{V} k_{ijk} t^k + 4 \tau_i \tau_j + \textrm{ (terms subleading in $\mc{V}$)}.
\ee
Thus in the limit described above we have
\be
\mc{K}^{T_s \bar{T}_s} = -\frac{4}{9} \mc{V} k_{ssk} t^k + \mc{O}(1),
\label{gss}
\ee
with
\be
\label{Vnp1Final}
V_{np1} \sim  \frac{(- k_{ssk}t^k) a_s^2 \vert A_s \vert^2 e^{-2 a_s \tau_s} e^{\mc{K}_{cs}}}{\mc{V}} +
\mc{O}\left(\frac{e^{-2a_s \tau_s}}{\mc{V}^2} \right).
\ee
We have dropped numerical prefactors.   Despite the minus sign in
(\ref{gss})
this component of the inverse metric will be positive since the
K\"ahler metric as a whole is
positive definite and this component computes the length squared of
the (dual) vector
$\partial_{T_s} W$.   In the limit we consider, so long as
we remain inside the
K\"ahler cone, the leading term must be positive.

We can perform a similar analysis for $V_{np2}$, from whence negative contributions to the potential arise. We have
\be
V_{np2} = - e^{\mc{K}}
\left( \mc{K}^{T_j \bar{T}_k} a_j A_j e^{-a_j T_j} \bar{W} \partial_{\bar{T}_k} \mc{K} + \mc{K}^{T_k \bar{T}_j}a_k \bar{A}_k e^{- a_k \bar{T}_k}
W \partial_{T_j} \mc{K} \right).
\ee
The only surviving exponential terms are again those involving $\tau_s$. $V_{np2}$ thus reduces to
\be
V_{np2} = - e^{\mc{K}} \left[ \mc{K}^{T_s \bar{T}_k} \left(a_s A_s e^{-
    a_s T_s} \bar{W} \partial_{\bar{T}_k} \mc{K} \right) +
\mc{K}^{T_k \bar{T}_s}\left(a_s \bar{A_s} e^{- a_s \bar{T}_s} W
\partial_k \mc{K} \right)\right].
\label{np2a}
\ee
The form of the inverse metric (\ref{inversemetric}) implies  $\mc{K}^{T_s \bar{T}_k} = \mc{K}^{T_k \bar{T}_s}$.
The sign of $V_{np2}$ is determined by the value of the axionic field $b_s = \textrm{Im}(T_s)$, which
will adjust to make $V_{np2}$ negative.    To see this, note
that at leading order
at large volume $W = W_0 + O(1/\mc{V})$,
and the only dependence on the
 axion $b_s$ is in $V_{np2}$.    Now write $V_{np2} = e^{ i a_s b_s} X
 + e^{- i a_s b_s} \bar{X}$, where
 we have collected all factors in (\ref{np2a}) except for the axion
 into $X$ and $\bar{X}$.  Extremizing the
potential with respect to $b_s$, it is easy to see that at a minimum
the axion will arrange its value to
cancel the overall phase from the prefactors and make $V_{np2}$
negative.

Thus we may without loss of generality simplify the calculation by replacing
$T_s$ by $\tau_s$ and assuming $A_s$ and $W$ to be both real.
Recall that
$$
\partial_{T_k} \mc{K} = \frac{t^k}{2 \mathcal{V} + \xi} \to \frac{t^k}{2 \mc{V}} + \mc{O}\left(\frac{1}{\mc{V}^2}\right).
$$
Therefore
\be
V_{np2} \sim - e^{\mc{K}} a_s A_s W e^{-a_s \tau_s} \mc{K}^{T_s \bar{T}_j} \frac{t^j}{\mc{V}} + \mc{O}
\left(\frac{e^{-a_s \tau_s}}{\mc{V}^2}\right).
\ee
We have introduced a minus sign as a reminder that this term will be negative.
Substituting in for $\mc{K}^{T_s \bar{T}_k}$ then gives
\bea
V_{np2} & \sim & - e^{\mc{K}_{cs}} a_s A_s W e^{-a_s \tau_s} \frac{- \frac{4}{9} \mc{V} k_{sjk} t^j t^k  + 4 \tau_s \tau_j t^j}
{\mc{V}^3} + \mc{O}\left(\frac{e^{-a_s \tau_s}}{\mc{V}^3}\right)\nonumber \\
\label{vnp2equation}
& \sim & - e^{\mc{K}_{cs}} a_s A_s W e^{-a_s \tau_s} \frac{ - \frac{8}{9} \mc{V} \tau_s + 4 \tau_s \tau_j t^j}{\mc{V}^3}
+ \mc{O}\left(\frac{e^{-a_s \tau_s}}{\mc{V}^3}\right)\nonumber.
\eea
As from the definition $\tau_j t^j \propto \mc{V}$,
we conclude that in the limit described above,
\be
\label{Vnp2General}
V_{np2} \sim - \frac{a_s \tau_s e^{-a_s \tau_s}}{\mc{V}^2} \vert A_s W_0 \vert e^{\mc{K}_{cs}}
+ \mc{O}\left(\frac{e^{-a_s \tau_s}}{\mc{V}^3}\right)\nonumber.
\ee

We may now study the full potential by combining equations (\ref{Valpha'}), (\ref{Vnp1Final}) and
(\ref{Vnp2General}).
\be
\label{fullpotential}
V \sim \left[ \frac{1}{\mc{V}} a_s^2 |A_s|^2 (-k_{ssk}t^k) e^{-2 a_s \tau_s} -
\frac{1}{\mc{V}^2} a_s \tau_s e^{-a_s \tau_s} \vert A_s W \vert
+ \frac{\xi}{\mc{V}^3} |W|^2 \right].
\ee
We have absorbed factors of $e^{\mc{K}_{cs}}$ into the values of $W$ and $A_s$.
We can now demonstrate the existence of a decompactification limit
in which this potential approaches zero from below.
This limit is given by
\be
\label{decomlimit}
\mc{V} \to \infty \quad \quad \textrm{ with } \quad \quad a_s \tau_s = \ln (\mc{V}).
\ee
In this limit the potential takes the following form
\be
\label{volpotential}
V \sim \left[ a_s^2 A_s^2 \frac{(-k_{ssj}t^j)}{\mc{V}^3} - \vert A_s W_0 \vert
\frac{(a_s k_{sjk} t^j t^k)}{\mc{V}^3} + \frac{\xi}{\mc{V}^3}
\vert W_0 \vert ^2 \right] + \mc{O}\left(\frac{1}{\mc{V}^4}\right).
\ee
As in this limit the non-perturbative corrections to $W$
are subleading by a power of $\mc{V}$, we have replaced $W$ by $W_0$.
 We have also written out $\tau_s = \frac{\ln \mc{V}}{a_s}$
in terms of 2-cycle volumes $t^i$.

All terms in equation (\ref{volpotential}) have the same volume dependence
and it is not immediately obvious which is dominant at large volume.
However, the numerator of the
second term of equation (\ref{volpotential}) is quadratic in the 2-cycle volumes,
whereas the others have at most a linear dependence.
As $\tau_i \to \infty$ for all $i$,
all 2-cycles must blow up to infinite volume.
The numerator of the second term is proportional to
$\tau_s$
and thus scales as $\ln \mc{V}$.  Thus, this term scales
as $\frac{\ln \mc{V}}{\mc{V}^3}$, and overcomes the first and third terms which scale schematically
as $\frac{\sqrt{\ln \mc{V}}}{\mc{V}^3}$ and $\frac{1}{\mc{V}^3}$.
In the limit (\ref{decomlimit}) the potential eventually behaves as
\be
V \sim - e^{\mc{K}_{cs}} \vert A_s W_0 \vert \frac{\ln \mc{V}}{\mc{V}^3},
\ee
and approaches zero from below.\footnote{
One concern in the above analysis may be our treatment of $A_s$, which we have treated as a constant.
If the $A_s$ were to depend on the K\"ahler moduli, this may invalidate the argument.
However, as a correction to the superpotential $A_s$ must be
holomorphic in the K\"ahler moduli and must also respect the axion
shift symmetry. This implies that $A_s$ cannot have a perturbative
dependence on the K\"ahler moduli. It may depend on complex structure
moduli, but these have been fixed by the fluxes. I thank Liam
McAllister for discussions on this point.}

Given this, it is straightforward to argue that there must exist a large volume AdS minimum.
At smaller volumes, the dominant term in the potential ($\ref{fullpotential}$) is either
the non-perturbative term $V_{np1}$ or the $(\alpha')^3$ term $V_{\alpha'}$, depending on the value of
$\tau_s$. Both are positive; the former because the metric on moduli space is positive definite and the
latter because we have required $h^{2,1} > h^{1,1}$.
Thus at small volumes the potential is positive, and so since
the potential must go to zero at infinity and is known to have
negative values at finite volume,
there must exist a local AdS minimum along the direction in K\"ahler moduli space where the volume changes.

One may worry that this argument involves the behaviour of the potential at small values of the volume where the
$\alpha'$ expansion cannot be trusted. However, for minima at very large volume, the `small' volumes
required in the above argument are extremely large in string units,
and so we can self-consistently neglect
terms of higher order in $\alpha'$. The relative strength of the $\alpha'$ correction varies between Calabi-Yaus
depending on the precise details of the geometry. For the explicit example studied in the next section,
the `small' volumes used in the above argument to establish the positivity of the potential
may easily be $\mc{O}(10^8) l_s^6$.

It remains to argue that the potential also has a minimum in the
remaining directions of the moduli space.   Imagine moving along the
surfaces in the moduli space that are of fixed Calabi-Yau volume, $\mc{V}$.
Then, as one approaches the walls of the K\"ahler cone the first term
in (\ref{fullpotential}) dominates, since it has the fewest powers of
volume in the denominator and the exponential contributions of
the moduli that are becoming small cannot be neglected. Only the
exponential contribution of $\tau_s$ is given in (\ref{fullpotential})
because of the assumed limit, but it is easy to convince oneself that
a similar term will appear for any modulus that is small while the
overall volume is large. Thus at large overall volume we expect the
potential to grow in the positive direction towards the walls of the
K\"ahler cone, provided all the moduli appear in the non-perturbative
superpotential.   All told, the potential is negative along the
special direction in moduli space that we have identified and
eventually rises to be positive or to vanish in all other directions, giving
an AdS minimum.
Since $V\sim O(1/\mc{V}^3)$ at the minimum, while $-3 e^{\mc{K}}|W|^2\sim
O(1/\mc{V}^2)$,
it is clear that this minimum is non-supersymmetric

We can heuristically see why
the minimum we are arguing for can be at exponentially large volume.  
The naive measure of the location of the minimum is the value of the volume at which the second term of
equation (\ref{volpotential}) becomes dominant.  As this only
occurs when $\ln \mc{V}$ is large, we expect to be able to find vacua
at large
values of $\ln \mc{V}$.  We will see this explicitly in the example %$P^4_{1,1,1,6,9}$ model
studied below.

We can also see that the gravitino mass for these vacua will, for fixed $g_s$, be independent of the flux choice. If the
minimum exists at large volume, it is found by playing off the three terms in equation (\ref{fullpotential})
against each other. If we write $\tilde \mc{V} = \frac{A_s \mc{V}}{W_0}$, then (\ref{fullpotential}) becomes
\be
\label{fullpotential2}
V \sim \left(\frac{A_s^3}{W_0}\right) \left[ \frac{1}{\tilde\mc{V}} a_s^2  (-k_{ssk}t^k) e^{-2 a_s \tau_s} -
\frac{1}{\tilde\mc{V}^2} a_s \tau_s e^{-a_s \tau_s}
+ \frac{\xi}{\tilde\mc{V}^3} \right].
\ee
The minimum of this potential as a function of $\tilde\mc{V}$ is independent of $A_s$ and $W_0$ and, given
$a_s$, depends only on the Calabi-Yau. We therefore have
\be
\mc{V} \sim \frac{W_0}{A_s} f(a_s, \mc{M}) + (\textrm{subleading corrections}),
\ee
where $f$ is a function of the geometry. The gravitino mass is then given by
\be
m_{\frac{3}{2}} = e^{\mc{K}/2}\vert W \vert \approx \frac{A_s}{2 f(a_s, \mc{M})}.
\ee

Although we have just considered the K\"ahler moduli in finding this minimum, it is straightforward to see that
it must actully be a minimum of the full potential.
Reinstating the dilaton-axion and complex structure moduli, the full potential can be written\cite{hepth0204254}
\bea
\label{Vcsdilaton}
V & = & e^{\mc{K}} (\mc{K}^{a\bar{b}} D_a W \bar{D}_b \bar{W} + \mc{K}^{\tau \bar{\tau}}
D_\tau W \bar{D}_\tau \bar{W}) + e^{\mc{K}} \frac{\xi}{2 \mc{V}} (W
\bar{D}_\tau \bar{W} + \bar{W} D_\tau W) \nonumber \\
& & + V_{\alpha'} + V_{np1} +
V_{np2}.
\eea
Recall that the moduli values found above give rise to a negative value of the potential of
$\mc{O}\left(\frac{1}{\mc{V}^3}\right)$.
The first term in (\ref{Vcsdilaton}) is positive definite and of $\mc{O}\left(\frac{1}{\mc{V}^2}\right)$.
This vanishes iff $D_\tau W = D_{\phi_i} W = 0$. Therefore, any movement of either the dilaton or
complex structure moduli away from their stabilised values would create a positive term
of $\mc{O}\left(\frac{1}{\mc{V}^2}\right)$, which the negative term cannot compete with. Thus this must increase
the potential, and so the solution above automatically represents a minimum of the full potential.

It is instructive to compare this with the behaviour of KKLT
solutions. The scalar potential here is
\be
V = e^{\mc{K}_{cs}} \Big(\frac{\mc{K}^{i \bar{j}} D_j W\bar{D}_{j} \bar{W}}{\mc{V}^2} -
\frac{ 3 \vert W \vert^2}{\mc{V}^2}\Big).
\ee
If $D_i W = 0$ for all moduli, the potential is negative and of magnitude $\mc{O}\left(\frac{1}{\mc{V}^2}\right)$.
However, if we move one modulus, for concreteness the dilaton, away from its stabilised value, the resulting
positive definite contribution $e^{\mc{K}_{cs}} \frac{\mc{K}^{\tau \bar{\tau}} D_\tau W \bar{D}_\tau \bar{W}}{\mc{V}^2}$
is only of the same order as the minimum. Moving the dilaton alters the value of $e^{\mc{K}_{cs}}$ and thus may increase
the numerator of the negative term. As the positive and negative contributions are of the same order, we see that
depending on the magnitude of $\mc{K}^{\tau \bar{\tau}} D_\tau W \bar{D}_\tau \bar{W}$, this may in general decrease
the overall value of the potential. Therefore it is necessary to check explicitly for each choice of fluxes that
the resulting potential has no minimum.

The above solution can be uplifted to a de Sitter vacuum through
the usual mechanisms of adding anti-D3 branes~\cite{hepth0301240}
or turning on magnetic fluxes on D7-branes~\cite{hepth0309187}. For
concreteness we take the uplift potential to be \be
\label{Vuplift} V_{uplift} = + \frac{\epsilon}{\mc{V}^2}. \ee
When
$\epsilon = 0$, the above minimum still exists and there are many
values of the moduli for which $V < 0$. For $\epsilon$
sufficiently large, the minimum is entirely wiped out and the
potential is positive for all values of the moduli.
At a critical value of $\epsilon$ the minimum will pass through zero.
By construction, this must still represent a minimum of the full potential.
After adding the uplift terms, the
total potential will once again go
to zero from above at large volumes because the scaling of (\ref{Vuplift}) will
overwhelm the $O(1/\mc{V}^3)$ negative terms even in the special limit
that we have been studying.  This then leads to the metastable de
Sitter vacua popular in the attempts to incorporate accelerating universes into
string theory.

\subsection{Explicit Calculations for $\mbb{P}^4_{[1,1,1,6,9]}$}
\label{DouglasModel}

We now illustrate the above ideas through
explicit
calculations for flux compactifications on an orientifold of
the Calabi-Yau manifold given by the degree 18 hypersurface in
$\mathbb{P}^4_{[1,1,1,6,9]}$. This has been studied by Denef, Douglas and
Florea\cite{hepth0404257} following earlier work in
\cite{hepth9403187}. The defining equation is
\be
\label{definingequation}
z_1^{18} + z_2^{18} + z_3^{18} + z_4^3 + z_5^2 - 18 \psi z_1 z_2 z_3 z_4 z_5 -3 \phi z_1^6 z_2^6 z_3^6
= 0,
\ee
with $h^{1,1}=2$ and $h^{2,1} = 272$.
The complex structure moduli $\psi$ and $\phi$ that have been written in (\ref{definingequation}) are the
two moduli invariant under the $\Gamma = \mbb{Z}_6 \ti \mbb{Z}_{18}$ action whose quotient gives
the mirror manifold \cite{GreenePlesser}. There are another 270
terms not invariant under $\Gamma$ which have not been written
explicitly, although some will be projected out by the
orientifold action.

We first stabilise the complex structure moduli through an explicit
choice of fluxes. If $W_{cs}$ denotes the flux superpotential, we must solve
\be
\label{StabilisingModuli}
D_\tau W_{cs} = 0 \quad \textrm{ and } \quad D_{\phi_i} W_{cs} = 0,
\ee
for the dilaton and complex structure moduli. There are two possibilities.
First, we may of course turn on fluxes along all relevant three-cycles
and solve (\ref{StabilisingModuli}) for all moduli. As we would need
to know all 200-odd periods this is impractical.
We however know of no theoretical reasons not to do it.
The easier approach is to turn on fluxes only along cycles
corresponding to $\psi$ and $\phi$, and then solve
\be
\label{SusyStabilising}
D_\tau W_{cs} = D_\psi W_{cs} = D_\phi W_{cs} = 0.
\ee
As first described in
\cite{hepth0312104} and reviewed in chapter \ref{StatisticsReview}, the invariance of ($\ref{definingequation}$)
under $\Gamma$ ensures that, at $\phi_k = 0$,
$D_{\phi_k} W = 0$ for all other moduli $\phi_k$. The necessary periods have been
computed in \cite{hepth9403187} and appropriate fluxes and solutions
to (\ref{SusyStabilising}) could be found straightforwardly as in
chapter \ref{StatisticsReview}
along the lines of \cite{hepth0312104,  hepth0404243, hepth0409215}.
This is not our focus here and we henceforth assume this to
have been done.

We now return to the K\"ahler moduli. Using the notation of
\cite{hepth0404257}, the K\"ahler geometry is specified by
\be
\mc{V} = \frac{1}{9\sqrt{2}}\left(\tau_5^{\frac{3}{2}} -
\tau_4^{\frac{3}{2}}\right), \nonumber
\ee
$$
\textrm{ with } \tau_4 = \frac{t_1^2}{2} \quad \textrm{ and } \quad
\tau_5 = \frac{\left(t_1 + 6t_5\right)^2}{2}.
$$
Here $\tau_4$ and $\tau_5$ are volumes of the divisors $D_4$ and
$D_5$,
corresponding to a particular set of 4-cycles, and $t_1$ and $t_5$
2-cycle volumes.
Generally the volume is only an implicit function of $\tau_i$, but here we are
fortunate and have an explicit expression.
As shown in \cite{hepth0404257}, both $D_4$ and $D_5$ correspond to divisors which
appear non-perturbatively in the superpotential.
We write this superpotential as
\be
W = W_0 + A_4 e^{- a_4 T_4} + A_5 e^{- a_5 T_5}.
\ee
We now take the limit described above,
in which $\mc{V} \to \infty$ (and hence $\tau_5\to\infty$) and $\tau_4\sim \log{\mc{V}}$. Note that the
alternative limit $\tau_4 \to \infty$ with $\tau_5\sim \log{\mc{V}}$ would not be
well-defined, as the volume of the
Calabi-Yau becomes formally negative.

\begin{figure}[ht]
\linespread{0.2}
\begin{center}
\epsfxsize=0.85\hsize \epsfbox{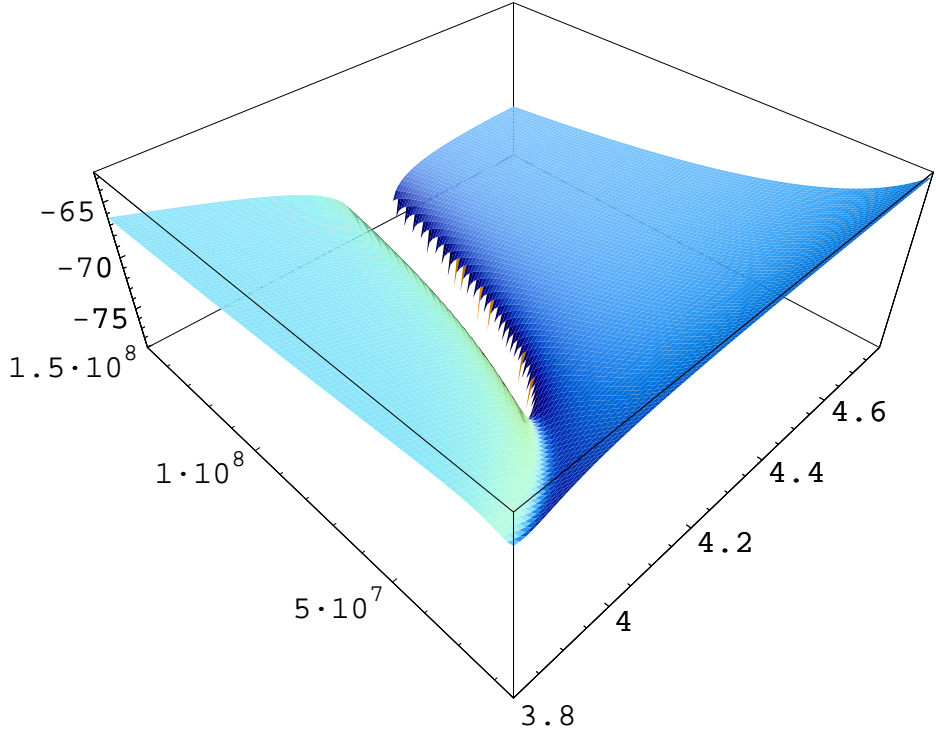}
\end{center}
\caption{ $\ln (V)$ for $\mbb{P}^4_{[1,1,1,6,9]}$ in the large volume
  limit, as a function of the divisors $\tau_4$ and $\tau_5$. The void
  channel corresponds to the region
  where $V$ becomes negative and $\ln (V)$ undefined. As $V \to 0$ at
  infinite volume, this immediately shows that a large-volume minimum
  must exist. Here the values $W_0 = 20,
  A_4 = 1$ and $a_4 = 2\pi$ have been used.}
\label{LargeVolumeLimit}
\end{figure}

The $\alpha'$ correction is given by equation (\ref{Valpha'}). For
$V_{np1}$ and $V_{np2}$ we must compute the inverse metric, which in
this limit is given by
\bea
\mc{K}^{T_4 \bar{T_4}} & = & 24 \sqrt{2} \sqrt{\tau_4} \mc{V} \sim \sqrt{\tau_4} \mc{V}, \\
\mc{K}^{T_4 \bar{T_5}} = \mc{K}^{T_5 \bar{T_4}} & = & 4 \tau_4 \tau_5 \sim \tau_4 \mc{V}^\frac{2}{3}, \\
\mc{K}^{T_5 \bar{T_5}} & = & \frac{4}{3} \tau_5^2 \sim \mc{V}^\frac{4}{3}.
\eea
We can then compute $V_{np1}$ and $V_{np2}$ with the result that the
full potential takes the schematic form
\be
\label{examplepotential}
V \sim \left[ \frac{1}{\mc{V}} a_4^2 |A_4|^2 \sqrt{\tau_4} e^{-2 a_4 \tau_4} -
\frac{1}{\mc{V}^2} a_4 \tau_4 e^{-a_4 \tau_4} \vert A_4 W \vert
+ \frac{\xi}{\mc{V}^3} |W|^2 \right].
\ee
where numerical coefficients have been dropped.  We have implicitly
extremised with respect to the axion $b_4$ to get a negative sign in
front of the
second term, as described below equation (\ref{np2a}).
It is obvious that in the limit
\be
\tau_5 \to \infty \quad \textrm{ with } \quad a_4 \tau_4 = \ln \mc{V},
\ee
the potential approaches zero from below as the middle term of equation (\ref{examplepotential})
dominates. This is illustrated in figure \ref{LargeVolumeLimit} where we plot the
numerical values of $\ln (V)$.

The location and properties of the AdS minimum may be found analytically.
To capture the form of equation (\ref{examplepotential}), we write
\be
\label{Vgeneral}
V = \frac{\lambda \sqrt{\tau_4} e^{-2 a_4 \tau_4}}{\mc{V}}
- \frac{\mu}{\mc{V}^2} \tau_4 e^{-a_4 \tau_4} + \frac{\nu}{\mc{V}^3}.
\ee
The axion field $b_5$ has been ignored as terms in which it appears are
exponentially suppressed.

We regard $V = V(\mc{V}, \tau_4)$, and solve
$$
\frac{\partial V}{\partial \mc{V}} = \frac{\partial V}{\partial \tau_4} = 0.
$$
One may easily check that the first of these equations may be
rearranged into a quadratic and solved for $\mc{V}$ to give
\be
\label{Vsoln}
\mc{V} = \frac{\mu}{\lambda} \sqrt{\tau_4} e^{a_4 \tau_4} \left(1 \pm
\sqrt{1 - \frac{3 \nu \lambda}{\mu^2 \tau_4 ^{\frac{3}{2}}}} \right)
\ee
We also have
\be
\frac{\partial V}{\partial \tau_4} = 0 \Rightarrow
\frac{\lambda \mc{V} e^{-a_4 \tau_4}}{\tau_4^{\half}} \left(\half - 2
a_4 \tau_4 \right) - \mu \left(1 - a_4 \tau_4 \right) = 0.
\ee
We then use (\ref{Vsoln}) to obtain an implicit equation for $\tau_4$,
\be
\label{tausoln}
\left(1 \pm \sqrt{1 - \frac{3 \nu \lambda}{\mu \tau_4^{\frac{3}{2}}}} \right)\left( \half - 2 a_4 \tau_4 \right)
= (1 - a_4 \tau_4).
\ee
We do not need to solve this fully; as we require
$a_4 \tau_4 \gg 1$ to be able to ignore higher instanton corrections, we can use this
to simplify (\ref{tausoln}) and solve for $\tau_4$ and $\mc{V}$, obtaining
\bea
\tau_4 & = & \left( \frac{4 \nu \lambda}{\mu^2} \right)^{\frac{2}{3}},
\nonumber \\
\mc{V} & = & \frac{\mu}{2 \lambda} \left( \frac{4 \nu \lambda}{\mu^2}
\right)^{\frac{1}{3}}
e^{a_4 \left( \frac{4 \nu \lambda}{\mu^2} \right)^{\frac{2}{3}}}.
\eea

For the potential of (\ref{Vgeneral}),
$$
\lambda \sim a_4^2 \vert A_4 \vert ^2, \quad \mu \sim a_4 \vert A_4
W_0 \vert, \quad \textrm{ and }
\nu \sim \xi \vert W_0 \vert^2.
$$
We then have
\be
\label{TauVolFinal}
\tau_4 \sim (4 \xi)^{\frac{2}{3}} \quad \textrm{ and }
\mc{V} \sim \frac{\xi^{\frac{1}{3}} \vert W_0 \vert}{a_4 A_4} e^{a_4 \tau_4}.
\ee
This formula justifies our earlier claim that these vacua can
generically be at exponentially large volume.

For the $\mbb{P}^4_{[1,1,1,6,9]}$ example,
$$
\xi = 1.31, \quad \lambda = 3\sqrt{2} a_4^2 \vert A_4 \vert^2, \quad
\mu = \half a_4 \vert A_4 W_0 \vert, \quad \nu = 0.123 \vert W_0
\vert^2.
$$
The analytic results derived here agree well with the exact locations of the
minima found numerically, with the small error almost entirely
due to the approximation made in solving equation (\ref{tausoln}).
As discussed above,
the values of the K\"ahler moduli found above combine with the flux-stabilised complex structure moduli
to give a minimum of the full potential (\ref{Vn=1sugra}).

We note that both the overall and divisor volumes are clearly larger
 than the string scale. The `large' modulus $\tau_5 \gg 1$, while for the
 `small' modulus $a_4 \tau_4 \sim \ln \mc{V} \gtrsim 1$. 
Putting back the factors of $g_s$ and $N$, we have $a_4$ =
 $\frac{2 \pi}{g_s N}$. We shall discuss scales and phenomenological
 features in more detail below, but shall now simply observe that
 small adjustments in $g_s$ and $N$ allow a very wide variety of
 scales to be realised. 

Let us summarise our results so far.
The argument above shows that there exists a decompactification direction in moduli space along
which nonperturbative effects dominate over
the perturbative $\alpha'^3$ corrections, leading to an exponentially large volume
minimum with all geometric moduli stabilised in a very general class of
compactifications (for which $h_{12}>h_{11}>1$).\footnote{If $h_{12} < h_{11}$, the $\alpha'$ corrections cause the
  fields to roll in towards small volume. In this limit it is very
  difficult to maintain control as all $\alpha'$ corrections become
  equally important.}

There are several very interesting features about this limit.
The mechanism described here stabilises all moduli and results in internal spaces that are
exponentially large in string units, which is the first time this has
been achieved in a string construction. While large internal volumes are
useful from the point of view of control, they also give the
interesting possibility of lowering the string scale, as the string
and Planck scales are related by $M_s \sim \frac{M_P}{\sqrt{\mc{V}}}$.
Furthermore, no tuning has been required to obtain this minimum. In
particular, the
volume scales as $\mc{V} \propto W_0$ and there is no requirement
that $W_0$ be small. This is
in contrast to the behaviour encountered
in the KKLT scenario, where very small values of $W_0$ are necessary
for consistency. 

As the volume is exponentially sensitive to (for example) the
stabilised dilaton, there
is no fine-tuning involved in adjusting the volume to give a TeV-scale
gravitino mass. The above moduli stabilisation mechanism therefore
gives a genuine solution, as opposed to simply a reinterpretation, 
of the supersymmetric hierarchy problem. 

\section{The $\lowercase{g}_{\lowercase{s}}$ and $\alpha'$ Expansions}
\label{PerturbativeExpansions}

In chapter 4, we argued that under almost all circumstances 
the neglect of $\alpha'$ corrections in IIB flux compactifications is
inconsistent. In the above section we have described how the inclusion
of the $\alpha'^3$ corrections of \cite{hepth0204254} gives a minimum
of the potential at exponentially large volumes. The natural worry is
that this is inconsistent: it is often felt that if some $\alpha'$
corrections are important all will be, and it is inconsistent to
include only a subset. Fortunately this is wrong and
the inclusion of  some $\alpha'$ corrections does not necessarily require full knowledge of the string theory. At
small volumes, the $\alpha'$ corrections do indeed appear democratically, and it would be difficult to extract reliable
results. However, the $\alpha'$ expansion is at heart an expansion in inverse volume. At very large volumes, the
expansion parameter for $\alpha'$ effects is $\frac{1}{\mc{V}}$, and a systematic inclusion of such effects will
give a controlled expansion. In discussing this, it will be useful to consider the origins of
perturbative effects from both ten- and four-dimensional perspectives.

\subsection*{Perturbative corrections in 10 dimensions}

It will be useful in this section to keep in mind the dimensional
reduction scalings analysed in chapter \ref{ChapterModuliAndFluxes}.
We will analyse perturbative corrections to the scalar potential
 by studying the volume scaling of terms arising from dimensional reduction
of the $\alpha'$-corrected type IIB supergravity action. Of course this action is not known fully, but our
arguments will only depend on the general form of the terms rather than the specific details of the tensor
structure. The supergravity action consists of bulk and localised terms and is \be \label{FullAction} S_{IIB} =
S_{b,0} + \alpha'^3 S_{b,3} + \alpha'^4 S_{b,4} + \alpha'^5 S_{b,5} + \ldots + S_{cs} + S_{l,0} + \alpha'^2
S_{l,2} + \ldots \ee The localised sources present are D3/D7 or O3/O7 planes. In string frame, we have
\cite{PolchinskiBook} \bea \label{iibaction} S_{b,0} & = & \frac{1}{(2 \pi)^7 \alpha'^4} \int d^{10}x \sqrt{-g}
\left\{ e^{-2 \phi}[\mc{R} + 4(\nabla \phi)^2] - \frac{F_1^2}{2} -
  \frac{1}{2 \cdot 3!} G_3 \cdot \bar{G}_3 - \frac{\tilde{F}_5^2}{4
    \cdot 5!} \right\},
  \nonumber \\
S_{cs} & = & \frac{1}{4 i (2 \pi)^7 \alpha'^4} \int e^\phi C_4 \wedge
G_3 \wedge \bar{G}_3, \nonumber \\
S_{l,0} & = & \sum_{sources} \left( - \int d^{p+1} \xi \, T_p \, e^{-\phi} \sqrt{-g}
 + \mu_p \int C_{p+1} \right).
\eea We may avoid the need to include D3-branes by taking the fluxes to saturate the $C_4$ tadpole. We shall
work throughout in the F-theory orientifold limit, in which the dilaton is constant: $\phi = \phi(y) = \phi_0$.

For flux compactifications with ISD 3-form fluxes, as reviewed in
chapter \ref{ChapterModuliAndFluxes}, the metric and fluxes take the form
\bea
\label{metric5flux}
ds_{10}^2 & = & e^{2A(y)}\eta_{\mu \nu} dx^\mu dx^\nu + e^{-2A(y)}\tilde{g}_{mn}dy^m dy^n, \nonumber \\
\tilde{F}_5 & = & (1 + *) \left[ d\alpha \wedge dx^0 \wedge dx^1
  \wedge dx^2 \wedge dx^3 \right], \nonumber \\
F_3, H_3 & \in & H^3(M, \mbb{Z}), \eea where $\alpha = e^{4A}$ parametrises both the magnitude of the warping
and the size of the 5-form flux. Here $\tilde{g}$ is a Calabi-Yau metric; the flux back-reacts to render the
compact space only conformally Calabi-Yau. The warp-factor transforms non-trivially under internal rescalings,
with warping effects suppressed at large volume. Specifically, under $\mc{V} \to \lambda^6 \mc{V}$, where
$\lambda \gg 1$, $\alpha = 1 + \mc{O}(\frac{1}{\lambda^4}) + \ldots $.

We first consider contributions tree level in $\alpha'$. The leading contribution to the 4-d scalar potential
arises from the flux term $\frac{1}{(2 \pi)^7 \alpha'^4} \int G_3
\cdot \bar{G}_3$ \cite{hepth0105097}. This has a topological part and a moduli-dependent
component. The former is pseudo-BPS and cancels against the tension of
localised sources. Dimensional reduction of the latter gives
\be
\label{Vflux}
V_{flux} \sim \mc{K}_{cs}^{a \bar{b}} \frac{D_a W
  D_{\bar{b}} W}{\mc{V}^2},
\ee
where the sum is
over dilaton and complex structure moduli. This term is positive
  semi-definite and vanishes at its minimum. 
As we will work in a limit $\mc{V} \gg 1$, the most important
  element of (\ref{Vflux}) is its volume scaling. This is 
understood as follows:
\be
V_{flux} \sim \overbrace{\mc{V}^{-2}}^{\textrm{conversion to
4d Einstein frame}} \ti \overbrace{\mc{V}}^{\textrm{internal integral}} \ti \overbrace{\mc{V}^{-1}}^{G_3 \cdot
\bar{G_3}} \sim \mc{V}^{-2}.
\ee
This volume scaling is valid for both topological and moduli-dependent parts.

In the absence of warping, (\ref{Vflux}) is the only $\mc{O}(\alpha'^0)$ contribution to the potential energy,
as $\tilde{F}_5 = \mc{R} = 0$ and dilaton gradients vanish. However, there is also a warping contribution. At
large volume, $\alpha \sim 1 + \frac{1}{\mc{V}^{\frac{2}{3}}}$ and thus
 $V_{F_5} = \int d^6 x \sqrt{g} F_5^2  \sim \int d^6 x \sqrt{g} (\textrm{d} \alpha)^2$ contributes
\be V_{F_5} \sim \overbrace{\mc{V}^{-2}}^{\textrm{conversion to Einstein frame}} \ti
\overbrace{\mc{V}}^{\textrm{internal integral}} \ti \overbrace{\mc{V}^{-\frac{5}{3}}}^{F_5^2} \sim
\mc{V}^{-8/3}. \ee As now $\mc{R} \neq 0$, the Einstein-Hilbert term $\int_{Y_6} \sqrt{-g} \mc{R}$ is also
important and in fact contributes identically with $V_{F_5}$. These terms may be related to the tree-level flux
term and in fact serve as an additional prefactor. The net result is \cite{hepth0208123}
$$
V_{0,unwarped} = \frac{1}{2 \kappa_{10}^2 \textrm{Im } \tau} \int_M G_3^{+} \wedge *_6 \bar{G}_3^{+} \to V_{0,
warped} = \frac{1}{2 \kappa_{10}^2 \textrm{Im } \tau} \int_M e^{4A} G_3^{+} \wedge *_6 \bar{G}_3^{+}.
$$
It is important that this potential remains no-scale, with no
potential generated for the  K\"ahler moduli \cite{hepth0208123}.

The IIB bulk effective action receives higher-derivative corrections starting at $\mc{O}(\alpha'^3)$, which is also
the order at which string loop corrections first appear; the tree level action is already $SL(2, \mbb{Z})$
invariant and receives no $g_s$ corrections. The discussion of loop corrections to $S_b$ is thus subsumed into
the discussion of higher-derivative corrections.

The bosonic fields are the metric, dilaton-axion, 3-form field strength $G_3$ and self-dual five form field
strength $F_5$. While its precise form is unknown, $S_{b,3}$ is expected to include all combinations of these
consistent with the required dimensionality. At $\mc{O}(\alpha'^3)$, the bosonic action takes the schematic form
\bea \label{Sb3} S_{b,3} & \sim & \frac{\alpha'^3}{\alpha'^4} \int d^{10}x \sqrt{-g} \Big[ \Big( \mc{R}^4 +
\mc{R}^3 \left( G_3 G_3 + \bar{G}_3 \bar{G}_3 + G_3 \bar{G}_3 + F_5^2 + \partial \tau \cdot \partial \tau +
\nabla^2 \tau
\right) \nonumber \\
& & + \mc{R}^2((DG_3)^2 + (DF_5)^2 + G^4 + \ldots ) + \mc{R}(G_3^6 + \ldots ) + (G_3^8 + \ldots)\Big) \Big].
\nonumber \eea Terms linear in the fluxes (e.g. $\mc{R}^3 DG_3$) are forbidden as the action must be invariant
under world-sheet parity. The tensor structure and modular behaviour of the majority of these terms is unknown.
A notable exception is the $\mc{R}^4$ term, the coefficient of which is known exactly to be an Eisenstein series
in the dilaton \cite{hepth9808061}. As stressed above however, our interest is in the volume scaling, which can be extracted on
merely dimensional grounds.

Let us first consider terms independent of warping. The $\mc{R}^3 (G^2 + c.c.)$ and $\mc{R}^3 G \bar{G}$ terms
are most easily understood. These are similar to the $\mc{O}(\alpha'^0)$ $G_3 \bar{G_3}$ term but with three
extra powers of curvature. Then \be V_{\mc{R}^3 G^2} \sim \overbrace{\mc{V}^{-2}}^{\textrm{ Einstein frame}} \ti
\overbrace{\mc{V}}^{\textrm{internal integral}} \ti \overbrace{\mc{V}^{-1}}^{\mc{R}^3} \ti
\overbrace{\mc{V}^{-1}}^{G_3^2} \sim \mc{V}^{-3}. \ee The same argument tells us that a similar scaling applies
for $\mc{R}^2 (DG_3)^2$ terms, whereas $\mc{R}^2 G^4$ terms contribute as $\sim \frac{1}{V^{11/3}}$, $\mc{R}
G^6$ terms as $\sim \frac{1}{V^{13/3}}$ and $G^8$ terms as $\sim \frac{1}{V^5}$. In the absence of warping, the
$\mc{R}^4$ term does not contribute to the potential energy; integrated over a Calabi-Yau, it vanishes. This
geometric result can be understood macroscopically; were this not to vanish, it would generate a potential for
the volume even in flux-less $\mc{N} = 2$ IIB compactifications. However, it is known there that tree level
moduli remain moduli to all orders in the $\alpha'$ and $g_s$ expansions, and thus higher derivative terms must
not be able to generate a potential for them. This term contributes indirectly by modifying the dilaton
equations of motion at $\mc{O}(\alpha'^3)$; we will return to this later.

There are also higher-derivative terms dependent on the warp factor. Examples are
$$
\frac{\alpha'^3}{\alpha'^4} \int \sqrt{g} \mc{R}^4, \quad \quad \frac{\alpha'^3}{\alpha'^4} \int \sqrt{g}
\mc{R}^3 F_5^2, \quad \quad \frac{\alpha'^3}{\alpha'^4} \int \sqrt{g} \mc{R}^2 (DF_5)^2.
$$
These are similar to the corresponding tree-level terms but with three extra powers of curvature. As $\mc{R}^3
\sim \frac{1}{\mc{V}}$, the contribution of such terms is no larger than $\mc{O}(\frac{1}{\mc{V}^{11/3}})$.

There are also potential contributions from internal dilaton gradients. As we have worked in the orientifold
limit of F-theory, we have thus far regarded such terms as vanishing. However, this is not quite true. In the
presence of higher derivative corrections, a constant dilaton no longer solves the equations of motion. Instead,
we have \be \label{newdilaton} \phi(y) = \phi_0 + \frac{\zeta(3)}{16} Q(y). \ee An explicit expression for
$Q(y)$ may be found in \cite{hepth0204254}, but for our purposes it is sufficient to note that $Q \sim \mc{R}^3
\sim \frac{1}{\mc{V}}$. It is then easy to see that terms such as
$$
\int (\partial \tau \cdot \partial \bar{\tau}) \mc{R}^2 G^2 \quad \quad \textrm{ or } \quad \quad \int (\nabla^2
\phi) (DF_5)^2 \mc{R}
$$
are suppressed compared to the terms considered above. Note that dilaton-curvature terms such as
$$
\int (\nabla^2 \phi) \mc{R}^3
$$
will not contribute to the potential energy; these exist in $\mc{N} = 2$ Calabi-Yau compactifications and so
must vanish either directly or by cancellation. Similar comments apply for the fact that, even without warping,
the internal space ceases to be Calabi-Yau at $\mc{O}(\alpha'^3)$.

There is one further effect associated with (\ref{newdilaton}). As the dilaton is no longer constant, under
dimensional reduction the four-dimensional Einstein-Hilbert term is renormalised. Rescaling this to canonical
form introduces a term $\frac{V_{tree}}{\mc{V}}$ of $\mc{O}(\frac{1}{\mc{V}^3})$. However, as $V_{tree}$ is
no-scale this correction does not break the no-scale structure and in particular vanishes at the minimum.

String loop corrections first appear at $\mc{O}(\alpha'^3)$ and are thus subsumed into the above analysis. While
it may be difficult to derive anything explicitly, we may conjecture their effect in the large volume limit. The
corrections to the K\"ahler potential arise from dimensional reduction of the 10-dimensional $\mc{R}^4$ term. In
Einstein frame, the string tree-level  $\alpha'^3$ corrected K\"ahler potential derived in \cite{hepth0204254}
takes the form \be \label{Kalpha2} \K = \mc{K}_{cs} -2 \ln \left(\mc{V} + \frac{\xi}{2 e^{3 \phi/2}}\right). \ee
Here $\xi  = - \frac{\chi(M) \zeta(3)}{2(2 \pi)^3}$. $\zeta(3)$ is distinctive as the tree-level coefficient of
the ten-dimensional $\mc{R}^4$ term, the coefficient of which is known exactly to be \be
f^{(0,0)}_{\frac{3}{2}}(\tau, \bar{\tau}) = \sum_{(m,n) \neq (0,0)} \frac{e^{-\frac{3 \phi}{2}}}{\vert m + n
\tau \vert^3}. \ee This has the expansion \be f^{(0,0)}_{\frac{3}{2}}(\tau, \bar{\tau}) = \frac{2
\zeta(3)}{e^{\frac{3
      \phi}{2}}} + \frac{2 \pi^2}{3} e^{\frac{\phi}{2}} + \textrm{ instanton
  terms }.
\ee Therefore, to incorporate $\mc{O}(\alpha'^3)$ string loop corrections to $S_{IIB}$, the natural conjecture
is that we should modify (\ref{Kalpha2}) to \be \K = \K_{cs} - 2 \ln \left(\mc{V} - \frac{\chi(M)}{8(2 \pi)^3}
f^{(0,0)}_{\frac{3}{2}}(\tau, \bar{\tau}) \right). \ee

Let us finally mention further higher derivative corrections at $\mc{O}(\alpha'^4)$ and above. At large volume
these are all subleading, with any terms generated being subdominant to the $\frac{1}{\mc{V}^3}$ terms present
at $\mc{O}(\alpha'^3)$. For example, an $\mc{O}(\alpha'^4)$ term $G^2 \mc{R}^4$ would give a
$\mc{V}^{-\frac{10}{3}}$ contribution to the potential. There are other terms that would naively give a
$\frac{1}{\mc{V}^3}$ contribution, such as a possible $\mc{O}(\alpha'^6)$ term $\mc{R}^6$. However, on a
Calabi-Yau such a term vanishes, either explicitly or through cancellation. This is for the reasons discussed
above: terms present even in pure $\mc{N} = 2$ Calabi-Yau compactification cannot generate potentials for the
moduli.

Our conclusion is therefore that the leading $\alpha'$ corrections breaking the no-scale structure appear at
$\mc{O}(\frac{1}{\mc{V}^3})$, coming from $\mc{R}^3 G^2$ and $\mc{R}^2
(DG)^2$ terms in ten dimensions.\footnote{One can indeed show that if
  $\mc{V} \gg 1$ any effect that is string scale and localised in the extra dimensions
  can only contribute to the F-term potential $V$ at $\mc{O}\left(\frac{1}{\mc{V}^3}\right)$.}
 This is
consistent with the result of \cite{hepth0204254}, where the corresponding correction (\ref{Kalpha2}) to the
K\"ahler potential was computed and the resulting scalar potential interpreted as descending from such
ten-dimensional terms. Although the scaling argument used above cannot reproduce the coefficient of the
correction, it makes it easier to see that other terms are subleading; in particular, the effects associated
with non-vanishing $F_5$ (which were not considered in \cite{hepth0204254}) do not compete.

Let us now consider higher derivative corrections to localised sources. The D3-brane action is \be S_{D3} = -T_3
\int d^4x \sqrt{-g} e^{-\phi} + \mu_3 \int C_4. \ee The D3-brane
world-volume is space-filling. Therefore, so long as higher derivative
corrections arise from pull-backs onto the brane world volume then
they necessarily involve space-time derivatives and so cannot give a
potential energy.
This is not so for D7-branes;
fortunately in that case such effects were already included in the analysis of \cite{hepth0105097}. We briefly
review the relevant results. The leading $\alpha'$ correction to the D7-brane Chern-Simons action is (we do not
turn on internal D7-brane fluxes) \be S_{D7, \alpha'^2} = \frac{\mu_7}{96}(2 \pi \alpha')^2 \int_{\mbb{M}^4 \ti
\Sigma} C_4 \wedge \textrm{Tr}( \mc{R}_2 \wedge \mc{R}_2), \ee whereas the leading correction to the wrapped
D7-brane tension arises from a term \be \frac{-\mu_7}{96}(2 \pi \alpha')^2 \int_{\mbb{M}^4 \ti \Sigma} \sqrt{-g}
\textrm{Tr}(\mc{R}_2 \wedge *\mc{R}_2). \ee These contribute effective D3-brane charge and tension. In F-theory,
the D3-brane charge from the wrapped D7-branes is \be \label{FD3charge} Q_{D3} = -\frac{\chi(X)}{24}, \ee where
$X$ is the Calabi-Yau fourfold. As the D7 branes are BPS, (\ref{FD3charge}) also gives the resulting effective
D3 tension.

We can then make a similar argument that higher $\alpha'$ corrections to the D7-brane action need not be considered.
As the branes are BPS, $\alpha'$ corrections to the tension are related to $\alpha'$ corrections to the charges.
However, these involve spacetime curvature, and so the corresponding
correction to the DBI action will not give
rise to a 4d potential energy.

The arguments above have considered contributions to the 4d
effective potential coming from dimension reduction of the 10d
effective action. This style of argument knows nothing about the
low-energy physics. It is well-known that the existence of
4-dimensional supersymmetry
places considerable constraints on the type of effective action
allowed. We now analyse allowed corrections from the viewpoint of
4-dimensional supergravity.
The
combination of 4d and 10d intuition leads to the most stringent
constraints on the possible perturbative corrections that may appear.

\subsection*{Perturbative corrections in 4 dimensions}

We now use the fact that the effective theory admits
a 4d supergravity description to constrain the perturbative $g_s$ and
$\alpha'$ corrections. We will confirm the self-consistency of the
effective field theory description below in section \ref{EFT}. The supergravity
structure implies all perturbative corrections to the potential should
manifest themselves through corrections to the K\"ahler potential, as
the superpotential is not renormalised in perturbation theory. To be
specific we shall illustrate the discussion with the K\"ahler
potential appropriate to the $\mbb{P}^4_{[1,1,1,6,9]}$ model studied
above.
At tree-level this is given by
\be \mc{K} = - 2 \ln \left((T_1 + \bar{T}_1)^{3/2} -
(T_2 + \bar{T}_2)^{3/2} \right). \ee 
We recall the
geometry has one overall K\"ahler mode ($T_1$) and one blow-up mode
($T_2$). We specialise to our limit of interest $\tau_1
\gg \tau_2 > 1$, with $\tau_i = \textrm{Re}(T_i)$. The
resulting K\"ahler metric has the form (neglecting terms subleading
in $\mc{V}$) \be \label{KMetric} \mc{K}_{i \bar{j}} = \left(
\begin{array}{cc} \vspace{0.2cm} \frac{3}{\mc{V}^{4/3}} & -\frac{9
\sqrt{\tau_2}}{2 \mc{V}^{5/3}} \\
\vspace{0.2cm} -\frac{9 \sqrt{\tau_2}}{2 \mc{V}^{5/3}} & \frac{3}{2
\sqrt{2 \tau_2} \mc{V}} \\
\end{array} \right)
\ee
The central point in this analysis is that valid perturbative corrections to
the K\"ahler potentials should also give perturbative corrections to
the K\"ahler metric. If a hypothetical correction term $\delta K$
generates a correction in the metric that dominates the tree-level
metric in an appropriate classical limit (weak coupling and large
volume), then it is unphysical and should not be allowed.

We can apply this condition to restrict the form of potential
corrections to $\mc{K}$. For example, consider the possible
correction \be \mc{K} + \delta \mc{K} = - 2 \ln (\mc{V}) +
\frac{\epsilon \sqrt{2 \tau_2}}{\mc{V}^{\alpha}}, \ee with $0 <
\alpha < 1$. The $2 \bar{2}$ component of the corrected K\"ahler
metric is \be \mc{K}_{2 \bar{2}} + \delta \mc{K}_{2 \bar{2}} =
\frac{3}{2 \sqrt{2 \tau_2} \mc{V}} - \frac{\epsilon}{8 \sqrt{2}
\tau_2^{3/2} \mc{V}^{\alpha}}. \ee As $\alpha < 1$, the correction
to the kinetic term would always dominate the tree-level term in the
limit $\mc{V} \gg 1$. This seems implausible as a large-volume limit
ought to make the correction less, rather than more, important.

A similar comment applies to a correction \be \mc{K} + \delta \mc{K}
= - 2 \ln (\mc{V}) + \frac{\epsilon \tau_2^2}{\mc{V}}, \ee which
leads to \be \mc{K}_{2 \bar{2}} + \delta \mc{K}_{2 \bar{2}} =
\frac{3}{2\sqrt{2 \tau_2} \mc{V}} + \frac{\epsilon}{4 \mc{V}}. \ee
In this case, as we take $\tau_2$ large, the correction becomes
increasingly dominant over the tree-level term. Given that large
$\tau_2$ reduces both the curvature of this cycle and the gauge
coupling on any brane wrapping it, we would again expect exactly
opposite behaviour to occur.

We could also consider the correction \be \mc{K} + \delta
\mc{K} = - 2 \ln (\mc{V}) + \frac{2 \epsilon
\tau_2}{\mc{V}^{\alpha}}, \ee with $0 < \alpha < 1$. In this case
$\delta K_{2 \bar{2}}$ is subleading to $\mc{K}_{2
  \bar{2}}$ in the classical limit. However, if we consider the $1
\bar{2}$ component we now have \be \mc{K}_{1 \bar{2}} + \delta
\mc{K}_{1 \bar{2}} = -\frac{9 \sqrt{\tau_2}}{2 \mc{V}^{5/3}} +
\frac{3 \epsilon}{2
  \mc{V}^{\alpha + 2/3}},
\ee and the correction again dominates in the large-volume limit.

Generally, what the above tells us is that, for $\alpha < 1$ and  nontrivial
$f(\tau_s)$, corrections of the form
$$
\delta K = - 2 \ln (\mc{V}) + \frac{f(\tau_s)}{\mc{V}^\alpha}
$$
are excluded. Furthermore, for a correction 
$$
\delta K = - 2 \ln (\mc{V}) + \frac{f(\tau_s)}{\mc{V}}
$$
to be permissible, we must have
\be
\lim_{\tau_s \to \infty} \frac{f''(\tau_s)}{\sqrt{\tau_s}} \to 0.
\ee

We note a correction
\be
\label{KCorrectionUsed} \mc{K} + \delta \mc{K} = - 2 \ln (\mc{V}) +
\frac{\epsilon \sqrt{\tau_2}}{\mc{V}} \ee 
is not excluded by the above arguments.
As a correction to the K\"ahler metric, this would give \bea \mc{K}_{2
\bar{2}} + \delta \mc{K}_{2 \bar{2}} & = & \frac{3}{2
  \sqrt{2 \tau_2} \mc{V}} -
\frac{\epsilon}{16 \tau_2^{3/2} \mc{V}}. \eea This correction is
suppressed compared to the tree-level term by a factor $\tau^{-1}$,
i.e. $g^2$ of the field theory on the brane. It is well-behaved (i.e. subdominant) in the
classical limit, as there does not exist a `bad' scaling limit in
which it dominates the tree-level term.
We shall consider such a possible correction further in chapter \ref{ChapterAxions}.

The above arguments show that for corrections which have a nontrivial
functional dependence on the `small' moduli, the correction to $\mc{K}$
- and hence the correction to $V$ - must be suppressed by a factor of
$\mc{V}$. This implies they can contribute to the scalar potential
at a maximal order of $1/\mc{V}^3$ in the volume expansion. This is a
similar order to the $\alpha'^3$ correction. Consequently, while these corrections
might modify the exact locus of the exponentially large volume
minimum, they cannot affect the main feature, namely the
exponentially large volume.

The above arguments do not allow us to say anything about corrections
to $\mc{K}$ that depend purely on the volume, such as
\be
\delta \mc{K} = - 2 \ln(\mc{V}) + \frac{1}{\mc{V}^\alpha},
\ee
with $0 < \alpha < 1$. No evidence exists for such corrections arising
from dimensional reduction of local ten-dimensional terms. 
They may still be generated by non-local effects. For example, 
there exists a string loop calculation of
corrections to the K\"ahler potential in IIB orientifold
backgrounds \cite{hepth0508043}. This found the existence of an effective correction
\be
\label{bhkcorrection}
\delta \mc{K} =- 2 \ln(\mc{V}) + \frac{1}{\mc{V}^{2/3}} + \mc{O}(\mc{V}^{-4/3}).
\ee
Naively such a correction will at large volume dominate in the scalar
potential over the $1/\mc{V}$
$\alpha'^3$ correction. However, in fact this is not the
case. Explicit calculation shows that for the correction (\ref{bhkcorrection})
\be
(\mc{K}^{i \bar{j}}\mc{K}_i \mc{K}_j - 3) = 0 + \frac{\alpha(2/3
  - 2/3)}{\mc{V}^{2/3}} + \frac{\beta}{\mc{V}^{4/3}},
\ee
and so the expected contribution at $\mc{O}(\mc{V}^{-2/3})$ is
absent. It is amusing that this cancellation occurs only for
$\mc{V}^{-2/3}$ corrections, which is precisely the case that is relevant.

We have used the above arguments to constrain perturbative
corrections to $\mc{K}$. A similar argument shows that the
non-perturbative correction
\be
\delta K = -2 \ln (\mc{V}) + e^{-2 \pi (T_s + \bar{T}_s)},
\ee
which naively would appear relevant, must also be absent as one can
again find a classical limit in which the `corrections' to the
K\"ahler metric would dominate the tree-level term.

We note that for the case where K\"ahler corrections have been
computed \cite{hepth0508043}, the corrections do fit into the
form above. We only focus on the K\"ahler moduli dependence: the full
expressions can be found in \cite{hepth0508043}. The correction
gives \be \mc{K} + \delta \mc{K} = - \ln (T_1 + \bar{T}_1) - \ln
(T_2 + \bar{T}_2) - \ln (T_2 + \bar{T}_3) + \sum_{i=1}^3
\frac{\epsilon_i}{T_i + \bar{T}_i}, \ee with for example \be
\mc{K}_{1 \bar{1}} + \delta \mc{K}_{1 \bar{1}} = \frac{1}{4
\tau_1^2} + \frac{\epsilon_1}{4 \tau_1^3}. \ee The loop-corrected
K\"ahler metric is suppressed by a factor $\tau_1^{-1} = g^2$.

The point of this discussion is that the form of possible
corrections to the K\"ahler potential can be heavily constrained by
the reasonable requirement that in a classical (large volume, weak
coupling) limit, corrections to the metric become increasingly
subdominant to tree-level terms: simply because K\"ahler corrections
are very hard to calculate does not make us entirely ignorant of
their form. In particular, denoting the `small', blow-up moduli by
$\tau_i$, these considerations exclude corrections of the form \be
\mc{K} + \delta K = - 2 \ln (\mc{V}) +
\frac{f(\tau_i)}{\mc{V}^{\alpha}}, \ee with $\alpha < 1$, as in a
classical, large volume limit there will be metric components whose
correction dominates the tree-level term. Corrections that are allowed
(with $\alpha > 1$) contribute terms at most of order
$\frac{1}{\mc{V}^3}$ to the scalar potential.

It is often said that, because the K\"ahler potential is
non-holomorphic, there is no control over its form. The arguments above
show that in certain cases we may restrict the form of possible
corrections without being able to do a quantum computation. This
is due to the existence of a classical limit in which quantum
corrections should be sub-dominant to the tree-level terms.

This justifies the neglect of higher perturbative and non-perturbative
corrections to the K\"ahler potential. The underlying reason why these
can be neglected is the exponentially large volume: the $\alpha'$
expansion is an inverse volume, and $\mc{V} \sim 10^{15}$ (for example)
implies that this expansion can be controlled.

\section{Moduli Spectroscopy}
\label{spectrosec}
\linespread{1.3}

We have shown above both that $\alpha'$ corrections must generally be
included to study K\"ahler moduli stabilisation, and that at
large volume we only need include the leading corrections of
\cite{hepth0204254}. We now commence our study of the phenomenological
properties of the models developed above by computing the moduli and modulino
spectra. It is easy (by small adjustments of $g_s N$) to stabilise
the overall volume at essentially any value. Our primary interest is
therefore in the scaling behaviour of the masses with the 
internal volume.

We focus on the 2-modulus $\mbb{P}^4_{[1,1,1,6,9]}$ model. As described above,
in the limit where $\tau_5 \gg \tau_4 > 1$, the scalar potential takes the form
\be
\label{LargeTau5Pot}
V = \frac{\lambda \sqrt{\tau_4} (a_4 A_4)^2 e^{-2 \frac{a_4
    \tau_4}{g_s}}}{\cal{V}} -
\frac{\mu W_0 (a_4 A_4) \tau_4 e^{- \frac{a_4 \tau_4}{g_s}}}{{\cal{V}}^2}
+ \frac{\nu \xi W_0^2}{{\cal{V}}^3},
\ee
where $\lambda$, $\mu$ and $\nu$ are model-dependent numerical
    constants.

\subsection{Bosonic Fields}
\label{MassesSection}
We set $\hbar = c = 1$ but will otherwise be pedantic on frames and factors of $2 \pi$ and $\alpha'$.
Our basic length will be $l_s = 2 \pi \sqrt{\alpha'}$ and our basic mass $m_s = \frac{1}{l_s}$.
These represent the only dimensionful scales, and unless specified
otherwise volumes are measured in units of $l_s$. We will furthermore require that
at the minimum the 4-dimensional metric is the metric the string worldsheet couples to.

We give notice here that this section unavoidably contains one messy
point. This involves the relations between string and Einstein frame
volumes. The origin of the messiness is that it is the string frame
volume that is the physical volume as seen by the string and which
determines the validity of string perturbation theory. However, it is
the Einstein frame volume - which includes factors of $g_s$ - which appears
in the supergravity K\"ahler potential and in defining the moduli.

We now study the mass spectrum.
Stringy excitations have
\be
m_S^2 = \frac{n}{\alpha'} \Rightarrow m_S \sim 2 \pi m_s.
\ee
To estimate Kaluza-Klein masses we recall toroidal
compactifications. A stringy ground state of Kaluza-Klein and winding integers
$n$ and $w$ has mass
\be
\label{KKmasses}
m_{KK}^2 = \frac{n^2}{R^2} + \frac{w^2 R^2}{\alpha'^2},
\ee
where $R$ is the dimensionful Kaluza-Klein radius.
Strictly (\ref{KKmasses}) only holds for toroidal compactifications, but it should
suffice to estimate the relevant mass scale.
If we write $R = R_s l_s$ and assume $R_s \gg 1$, we have
\be
\label{NewKKmasses}
m_{KK} \sim \frac{m_s}{R_s} \quad \textrm{ and } \quad m_W \sim (2 \pi)^2 R_s m_s.
\ee
It is conceivable that the geometry of the internal space is elongated such
that the Kaluza-Klein radius $R_s$ is
uncorrelated with the overall volume, giving
KK masses of order $m_{KK}^4\sim 1/\tau_i$ for
the different cycles.
However, in absence of evidence to the contrary we assume
the simplest scenario in which $(2 \pi R_s)^6 = \mc{V}_s$.
Then
\be
\label{KKMass}
m_{KK} \sim \frac{2 \pi m_s}{{\cal V}_s^{\frac{1}{6}}}. %\ quad \textrm{ and } \quad
%m_W \sim (2 \pi) {\cal V}_s^{\frac{1}{6}} m_s.
\ee
Here $m_{KK}$ refers only to the lightest KK mode, as in our
situation the overall volume is large but there are
relatively small internal cycles. Therefore, while there may be many
KK modes, there is a hierarchy  with the others being naturally
heavier than the scale
of (\ref{KKMass}).

We next want to determine the masses of the complex structure and K\"ahler moduli.
This requires an analytic expression for the potential
in terms of canonically normalised fields.
The dimensional reduction of the 10-dimensional action into this
framework is carried out in more detail in section \ref{appendixsec1} of
the Appendix; here we shall just state results.

An $\mc{N} = 1$ supergravity is completely specified by a K\"ahler potential, superpotential and gauge kinetic function.
Neglecting the gauge sector to focus on moduli dynamics, the action is
\be
S_{\mc{N}=1} = \int d^4 x \sqrt{-G} \left[ \frac{M_P^2}{2} \mc{R} - \K_{i\bar{j}} D_\mu \phi^{i} D^\mu \bar{\phi}^j
- V(\phi, \bar{\phi}) \right],
\ee
where
\be
\label{ScalarPotential2}
V(\phi, \bar{\phi}) = e^{\K/M_P^2} \left(\K^{i \bar{j}} D_i \hat{W} D_{\bar{j}} \bar{\hat{W}} - \frac{3}{M_P^2} \hat{W}
\bar{\hat{W}} \right) + \textrm{ D-terms}.
\ee
$\K$ is the K\"ahler potential, which has mass dimension 2, and $\hat{W}$ the superpotential, with mass dimension 3.
$M_P$ is the reduced Planck mass $M_P = \frac{1}{(8 \pi G)^\half} =
2.4 \ti 10^{18} \textrm{GeV}$ and the Planck and string scales are related by
\be
M_P^2 = \frac{4 \pi \mc{V}_s^0}{g_s^2 l_s^2} \quad \textrm{ or } \quad
m_s = \frac{g_s}{\sqrt{4 \pi \mc{V}_s^0}} M_P.
\ee
Here $\mc{V}_s^0 = \langle \mc{V}_s \rangle$ is the string-frame
volume at the minimum. $\mc{V}_s$ is defined as
\be
\mc{V}_s = \int_X d^6 x \sqrt{\tilde{g}}.
\ee
It measures the volume of the internal space using the metric
$\tilde{g}$ defined by
\be
\tilde{g}_{MN} = e^{(\phi_0 - \phi)/2} g_{MN},
\ee
where $g_{MN}$ defines the ten-dimensional string frame metric. The
factor of $e^{-\phi/2}$ in $\tilde{g}$ ensures it is an Einstein-frame
metric, but the factor $e^{\phi_0 / 2}$ also ensures that for the
vacuum solution volumes measured with $\tilde{g}_{MN}$ agree with those
measured with $g_{MN}$.

As described in the Appendix, including the $\alpha'$ and non-perturbative corrections, the
K\"ahler and superpotentials are
\bea
\label{Potentials}
\frac{\mc{K}}{M_P^2} & = & - 2 \ln \left(\mc{V}_s + \frac{ \xi g_s^{\frac{3}{2}}}{2 e^{\frac{3 \phi}{2}}} \right)
- \ln(S + \bar{S}) - \ln \left(-i \int_{CY} \Omega \wedge \bar{\Omega}\right), \nonumber \\
\hat{W}  & = & \frac{g_s^{\frac{3}{2}} M_P^3}{\sqrt{4 \pi} l_s^2} \left( \int_{CY}
 G_3 \wedge \Omega + \sum A_i e^{\frac{- 2 \pi}{g_s} T_i} \right)
\equiv \frac{g_s^{\frac{3}{2}} M_P^3}{\sqrt{4 \pi}} W.
\eea
Here $\xi = -\frac{\zeta(3) \chi(M)}{2 (2 \pi)^3}$ and
$T_i = \tau_i + i b_i$, where $\tau_i = \int_{\Sigma_i} d^4 x
\sqrt{\tilde{g}}$ is a  4-cycle volume and $b_i$ its axionic
partner arising from the RR 4-form;
these are good K\"ahler coordinates for IIB orientifold compactifications.

The gravitino mass can be read off immediately from
(\ref{Potentials}): 
\be
\label{GravitinoMass}
m_{\frac{3}{2}} = e^{\K/2} \vert \hat{W} \vert = \frac{g_s^2
  e^{\frac{\K_{cs}}{2}} \vert W_0 \vert}{{\cal V}_s^0 \sqrt{4 \pi}} M_P.
\ee
It will be useful to relate the scale of the bosonic masses to (\ref{GravitinoMass}).

To calculate scalar masses, we must express the potential (\ref{ScalarPotential2}) in terms of
canonically normalised fields.
The K\"ahler metric for the complex
structure moduli is given by
\be
\label{CSmetric}
\mc{K}_{i \bar{j}} = \partial_i \partial_{\bar{j}} \mc{K}_{cs} = \partial_i
\partial_{\bar{j}} \ln \left( - i \int_{CY} \Omega \wedge \bar{\Omega} \right).
\ee
In the no-scale approximation, which holds to leading order, the potential for the complex structure moduli is
\be
V = \frac{g_s^4 M_P^4}{8 \pi (\mc{V}_s^0)^2} \int d^4 x \sqrt{-g_E} e^{\mc{K}_{cs}} \left[ G^{a \bar{b}}
D_a W D_{\bar{b}} \bar{W}  \right],
\ee
where the sum runs over complex structure moduli only.
The inverse of (\ref{CSmetric}) is hard to make explicit, as to do so
would require knowledge of all the Calabi-Yau
periods.
However, as (\ref{CSmetric}) is independent of dilaton and K\"ahler
moduli, this process will not introduce
extra factors of $\mc{V}_s$ or $g_s$. Thus if we assume numerical factors
to be $\mc{O}(1)$, we find
\be
m_{cs}^2 = \mc{O}(1) \frac{g_s^4 N^2 M_P^2}{4 \pi ({\cal V}_s^0)^2},
\ee
where $N \sim \mc{O}(\sqrt{\frac{\chi}{24}})$ is a measure of the typical number of flux quanta
and arises from the $D_a W$ terms.
We therefore have
\be
m_{cs} = \mc{O}(1) \frac{g_s N m_s}{\sqrt{\mc{V}_s^0}}.
\ee
As emphasised in \cite{hepth0309170}, one requires a clear separation between
Kaluza-Klein and complex structure masses to trust the supergravity analysis.
We have
\be
\label{csKK}
\frac{m_{cs}}{m_{KK}} \sim \frac{g_s N}{2 \pi ({\cal V}_s^0)^{\frac{1}{3}}}.
\ee
At large volumes, this ratio is much less than one, which is reassuring.

For the concrete $\mbb{P}^4_{[1,1,1,6,9]}$ example,
the K\"ahler moduli may be treated more explicitly. For this model we
shall give a completely explicit analysis in chapter \ref{ChapterAxions}. Here we focus
on scaling properties, as these generalise to multi-modulus models in
a way that numerical factors will not.

It is nonetheless hard to normalise the fields canonically across the entirety of
moduli space. However, to compute the spectrum we only need
normalise the moduli at the physical minimum.
It turns out (see section \ref{appendixsec2} of the Appendix) that the appropriately normalised fields are
\bea
\label{CanonicalFields}
\tau_5^{c} = \sqrt{\frac{3}{2}} \frac{\tau_5}{\tau_5^0} M_P, & &
b_5^{c} = \sqrt{\frac{3}{2}} \frac{b_5}{\tau_5^0} M_P, \nonumber \\
\tau_4^{c} = \sqrt{\frac{3}{4}} \frac{\tau_4}{(\tau_5^0)^{\frac{3}{4}}
  (\tau_4^0)^{\frac{1}{4}}} M_P, & &
b_4^{c} = \sqrt{\frac{3}{4}} \frac{b_4}{(\tau_5^0)^{\frac{3}{4}} (\tau_4^0)^{\frac{1}{4}}} M_P.
\eea
Here $\tau_5^0 = \langle \tau_5 \rangle$, etc. The bosonic mass matrix follows
by taking the second derivatives of the scalar potential with respect to $\tau_i^c$ and $b_i^c$.
In the vicinity of the large volume minimum, the scalar potential takes the form
\be
\label{Vgeneral2}
V = g_s^4 M_P^4 \left(\frac{\lambda' \sqrt{\tau_4} e^{-2 \frac{a_4 \tau_4}{g_s}}}{\tau_5^{\frac{3}{2}}}
+ \frac{\mu'}{\tau_5^3} \tau_4 e^{- \frac{a_4 \tau_4}{g_s}}\cos \left(
\frac{a_4 b_4}{g_s} \right) + \frac{\nu'}{\tau_5^{\frac{9}{2}}} \right).
\ee
As discussed above we have
$$
\lambda' \sim \frac{a_4^2 \vert A_4 \vert^2}{g_s^2}, \quad \mu' \sim \frac{a_4 \vert A_4 W_0 \vert}{g_s},
\textrm{ and } \nu' \sim \xi \vert W_0 \vert^2.
$$
The $b_5$ axion appears only in terms suppressed by $e^{-a_5 \tau_5}$ and has not been
written explicitly. As $\tau_5^0 \gg 1$, it follows that this field is
essentially massless.
In terms of the canonical fields, the scalar potential
(\ref{Vgeneral2}) becomes
\be
V = \frac{\lambda \sqrt{\tau_4^{c} \beta}
e^{-2 a_4 \beta
  \tau_4^{c}}}{(\tau_5^0)^{\frac{3}{2}}
(\tau_5^{c})^{\frac{3}{2}}}
 + \frac{\mu \tau_4^{c} \beta
e^{-a_4 \beta \tau_4^{c}}
\cos(a_4 \beta b_4)}{(\tau_5^0)^3 (\tau_5^{c})^3} +
\frac{\nu}{(\tau_5^0)^{\frac{9}{2}}(\tau_5^{c})^{\frac{9}{2}}},
\ee
where $\tau_4 = \beta g_s \tau_4^c$.
The mass matrix is
\be
\textrm{d}^2 V =
\left( \begin{array}{ccc} \frac{\partial^2 V}{\partial \tau_5^{c} \partial \tau_5^{c}} & \frac{\partial^2 V}{\partial
    \tau_5^{c} \partial \tau_4^{c}} & \frac{\partial^2 V}{\partial
    \tau_5^{c} \partial b_4^{c}} \\   \frac{\partial^2 V}{\partial
    \tau_4^{c} \partial \tau_5^{c}} &  \frac{\partial^2 V}{\partial
    \tau_4^{c} \partial \tau_4^{c}} & \frac{\partial^2 V}{\partial
    \tau_4^{c} \partial b_4^{c}} \\ \frac{\partial^2 V}{\partial
    b_4^{c} \partial \tau_5^{c}}& \frac{\partial^2 V}{\partial
    b_4^{c} \partial \tau_4^{c}} & \frac{\partial^2 V}{\partial
    b_4^{c} \partial b_4^{c}} \end{array}
\right).
\ee
At the minimum this
mass matrix takes the schematic form
$$
\frac{M_P^2 g_s^4}{({\cal V}_s^0)^2} \left( \begin{array}{ccc}
  \frac{a}{{\cal V}_s^0} & \frac{b}{({\cal V}_s^0)^{\half} g_s} & 0 \\
\frac{b}{({\cal V}_s^0)^{\half} g_s} & \frac{c}{g_s^2} & 0 \\ 0 & 0 & d \end{array} \right).
$$
Cross terms involving the axion decouple and we obtain
\bea
\label{Masses}
m_{\tau_5^{c}} = \mc{O}(1) \frac{g_s^2 W_0}{\sqrt{4 \pi}
  ({\cal V}_s^0)^{\frac{3}{2}}} M_P, & &  m_{b_5^{c}} \sim \exp
  (-\tau_5^0) M_P,
\nonumber \\
m_{\tau_4^{c}} = \mc{O}(1) \frac{a_4 g_s W_0}{\sqrt{4 \pi} {\cal V}_s^0}
M_P, & &  m_{b_4^{c}} = \mc{O}(1)
\frac{a_4 g_s W_0}{\sqrt{4 \pi}{\cal V}_s^0} M_P.
\eea
This division of scales between the large modulus ($\tau_5$) and the small
modulus ($\tau_4$) is a general feature of these models.
The $\mc{O}(1)$ factors depend on the detailed geometry of the
particular Calabi-Yau and are therefore not written explicitly, although
given the K\"ahler potential their numerical
computation is straightforward.

Comparing formulae (\ref{Masses}) with (\ref{GravitinoMass}) suggests
the existence of a small hierarchy between the small modulus $\tau_4$
and the gravitino mass $m_{3/2}$ set by $\frac{a_4}{g_s}$. The above
analysis has focused on the scaling of masses with volume
$\mc{V}_s^0$. We will see in chapter \ref{ChapterAxions} that a precise treatment
indeed gives a small hierarchy between $m_{\tau_4}$ and $m_{3/2}$,
which is in fact set by $\left(\frac{a_4 \tau_4}{g_s}\right) \sim \ln
\left( \frac{M_P}{m_{3/2}} \right)$.
$$
\frac{m_{\tau_4}}{m_{3/2}} \sim \frac{a_4 \tau_4}{g_s} \sim \ln\left(\frac{M_P}{m_{3/2}}\right).
$$

The $\tau_4$ and $b_4$ moduli have masses similar to - indeed
marginally heavier than - the dilaton and
complex structure moduli.
We may worry that our above treatment was inconsistent, as
we first integrated out complex structure moduli
and only then considered K\"ahler moduli. However, as argued in
section \ref{sseLVL},
we see from the form of the full potential
that the solution derived by first integrating out
the complex structure moduli
remains a minimum of the full potential. Physically, the reason why we
can get away with `integrating out' light fields is that the two
sectors - K\"ahler and complex structure moduli - decouple, as is
evident from the K\"ahler potential (\ref{Potentials}).

\subsection{Fermionic Fields}

The fermions divide into the gravitino and the fermionic partners of
the chiral superfields. The gravitino mass is given by
\be
m_{\frac{3}{2}} = e^{\K/2} \vert \hat{W} \vert = \frac{g_s^2
  e^{\frac{\K_{cs}}{2}} \vert W_0 \vert}{{\cal V}_s^0 \sqrt{4 \pi}} M_P.
\ee
with $\K$ and $\hat{W}$ given by (\ref{Potentials}).

We use expressions appropriate for $V=0$ and so
assume we have included lifting terms to restore a Minkowski minimum.
The mass matrix for the other fermions is then $[M_\psi]_{ij}
\bar{\psi}_{Li} \psi_{Lj}$, where
$[M_\psi]_{ij} = \sum_{n=1}^4 [M_\psi^n]_{ij}$, with
\bea
\label{FermionMasses}
\left[ M_{\psi}^{1} \right]_{ij} & = & -e^{\K/2} \vert \hat{W} \vert \left\{ \K_{ij} + \frac{1}{3}
  \K_i \K_j \right\}, \nonumber \\
\left[ M_{\psi}^{2} \right]_{ij} & = & -e^{\K/2} \vert \hat{W} \vert \left\{ \frac{\K_i \hat{W}_j +
     \K_j \hat{W}_i}{3 \hat{W}} - 2 \frac{\hat{W}_i \hat{W}_j}{3 \hat{W}^2} \right\}, \nonumber \\
\left[ M_{\psi}^{3} \right]_{ij} & = & - e^{\K/2} \sqrt{\frac{\bar{\hat{W}}}{\hat{W}}} \hat{W}_{ij}, \nonumber \\
\left[ M_{\psi}^{4} \right]_{ij} & = & e^{\mc{G}/2} \mc{G}_l (\mc{G}^{-1})_k^l \mc{G}_{ij}^k,
\eea
where $\mc{G} = \K + \ln(\hat{W}) + \ln(\bar{\hat{W}})$, $\hat{W}_i = \partial_i \hat{W}$, $\K_i
  = \partial_i \K$, $\mc{G}^l = \partial_{\bar{l}} \mc{G}$,
  etc. Here derivatives are with respect to the canonically normalised
  fields (\ref{CanonicalFields}).

There is one massless fermion corresponding to the goldstino,
eaten by the gravitino in the superHiggs effect. This corresponds to
the fermionic partner of the field breaking supersymmetry, which is essentially 
$\tilde{\tau}_5$, although there is some small mixing with $\tilde{\tau}_4$.
The mass of $\tilde{\tau}_4$ can be estimated from
(\ref{FermionMasses}); we find
\be
m_{\tilde{\tau_4}} \approx \frac{g_s^2 a_4 W_0}{{\cal V}_s} M_P \approx m_{\tau_4} \approx m_{3/2}.
\ee
As with the bosonic spectrum, it is hard to obtain explicit expressions for modulino masses
for the complex structure moduli. However, there is no explicit volume dependence in $\K_{cs}$ or $W_{flux}$,
and so the volume dependence of $m_{\tilde{\phi}}$ is determined by the $e^{\frac{\K}{2}}$ terms. Therefore
\be
m_{\tilde{\phi}} \sim \frac{g_s^2 W_0}{{\cal V}_s} M_P, \qquad m_{\tilde{\tau}}
\sim \frac{g_s^2 W_0}{\mc{V}_s} M_P.
\ee
and modulino masses have a scale
set by the gravitino mass. Thus, as expected, $m_{3/2}$ determines the scale of Bose-Fermi
splitting. As at large volume $m_{3/2} \ll m_s, m_{KK}$, the moduli and
modulino physics should decouple from that associated with stringy or
Kaluza-Klein modes.

\subsection{Properties of the Spectrum}

At large volume, the single most important factor in determining the moduli masses is
the stabilised internal volume. The different scales are suppressed compared to the 4-dimensional
Planck scale by various powers of the internal volume. In table \ref{ModuliTable1} we show this
scaling explicitly for the various moduli. There are also
model-dependent $\mc{O}(1)$ factors, which we do not show explicitly.
\begin{table}
\caption{Moduli spectrum for $\mbb{P}^4_{[1,1,1,6,9]}$ in terms of
  $\mc{V}_s^0 = \langle \mc{V}_s \rangle$}
\label{ModuliTable1}
\centering
\vspace{3mm}
\begin{tabular}{|c|c|}
\hline
Scale & Mass \\
\hline
\textrm{4-dimensional Planck mass} & $\frac{4 \pi \mc{V}_s^0}{g_s}
m_s = M_P$ \\
\textrm{String scale } $m_s$ & $m_s = \frac{g_s}{\sqrt{4 \pi \mc{V}_s^0}} M_P$ \\
\textrm{Stringy modes} $m_S$ & $2 \pi m_s = \frac{g_s \sqrt{\pi}}{\sqrt{\mc{V}_s^0}} M_P$ \\
\textrm{Kaluza-Klein modes} $m_{KK}$ & $\frac{2 \pi}{({\cal
    V}_s^0)^{\frac{1}{6}}} m_s =
\frac{g_s \sqrt{\pi}}{(\mc{V}_s^0)^{\frac{2}{3}}} M_P$ \\
\textrm{Gravitino} $m_{3/2}$ & $\frac{g_s W_0}{\sqrt{\mc{V}_s^0}} m_s =
\frac{g_s^2 W_0}{\sqrt{4 \pi} \mc{V}_s^0} M_P$ \\
\textrm{Dilaton-axion } $m_{\tau}$
& $\frac{g_s N}{\sqrt{\mc{V}_s^0}} m_s = \frac{g_s^2 N}{\sqrt{4 \pi} \mc{V}_s^0} M_P$
\\
\textrm{Complex structure moduli } $m_{\phi}$ &
$\frac{g_s N}{\sqrt{\mc{V}_s^0}} m_s = \frac{g_s^2 N}{\sqrt{4 \pi} \mc{V}_s^0} M_P$
\\
\textrm{`Small' K\"ahler modulus } $m_{\tau_4}, m_{\tilde{\tau_4}}$ & $\frac{
  a_4 \tau_4 W_0}{\sqrt{\mc{V}_s^0}} m_s =
\frac{a_4 \tau_4 g_s W_0}{\sqrt{4 \pi} \mc{V}_s^0} M_P$
\\
\textrm{Modulinos } $m_{\tilde{\tau}}, m_{\tilde{\phi}}$ & $\frac{g_s
  W_0}{\sqrt{\mc{V}_s^0}} m_s =
\frac{g_s^2 W_0}{\sqrt{4 \pi} \mc{V}_s^0} M_P$
\\
\textrm{`Large' volume modulus} $m_{\tau_5}$ & $\frac{g_s
  W_0}{\mc{V}_s^0} m_s = \frac{g_s^2 W_0}{\sqrt{4 \pi} (\mc{V}_s^0)^{\frac{3}{2}}} M_P$ \\
\textrm{Volume axion} $m_{b_5}$ & $\exp( - (\mc{V}_s^0)^{\twothirds}) M_P \sim 0 $ \\
\hline
\end{tabular}
\end{table}

This spectrum has several characteristic features. First, the string scale is hierarchically
smaller than the Planck scale. The internal volume depends exponentially on the inverse string coupling
and thus very small changes in the stabilised dilaton lead to large
effects in the compact space.
The above construction therefore represents the first stringy
construction of large extra dimensions.\footnote{It is good to keep in
  mind that `large' does not mean macroscopic. With six extra
  dimensions and $m_s \sim 1 \hbox{TeV}$, the radii of the extra
  dimensions are given by $r \sim \mc{V}^{\frac{1}{6}} m_s^{-1} \sim
  10^{-14}\hbox{m}$.} We note that in this framework
the extra dimensions are isotropic, with all six dimensions having
comparable radii.

The majority of moduli masses are stabilised at a high scale $\mc{O}\left(\frac{M_P}{\mc{V}_s^0}\right)$,
comparable to $m_{\frac{3}{2}}$ but below the scale of stringy and Kaluza-Klein modes.
There are two light moduli; the radion and its associated axion.
The latter has an extremely small mass that is light in any units. As an axion, one would like to
use this as a solution to the strong CP problem. Unfortunately this
axion corresponds precisely to the D7 gauge theory with gauge
coupling determined by the inverse size of the large 4-cycle. As this is 
extremely small, we do not expect the Standard Model to
live on such branes. It may be possible to use axions associated
with a small cycle as a QCD axion: we shall develop this idea further in
chapter \ref{ChapterAxions}. The radion mass is hierarchically lighter than the
gravitino mass: this may have cosmological implications
which it would also be interesting to explore further.

The principal factor entering the scales is the internal volume, and we present in table
\ref{ModuliTable2} possible spectra arising from various choices of the
internal volume. We reemphasise here that the reason we can
talk about `choices' of the volume is that its stabilised value is
exponentially sensitive to $\mc{O}(1)$ parameters such as the string
coupling, and thus it may be dialled freely from the Planck to TeV scales.
The moduli spectra are shown for GUT, intermediate
and TeV string scales.
If the fundamental scale is at the GUT scale, all moduli are heavy
with the exception of the
light $b_5$ axion. For the intermediate scale, the volume modulus $\tau_5$ is relatively light. Its mass does not
present a problem with fifth force experiments but may be problematic in a cosmological context
\cite{hepph9308292, hepph9308325}. Finally, for TeV scale gravity all
moduli are very light.
In particular, $\tau_5$ is now so light ($10^{-17}$ eV) as
to be in conflict with fifth force experiments. It would then be
difficult to realise this scenario unless either observable matter
did not couple to $\tau_5$ or its mass received extra corrections.
\begin{table}
\caption{Moduli spectra for typical GUT, intermediate and TeV string scales}
\label{ModuliTable2}
\centering
\vspace{3mm}
\begin{tabular}{|c|c|c|c|c|}
\hline
Scale & Mass & GUT & Intermediate & TeV \\
\hline
$M_P$ & $M_P$ & $2.4 \ti 10^{18}$ GeV &  $2.4 \ti 10^{18}$ GeV &  $2.4 \ti 10^{18}$ GeV\\
$m_s$ & $\frac{g_s}{\sqrt{4 \pi \mc{V}_s^0}} M_P$ & $1.0 \ti 10^{15}$ GeV & $1.0 \ti 10^{12}$ GeV& $1.0 \ti 10^{3}$ GeV\\
$m_S$ & $2 \pi m_s = \frac{g_s \sqrt{\pi}}{\sqrt{\mc{V}_s^0}} M_P$ & $6 \ti 10^{15}$ GeV
& $6 \ti 10^{12}$ GeV& $6 \ti 10^{3}$ GeV\\
$m_{KK}$ & $\frac{2 \pi m_s}{(\mc{V}_s^0)^{\frac{1}{6}}} =
\frac{g_s \sqrt{\pi}}{(\mc{V}_s^0)^{\frac{2}{3}}} M_P$ & $1.5 \ti 10^{15}$ GeV & $1.5 \ti 10^{11}$ GeV
& $0.15$ GeV \\
$m_{3/2}$ & $\frac{g_s^2 W_0}{\sqrt{4 \pi} \mc{V}_s^0} M_P$ &
$1.5\ti 10^{12}$ GeV & $1.5\ti 10^6$GeV & $1.5\ti 10^{-12} $GeV\\
$m_\tau$ & $\frac{g_s N m_s}{\sqrt{\mc{V}_s^0}} = \frac{g_s^2 N}{\sqrt{4 \pi} \mc{V}_s^0} M_P$ &
$1.5\ti 10^{12}$ GeV & $1.5\ti 10^6$GeV & $1.5\ti 10^{-12} $GeV\\
$m_{cs}$ & $\frac{g_s N m_s}{\sqrt{\mc{V}_s^0}} = \frac{g_s^2 N}{\sqrt{4 \pi} \mc{V}_s^0} M_P$ &
$1.5\ti 10^{12}$ GeV & $1.5\ti 10^6$GeV & $1.5\ti 10^{-12} $GeV\\
$m_{\tau_4}, m_{b_4}$ & $\frac{a_4 g_s W_0}{\sqrt{4 \pi} \mc{V}_s^0} M_P$  &
$7.5\ti 10^{12}$ GeV & $1.5\ti 10^7$GeV & $1.5\ti 10^{-10} $GeV\\
$m_{\tau_5}$ & $\frac{g_s^2 W_0}{\sqrt{4 \pi} (\mc{V}_s^0)^{\frac{3}{2}}} M_P$
& $2.2 \ti 10^{10}$ GeV & 22 GeV &  $2.2\ti 10^{-26}$ GeV\\
$m_{b_5}$ & $\exp( - a_5 \tau_5) M_P \sim 0 $ & $\sim 10^{-300}$ GeV &
$\exp( -10^6)$ GeV&
$\exp(-10^{18})$ GeV\\
\hline
\end{tabular}
\end{table}

Overall, the clearest feature of the spectra in tables
\ref{ModuliTable1} and \ref{ModuliTable2} is the
presence of significant hierarchies governed by the stabilised
volume. It is a very interesting that a single quantity (the overall
volume) is capable of generating such large hierarchies. In addition
to the large hierarchies, there is also a small hierarchy between $m_{\tau_4}$ 
and $m_{3/2}$. We shall discuss this small hierarchy in more detail in chapter \ref{chapterSoftSusy}.

\section{On The Validity of Effective Field Theory}
\label{EFT}

Given these results, we can return and check that the
four-dimensional effective field theory is self-consistent.
Our use of an $\mc{N} =1 $ supergravity framework
should be valid so long as there is a separation of scales in which
the 4-dimensional physics decouples from the high-energy physics.
Let us enumerate the consistency conditions that a candidate minimum must satisfy
in order to trust the formalism.
First,
\be
\label{Constraint1}
\langle \mc{V}_s \rangle = \langle \mc{V}_E g_s^{\frac{3}{2}} \rangle
\gg 1.
\ee
To control the $\alpha'$ expansion, the string-frame compactification
volume must be much greater than unity. It is important to keep track
of frames here;
this condition is often incorrectly stated with an Einstein frame
$(g_{\mu \nu, s} = e^{\phi/2} g_{\mu \nu, E})$ volume. However, in Einstein frame the $\alpha'$
corrections come with inverse powers of $g_s$ (as in (\ref{alpha'correction})) and
thus the string frame volume is the correct measure of the reliability
of the perturbative expansion.

Secondly, we require
\bea
\label{Constraint2}
 \langle V \rangle & \ll & m_s^4, \nonumber \\
\langle V \rangle & \ll & m_{KK}^4.
\eea
The vacuum energy must be suppressed compared to the string and
compactification scales. Otherwise, the neglect of stringy and Kaluza-Klein modes
in the analysis is untrustworthy. To trust the moduli potential we likewise
require the decoupling of the moduli masses, namely 
\be
\label{Constraint3}
m_{\frac{3}{2}}, m_{\phi}, m_{\tilde{\phi}} \ll m_s, m_{KK},
\ee
where $\phi$ and $\tilde{\phi}$ are generic moduli and modulini.

There is one further potential constraint we wish to discuss. The
$\mc{N}=1$ SUGRA potential can be written as
\be
V_{\mc{N}=1} = \underbrace{e^\K\left[\mc{K}^{i \bar{j}}D_i W D_{\bar{j}}
    \bar{W}\right]}_{\textrm{F-term energy}}
- \underbrace{3e^\K \vert W \vert^2}_{\textrm{gravitational energy}},
\ee
and we may thus define a susy breaking energy $m_{susy}$, by $m_{susy}^4 =
e^\K\left[\mc{K}^{i \bar{j}} D_i W D_{\bar{j}} \bar{W}\right]$. In no-scale models, $m_{susy}^4$ cancels against
$3 e^\K \vert W \vert^2 $ to give a vanishing cosmological constant. Is it necessary to
require $m_{susy} \ll m_{KK}, m_s$? As $m_{susy} \sim
W_0 m_s$, imposing this would lead to the constraints
\bea
\label{MSusyConstraints}
W_0 & \ll & 1 \nonumber \\
W_0 \langle \mc{V}_s \rangle^{\frac{1}{3}} & \ll & 1.
\eea
Normally, there is a twofold reason for requiring $m_{susy} \ll m_s$.
First, once susy is broken the vacuum energy is of $\mc{O}(m_{susy}^4)$
and secondly, the boson-fermion mass splittings are $\mc{O}(m_{susy})$. However, neither
reasons are valid here. In a no-scale model, irrespective of the value of $m_{susy}$, the vacuum
energy vanishes. Furthermore, the boson associated with supersymmetry breaking is
massless in the no-scale approximation (as the potential is flat), whereas the associated fermion
is the goldstino that is eaten by the gravitino. It is the gravitino that sets the scale of
Bose-Fermi splitting, but this has mass $m_{3/2} = e^{\frac{\K}{2}}
\vert W \vert \sim \frac{m_{susy}}{\sqrt{{\cal{V}}}}$
at large volumes.
Requiring $m_{3/2} \ll m_{KK},
m_s$ leads to the much weaker constraint
\be
\label{WeakConstraint}
W_0 \ll \langle \mc{V}_s \rangle^{\frac{1}{3}}.
\ee
Thus $m_{susy}$ as defined above is an imaginary
scale; it sets neither the scale of Bose-Fermi splitting
nor the vacuum energy. Thus we shall only consider (\ref{Constraint1}),
(\ref{Constraint2}) and (\ref{Constraint3}) as relevant constraints.
We should note that this distinction only arises in a
large extra dimensions scenario. If $\langle \mc{V}_s \rangle \sim
\mc{O}(1)$, the conditions (\ref{MSusyConstraints}) and
(\ref{WeakConstraint}) coalesce.

There is finally a consistency condition peculiar to IIB
flux compactifications. If $\tau_i$ are the divisor volumes, we
require
\be
\label{Constraint4}
\frac{a_i \tau_{i,s}}{g_s} = a_i \tau_{i,E} \gg 1.
\ee
This allows us to neglect
multi-instanton contributions.

Let us now consider these constraints as applied to our model.
As the stabilised volume is exponentially large, (\ref{Constraint1}) is trivially
satisfied. The conditions (\ref{Constraint2}) depend on the vacuum energy
at the minimum. After the de Sitter uplift, these are satisfied by construction.
Before the uplift, we recall that the vacuum energy at the minimum is
$\mc{O}(\frac{W_0^2 M_P^4 g_s^4}{4 \pi {\cal{V}}^3})$,
whereas $m_{KK}^4 \sim \frac{\pi^2 g_s^4
  M_P^4}{{\cal{V}}^{\frac{8}{3}}}$. This gives a restriction
\be
W_0 \ll \pi^{\frac{3}{2}} \langle \mc{V}_s \rangle ^{\frac{1}{6}}.
\ee
At large volumes this is not an onerous condition to satisfy.

The particle mass constraints are likewise satisfied. The most dangerous of these is the requirement
$m_{3/2}, m_{\phi} \ll m_{KK}$. From (\ref{csKK}) we require
\be
\frac{g_s N}{2 \pi} \ll \langle \mc{V} \rangle^{\third},
\ee
were $N$ is a measure of the typical number of flux
quanta\footnote{Not to be
confused with $N$ measuring the rank of the
  hidden sector group that enters in the coefficients $a_i$ for
  gaugino condensation superpotentials. } and can be
taken
to be $\mc{O}(\frac{\chi}{24}) \sim 30$. At the large volumes we work
at this constraint is satisfied comfortably.
The constraints on the divisor volumes are also satisfied at large
volume, as for the
`small' divisor $\tau_4$, $a_4 \tau_4 \sim \ln {\cal V}$ at the minimum.

Thus all consistency conditions are satisfied and we see no reason to
regard the use of $\mc{N} = 1$ supergravity as inconsistent.
An important point is that we obtain no strong constraints on the value of $W_0$;
this is in contrast to KKLT-type solutions, for which
very small values of $W_0$ are essential. As large values of $W_0$ are
preferred by the statistical results of Douglas and collaborators
\cite{hepth0307049, hepth0404116, hepth0411183}, we expect a
typical solution to have large $W_0$. The maximum value of $W_0$ is
determined by the fluxes satisfying the tadpole conditions and can 
in general be $\mc{O}(10-100)$.

In KKLT constructions the constraint (\ref{Constraint1}) leads to the
requirement $W_0 \ll 1$ and $g_s$ not too small. To be more precise,
as $A_i e^{- \frac{a_i \tau_i}{g_s}} \sim W_0$, this gives
$$
-\frac{g_s \ln (W_0)}{a_i} \gg 1.
$$
If this is satisfied then satisfying (\ref{Constraint2}) is automatic.
As in such compactifications $\langle \mc{V}_s \rangle^{\third} \sim
\mc{O}(1)$, (\ref{Constraint3}) gives
$$
\frac{g_s N}{2 \pi} \ll \mc{O}(1).
$$
This is in general hard to satisfy (as also pointed out in \cite{hepth0309170}).

\section{The Many-Moduli Case}
\label{GenSection}
Our discussion of cycle sizes and moduli masses
has so far focused on a particular model,
${\mathbb{P}}^4_{[1,1,1,6,9]}$. We now argue that the above framework
will extend 
to more general Calabi-Yaus with $h^{2,1} > h^{1,1}$, so long as we assume 
the existence of appropriate non-perturbative superpotentials.

In the ${\mathbb{P}}^4_{[1,1,1,6,9]}$ example, we found that of
the two K\"ahler moduli one ($\tau_5$) was large,
whereas the other ($\tau_4$) was stabilised
at a small value. Our claim is that this behaviour will persist
in the general case,
with one K\"ahler modulus being large and all others being
small. It is the one overall modulus that is responsible for the
large volumes obtained.

To show this, let us write the large-volume expression for the scalar
potential (\ref{potential}) as:
\be
\label{LargeToSmallEquation}
V = \ \sum_{i,j} \frac{C_1 e^{-{a_i\tau_i}-
a_j\tau_j}}{\cal V}\ - \ \sum_i \frac{C_2
  e^{-a_i\tau_i}}{{\cal V}^2}\ +\ \frac{C_3}{{\cal V}^3},
\ee
with $C_1 \sim \sqrt{\tau}$, $C_2 \sim \tau$ and $C_3 \sim -
\chi(M)$. $C_3$
is therefore positive so long as $h^{2,1} > h^{1,1}$.
We shall not quibble here over frames or the factors of $g_s$ in the
exponent
as they do not affect the argument.

Let us start in a position where $\mc{V} \gg 1$, $V < 0$ and there are
many large moduli $\tau_i \gg 1$. This is the limit identified above
as leading to a decompactification direction with $V < 0$.
We may investigate the behaviour of the potential as one of the
$\tau_i$ fields change. Originally, all terms non-perturbative in the
large $\tau_i$ can be neglected. As the second term is the only negative term,
it wants to increase its magnitude in order to
minimise the value of the potential.
This will naturally reduce the value of the corresponding $\tau_i$
to make the exponential more relevant. As
long as the other large moduli are adjusted to keep $\mc{V}$ constant,
this reduces the value of the potential. This continues until  $e^{-a_i \tau_i}
\sim \frac{1}{\mc{V}}$, when the (positive) double exponential term in
(\ref{LargeToSmallEquation}) becomes important and the modulus
will be stabilised.

We may carry on doing this with all the large moduli, but since we are holding
$\cal V$ constant, one (combination)
of the fields must remain large. This finally leaves us with a picture of a
manifold with one large 4-cycle (and corresponding 2-cycle), but all
other 4-cycles of size close to the string scale.

We can now identify the locus of the minimum.
In the limit $\tau_b \to \infty$ with $\tau_{s,i} \textrm{ small }$,
the scalar potential takes the form
\be
V = e^{\mc{K}} \left[ \mc{K}^{i \bar{\jmath}} \partial_i W \partial_{\bar{\jmath}} \bar{W}
+ \mc{K}^{i \bar{\jmath}} ((\partial_i \mc{K}) W \partial_{\bar{\jmath}} \bar{W} + c.c.)
+ \frac{3 \xi}{4 \mc{V}} \right]
\ee
We take $W$ to be
\be
W = W_0 + \sum_{s,i} A_i e^{- a_i T_i}.
\ee
We may include an exponential dependence on $\tau_b$ in $W$; as
$\tau_b \gg 1$ this is in any case insignificant. Now,
\be
\mc{K}^{i \bar{\jmath}} \partial_i \K \propto \tau_j,
\ee
and we may also verify that, so long as $i$ and $j$ both correspond to small moduli,
\be
\mc{K}^{i \bar{\jmath}} \sim \mc{V} \sqrt{\tau_s^{'}}
\ee
where $\tau_s^{'}$ is a complicated function of the $\tau_{s,i}$ that scales
linearly under $\tau_{s,i} \to \lambda \tau_{s,i}$. The scalar potential then takes the form
\be
V = \frac{\sqrt{\tau_s^{'}} \partial_i W \cdot \partial_j W}{\mc{V}} -
\frac{\tau_i \cdot \partial_i W}{\mc{V}^2} + \frac{3 \xi}{4 \mc{V}^3}.
\ee
If we then take the decompactification limit $\tau_b \to \infty$ with
$\partial_i W = \frac{1}{\mc{V}}$ and $\tau_i \sim \mc{O}(\ln \mc{V})$,
then the potential goes to zero from below.
As this result is independent of the strength of the non-perturbative corrections,
which always eventually come to dominate the positive $\alpha'^3$ terms, there is an associated
minimum at large volume.

In order to get a clearer picture
let us make a brief geometric digression. The volume of a Calabi-Yau
can be expressed either in terms of 2-cycles, $\mc{V} = \frac{1}{6} k_{ijk}t^i t^j
t^k$, or 4-cycles $\mc{V} = \mc{V}(\tau_i)$. It is a fact that the matrix
\be
\label{d2Vrelation}
M_{ij} = \frac{\partial^2 \mc{V}}{\partial \tau_i \partial \tau_j}
\ee
has signature $(1,h^{1,1} - 1)$ (one plus, the rest minus). This
follows from the result that $\tau_i = \tau_i (t^j)$ is simply a
coordinate
change on K\"ahler moduli space, and it is a standard result
\cite{CandelasDeLaOssa} that
\be
M_{ij}^{'} = \frac{\partial^2 \mc{V}}{\partial t^i \partial t^j}
\ee
has signature $(1, h^{1,1} -1)$. The coordinate change cannot alter
the signature of the metric.

This signature manifests itself in explicit models. In both
the ${\mathbb{P}}^4_{[1,1,1,6,9]}$ example studied above and an $\mc{F}_{11}$
model also studied in \cite{hepth0404257}, the volume may in fact be
written explicitly in terms of the divisor volumes. With the $\tau_i$ as
defined in \cite{hepth0404257}\footnote{Note that in the notation of
  \cite{hepth0404257} we do not have $\tau_i \neq
\frac{\partial \mc{V}}{\partial t_i}$, but rather linear
combinations thereof.}, we have
\bea
\label{ModelVolumes}
{\mathbb{P}}^4_{[1,1,1,6,9]} & & {\cal{V}} = {1\over{9\sqrt{2}}}\left(\tau_5^{\frac{3}{2}} -
\tau_4^{\frac{3}{2}}\right), \nonumber \\
& & \tau_4 = \frac{t_1^2}{2} \textrm{ and } \tau_5 = \frac{(t_1 + 6t_5)^2}{2}.
\nonumber \\
\mc{F}_{11} & & {\cal{V}} =
2\left(\tau_1+\tau_2 +2\tau_3\right)^{3/2}-\left(\tau_2+2\tau_3\right)^{3/2} - \tau_2^{3/2},
\nonumber \\
& & \tau_1 = \frac{t_2}{2}\left(2 t_1 + t_2 + 4 t_3\right), \,
\tau_2 = \frac{t_1^2}{2}, \textrm{ and } \tau_3 = t_3 \left(t_1 + t_3\right).
\eea
For the $\mc{F}_{11}$ model, we see from the expressions for
$\tau_i$ in terms of 2-cycles that it is consistent to have
$\tau_1$ large and $\tau_2, \tau_3$ small but not otherwise. The signature of
${\rm{d}}^2 V$ is manifest in (\ref{ModelVolumes}); each expression contains
$h^{1,1} - 1$ minus signs. There is another important point. In each case, there is
a well-defined limit in which the overall volume goes to infinity and all but one divisors remain
small. These limits are given by $(\tau_5 \to \infty, \tau_4 \textrm{ constant })$ and
$(\tau_1 \to \infty, \tau_2, \tau_3 \textrm{ constant })$ respectively. Furthermore,
in each case this limit is unique: e.g. the alternative limit $(\tau_2 \to \infty, \tau_1, \tau_3
\textrm{ constant })$ is not well-defined.

\begin{figure}[ht]
\begin{center}
\makebox[10cm]{
\epsfxsize=12cm
\epsfysize=8cm
\epsfbox{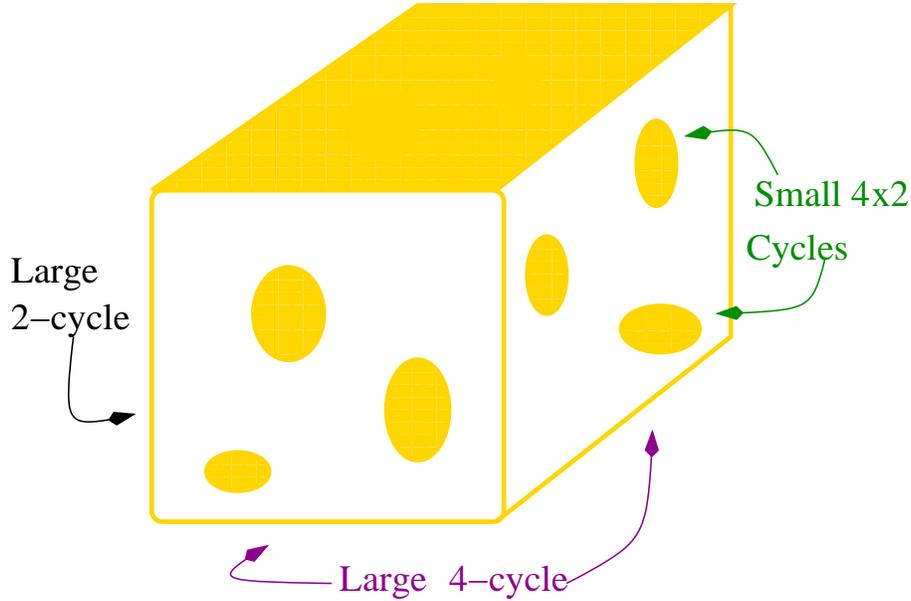}}
\end{center}
\caption{A Swiss cheese picture of a Calabi-Yau. There is one 
large 4-cycles - increasing this cycle volume increases the overall
volume. The other cycles are such that increasing the cycle volume decreases the
overall volume.}
\label{SwissCheese}
\end{figure}

This motivates a `Swiss-cheese' picture of the Calabi-Yau, illustrated in figure \ref{SwissCheese}.
A Swiss cheese is a 3-manifold with 2-cycles. Of these 2-cycles, one
($t_b$) is large and controls the size of the cheese,
while the others ($t_{s,i}$) are small and control the size of the
holes.\footnote{This distinction between large $t_b$ and small
  $t_{s,i}$ only holds for reputable cheesemongers!} 
  The volume of the cheese can be written
\be
\mc{V} = t_b^{3/2} - \sum_i t_{s,i}^{3/2},
\ee
and $\frac{\partial^2 \mc{V}}{\partial t_i \partial t_j}$ has signature $(1, h^2 - 1)$.
The small cycles are internal; increasing their volume decreases the overall volume of the manifold.
There is one distinguished cycle that controls the overall volume; this cycle may be
made arbitrarily large while holding all other cycles small, and controls the overall volume.
For all other cycles, an arbitrary increase in their volume decreases the overall volume and eventually
leads to an inconsistency. The small cycles may be thought of as local
blow-up effects; if the bulk
cycle is large, the overall volume is largely insensitive to the size of the small cycles.

To capture the above, let us consider a Calabi-Yau with divisors $\tau_b, \tau_{s,i}$ such that the
volume can be written
\be
\label{CYLvol}
{\cal{V}} = \left(\tau_b + \sum a_i \tau_{s,i}\right)^{\frac{3}{2}} -
 \left(\sum b_i \tau_{s,i}\right)^{\frac{3}{2}}
- .... - \left(\sum k_i \tau_{s,i}\right)^{\frac{3}{2}}.
\ee
We assume that a limit $\tau_b \gg\tau_{s,i}$ is well-defined.
The minus signs are necessary to ensure (\ref{d2Vrelation}) is satisfied. The form given above is
valid globally for both ${\mathbb{P}}^4_{[1,1,1,6,9]}$ and
$\mc{F}_{11}$ models.
The form (\ref{CYLvol}) is illustrative and it is not important for
our argument that it hold generally; the important assumption is that
there exists a well-defined limit $\tau_b \gg \tau_{s,i}$. The effect
 of the $\alpha'$-corrected potential is then to drive the moduli to a limit where $\tau_b \gg
 1$ and $\tau_s \sim \ln(\tau_b)$.

The geometric interpretation of this is that the `external' cycle controlling the
overall volume may be very large, whereas the small, `internal' cycles are
always stabilised at small volumes. As in section \ref{spectrosec}, the
masses associated with the
moduli parametrising the small cycles have a naive scaling
$m_{\tau_i} \sim \mc{O}(\frac{g_s^2 W_0}{\mc{V}})$, comparable to the masses
of the dilaton and complex structure moduli. Thus the features of the resulting spectrum are largely
model-independent - in particular, the scales of moduli masses are set
by the overall volume irrespective of the detailed model. As before the
`volume' modulus, which is distinguished by its large size, is
relatively light, being suppressed by a factor of $\mc{V}^{3/2}$.

We caution here that the `Swiss Cheese' picture is only a picture. The
signature (\ref{d2Vrelation}) is a feature of the second derivatives
of the volume rather than the first derivatives. Nonetheless, it captures the
outcome of the moduli stabilisation results and is a useful way to 
visualise and understand them.

We also comment here that in order for the above moduli stabilisation
mechanism to work, we do need at least one `blow-up' modulus that can sensibly
be kept small while the overall volume is made arbitrarily large. For
example, this approach cannot be made to work on tori with $h^{1,1} =
3$. This is because the limit in which one 4-cycle is kept small while
the overall volume is made extremely large is very anisotropic, with two
large dimensions and four small ones. The assumptions made above are
then no longer valid and the resulting
structure of the scalar potential is quite different from that
analysed above.

\section{Comparison with KKLT Vacua}

This chapter has described the construction of new large-volume
non-supersymmetric minima of the scalar potential. The third part of
this thesis will study phenomenological applications of this
construction. Before moving on to this, let us first compare this
construction with the more standard KKLT vacua

\begin{enumerate}

\item First, there is a sense in which the above construction
  completes KKLT across the entire range of moduli space. As shown in
  chapter \ref{ChapterKahlerModuli}, the regime of validity of the KKLT potential is extremely limited,
being restricted both in the values of $W_0$ allowed and the regime
  of moduli space over which the potential can be trusted. In
  particular, the KKLT potential is always invalid in the asymptotically large
  volume limit. This can be traced to the neglect of the 
perturbative corrections which dominate at large volume. By including
  these corrections, the resulting
potential includes both the large-volume minimum discussed above and,
for very small values of $W_0$, the KKLT minimum. Unlike KKLT, the
  large-volume minimum has no restrictions on the permissible values
  of $W_0$. It is a property of
  the large-volume minimum that $\mc{V} \propto W_0$. When $W_0 \ll 1$ the
two minima coexist, and gradually decreasing $W_0$ causes the two minima to
approach each other and to eventually merge. This behaviour is illustrated in figures
\ref{kkltbbcq1} and \ref{kkltbbcq2}.
\begin{figure}[ht]
\linespread{0.2}
\begin{center}
\epsfxsize=0.75\hsize \epsfbox{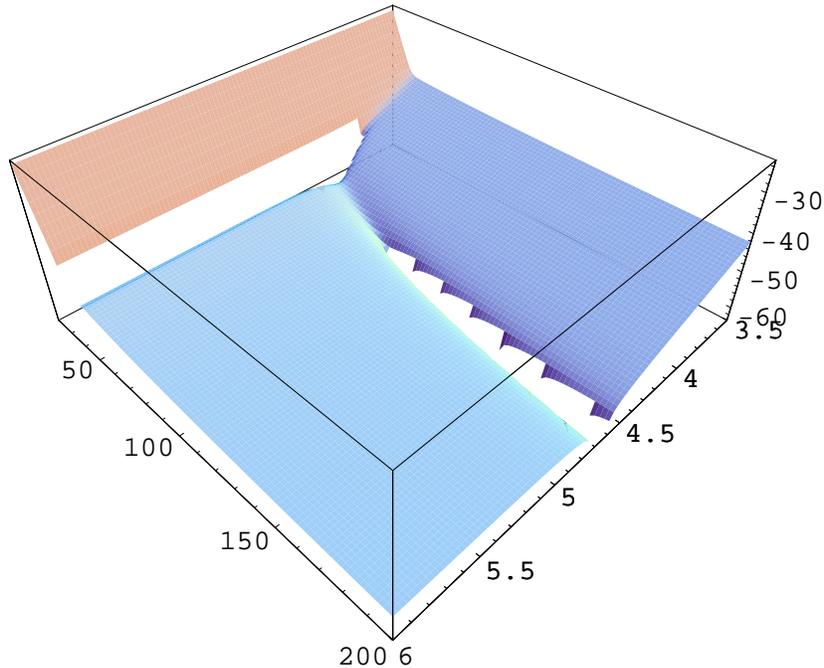}
\end{center}
\caption{ A plot of $\ln{V}$, showing the region of the scalar potential in which the large volume
  minimum coexists with a KKLT minimum at smaller volume. This picture
  is valid for very small $W_0$ ($\sim 10^{-10}$)}
\label{kkltbbcq1}
\end{figure}
\begin{figure}[ht]
\linespread{0.2}
\begin{center}
\epsfxsize=0.75\hsize \epsfbox{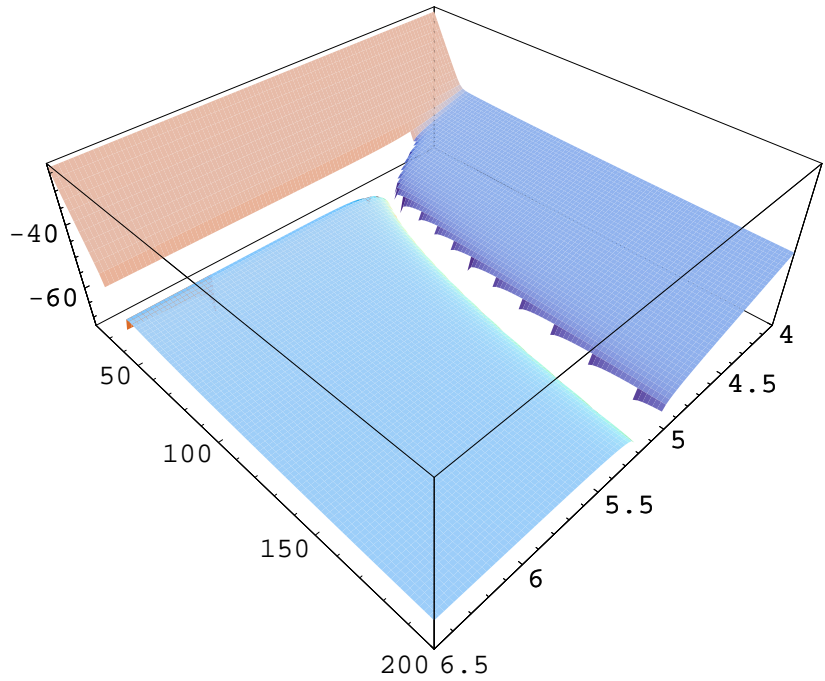}
\end{center}
\caption{ A plot of $\ln{V}$, showing how as $W_0$ is further
  decreased the large volume
  minimum merges with the KKLT minimum.}
\label{kkltbbcq2}
\end{figure}

\item The most obvious difference is in the volume of the extra
  dimensions. In KKLT the stabilised volume is always relatively
  small, going as $\ln (W_0)$, and the string and Planck scales are
  comparable. In the scenario constructed above, the volume is
  naturally exponentially large and is in fact exponentially sensitive
  to the stabilised dilaton. As $m_{3/2} \sim \frac{M_P W_0}{\mc{V}}$,
  this large volume allows a natural generation of hierarchies. In
  particular it has the distinct advantage that the hierarchy between
  the
weak and Planck scales may be
  explained without the fine-tuning of $W_0$ that is necessary in KKLT.

\item A second obvious difference is in supersymmetry breaking.
In KKLT, the AdS minimum is supersymmetric.
and the principal source of supersymmetry breaking terms comes from the
uplift mechanism. The lifting term
dominates the soft supersymmetry breaking terms in the sense that
all $F$-terms vanish if this term is absent. In the above scenario the
original minimum is already non-supersymmetric AdS, and
the sources of supersymmetry
breaking are primarily the K\"ahler moduli F-terms. This arises from
the underlying no-scale structure.
These F-terms give the dominant contribution
to soft terms irrespective of the uplift mechanism (as we shall
see in chapter \ref{chapterSoftSusy}).

\item In the KKLT scenario the potential may
 develop tachyonic
directions after fixing the K\"ahler moduli. This is not so for the
 large-volume minimum,
for which the minimum of the potential
is at order $-\frac{e^{\K_{cs}} \vert W \vert^2 (S,U)}{{\cal V}^3}$. As the
contribution from dilaton and complex structure F-terms is always
positive and $\mc{O}(\frac{1}{\mc{V}^2})$, any displacement of these moduli
from $D_i W = 0$
increases
the potential due to the different volume scalings present.
Thus in this scenario there are
no tachyonic directions in the geometric moduli. This
is the main reason these models are far simpler to analyse regarding
soft supersymmetry breaking. In the KKLT scenario, it
is a hard problem to minimise the full potential without following the two step procedure
in which dilaton and complex structure moduli are
fixed by the fluxes and then integrated out. 
This procedure
fails in many cases giving rise to tachyonic directions.
This was explicitly seen in \cite{hepth0411066, ciqs} for the simplest
cases with no complex structure moduli. For more complicated examples
 with several complex structure moduli minima can be found, but this
 is model-dependent\cite{hepth0404116, hepth0411183, hepth0506090}.

The above construction of the non-supersymmetric large-volume minimum
concludes Part II of this thesis. In Part III we will investigate the
phenomenological properties of this scenario.

\end{enumerate}

\part{Applications: Towards String Phenomenology}

\chapter{Soft Supersymmetry Breaking}
\label{chapterSoftSusy}

This chapter is based on part of the paper \cite{hepth0505076} and the
paper \cite{hepth0605141}. Sections \ref{gf} and \ref{xyz} are based
on a part of the paper \cite{hepth0505076} primarily due to Kerim Suruliz.

The second part of this thesis described the construction of a
moduli potential with a minimum at exponentially large values of the
internal volume. The minimum found was an AdS minimum,
non-supersymmetric even before uplifting. As the minimum is
non-supersymmetric, the supersymmetry is spontaneously broken. The
moduli whose potential we studied are gravitationally coupled to matter
fields. As matter fields are assumed to have vevs that are either
vanishing or hierarchically suppressed (as with the Higgs), it
was consistent to neglect them when considering the moduli potential.

If supersymmetry is realised in nature, it is realised as a broken
symmetry. The breaking of supersymmetry by the moduli potential is
transmitted to the (observable) matter sector through
Planck-suppressed interactions - this is the `gravity mediation'
scenario. For phenemomenological purposes, we
are interested in the matter Lagrangian and in particular the terms
arising from supersymmetry breaking.
In this context 
it is a well-known result that 
\emph{spontaneous} breaking in the locally supersymmetric supergravity
Lagrangian gives rise to
\emph{explicit} soft breaking in the globally supersymmetric matter
Lagrangian. If the LHC sees low-energy supersymmetry, it is these soft
breaking terms that phenomenologists will seek to extract from the data.
 
The problem of how to go from a spontaneously broken supergravity
theory to an explicitly but softly broken supersymmetric field theory
is a classic one and the general formalism has existed for some time
\cite{SoniWeldon, hepth9303040, hepph9308271}.
While the formalism is general, the matter Lagrangian is
model-dependent. In particular there are various choices as to where to embed the
Standard Model and how to realise the matter content. In the context
of the IIB flux compactifications considered in this thesis, the main
choice is whether the Standard Model gauge groups should be supported
on D3 or D7 branes.\footnote{A detailed realisation of the Standard
  Model may involve more choices, but this is the most fundamental one.}

We start by reviewing the general formalism before estimating the magnitude of
the soft parameters for the specific large-volume models studied in the previous chapter.

\section{General Formulae}
\label{gf}

\subsection*{Matter-Moduli Couplings}

In order to calculate soft scalar masses we need 
to know the matter K\"ahler potential. 
A full account of matter-moduli couplings would require a full
embedding of the Standard Model and is not feasible. In general,
computing matter K\"ahler potentials on Calabi-Yau spaces is a
notoriously hard problem, the difficulty stemming from the fact that
the K\"ahler potential is non-holomorphic and so not
protected. Standard Model matter belongs to bifundamental representations
and corresponds to strings stretching between brane stacks. It is
not known how to compute the K\"ahler potential for such strings on
arbitrary backgrounds even at tree-level. There do exist explicit orbifold
computations \cite{hepth0404134}, but these are restricted to one very special
locus in moduli space.

Instead of using bifundamental matter to estimate soft terms we
therefore use adjoint matter, corresponding physically to the motion of branes in
the internal space or mathematically to deformations of the calibrated
cycle wrapped by the brane world-volume. In this case progress has
been made in a Calabi-Yau context and formulae exist for the combined matter-moduli K\"ahler
potential for both D3 and D7 adjoint matter.
As reviewed in section \ref{seCCS}, in \cite{hepth0312232} the general K\" ahler potential for Calabi-Yau
orientifolds with D3 branes was derived by a dimensional reduction. The result is
\begin{equation}
\label{kahlerpot}
K(S, T, U,\phi) = -\log (S+\bar{S}) -\log \left(-i\int
\Omega\wedge \bar{\Omega} \right) - 2 \log ({\cal{V}} (T, U, \phi)).
\end{equation}
Here $U$ are the complex structure moduli
and $\phi^i, i=1,2,3,$ are scalar fields corresponding to the
positions of D3 branes on the Calabi-Yau. The Calabi-Yau volume $\cal{V}$
is to be understood
as a function of the complexified K\" ahler moduli $T^i$, the expression
for which is:
\begin{equation}
\label{complexified}
T_\alpha = \tau_\alpha + i\rho_\alpha + i\mu_3 l^2
(\omega_\alpha)_{\imath{\bar{\jmath}}} {\rm Tr} \phi^i
\left( \bar{\phi}^{\bar{\jmath}}
- {i\over2} \bar{U}^{\hat{a}} (\bar{\chi_{\hat{a}}})^{\bar{\jmath}}_l \phi^l\right).
\end{equation}
$\omega_\alpha$ are a basis for (1,1) forms on the Calabi-Yau
surviving the orientifold projection, while the $\chi_{\hat{a}}$ are a basis
of $(2,1)$ forms with negative sign under the orientifold
projection - these correspond to the complex structure
moduli $U^{\hat{a}}$ surviving the orientifold projection. Also
$$
\mu_3 =\frac{1}{(2\pi)^{3} \alpha'^{2}}, \qquad l = 2 \pi \alpha' \qquad
\hbox{ and } \qquad \mu_3 l^2 = \frac{1}{2 \pi}.
$$

If both D3 and D7 branes are present, the K\" ahler potential was derived
in \cite{hepth0409098} and has the form
\be
\label{KahlerWithD7}
\K(S,T, U, \zeta, \phi) = -\log \left( S+\bar{S}+2i \mu_7 {\cal{L}}_{AB} \zeta^A
\bar{\zeta}^{\bar{B}} \right) -\log \left( -i\int\Omega\wedge
\bar{\Omega} \right) -2 \log \left( \mc{V} \right),
\ee
where ${\cal{L}}_{AB}$ are certain geometric quantities, $\zeta^A$ are
the moduli describing the position of the D7 brane and $\mc{V}$ is
understood as a function of $T$, $U$ and $\phi$. In a similar fashion to
(\ref{complexified}) the dilaton modulus $S$ undergoes a redefinition,
such that the K\"ahler potential remains
\be
\mc{K}(S, T, U, \zeta, \phi) = - \ln \left( \frac{2}{g_s} \right) -
\ln \left( -i \int \Omega \wedge \bar{\Omega} \right) - 2 \ln \left(
\mc{V} \right).
\ee
We neglect in the above the possibility
of Wilson line moduli.

The main point we take from these expressions is not so much the
detailed form as how the matter fields appear. For D3 branes, the
matter fields couple to the K\"ahler moduli, leading to a redefinition
of $T_\alpha$. For D7 branes, the matter fields couple to the dilaton
leading to a redefinition of $S$. This can be understood naturally
from F-theory, where both the dilaton and the D7 moduli are complex
structure moduli of the F-theory fourfold.

\subsection*{F-terms}

We use the standard formalism for calculating soft supersymmetry breaking terms,
as described for example in \cite{hepph9707209}. We proceed by expanding
the K\" ahler potential and superpotential in terms of the visible sector fields $\varphi$
\begin{eqnarray}
\K &=& \hat{\K} + (\tilde{\K}_{i\bar{\jmath}}) \varphi^i
\bar{\varphi}^{\bar{\jmath}} + Z_{ij} \varphi^i \varphi^j + \cdots,
\nonumber\\
W &=& \hat{W} + \mu_{ij} \varphi^i \varphi^j + Y_{ijk} \varphi^i \varphi^j \varphi^k +\cdots,
\end{eqnarray}
where $\hat{\K}, \tilde{\K}_{i\bar{\jmath}}, Z_{ij}, \mu_{ij}$ and $Y_{ijk}$
depend on the hidden moduli only.
Hidden sector F-term supersymmetry breaking is
characterised by nonvanishing expectation values for the auxiliary fields of the hidden
sector chiral superfields. These F-terms are fundamental
for the calculation of soft masses and may be written as
\begin{equation}
\label{fterms}
{\bar{F}}^{\bar{m}} = e^{{\hat{\K}/{2M_P^2}}} {\hat{\K}}^{\bar{m}n} {{D_n \hat{W}}\over{M_P^2}}.
\end{equation}
In this and all subsequent formulae, $m$ and $n$ range over the
hidden moduli - the dilaton, complex structure and K\" ahler moduli
in our case. These moduli are called `hidden' because, being
uncharged, they have no gauge couplings and so cannot be seen
in collider experiments.
We henceforth work in Planck mass units and do not include explicit
factors of $M_P.$

Given the F-terms, the various soft parameters can be calculated.
Perhaps the simplest of all quantities to compute are the canonically
normalised gaugino masses $M_a$. For a sector with gauge kinetic
function $f_a$, these are
\begin{equation}
\label{gauginomass}
M_a = {1\over2} ({\rm Re} f_a)^{-1} F^m \partial_m f_a.
\end{equation}
Once we know $f_a$ the gaugino masses can be computed directly from the
moduli vevs.
There is also a general formula for scalar masses.
Assuming a diagonal matter field metric
$\tilde{\K}_{i\bar{\jmath}}= \tilde{\K}_i
\delta_{i\bar{\jmath}}$,\footnote{Note that there is no sum over
  $i$. Furthermore, $\tilde{\K}_i$ is not to be confused with the derivative of
the K\" ahler potential with respect to modulus $i.$} the squared
soft masses of
canonically normalised matter fields $\varphi^i$  can be written as
\begin{equation}
\label{softmass}
m_i^2 = m_{3/2}^2 +V_0 - F^{m} {\bar{F}}^{\bar{n}} \partial_m \partial_{\bar{n}} \log \tilde{\K}_i,
\end{equation}
where $V_0$ denotes the value of the cosmological constant.
The A-terms of the normalised matter fields $\hat{\varphi}^i$ are
defined by $A_{ijk}
\hat{Y}_{ijk} \hat{\varphi}^i \hat{\varphi}^j \hat{\varphi}^k$ with
\begin{eqnarray}
\label{aterms}
A_{ijk} &=& F^m (\hat{\K}_m +\partial_m \log Y_{ijk}- \partial_m \log (\tilde{\K}_i
\tilde{\K}_j\tilde{\K}_k)),\nonumber\\
\hat{Y}_{ijk} &=& Y_{ijk} {{\hat{W}^*}\over{|\hat{W}|}} e^{\hat{\K}/2}
(\tilde{\K}_i \tilde{\K}_j \tilde{\K}_k)^{-1/2}.
\end{eqnarray}
Finally, if $Z_{ij} = \delta_{ij} Z$ and $\mu_{ij} = \mu \delta_{ij}$,
the B-term $\hat{\mu} B \hat{\varphi}^i \hat{\varphi}^i$ for the field
$\varphi^i$ can be written as
\begin{eqnarray}
\hat{\mu} B &=& (\tilde{\K}_i)^{-1} \Bigl\{ {\hat{W}^*\over{|\hat{W}|}} e^{\hat{\K}/2}
\mu \Bigl[F^m (\hat{\K}_m +\partial_m \log\mu - 2\partial_m \log
(\tilde{\K}_i))\nonumber\\
& &{}-m_{3/2}\Bigr]+(2m_{3/2}^2+V_0) Z - m_{3/2} \bar{F}^{\bar{m}}
\partial_{\bar{m}} Z \nonumber\\
& &{}+ m_{3/2} F^m (\partial_m Z - 2 Z\partial_m \log
(\tilde{\K_i}))\nonumber\\
& &{}- \bar{F}^{\bar{m}} F^n (\partial_{\bar{m}} \partial_n
Z - 2\partial_{\bar{m}} Z \partial_n \log (\tilde{\K}_i))\Bigr\}.
\end{eqnarray}
where the effective $\mu$-term is given by
\begin{equation}
\hat{\mu} = \left( {\hat{W}^*\over{|\hat{W}|}} e^{\hat{\K}/2} \mu + m_{3/2} Z
 - F^{\bar{m}} \partial_{\bar{m}} Z \right) (\tilde{\K}_i)^{-1}.
\end{equation}
We shall set $V_0=0$ since soft masses are to be evaluated after lifting
the vacuum energy.

It is important to have an idea of the approximate sizes of the various
F-terms used in the computation of soft terms. We continue to work in the
context of the ${\mathbb{P}}^4_{[1,1,1,6,9]}$ model with two K\" ahler moduli $T_4,T_5$,
although we expect our results to extend to other models in this
framework (but not to KKLT-style models).

At large volume, the volume scaling of the relevant parts of the inverse metric are
\begin{eqnarray}
\label{invmetric}
\K^{\bar{T}_4 T_4} &\sim& {\cal{V}},\nonumber\\
\K^{\bar{T}_4 T_5} &\sim& {\cal{V}}^{2/3},\nonumber\\
\K^{\bar{T}_5 T_5} &\sim& {\cal{V}}^{4/3},\nonumber\\
\K^{\bar{S}   T_4} &\sim& {1\over{{\cal{V}}}},\nonumber\\
\K^{\bar{S}   T_5} &\sim& {1\over{{\cal{V}}^{1/3}}}.
\end{eqnarray}
The derivatives of the K\" ahler potential are
\bea
\partial_4 \K \equiv \partial_{T_4} \K & \sim & \frac{1}{\mc{V}} \nonumber \\
\partial_5 \K \equiv \partial_{T_5} \K & \sim & \frac{\tau_5^{1/2}}{\mc{V}} \sim \frac{1}{\mc{V}^{2/3}}. \nonumber 
\eea
Since at the minimum of the potential $D_i W=0$ for
the dilaton and complex structure moduli, the volume dependence
of the relevant F-terms is given by
\begin{eqnarray}
F^4 &\sim& {1\over\cal{V}}\left(
{\cal{V}\over\cal{V}} +
{{\cal{V}}^{2/3}\over {\cal{V}}^{2/3}}\right) \sim {1\over{\cal{V}}} \nonumber\\
F^5 &\sim& {1\over{\cal{V}}}\left( {{\cal{V}}^{4/3}\over {\cal{V}}^{2/3}} +
{{\cal{V}}^{2/3}\over {\cal{V}}} \right)\sim {1\over{\cal{V}}^{1/3}}\nonumber\\
F^S &\sim& {1\over{\cal{V}}} \left( {1\over{{\cal{V}}}} +
{1\over{{\cal{V}}^2}}\right) \sim {1\over{\cal{V}}^2}.
\end{eqnarray}
We see that at large volume, 
$${\cal{V}}^{-1/3} \sim F^5\gg F^4 \sim {\cal{V}}^{-1}\gg F^S
\sim {\cal{V}}^{-2}.$$

The F-terms corresponding to complex structure moduli vanish since for $i$ ranging over dilaton and K\" ahler moduli
$\K^{\bar{U}i}=0$,
even after the inclusion of the $\alpha'$ corrections of
\cite{hepth0204254}. Furthermore,
 $D_U W=0$ at the minimum of the scalar potential so long as we only turn on ISD
fluxes. If we use IASD fluxes for the uplift, the complex structure
F-terms will be nonvanishing but still suppressed, and we do not
anticipate any significant modifications to the results below.

\section{Soft Parameters: Basic Estimates}
\label{xyz}

We now use the above formalism to estimate the
magnitude of the soft parameters for matter fields living on either D3
or D7 branes.

\subsection*{Soft Parameters for D3 branes}

To calculate the masses of D3 moduli it is sufficient to work with
a low energy theory only containing the K\" ahler moduli and the D3 fields.
This is because ISD fluxes do not give masses to D3 moduli 
but do give large masses $m \sim {\cal{O}} (m_{3/2})$ to dilaton, complex
structure moduli and D7 brane moduli
\cite{hepth0105097,hepth0312232, hepth0311241}.
We also restrict to a single D3-brane rather than a stack, and
assume for simplicity that the D3 moduli metric is diagonal. Of course
all these assumptions are naive, but since we are primarily concerned
with the volume scaling of the soft parameters, we hope the features we obtain
survive this approximation.

Equations (\ref{kahlerpot}) and (\ref{complexified}) describe how D3 moduli
enter the K\" ahler potential.
Concentrating on a single D3 modulus $\phi$, the K\" ahler potential becomes
\begin{equation}
\K = -2\log \Bigl[
(T_5 + \bar{T}_5 - c|\phi|^2)^{3/2} - (T_4+\bar{T}_4 - d|\phi|^2)^{3/2}
+ {\xi'\over2}\Bigr] + \K_{cs} +2\log(36).
\end{equation}
This is obtained from the original
K\" ahler potential $\K = -2\log \left({\cal{V}} +{\xi\over2}\right) + \K_{cs}$.
Here $\xi'=36\xi$ and  $c, d \sim \mc{O}(1)$ parametrise our ignorance of the forms
$\omega_\alpha$. The factors of $36$ come from $9 \sqrt{2} \ti 2
\sqrt{2}$ $\Big($recall $\mc{V} = \frac{1}{9\sqrt{2}}
\left(\tau_5^{3/2} - \tau_4^{3/2}\right) \Big)$.

The superpotential is
\begin{equation}
\label{wnodilaton}
W = \hat{W} = {g_s^2\over{\sqrt{4\pi}}} \left(W_0 + A_4 e^{-i {a_4\over g_s} T_4} + A_5 e^{-i {a_5\over
    g_s} T_5}\right),
\end{equation}
there being no supersymmetric $\mu$-term for D3 brane scalars. In
another simplifying assumption we also assume that $W$ is real.

After expanding $\K$ around $\phi=0$ we obtain the following
expressions for $\tilde{\K}_i$ and $\hat{\K}$:
\begin{eqnarray}
\label{ktilde}
\tilde{\K}_i &=& 3{{{c(T_5+\bar{T}_5)^{1/2} - d(T_4+\bar{T}_4)^{1/2}}}\over
{(T_5+\bar{T}_5)^{3/2} - (T_4+\bar{T}_4)^{3/2}+{\xi'\over2}}},\nonumber\\
\hat{\K} &=& -2\log \left[ (T_5+\bar{T}_5)^{3/2} - (T_4+\bar{T}_4)^{3/2}+{\xi'\over2}\right].
\end{eqnarray}
We now introduce the variables $X=(T_4+\bar{T}_4)^{1/2} = \sqrt{2 \tau_4},
Y=(T_5+\bar{T}_5)^{1/2} = \sqrt{2 \tau_5}$ to simplify various expressions appearing
in the rest of this section.

It is important to note that in the no-scale approximation (obtained
by setting $\xi=0$ in $\K$ and $A_4=A_5=0$ in $W$), the nonvanishing F-terms
are $F^4 = -e^{\hat{\K}/2} X^2 W$,
$F^5 = -e^{\hat{\K}/2} Y^2 W$ (these are derived in the
Appendix). We also have, after a redefinition of $c$ and $d$,
\begin{equation}
\log (\tilde{\K}_i) = \log (cX+dY)-\log ({\cal{V}}) + {const}.
\end{equation}
It is easy to check that $F^m \bar{F}^{\bar{n}}
\partial_m \partial_{\bar{n}} \log (cX+dY) = -e^{\hat{\K}}|W|^2/2.$ Also
\begin{equation}
F^m \bar{F}^{\bar{n}}\partial_m
  \partial_{\bar{n}} \log ({\cal{V}}) =
- \half \hat{\K}_{m\bar{n}} F^m \bar{F}^{\bar{n}} =
- \frac{3}{2} e^{\hat{\K}}|W|^2
\end{equation}
in this approximation, so that
\begin{equation}
F^m \bar{F}^{\bar{n}} \partial_m \partial_{\bar{n}} \log {\tilde{\K}_i}
= e^{\hat{\K}}|W|^2.
\end{equation}
This cancels against $m_{3/2}^2 = e^{\hat{\K}} |W|^2$ in the
expression for soft masses (\ref{softmass}) giving the no-scale result
$m_i^2=0.$

Let us now estimate the size of soft scalar masses in the ${\mathbb{P}}^4_{[1,1,1,6,9]}$
model.
Consider the expression (\ref{softmass}). The inclusion of $\alpha'$
and nonperturbative effects alters the F-terms (through $\hat{\K}$, $\hat{\K}^{\bar{T}_i T_j}$
and $\partial_i W$) and the expression for $\tilde{\K}_i.$
After including nonperturbative contributions (but temporarily neglecting $\alpha'$
corrections) the F-terms are
\begin{eqnarray}
F^4 &\sim& {1\over{\cal{V}}} + {1\over{\cal{V}}} \K^{T_4 \bar{T}_4}
(\partial_4 W)
\nonumber\\
F^5 &\sim& {1\over{{\cal{V}}^{1/3}}} + {1\over{\cal{V}}} \K^{T_4 \bar{T}_5}
(\partial_4 W).
\end{eqnarray}
By construction, the modulus $\tau_4$ is small at the minimum,
while $\tau_5$ is exponentially large, so we are justified in including
only the nonperturbative contribution from $\tau_4.$ Moreover, $\tau_4$
is such that at the (AdS) minimum we have
\begin{equation}
-\partial_{\tau_4} W = {{a_4 A_4}\over{g_s}} e^{-{a_4\over{g_s}} \tau_4}\sim {\xi^{1\over3}
|W_0|\over{\langle{\cal{V}}_s\rangle}},
\end{equation}
and so $(\partial_4 W)\sim 1/{\cal{V}}.$ Therefore, we may use the
expressions for the inverse metric $\K^{\bar{T}_i T_j}$ given in
(\ref{invmetric}) to write
\begin{eqnarray}
\label{Ftermsfirststab}
F^4 &\sim& {1\over{\cal{V}}} + {1\over{\cal{V}}},
\nonumber\\
F^5 &\sim& {1\over{{\cal{V}}^{1/3}}} + {1\over{{\cal{V}}^{4/3}}}.
\end{eqnarray}
In section \ref{cqpap} we shall see that there is actually a cancellation
in $F^4$ and the above overestimates this F-term by a factor $\ln
(\mc{V}) \sim \ln (m_{3/2})$.
In the expressions (\ref{Ftermsfirststab}) the second term corresponds to the modification coming
from the nonperturbative addition to the superpotential.
The dominant contributions to scalar masses will come from
terms of the form
$F^m_{np} \bar{F}^{\bar{n}} \partial_m \partial_{\bar{n}} \log (\tilde{\K}_i)$
where $F^m_{np}$ is the nonperturbative contribution to the F-term.
Finally we have to see how $\partial_m \partial_{\bar{n}} \log
\tilde{\K}_i$ scales with the volume, for $m,n \in \{ 4,5\}.$
Using the explicit expressions (\ref{logcu}) derived in the
appendix we see that
\begin{eqnarray}
\label{logcunew}
\partial_4 \partial_4 \log (cX+dY) &\sim& {\cal{V}}^{-1/3},
\nonumber\\
\partial_4 \partial_5 \log (cX+dY) &\sim& {1\over{\cal{V}}},
\nonumber\\
\partial_5 \partial_5 \log (cX+dY) &\sim& {\cal{V}}^{-4/3}.
\end{eqnarray}
Using the above expressions for F-terms and (\ref{logcunew}), we have
\begin{eqnarray}
F^4_{np} \bar{F}^{\bar{4}} \partial_4 \partial_{\bar{4}} \log (cX+dY)
&\sim& {1\over{\cal{V}}} {1\over{\cal{V}}}  {\cal{V}}^{-1/3}
= {\cal{V}}^{-7/3},\nonumber\\
F^4_{np} \bar{F}^{\bar{5}} \partial_4 \partial_{\bar{5}} \log (cX+dY)
&\sim& {1\over{\cal{V}}} {1\over{{\cal{V}}^{4/3}}}  {\cal{V}}^{-1}
= {\cal{V}}^{-10/3},\nonumber\\
F^5_{np} \bar{F}^{\bar{5}} \partial_5 \partial_{\bar{5}} \log (cX+dY)
&\sim& {1\over{{\cal{V}}^{4/3}}} {1\over{{\cal{V}}^{1/3}}}  {\cal{V}}^{-4/3}
= {\cal{V}}^{-3}.
\end{eqnarray}
We therefore expect the masses squared to be ${\cal{O}}(\mc{V}^{-7/3}).$
A similar analysis can be done for the terms of type
$F^m_{np} \bar{F}^{\bar{n}} {\hat{\K}}_{m\bar{n}}$ and also including $\alpha'$
corrections - these turn out to be subleading compared to
the contribution from nonperturbative corrections to the superpotential.
Explicit expressions can be found in  
appendix C.

Our conclusion is that the masses of D3 moduli are suppressed compared
to the gravitino mass by a factor ${\cal{V}}^{-1/6}$ (the factors
of $g_s$ and $W_0$ also present are derived in appendix C):
\begin{equation}
\label{D3masssca}
m_i = {\cal{O}} (1) {g_s^2 W_0\over{\sqrt{4\pi} ({\cal V}^0_s)^{7/6}}}
\sim \frac{m_{3/2}}{\mc{V}^{1/6}}.
\end{equation}

We next examine D3 gaugino masses.
The gauge kinetic function is $f=\mu_3 l^2 S= \frac{S}{2\pi}$ and the normalised
gaugino masses are
\begin{equation}
\label{D3gaugmass}
M_{D3} = \frac{2\pi}{2} ({\rm Re } S)^{-1} F^S = {{\cal{O}}} \left(
{1\over{{\cal{V}}^2}}\right).
\end{equation}
The dilaton and K\"ahler moduli mix in the metric through the
$\alpha'^3$ correction.
$F^S$ must therefore be calculated using the full K\" ahler
potential, before integrating out the dilaton and complex structure moduli.
$M_{D3}$ turns out to be proportional to $g_s^2$ and
$W_0$ in the same way scalar masses are---this can be deduced from the
$1/g_s$ prefactor in the inverse metric $\K^{\bar{S} T_i}$
(as shown in \cite{hepth0412239}) and the factor of $g_s^{3/2}$ in
$W$, which can be observed in (\ref{Potentials}).

We can also estimate the magnitude of the A-terms, which
again vanish in the no-scale approximation.
If we use (\ref{ktilde}) in the A-term expression (\ref{aterms}), with
constant $Y_{ijk}$, we obtain
\begin{eqnarray}
A_{\phi\phi\phi} &=& e^{\hat{\K}/2}\Bigl\{ -{{3\xi}\over{4\cal{V}}}W + (\partial_4 W)
\Bigl[
{X^2\over{36{\cal{V}}}} (2Y^3 + X^3 - {\xi'\over2}) - {3Y\over{\cal{V}}} X^2 Y^2
 \nonumber\\
& & +{3\over{2(cX+dY)}}\left( {c\over3} (2Y^3 + X^3 -{\xi'/2}) +
d X^2 Y\right)\Bigr]\Bigr\}.
\end{eqnarray}
As $\partial_4 W \sim \mc{V}^{-1}$,
$A\sim Y^2/{{\cal{V}}^2} \sim {\cal{V}}^{-4/3}.$
As with the scalar and gaugino masses, the dependence of $A$ on $g_s$ and $W_0$ is given by
by $A\propto g_s^2 W_0.$

Let us finally consider $\mu$ and B-terms.
For D3 branes, the supersymmetric $\mu$ term vanishes, but there is an effective $\mu$-term
generated by the Giudice-Masiero mechanism \cite{GiudiceMasiero}. This
is due to the appearance of
a bilinear in the K\" ahler potential dependent on complex
structure moduli, as follows from formulae (\ref{kahlerpot}) and
(\ref{complexified}).
The prefactor of the bilinear $\phi^i \phi^j$ in the expansion of the
K\" ahler potential is
\begin{equation}
Z_{ij} = {{3\mu_3 l^2}\over{ {\cal{V}} +\xi/2}} t^{\alpha}
(\omega_\alpha)_{(j|\bar{s}} (\bar{\chi}_{\hat{a}})^{\bar{s}}_{|i)}
\bar{U}^{\hat{a}}.
\end{equation}
For simplicity we consider only one $Z$ with $Z=(c'X+d'Y)(a_i U^i)/(
  {\cal{V}} + \xi/2).$
The complex structure moduli dependence of $Z$ is in fact
unimportant since the F-terms corresponding to these moduli are
vanishing.
The calculation of the effective $\mu$-term will be very similar to the
computation of A-terms; we do not need to differentiate by $U$, and
can absorb $a_i U^i$
into $c'$ and $d'$ to write $Z=(c'X+d'Y)/({\cal{V}} + \xi/2).$
Then
\begin{eqnarray}
\partial_4 Z &\sim& \frac{1}{\mc{V}},\nonumber\\
\partial_5 Z &\sim&  \frac{1}{\mc{V}^{4/3}},
\end{eqnarray}
so that $F^m \partial_m Z$ gives rise to terms
\begin{eqnarray}
e^{\K/2} \K^{44} (\partial_4 W)\partial_4 Z &\sim& \mc{V}^{-2},\nonumber\\
e^{\K/2} \K^{54} {\partial_4 W}\partial_5 Z &\sim& \mc{V}^{-8/3}.
\end{eqnarray}
Assuming the complex structure moduli to be fixed at ${\cal{O}}
(1)$ values, it is easy to confirm that the $\hat{\mu}$ term scales as
${\cal{O}} (\mc{V}^{-4/3}).$

As $Z$ can be treated as
$(c'X+d'Y)/ ({\cal{V}}+\xi/2)$ and by analogy with the calculation
of the masses squared, it is easy to see that the expression
for $\hat{\mu} B$ behaves like $\mc{O} (\mc{V}^{-7/3}).$
If we do not include anomaly mediated contributions, the dependence of D3
brane soft terms
on ${\cal{V}}, W_0$ and $g_s$ is summarised in table \ref{d3softtable}
for GUT, intermediate and TeV string scales.
\begin{table}
\label{d3softtable}
\caption{Soft terms for D3 branes
  (AMSB contributions not included)}
\centering
\vspace{3mm}
\begin{tabular}{|c|c|c|c|c|}
\hline
Scale & Mass & GUT & Intermediate & TeV \\
\hline
\textrm{Scalars} $m_i$ & $\frac{g_s^2}{({\cal{V}}_s^0)^{7/6}}W_0 M_P$ &
$3.6\ti 10^{11}$ GeV & $3.6\ti 10^4$GeV & $3.6\ti 10^{-17}$GeV\\
\textrm{Gauginos} $M_{D3}$ & $\frac{g_s^2}{({\cal{V}}_s^0)^2}W_0 M_P$ &
$3.6\times 10^9$GeV & $3.6\ti 10^{-3}$GeV & $3.6\ti 10^{-39}$ GeV\\
\textrm{A-term} $A$ & $\frac{g_s^2}{({\cal{V}}_s^0)^{4/3}}W_0 M_P$ &
$3.2\times 10^{11}$GeV
& $3.2 \ti 10^3$ GeV  & $3.2\ti 10^{-21}$GeV\\
$\mu$-\textrm{term} $\hat{\mu}$ & $\frac{g_s^2}{({\cal{V}}_s^0)^{4/3}}W_0
M_P$ & $3.2\times 10^{11}$GeV
& $3.2 \ti 10^3$ GeV  & $3.2\ti 10^{-21}$GeV\\
\textrm{B term} $\hat{\mu}B$ &
 $\frac{g_s^2}{({\cal{V}}_s^0)^{7/6}}W_0 M_P$ &
$3.6\ti 10^{11}$ GeV & $3.6\ti 10^4$GeV & $3.6\ti 10^{-17}$GeV\\
\hline
\end{tabular}
\end{table}

\subsection*{Soft Parameters for D7 Branes}

Standard Model matter may also be
supported on D7 branes.
The open string sector can give rise to several different
types of moduli in the low energy theory.
There are geometric moduli corresponding to deformations of the
internal 4-cycle $\Sigma$ that the D7 brane wraps.
As discussed in \cite{hepth0409098}, the number of such moduli is
related to the $(2,0)$ cohomology of
the cycle $\Sigma.$ There are also Wilson line moduli $a_I$,
which are present if the cycle $\Sigma$ possesses harmonic $(1,0)$
forms. There are no harmonic $(1,0)$ forms on Calabi-Yaus as $h^{1,0}
= 0$, but they may exist when we restrict to the submanifold $\Sigma$.
If
present, these enter the K\" ahler potential through the complexified K\"
ahler moduli, which are further redefined from (\ref{complexified}) to
\begin{equation}
\label{complexified2}
T_\alpha = \tau_\alpha + i\rho_\alpha + i\mu_3 l^2
(\omega_\alpha)_{\imath{\bar{\jmath}}} {\rm Tr} \phi^i
\left( \bar{\phi}^{\bar{\jmath}}
- {i\over2} \bar{U}^{\hat{a}} (\bar{\chi_{\hat{a}}})^{\bar{\jmath}}_l
\phi^l\right)
+ \mu_7 l^2 C^{I\bar{J}} a_I \bar{a}_{\bar{J}},
\end{equation}
for geometry dependent coefficients $C^{I\bar{J}}.$
The D7 geometric moduli generically obtain  large, ${\cal{O}} (m_{3/2})$
masses after turning on fluxes.
This is most easily seen from the F-theory perspective, where the
D7 moduli are among the complex structure moduli of the Calabi-Yau
4-fold $M_8$. 
The relevant Gukov-Vafa-Witten
superpotential is then
\begin{equation}
W = \int_{M_8} G_4\wedge\Omega,
\end{equation}
which generically induces a nontrivial potential for the D7 moduli.
These are analogous to dilaton and complex structure moduli and we expect the $\alpha'$ and
non-perturbative effects to only have small effects.

For the case of a single geometric D7 brane modulus, the K\" ahler
potential is
\be
\mc{K} = -\log (S+\bar{S} - {\cal{L}} |\zeta|^2),
\ee 
giving $\tilde{\K} = {\cal{L}}/(S+\bar{S}).$
As $F^S$ vanishes before breaking
the no-scale structure, for D7 branes
$F^m \bar{F}^{\bar{n}} \partial_m \partial_{\bar{n}} \log {\tilde{\K}}$ vanishes and
the flux-induced soft mass is ${\cal{O}} (m_{3/2})$.
Including $\alpha'$ corrections, $F^S$ is no longer zero and
\begin{equation}
F^S \bar{F}^{\bar{S}} \partial_S \partial_{\bar{S}} (-\log (S+\bar{S}))
= {1\over{(S+\bar{S})^2}} F^S \bar{F}^{\bar{S}} =
{\cal{O}} \left( {1\over{{{\cal{V}}^4}}}\right).
\end{equation}
This is a manifestly tiny correction to $m_{\zeta}$, and so
\begin{equation}
m_{\zeta} \approx m_{3/2} = {\cal{O}} (1) {g_s^2 W_0 \over{\sqrt{4\pi}{\cal V}^0_s}}M_P.
\end{equation}
Note that D7 moduli receive both supersymmetric and soft-breaking masses.

The masses of Wilson line moduli for D7 branes 
may be found through the modified K\" ahler coordinates (\ref{complexified2}).
These moduli appear in the K\" ahler potential in a similar
fashion to D3 moduli and
the calculation of their masses squared will be parallel.
We therefore expect the Wilson line moduli to obtain ${\cal{O}}
({\cal{V}}^{-7/6})$
masses due to nonperturbative effects:
\begin{equation}
m_{Wilson} = {\cal{O}} (1) {g_s^2 W_0\over{\sqrt{4\pi} ({\cal V}^0_s)^{7/6}}} M_P.
\end{equation}

For D7 branes, the gauge kinetic function $f_a$ is given by the K\"
ahler modulus, $T_a$, of the cycle the D-branes wrap:
\begin{eqnarray*}
f_a = {T_a\over{2\pi}}.
\end{eqnarray*}
In general $F^{T^a}\ne0$ and the gaugino masses are nonvanishing.
For the $\mathbb{P}^4_{[1,1,1,6,9]}$ example, we have
\bea
\label{NaiveM4}
M_4 = \frac{F^4}{2 \tau_4} & \sim & {\cal{V}}^{-1}, \\
M_5 = \frac{F^5}{2 \tau_5} & \sim & {\cal{V}}^{-1/3}/ {\cal{V}}^{2/3} = {\cal{V}}^{-1}.
\eea
In either case
\begin{equation}
M_{D7} = {\cal{O}} (1) {g_s^2 W_0 \over{\sqrt{4\pi} {\cal V}^0_s}}
\sim m_{3/2}.
\end{equation}
As noted, in fact there is a suppression in $F^4$. We shall see
shortly in section \ref{cqpap} that actually $M_4 \sim \frac{m_{3/2}}{\ln (m_{3/2})}$.

We note a crucial difference between D3 and D7 branes is that a supersymmetric
$\mu$-term is induced for geometric D7 moduli with only ISD fluxes;
in fact, it was shown in \cite{hepth0408036} that for vanishing magnetic fluxes on the
D7 brane, the $\mu$-term corresponds to the $(2,1)$ component of flux
$G_3.$ $\mu$ terms do not exist for Standard Model fermion multiplets, and so in
estimating the soft masses we only use the soft breaking contributions.

Let us now consider the A-term corresponding to a D7 scalar field $\phi$
with $\tilde{\K}_i=\tilde{\K}_j=\tilde{\K}_k=\frac{\cal{L}}{S+\bar{S}}.$
We have
\begin{equation}
A_{\phi\phi\phi} = F^4 (\partial_4 \hat{\K}) + F^5 (\partial_5 \hat{\K}) + F^S
(\partial_S \hat{\K}) + {3F^S\over{S+\bar{S}}}.
\end{equation}
As $\partial_S \hat{\K} = -1/(S+\bar{S}) + {\cal{O}} ({\cal{V}}^{-1})$ and
$F^S\sim {1/{\cal{V}}^2}$, the largest contribution to
the D7 A-term comes from $F^5 (\partial_5 \K)$,and so
\begin{equation}
A\sim F^5 (\partial_5 K) \sim {1\over{{\cal{V}}^{1/3}}}\cdot {1\over{{\cal{V}}^{2/3}}} = {1\over{{\cal{V}}}}.
\end{equation}
Thus the A-term is ${\cal{O}} (m_{3/2}).$

If we try and realise the Standard Model on D7 branes, there is one
potential worry. The Yang-Mills gauge coupling is determined by
$g_{YM,a}^{-2} = \textrm{Re } f_a$, and thus if $f_a = T^5$ then $g_{YM}$
would be unacceptably small. However, as shown above,
in general we have $T_b \gg 1$ but $T_{s,i}$ relatively small. Thus so long as
the Standard Model is realised on branes wrapping the smaller cycles,
the resulting gauge kinetic function will have phenomenologically
acceptable values. We show in table \ref{d7softtable} soft masses for
GUT, intermediate and TeV string scales.
\begin{table}
\label{d7softtable}
\caption{Soft terms for D7 branes (AMSB not included)}
\centering
\vspace{3mm}
\begin{tabular}{|c|c|c|c|c|}
\hline
Scale & Mass & GUT & Intermediate & TeV \\
\hline
\textrm{Scalars} $m_\zeta$ & $m_{3/2}$ &
$1.5\ti 10^{12}$ GeV & $1.5\ti 10^3$GeV & $1.5\ti 10^{-12} $GeV\\
\textrm{Gauginos} $M_4,M_5$ & $m_{3/2}$ &
$1.5\ti 10^{12}$ GeV & $1.5\ti 10^3$GeV & $1.5\ti 10^{-12} $GeV\\
\textrm{A-term} $A$ & $m_{3/2}$ &
$1.5\ti 10^{12}$ GeV & $1.5\ti 10^3$GeV & $1.5\ti 10^{-12} $GeV\\
$\mu$-\textrm{term} $\hat{\mu}$ & $m_{3/2}$ &
$1.5\ti 10^{12}$ GeV & $1.5\ti 10^3$GeV & $1.5\ti 10^{-12} $GeV\\
\textrm{B term} $\hat{\mu}B$ &
$m_{3/2}$ &  $1.5\ti 10^{12}$ GeV & $1.5\ti 10^3$GeV & $1.5\ti 10^{-12} $GeV\\
\hline
\end{tabular}
\end{table}

\subsection*{D3-D7 States}
The intersections of D3 and D7 branes can also give rise to massless open
string states. Unfortunately their appearance in the K\" ahler
potential of the low energy theory cannot be deduced from the dimensional
reduction of Dirac-Born-Infeld and Chern-Simons actions for stacks for
branes, and thus their soft terms are difficult to analyse from
the four-dimensional point of view. In \cite{hepth0408036},
the soft masses for 3-7 scalar fields due to a general
flux background were computed using various symmetry arguments.
It was found that no scalar or fermion masses are generated.
In some particular
examples where it is known how the 3-7 scalars enter the 4D K\" ahler
potential this can be seen from the four dimensional perspective. For
example, in
\cite{hepph9812397, hepth0406092} the K\" ahler potential was obtained
for D3/D7 compactifications on $T^2\ti T^2\ti T^2$. For a D7-brane
wrapping the first two tori, 
the dependence on 3-7 fields $\phi_{37}$ is
\begin{equation}
\K = { {|\phi_{37}|^2}\over{(T_1+T_1^*)^{1/2} (T_2+T_2^*)^{1/2}}} +\cdots.
\end{equation}
If there is only one overall K\" ahler
modulus $T=T_1=T_2=T_3$ this becomes
\begin{equation}
\K = {|\phi_{37}|^2 \over{T+\bar{T}}}+\cdots,
\end{equation}
and we obtain the same no-scale cancellation argument for $\phi_{37}$
that applied previously for D3 matter. We therefore likewise
expect that after including no-scale breaking effects, 
\begin{equation}
m_{37} \sim m_{D3} = {\cal{O}} (1) {g_s^2 W_0\over{\sqrt{4\pi} ({\cal V}^0_s)^{7/6}}}.
\end{equation}

\subsection*{D-terms and de Sitter Uplifting}

To perform a semirealistic computation of the soft supersymmetry
breaking terms, we must uplift the nonsupersymmetric AdS vacuum
obtained by fixing the moduli using $\alpha'$ corrections and
nonperturbative effects. This can be done in several ways. The
original option, 
proposed in \cite{hepth0301240}, involves the use of an anti-D3 brane
at the bottom of a highly warped Klebanov-Strassler throat. Although the resulting uplifting term is
reminiscent of a D-term in the low energy theory, supersymmetry is
broken explicitly, albeit by a small amount.  

An alternative method of uplift 
was proposed in \cite{hepth0309187}: turning on
magnetic fluxes on a D7 brane wrapping a compact 4-cycle in the
Calabi-Yau.
The advantage of this is that
the uplifting term generated can indeed be interpreted as a
D-term in the low energy theory. For fine-tuning the cosmological
constant, one again needs the brane to be in a
highly warped region.
A further mechanism was proposed
in \cite{hepth0402135}: instead of using a strongly warped region, one can
look for local minima of the no-scale potential with
$V_0\ne 0.$ This will happen for certain non-ISD choices of fluxes;
for example, in a model with only one K\" ahler modulus $T$, setting
$W=\int G_3\wedge \Omega$ gives a source term for $T$
\begin{equation}
V_0 = {1\over{(S+\bar{S})(T+\bar{T})^3}}\left| \int G_3^* \wedge \Omega \right|^2.
\end{equation}
If $G_3$ contains a (3,0) component $V_0$ will be non-vanishing.
The large number of flux choices generically present in Calabi-Yau
compactifications suggests there should exist non-supersymmetric
minima with sufficiently
small $\int G_3^* \wedge \Omega.$ The existence of non-supersymmetric
minima of the flux ensemble has been investigated statistically in \cite{hepth0411183}.

Irrespective of the details of the uplift mechanism,
there will be extra contributions to scalar masses.
For concreteness let us use an uplift appropriate to IASD fluxes
\begin{equation}
V_{uplift}={\epsilon\over{{\cal{V}}^2}} =
{\epsilon\over\bigl[
(T_5+\bar{T}_5-c|\phi|^2)^{3/2} - (T_4+\bar{T}_4-d|\phi|^2)^{3/2}\bigr]^2},
\end{equation}
expressed in terms of the complexified K\" ahler moduli $T_4$ and $T_5.$
This may be expanded around $\phi=0$ as
\begin{equation}
V_{uplift} = {\epsilon\over\bigl[(T_5+\bar{T}_5)^{3/2}-(T_4+\bar{T}_4)^{3/2}\bigr]^2}
\left( 1+{3|\phi|^2(c(T_5+\bar{T}_5)^{3/2}-d(T_4+\bar{T}_4)^{3/2})
\over{(T_5+\bar{T}_5)^{3/2}-(T_4+\bar{T}_4)^{3/2}}}\right).
\end{equation}
Ignoring $\alpha'$ corrections, the coefficient of $|\phi|^2$ in the brackets can be
identified with the kinetic term prefactor for the $\phi$ fields
(\ref{ktilde}). 
The uplift gives a contribution $\epsilon/{\cal{V}}^2$ to scalar
masses squared.
To uplift to Minkowski space we require $\langle V_{uplift} \rangle = \langle
V_{min} \rangle\sim {\cal{O}} (1/{{\cal{V}}^3})$ and thus
$\epsilon\sim 1/{\cal{V}}.$
Consequently the uplift contribution to scalar masses squared is ${\cal{O}}
(1/{\cal{V}}^3)$,
which is much smaller than the ${\cal{O}} ({\cal{V}}^{-7/3})$
contribution obtained from
nonperturbative and $\alpha'$ effects.

If IASD fluxes are present, we no longer have $F^U = 0$.
However, if the IASD fluxes are to be used as a lifting term, we can
  estimate the magnitude of the complex structure F-terms. We require
$$
\frac{e^{\K_{cs}} \K_{cs}^{\bar{\imath}\jmath} (D_{\bar{\imath}} W)
  (D_j W)}{\mc{V}^2}
\sim \frac{1}{\mc{V}^3},
$$
and so $D_i W\sim {\cal{V}}^{-1/2}$. It then follows that $F^U \sim
\mc{V}^{-\frac{3}{2}}$
and the resulting effect on the masses is subleading to the F-terms
associated with the K\"ahler moduli. 

The conclusion is that the uplift has negligible effect on the soft terms.
This result is in fact intuitive. Because of the underlying no-scale
structure of the scalar potential, the magnitude of the potential is
far smaller than would be naively expected from the size of the F-terms. In
particular, $V \sim \mc{V}^{-3}$ while $\vert F \vert^2
\sim \mc{V}^{-2}$. Thus the `extra' supersymmetry breaking that must
be introduced to uplift to de Sitter space is much smaller than the
susy breaking already present in the AdS vacuum, and so does not alter
the soft terms already computed for the AdS minimum.

\subsection*{Comparison with Anomaly-Mediated Contributions}

In a general model with hidden sector supersymmetry breaking,
scalar masses, gaugino masses and A-terms are generated through loop effects
as a consequence of the super-Weyl anomaly \cite{hepth9810155, hepph9810442}. Assuming that soft terms
are generated solely through anomaly mediation, their values are
\begin{eqnarray}
M &=& {\beta_g\over{g}} m_{3/2},\nonumber\\
m_i^2 &=& -{1\over4}\left( {\partial \gamma_i \over \partial g}\beta_g
+ {\partial \gamma_i \over \partial y}\beta_y\right) m_{3/2}^2,\nonumber\\
A_y &=& -{\beta_y\over y}m_{3/2},
\end{eqnarray}
for gaugino mass $M$ and scalar masses $m_i.$
Here $\beta$ are the relevant $\beta$ functions, $\gamma_i$ 
the anomalous dimensions of the chiral superfields, and $g$ and $y$
respectively denote the gauge and Yukawa couplings, respectively.
One can alternatively use the compensator field formalism to write
\begin{eqnarray}
M &=& {{b g^2}\over{8\pi^2}} {F^C\over C_0},\nonumber\\
A_{ijk} &=& -{\gamma_i+\gamma_j+\gamma_k\over{16\pi^2}} {F^C\over C_0},
\nonumber\\
m_i^2 &=& {1\over{32\pi^2}} {d\gamma_i\over{d\log\mu}}\left|{F^C\over C_0}
\right|^2
+ {1\over{16\pi^2}} \left( \gamma^i_m F^m \left({F^C\over C_0}\right)^*\right),
\end{eqnarray}
where $C$ is the compensator field and $\gamma^i_m \equiv
\partial_m\gamma^i.$
In this case $F^C/C_0\sim m_{3/2}$.
The contribution to scalar masses from anomaly mediation is then (see \cite{hepth0208123})
\begin{equation}
\label{amsbcont}
m \sim b_0 \left( {g^2\over{16\pi^2}}\right) m_{3/2},
\end{equation}
where $b_0$ is the one loop beta function coefficient. Note that
(\ref{amsbcont})
is suppressed with respect to the gravitino mass by a factor
$1/(16\pi^2).$
In a similar fashion, the gaugino masses are
suppresed compared to $m_{3/2}$ by $1/(8\pi^2).$
For D3 branes, the large volume suppression in the gaugino mass
formula (\ref{D3gaugmass}) implies that anomaly mediation ought to dominate gaugino
mass generation. As the D3 scalar masses are also suppressed compared
to the gravitino mass in (\ref{D3masssca}), the anomaly-mediated contribution will
also be important at large volume. A full phenomenological analysis of
the D3-brane scenario should therefore include anomaly-mediated contributions.

For D7 branes, a naive application of the above results suggests that
anomaly mediation will be irrelevant as all soft masses are
comparable to the gravitino mass. However we shall now see that this is
not the case as gaugino masses are in fact suppressed compared to the
gravitino mass. In this case anomaly mediation contributes to gaugino
masses but not scalar masses (which will have $m_i^2 \sim m_{3/2}^2$).

We now justify these claims by analysing the fine structure of the
soft terms.
 
\section{Soft Parameters: Fine Structure}
\label{cqpap}

This thesis has mostly focused on the large-volume models developed in
the previus chapter. However we now show a more general result, namely that whenever a modulus $T_a$ in the IIB
landscape is stabilised by nonperturbative effects there is a
small hierarchy between the masses of the gravitino $m_{3/2}$ and the
associated D7 gaugino $M_a$,
\be
\label{GauginoRelation}
M_a^2 \sim \frac{m_{3/2}^2}{\ln(m_{3/2})^2}.
\ee
The large-volume models of chapter \ref{chapterLargeVol} are a
particular case of this.
This small hierarchy was first identified in single-modulus
KKLT models \cite{hepth0411066} (with a two-modulus example also studied in
\cite{hepph0511162}). For the KKLT case this suppression of $M_a$ leads to mixed
modulus-anomaly mediation and
the phenomenology of this scenario
has been analysed in \cite{hepth0503216, hepph0504036, hepph0504037,
  hepph0507110, hepph0508029, hepph0509039, hepth0509158,
  hepph0511320, hepth0603047, hepph0604192, hepph0604253}.\footnote{In these
  references this hierarchy is present  for all soft supersymmetry
  breaking terms. We can establish it in generality only for
  gaugino masses. As discussed below, in the
  large volume scenario scalar masses are generically $\mc{O}(m_{3/2})$.} 
Here
we show this relation to be a truly
generic feature of the landscape, by showing it to hold for arbitrary
multi-modulus models and to be independent of
the precise details of the scenario used to stabilise the
moduli. 

In section \ref{secSM} I also demonstrate soft scalar mass
universality in the large volume scenario and estimate the fractional
non-universality to be $1/\ln (m_{3/2})^2 \sim 1/1000$.
This is an interesting result as flavour non-universality is usually a
serious problem for gravity-mediated models. I briefly discuss the
phenomenology but do not anticipate the detailed discussion that will
appear in \cite{ACQSinprogress}.

\subsection*{Gaugino Masses}

For IIB flux compactifications
the appropriate K\"ahler potential and superpotential are given by
\bea
\label{KahlerPot}
\mc{K} & = & - 2 \ln \left( \mc{V} + \frac{\hat{\xi}}{2} \right),
\nonumber \\
\label{WDef}
W & = & W_0 + \sum A_i e^{-a_i T_i}.
\eea
Here $\mc{V} = \frac{1}{6}k_{ijk}t^i t^j t^k$ is the Calabi-Yau
volume, $\hat{\xi} = - \frac{\chi(M) \zeta(3)}{2(2 \pi)^3
g_s^{3/2}}$
and $T_i = \tau_i + i b_i$ are the K\"ahler moduli, corresponding to
the volume $\tau_i = \frac{\partial \mc{V}}{\partial t^i}$ of a
4-cycle $\Sigma_i$, complexified by the axion
$b_i = \int_{\Sigma_i} C_4$. $W_0 = \left\langle \int G_3 \wedge \Omega \right\rangle$ is the flux-induced superpotential
that is constant after integrating out dilaton and complex structure moduli.
In general there does not exist an explicit expression for $\mc{V}$ in
terms of the $T_i$.
We have included the perturbative K\"ahler corrections of
\cite{hepth0204254}. These are crucial in the exponentially large volume 
scenario but are not important for KKLT. We use single exponents in
the superpotential and do not consider
racetrack scenarios.

For a D7-brane wrapped on a cycle $\Sigma_i$, the gauge
kinetic function is given by
\be
f_i = \frac{T_i}{2\pi}.
\ee
As above, the gaugino masses are
computed through the general expression (\ref{gauginomass})
\be
M_a = \half \frac{1}{\hbox{Re } f_a} \sum_\alpha F^\alpha \partial_\alpha f_a,
\ee
where $a$ runs over gauge group factors and  $\alpha$ over the moduli fields.
The F-term $F^\alpha$ is 
\bea
F^\alpha & = & e^{\mc{K}/2} \sum_{\bar{\beta}} \mc{K}^{\alpha \bar{\beta}} D_{\bar{\beta}}
\bar{W} \nonumber \\
& = & e^{\mc{K}/2} \sum_{\bar{\beta}} \mc{K}^{\alpha \bar{\beta}} \left( \partial_{\bar{\beta}}
\bar{W} + (\partial_{\bar{\beta}} \mc{K}) \bar{W} \right),
\eea
where we have expanded the covariant derivative
$D_{\bar{\beta}}
\bar{W} = \partial_{\bar{\beta}} \bar{W} + (\partial_{\bar{\beta}} \mc{K})
\bar{W}$.

Thus for a brane wrapping cycle $k$ we have
\be
\label{GauginoMass}
M_k = \half \frac{F^k}{\tau_k}.
\ee
It is a property of the K\"ahler potential $\mc{K} = - 2 \ln (\mc{V} +
\frac{\hat{\xi}}{2})$
that
\be
\label{Relation}
\sum_{\bar{j}} \mc{K}^{k \bar{j}} \partial_{\bar{j}} \mc{K} = -2
\tau_k \left(1 + \frac{\hat{\xi}}{4\mc{V}}\right) \equiv -2 \hat{\tau}_k.
\ee
We therefore obtain
\be
F^k = e^{\mc{K}/2} \left( \sum_{\bar{j}} \mc{K}^{k \bar{j}} \partial_{\bar{j}}
\bar{W} - (2 \hat{\tau}_k) \bar{W} \right).
\ee
From the superpotential (\ref{WDef}), we see that $\partial_{\bar{j}} \bar{W} = -a_j \bar{A_j} e^{-a_j \bar{T}_j}$,
and so
\be
\label{FTerm}
F^k = e^{\mc{K}/2} \left( \sum_{\bar{j}} - \mc{K}^{k \bar{j}} a_j \bar{A_j}
e^{-a_j \bar{T}_j} - 2 \hat{\tau}_k \bar{W} \right).
\ee
We now show that if the modulus $T_k$ is stabilised by
nonperturbative effects, the two terms in equation (\ref{FTerm})
cancel to leading order.
To see this, we start with the full F-term supergravity potential,
\be
\label{Ftermpott}
V_F = e^{\mc{K}} \mc{K}^{i \bar{j}} \partial_i W \partial_{\bar{j}} \bar{W}
+ e^{\mc{K}} \mc{K}^{i \bar{j}} \left( (\partial_i K) W
\partial_{\bar{j}} \bar{W} + (\partial_{\bar{j}} \mc{K}) \bar{W}
\partial_i W \right) + e^{\mc{K}} (\mc{K}^{i \bar{j}} \mc{K}_i
\mc{K}_{\bar{j}} - 3) \vert W \vert^2.
\ee
As in chapter \ref{chapterLargeVol}, this becomes
\be
\label{ExpandedPotential}
V = \sum_{i \bar{j}} \frac{\mc{K}^{i \bar{j}} (a_i A_i) (a_j \bar{A}_j)
e^{-a_i T_i - a_j \bar{T}_j}}{\mc{V}^2} + \sum_j \frac{ 2 \hat{\tau}_j \left( a_j
\bar{A}_j e^{-a_j \bar{T}_j} W + a_j A_j \bar{W} e^{-a_j T_j}
\right)}{\mc{V}^2}
+ \frac{3 \hat{\xi} \vert W \vert^2}{4 \mc{V}^3}.
\ee
The perturbative K\"ahler corrections of (\ref{KahlerPot}) break no-scale,
giving the third term of (\ref{ExpandedPotential}).
We have only displayed the leading large-volume behaviour of these
corrections. This is reasonable as in KKLT such corrections are
not important, while in the exponentially large volume 
scenario $\mc{V} \gg 1$ and the higher volume-suppressed terms are
negligible.
We shall also assume throughout that $m_{3/2} \ll M_P$. This is
motivated by phenomenological applications and is any case necessary
to make sense of a small hierarchy governed by $\ln (M_P/m_{3/2})$.

We assume the modulus $T_k$ is stabilised by effects non-perturbative
in $T_k$
and locate the stationary locus by extremising
with respect to its real and imaginary parts. We first perform the
calculation keeping the dominant terms to demonstrate the cancellation
in (\ref{FTerm}). We subsequently show that the subleading terms are
indeed subleading and estimate their magnitude.

\subsubsection*{Leading  Terms}

We first
solve for the axionic component,
$\partial V / \partial b_k = 0$. The axion only appears in the
superpotential and we have
\bea
\label{LongEquation}
\frac{\partial V}{\partial b_k} & = & \frac{i}{\mc{V}^2} \Bigg[
  \sum_{i \bar{j}} \mc{K}^{i \bar{j}} (a_i A_i)(a_j \bar{A}_j) \left[
    -a_i \delta_{i k} + a_j \delta_{j k} \right] e^{-(a_i T_i + a_j
    \bar{T}_j)} + \nonumber \\
& &
+ \sum_j 2 a_j \hat{\tau}_j \left( \bar{A}_j W a_j \delta_{jk} e^{-a_j
  \bar{T}_j} - A_j \bar{W} a_j \delta_{jk} e^{-a_j T_j} \right) \Bigg]
+ \\
& &  \sum_j \frac{ 2 \hat{\tau}_j \left( a_j
\bar{A}_j e^{-a_j \bar{T}_j} (\partial_{b_k} W) + a_j A_j (\partial_{b_k}
  \bar{W}) e^{-a_j T_j}
\right)}{\mc{V}^2} + \frac{3 \hat{\xi} ((\partial_{b_k} W) \bar{W} + W
  (\partial_{b_k} \bar{W}))}{4 \mc{V}^3}  \nonumber \\
& = & \frac{i}{\mc{V}^2} \Bigg[ - \sum_{j} \mc{K}^{k \bar{j}} (a_k^2
  A_k) (a_j \bar{A}_j) e^{-(a_k T_k + a_j \bar{T}_j)} + \sum_i
  \mc{K}^{i \bar{k}} (a_i A_i) (a_k^2 \bar{A}_k) e^{-(a_i T_i + a_k
    \bar{T}_k )} + \nonumber \\
& & + 2 a_k^2 \hat{\tau}_k \left( \bar{A}_k W e^{-a_k
    \bar{T}_k} - A_k \bar{W} e^{-a_k T_k} \right) \Bigg].
\label{ShorterEquation}
\eea
In going from (\ref{LongEquation}) to (\ref{ShorterEquation}) we have
dropped the third line of (\ref{LongEquation}) as subleading. We will
estimate the magnitude of these subleading terms below.

We now change the dummy index in (\ref{ShorterEquation}) from $j$ to $i$, and use $\mc{K}^{k \bar{i}}
= \mc{K}^{i \bar{k}}$ together with $\frac{\partial V}{\partial b_k} =
0$ to obtain
\bea
\label{AxionEquation}
2 \hat{\tau}_k (\bar{A}_k W e^{-a_k \bar{T}_k} - A_k \bar{W} e^{-a_k
  T_k}) & = &
\sum_i \mc{K}^{k \bar{i}} \left( (a_i \bar{A}_i) A_k  e^{-(a_k T_k +
  a_i \bar{T}_i)} - (a_i A_i) \bar{A}_k e^{-(a_k \bar{T}_k +a_i T_i)}
\right)  \nonumber \\
& & + (\hbox{subleading terms}).
\eea
As the axion does not appear (at least in perturbation theory) in the
K\"ahler potential, its stabilisation is always entirely due to
  nonperturbative superpotential effects.

We next consider the stabilisation of $\tau_k = \hbox{Re}(T_k)$.
As stated above, our main assumption is that $T_k$ is stabilised by
superpotential effects nonperturbative in $T_k$. Another way of stating this
is to say that, when computing $\frac{\partial V}{\partial \tau_k}$,
the dominant contribution must arise from the superpotential term $A_k
e^{-a_k T_k}$: if this were not the case, our assumption about how
$T_k$ is stabilised is invalid.
In evaluating $\frac{\partial V}{\partial \tau_k}$,
we therefore focus on such terms as dominant and neglect terms
arising from e.g. $\frac{\partial}{\partial \tau_k} \left( \frac{\mc{K}^{i
  \bar{j}}}{\mc{V}^2}\right)$ as subdominant. We show below
 that the magnitude of the subdominant terms is suppressed by factors of
$\ln \left( m_{3/2} \right)$.

If we only consider superpotential terms, the calculation of
$\frac{\partial V}{\partial \tau_k}$
exactly parallels that of $\frac{\partial V}{\partial b_k}$ above. The only
differences lie in the signs, as
$$
\frac{\partial T_k}{\partial \tau_k} = \frac{\partial \bar{T}_k}{\partial \tau_k}
= 1, \hbox{ whereas } \frac{\partial T_k}{\partial b_k} =
-\frac{\partial \bar{T}_k}{\partial b_k} = i.
$$
In a similar fashion to (\ref{ShorterEquation}) we therefore obtain
\bea
\label{Tderivative}
\frac{\partial V}{\partial \tau_k} & = & \frac{a_k^2}{\mc{V}^2} \Big[
  \sum_i \mc{K}^{k \bar{i}} \Big( - (a_i \bar{A}_i) A_k e^{-(a_k T_k
    +a_i \bar{T}_i)} -  (a_i A_i) \bar{A}_k e^{-(a_k \bar{T}_k +a_i
    T_i)} \Big) \nonumber \\
& &
- 2 \hat{\tau}_k \left( \bar{A}_k W e^{-a_k \bar{T}_k} +
  A_k \bar{W} e^{-a_k T_k} \right) \Big] + (\hbox{subleading terms}).
\eea
Setting $\frac{\partial V}{\partial \tau_k}=0$ then implies
\bea
\label{TauEquation}
2 \hat{\tau}_k \left( \bar{A}_k W e^{-a_k \bar{T}_k} + A_k \bar{W} e^{-a_k
  {T}_k} \right) & = & - \sum_i \mc{K}^{k \bar{i}} \left( a_i \bar{A}_i A_k
e^{-(a_k T_k + a_i \bar{T}_i)} + a_i A_i \bar{A}_k e^{-(a_k \bar{T}_k
  + a_i T_i)} \right) \nonumber  \\
& & + (\hbox{subleading terms}).
\eea
We now sum (\ref{AxionEquation}) and (\ref{TauEquation}) to obtain
\bea
4 \hat{\tau}_k \bar{A}_k W e^{-a_k \bar{T}_k} & = & -2 \sum_i \mc{K}^{k
  \bar{i}} a_i A_i \bar{A}_k e^{-(a_k \bar{T}_k + a_i T_i)} \nonumber \\
\Rightarrow -2 \hat{\tau}_k W & = & \sum_i \mc{K}^{k \bar{i}} a_i A_i
  e^{-a_i T_i} \nonumber \\
\label{LastEquation}
\Rightarrow - 2 \hat{\tau}_k \bar{W} & = & \sum_i \mc{K}^{\bar{k} i} a_i \bar{A}_i
  e^{-a_i \bar{T}_i}.
\eea
Comparison with equations (\ref{GauginoMass}) and (\ref{FTerm}) shows
  that there exists a leading-order cancellation in the computation of
  the gaugino mass. This cancellation has followed purely from the
  assumption that the modulus $\tau_k$ was stabilised by non-perturbative
  effects: we have only required $\frac{\partial V}{\partial \tau_k} =
  0$ and not $D_{T_k} W = 0$. 

In deriving equations (\ref{ShorterEquation}) and (\ref{Tderivative}) we dropped
  subleading terms suppressed by $\ln \left( \frac{M_P}{m_{3/2}}
  \right)$. We then expect the cancellation from (\ref{LastEquation})
  and (\ref{FTerm}) to
  fail at this order, giving
\be
F^k \sim - 2 \frac{\hat{\tau}_k e^{\mc{K}/2} \bar{W}}{\ln(m_{3/2})}.
\ee
Equation (\ref{GauginoMass}) then gives
\be
M_k \sim \frac{e^{\mc{K}/2}\bar{W}}{\ln(m_{3/2})} = \frac{m_{3/2}}{\ln (m_{3/2})},
\ee
the relation we sought.

The above argument is general and model-independent. We have used the
K\"ahler potential appropriate to an arbitrary compactification,
making no assumptions about the number of moduli. Furthermore, as the
result comes from directly extremising the scalar potential it is
independent of whether the moduli stabilisation is approximately
supersymmetric or not.
Indeed, the argument above has
  not depended on finding a global minimum of the potential, or even on
  extremising the potential with respect to any of the moduli except $T_k$.
This result shows that the small hierarchy of (\ref{GauginoRelation})
will exist in both
KKLT and exponentially large volume approaches to moduli
  stabilisation. In the latter case this is possibly
  surprising as
  the minimum is in no sense approximately susy: each contribution to
  the sum in (\ref{FTerm}) individually gives $M_a \sim m_{3/2}$: it
  is only when summed the mass suppression is obtained.

We note, as an aside, that if \emph{all} moduli are
  stabilised by non-perturbative effects then by contracting
  (\ref{LastEquation}) with $\mc{K}_{j \bar{k}}$ we obtain
\be
-2 \sum_k \mc{K}_{j \bar{k}} \hat{\tau}_k \bar{W} = a_j \bar{A}_j
  e^{-a_j \bar{T}_j}.
\ee
Now, $\mc{K}_{j}$ is homogeneous of degree -1 in $\tau_k$, so
  recalling that $\frac{\partial}{\partial \tau_k} = 2
  \frac{\partial}{\partial T_k}$,
$$
\sum_k -2 \mc{K}_{j \bar{k}} \tau_k = \sum_k - \frac{\partial
  \mc{K}_j}{\partial \tau_k} \tau_k = \mc{K}_j,
$$
and therefore to leading order
\be
\label{SusyStab}
\partial_j W + (\partial_j \mc{K}) \bar{W} = 0.
\ee
Consequently if \emph{all} moduli are stabilised by nonperturbative
effects then the stabilisation is always approximately supersymmetric.

\subsubsection*{Subleading terms}

We now want to show that the terms neglected in computing
$\frac{\partial V}{\partial \tau_k}$ are all suppressed, under the
assumption that the modulus is solely stabilised by nonperturbative
effects. For concreteness we focus on the term in the potential
\be
\label{PotentialFirstTerm}
\sum_{i, \bar{j}} \left( \frac{\mc{K}^{i \bar{j}}}{\mc{V}^2} \right) (a_i A_i)(a_j \bar{A}_j)
  e^{-(a_i T_i + a_j \bar{T}_j)}.
\ee
The argument used for this term will also apply to the other terms of (\ref{ExpandedPotential}).
$ \frac{\mc{K}^{i \bar{j}}}{\mc{V}^2}$ is homogeneous in the $\tau_k$
of degree $-1$. To see this, we note that as by dimensional analysis
$\mc{V}$ is homogeneous in the $\tau_i$ of degree 3/2, $\mc{K}_{i
  \bar{j}} = \frac{\partial}{\partial T_i} \frac{\partial}{\partial
  \bar{T}_j} ( -2 \ln (\mc{V}))$
  is homogeneous in the $\tau_i$ of degree -2, and so $\mc{K}^{i
    \bar{j}}$ is homogeneous in the $\tau_i$ of degree 2. Therefore,
  summing over $k$,
\be
\sum_k \tau_k \frac{\partial}{\partial \tau_k} \left( \frac{\mc{K}^{i
    \bar{j}}}{\mc{V}^2} \right) = - \frac{\mc{K}^{i
    \bar{j}}}{\mc{V}^2},
\ee
and so
\be
\frac{\partial}{\partial \tau_k} \left( \frac{\mc{K}^{i
    \bar{j}}}{\mc{V}^2}\right) \lesssim \frac{\mc{K}^{i
    \bar{j}}}{\tau_k \mc{V}^2}.
\ee
Cosequently, differentiating (\ref{PotentialFirstTerm}) w.r.t $\tau_k$ gives
$$
\mc{O}\left( \frac{1}{\tau_k} \sum_{i,j} \frac{\left(\mc{K}^{i \bar{j}} (a_i A_i) (a_j
  \bar{A}_j) e^{-(a_i T_i + a_j \bar{T}_j)} \right)}{\mc{V}^2} \right)
+ a_k \sum_j \frac{\left(\mc{K}^{k \bar{j}} (a_k A_k) (a_j \bar{A}_j)
  e^{-(a_k T_k + a_j \bar{T}_j)} + c.c\right)}{\mc{V}^2}.
$$
The basic assumption we make is that the location of the minimum for
$\tau_k$ is dominantly determined by the effects nonperturbative in
$\tau_k$. Therefore in the first sum we should only include the terms which
depend nonperturbatively on $a_k T_k$. This gives
\be
\label{Rewrite}
\mc{O}\left( \frac{1}{\tau_k} \sum_j \frac{\left(\mc{K}^{k \bar{j}}
  a_j a_k \bar{A}_j A_k
  e^{-(a_k T_k + a_j \bar{T}_j)} + c.c\right)}{\mc{V}^2} \right)
+ \sum_j \frac{\left(\mc{K}^{k \bar{j}} (a_k^2 A_k) (a_j \bar{A}_j)
  e^{-(a_k T_k + a_j \bar{T}_j)} + c.c\right)}{\mc{V}^2}.
\ee
We then see that the first
term of (\ref{Rewrite}) is suppressed compared to the second by a factor $a_k
\tau_k$.

We note that there can exist moduli $\tau_k$ not stabilised
by effects nonperturbative in $\tau_k$ - this certainly holds for the
volume modulus for the large volume models. 
Furthermore, one can argue that in order to avoid
generating a potential for the QCD axion, the modulus $\tau_k$ associated with the QCD cycle should
be stabilised without using effects nonperturbative in $\tau_k$. This
point will be discussed further in chapter \ref{ChapterAxions}.
The argument above is restricted to the
case where the modulus $\tau_k$ is stabilised by effects
nonperturbative in $\tau_k$.

An identical analysis applies to the other two terms of
equation (\ref{ExpandedPotential}). As the K\"ahler dependent terms
are polynomials in $\tau_k$, derivatives of these with respect to
$\tau_k$ also give a suppression factor of $\tau_k$, while the derivatives
of superpotential exponents are enhanced by a factor $a_k$. The latter (which we
keep) are therefore larger by a factor $a_k \tau_k$ than the terms discarded.

In passing from (\ref{LongEquation}) to (\ref{ShorterEquation}) we
dropped the last line of (\ref{LongEquation}). This is self-consistent
so long as $\sum_j A_j e^{-a_j T_j}$ is suppressed compared to $W$. In
the models of chapter \ref{chapterLargeVol} this is trivial as $e^{-a_k
  \tau_k} \sim \frac{1}{\mc{V}}$ while $W \sim \mc{O}(1)$. In KKLT
models, as 
$$
\partial_{T_i} \mc{K} = -2 \frac{\partial_{T_i} \mc{V}}{\mc{V}} \lesssim
\frac{2}{\tau_i},
$$
we can use (\ref{SusyStab}) to see that
\be
\bar{W} \gtrsim (a_k \tau_k) A_k e^{-a_k \tau_k},
\ee
and so the third line of (\ref{LongEquation}) is suppressed compared
to the second by a factor of (at least) $a_k \tau_k$. 

The above arguments imply that the terms dropped in our evaluation
of $\frac{\partial V}{\partial \tau_k}$ are either 
suppressed by a factor of $a_k \tau_k \sim \ln(m_{3/2})$ or
have no non-perturbative dependence on $\tau_k$.
Consequently the gaugino mass suppression found above will hold at
leading order in $\frac{1}{\ln (m_{3/2})}$.

\subsubsection*{The uplift term}

In order to attain almost vanishing but positive vacuum energy, an
uplift term must also be
included. For KKLT models this is in a sense responsible for the soft
masses, as the original AdS minimum is supersymmetric. In the
large volume models the AdS minimum is already
non-supersymmetric and the contribution of the uplift to soft terms
is less relevant. 

We take as
uplift 
\be V_{uplift} = \frac{\epsilon}{\mc{V}^\alpha}, \ee 
where the power $4/3 \le \alpha \le 2$ depends on the uplift mechanism
\cite{hepth0301240, hepth0309187, hepth0402135}.
Including
this phenomenological uplift term, the full potential is \be
\label{UpliftedPotential} V_{full} = V_{SUGRA} + V_{uplift}. \ee At
the minimum $\langle V_{full} \rangle = 0$ and so we must have
$$
\langle V_{SUGRA} \rangle = - \langle V_{uplift} \rangle.
$$
$\mc{V}^{-\alpha}$ is homogeneous of degree $-3\alpha/2$ in the $\tau_i$ and so
\be
\sum_k \tau_k \frac{\partial}{\partial \tau_k} \mc{V}^{-\alpha} =
-\frac{3\alpha}{2}  \mc{V}^{-\alpha},
\ee
implying
\be
\frac{\partial}{\partial \tau_k} \mc{V}^{-\alpha} \lesssim - \frac{3\alpha}{2\tau_k}
\mc{V}^{-\alpha}.
\ee
Thus $\frac{\partial V_{uplift}}{\partial \tau_k}$ is
suppressed compared to $V_{uplift}$ by a factor of $\tau_k$.
In contrast, the derivatives of $V_{SUGRA}$ involve an enhancement by
$a_k$ due to the exponentials. As the two terms of
(\ref{UpliftedPotential}) are by definition equal at the minimum, we
see that
\be
\frac{\partial V_{SUGRA}}{\partial \tau_k} \gtrsim a_k \tau_k
\frac{\partial V_{uplift}}{\partial \tau_k}.
\ee
This fits in with
our previous analysis: the terms giving rise to the cancellation in (\ref{FTerm})
are the leading ones, with subleading terms suppressed by $a_k
\tau_k$. We do not control the subleading terms and they will
generically be non-vanishing, giving further contributions to the gaugino
masses at $\mc{O}\left(\frac{m_{3/2}}{\ln(m_{3/2})}\right)$.

As an aside, we note that for the large volume models
\be
\frac{\partial V_{uplift}}{\partial \tau_k} \sim \frac{1}{\mc{V}}
V_{uplift},
\ee
and so the presence of the uplift term does not significantly affect
the stabilisation of $\tau_k$. 

\section{Explicit Calculations for $\mbb{P}^4_{[1,1,1,6,9]}$}

The above has established that a suppression of gaugino masses
compared to the gravitino mass is generic in the landscape.
We now illustrate the above with explicit calculations for
the $\mbb{P}^4_{[1,1,1,6,9]}$ model analysed in chapter \ref{chapterLargeVol}.

\subsection{Gaugino Masses}

We first briefly recall the relevant properties of this model from
chapter \ref{chapterLargeVol}. As the manifold has $h^{1,1}=2$, there are two moduli, $T_b$ and
$T_s$. $T_b$ controls the overall volume and $T_s$ is a blow-up mode.
At the minimum $\tau_b = \hbox{Re}(T_b) \gg \tau_s = \hbox{Re}(T_s)
\sim \ln(\tau_b)$.
The K\"ahler and superpotential are
given by\footnote{For simplicity we drop a factor of
  $\frac{1}{9\sqrt{2}}$ from $\mc{V}$: this does not affect the results.}
\bea
\mc{K} & = & - 2 \ln \left( \mc{V} + \frac{\hat{\xi}}{2} \right) \equiv -
2 \ln \left( \tau_b^{3/2} - \tau_s^{3/2} + \frac{\hat{\xi}}{2}
\right), \\
W & = & W_0 + A_s e^{-a_s T_s} + A_b e^{-a_b T_b}.
\eea
The resulting K\"ahler metric is (dropping terms subleading in powers
of $\mc{V}$)
\be
\label{Metric}
\mc{K}_{i \bar{j}} = \left( \begin{array}{cc} \mc{K}_{b \bar{b}} &
  \mc{K}_{b \bar{s}} \\ \mc{K}_{s \bar{b}} & \mc{K}_{s \bar{s}}
\end{array} \right) = \left( \begin{array}{cc} \frac{3}{4 \mc{V}^{4/3}} &
  -\frac{9 \tau_s^{\half}}{8 \mc{V}^{5/3}} \\
-\frac{9 \tau_s^{\half}}{8 \mc{V}^{5/3}} & \frac{3}{8 \sqrt{\tau_s}
  \mc{V}} \end{array} \right),
\ee
with inverse metric
\be
\label{InverseMetric}
\mc{K}^{i \bar{j}} =  \left( \begin{array}{cc} \mc{K}^{b \bar{b}} &
  \mc{K}^{b \bar{s}} \\ \mc{K}^{s \bar{b}} & \mc{K}^{s \bar{s}}
\end{array} \right) = \left( \begin{array}{cc} \frac{4 \mc{V}^{4/3}}{3} &
  4 \tau_s \tau_b \\
4 \tau_s \tau_b & \frac{8 \sqrt{\tau_s} \mc{V}}{3} \end{array} \right).
\ee

In a limit $\mc{V} \equiv \tau_b^{3/2} - \tau_s^{3/2} \gg 1$ with
$\tau_s \sim \mc{O}(1)$, direct
evaluation of the scalar potential gives (dropping terms
subleading in $\mc{V}$),
\be
\label{BBCQPotential}
V = \frac{\lambda a_s^2 A_s^2 \sqrt{\tau_s} e^{-2 a_s \tau_s}}{\mc{V}}
- \frac{\mu a_s A_s \tau_s \vert W_0 \vert e^{-a_s \tau_s}}{\mc{V}^2} +
\frac{\nu \vert W_0 \vert^2}{\mc{V}^3}.
\ee
Explicitly, $\lambda = \frac{8}{3}$ and $\mu = 4$. The minus sign in
(\ref{BBCQPotential}) comes from minimising the potential with respect
to the axion $b_s$. As
$\tau_b \gg 1$ all terms nonperturbative in $\tau_b$ vanish. Noting
that
 $\frac{\partial}{\partial \tau_s}\left( \mc{V}^{-1} \right) \sim
\mc{O}\left(\frac{1}{\mc{V}^2}\right)$, we obtain
\be
\label{aaa}
\frac{\partial V}{\partial \tau_s} = \frac{\lambda a_s^2 A_s^2 \sqrt{\tau_s}
  e^{-2 a_s \tau_s}}{\mc{V}} \left( -2 a_s + \frac{1}{2 \tau_s}
\right) - \frac{\mu a_s A_s e^{-a_s \tau_s} \vert W_0 \vert}{\mc{V}^2}
\left( -a_s \tau_s +1 \right) + \mc{O}\left( \frac{1}{\mc{V}^2}
\right).
\ee
Imposing $\frac{\partial V}{\partial \tau_s} = 0$ and rearranging (\ref{aaa})
gives
\be
\label{SmallModSolution}
e^{-a_s \tau_s} = \left( \frac{\mu}{2 \lambda}\right)\frac{\vert W_0
  \vert}{\mc{V} a_s} \sqrt{\tau_s} \left(1 - \frac{3}{4 a_s \tau_s}
  \right) + \mc{O}\left(\frac{1}{(a_s \tau_s)^2} \right).
\ee
If we also solve $\frac{\partial V}{\partial \mc{V}} = 0$, we obtain
$$
\mc{V} \sim \left\vert \frac{W_0}{A_s} \right\vert e^{a_s \tau_s} \qquad \textrm{ with }
\tau_s \sim \hat{\xi}^{\frac{2}{3}}.
$$
$\mc{V}$ is exponentially sensitive to $\hat{\xi}$ (which includes
$g_s$) and $a_4$ and so can
take on essentially any value. A TeV scale gravitino mass requires
$\mc{V} \sim 10^{14}$, which we assume.

There may be gaugini associated with each of the moduli $\tau_b$ and
$\tau_s$. From (\ref{GauginoMass}) these have masses $\frac{F^b}{2 \tau_b}$ and $\frac{F^s}{2
  \tau_s}$ respectively. We calculate both masses: however we note
that the small cycle is the only cycle appropriate for Standard Model
matter, as a brane wrapped on the large cycle would have too small a
gauge coupling.
First,
\bea
F^b & = & e^{\mc{K}/2} \left(\mc{K}^{b \bar{b}} D_{\bar{b}} \bar{W} + \mc{K}^{b
  \bar{s}} D_{\bar{s}} \bar{W} \right) \\
& = & e^{\mc{K}/2} \left( -2 \tau_b \bar{W} + \mc{K}^{b \bar{b}}
\partial_{\bar{b}} \bar{W} + \mc{K}^{b \bar{s}} \partial_{\bar{s}}
\bar{W} \right),
\eea
where we have used $\mc{K}^{b \bar{b}} \partial_{\bar{b}} \mc{K} +
\mc{K}^{b \bar{s}}\partial_{\bar{s}} \mc{K} = -2 \tau_b$ from (\ref{Relation}).
However, as $\tau_b \sim \mc{V}^{2/3} \gg 1$, $\partial_{\bar{b}}
\bar{W} \sim \exp (-a_b \tau_b) \sim 0$. Furthermore,
$$
\mc{K}^{b \bar{s}} \partial_{\bar{s}} \bar{W} \sim (4 \tau_b \tau_s) \times
(a_s A_s \exp(-a_s \tau_s)) \sim \mc{V}^{-1/3},
$$
and so
\be
F^b = \frac{1}{\mc{V}} \left( -2 \tau_b W_0 +
\mc{O}\left(\mc{V}^{-1/3}\right)\right),
\ee
implying
\be
\vert M_b \vert = \left\vert \frac{F^b}{2 \tau_b} \right\vert = m_{3/2} + \mc{O}(\mc{V}^{-1/3}).
\ee
This is an identical relation to that of the fluxed MSSM \cite{hepph0408064}.

We now calculate $M_s$. Using $\mc{K}^{i \bar{j}} \mc{K}_{\bar{j}} =
- 2 \tau_i$ and $\mc{K}^{s \bar{b}} \partial_{\bar{b}} W \sim 0$, we get
\bea
F^s & = & e^{\mc{K}/2} \left( \mc{K}^{s \bar{s}} \partial_s \bar{W} - 2
\tau_s \bar{W} \right) \\
& = & e^{\mc{K}/2} \left( \mc{K}^{s \bar{s}} (-a_s A_s e^{-a_s T_s}) - 2
\tau_s \bar{W} \right).
\eea
From (\ref{InverseMetric}), $\mc{K}^{s \bar{s}} = \frac{8 \sqrt{\tau_s} \mc{V}}{3}$. Using
(\ref{SmallModSolution}) we then have
\be
F^s = \frac{2 \tau_s \bar{W}}{\mc{V}} \left( \left(1 - \frac{3}{4 a_s
  \tau_s} \right) - 1 \right).
\ee
We therefore obtain
\be
\label{SmallGauginoMass}
\vert M_s \vert = \frac{3 m_{3/2}}{4 a_s \tau_s}\left( 1+
\mc{O}\left(\frac{1}{a_4 \tau_4}\right)\right) = \frac{3 m_{3/2}}{4
  \ln (m_{3/2})} \left(1 + \mc{O}\left( \frac{1}{\ln(m_{3/2})}\right)\right),
\ee
with the expected small hierarchy. We therefore conclude that the gaugino mass associated to
 the exponentially large modulus $\tau_b$ is equal to the gravitino
 mass, whereas the gaugino mass associated to the small modulus
 $\tau_s$ is suppressed by $\ln (m_{3/2})$. While in the
 $\mbb{P}^4_{[1,1,1,6,9]}$ case there is only one
 small modulus, as shown above this result will extend
 to more realistic multi-modulus examples.
This confirms the logarithmic suppression of gaugino masses compared to the naive
estimates of section \ref{xyz}.

\subsection{Moduli Masses}

We also here prove the existence of an enhancement by
$\ln(m_{3/2})$ in the mass of the small modulus $\tau_s$ compared to
$m_{3/2}$. This is similar behaviour as found in KKLT solutions
\cite{hepth0503216}.
This again extends the simple
volume scaling arguments of chapter \ref{chapterLargeVol} which gave $m_s \sim \frac{M_P}{\mc{V}}$ and $m_b
\sim \frac{M_P}{\mc{V}^{3/2}}$. Focusing purely on the field
$\tau_s$, its Lagrangian is
\be
\int d^4 x \, \mc{K}_{s \bar{s}} \partial_\mu \tau_s \partial^\mu \tau_s + V(\tau_s).
\ee
As $\tau_s$ is the heavier of the two moduli $\tau_s$ and $\tau_b$, a lower bound on its
mass is given by
\be
m_s^2 \gtrsim \frac{1}{2 \mc{K}_{s\bar{s}}} \left\langle \frac{\partial^2
  V}{\partial \tau_s^2} \right\rangle.
\ee
Now, $\mc{K}_{s \bar{s}} = \frac{3}{8 \sqrt{\tau_s} \mc{V}}$. If we
evaluate $\frac{\partial^2 V}{\partial \tau_s^2}$, we obtain
\be
\frac{\partial^2 V}{\partial \tau_s^2} = \frac{4 \lambda a_s^4
  \sqrt{\tau_s} e^{-2 a_s \tau_s}}{\mc{V}} - \frac{\mu a_s^3 \tau_s
  e^{-a_s \tau_s} \vert W_0 \vert}{\mc{V}^2} + \left(\hbox{terms suppressed
  by }\frac{1}{a_s \tau_s}\right).
\ee
Substituting in our evaluation of $\langle e^{-a_s \tau_s} \rangle $
from (\ref{SmallModSolution}), we
obtain
\be
\frac{\partial^2 V}{\partial \tau_s^2} = \left( \frac{\mu^2}{2
  \lambda} \right) \frac{a_s^2 \tau_s^{3/2} \vert W_0
  \vert^2}{\mc{V}^3} \left(1 + \mc{O}\left(\frac{1}{a_s \tau_s}\right)\right)
\ee
Consequently
\bea
m_{\tau_s}^2 & \gtrsim & \left( \frac{4 \sqrt{\tau_s}\mc{V}}{3}\right)
\left( \frac{\mu^2}{2 \lambda} \right) \frac{a_s^2 \tau_s^{3/2} \vert
  W_0 \vert^2}{\mc{V}^3} \nonumber \\
& = & \left( \frac{2 \mu^2}{3 \lambda} \right) \frac{a_s^2 \tau_s^2
  \vert W_0 \vert^2}{\mc{V}^2}.
\eea
This gives
\be
\label{SmallModMass}
m_{\tau_s} \gtrsim 2 \ln \left( \frac{M_P}{m_{3/2}} \right) m_{3/2}
\left(1 + \mc{O}\left(\frac{1}{\ln(m_{3/2})}\right)\right).
\ee
While technically a lower bound, this is actually a very good estimate
of $m_{\tau_s}$, as the canonically normalised heavy modulus has only a very
small admixture of $\tau_b$.
This is confirmed by explicit numerical evaluation, which shows the
formulae  (\ref{SmallGauginoMass}) and (\ref{SmallModMass}) to be
accurate to within a couple of per cent.

\subsection{Scalar Masses}
\label{secSM}

The suppressed values for the
gaugino masses are a direct consequence of a cancellation in the 
calculation for the F-terms for the `small' moduli. Having suppresed
F-terms could naively lead to the conclusion that the
other soft terms must also be suppressed. 
However, consider the expression for scalar masses\footnote{This form assumes the matter metric is diagonal: the
  results below are unaffected if we use the fully general expressions \cite{ACQSinprogress}.}
\be
\label{scalarmass}
m_{\vphi}^2 \ =\ m_{3/2}^2 + V_0 - F^m\bar{F}^{\bar n} \partial_m\partial_{\bar
  n}  (\ln \tilde{\cal K}).
\ee 
In order to get suppressed values for $m_{\vphi}^2$ there must be a
contribution cancelling the leading $m_{3/2}^2$
contribution (assuming a negligible vacuum energy
$V_0$). In the KKLT models, this is provided by the anti-D3 brane \cite{hepth0503216}.
However, this cancellation is not generic.

For the large-volume models the uplift term is subdominant in susy
breaking and so has no significant effect on (\ref{scalarmass}). As
the F-terms are suppressed by $\ln (m_{3/2})$ for all small moduli,
the only F-term that can cancel the gravitino mass contribution is
that associated to the large volume modulus. The dependence of
$\tilde{K}$ on $\mc{V}$ varies depending on the type of scalar field considered.
To leading order we can write
\be
\label{MatterMetric}
{\tilde K} = h(\tau_s) {\cal V}^{-a},
\ee
with the exponent $a\geq 0$ taking different values for the different
kinds of matter fields in the model. Here $h$ is a flavour dependent function
of the smaller moduli that in general will be very hard to compute.

One can show that the F-term contribution only cancels the leading
$m_{3/2}^2$, giving scalars suppressed by $\ln (m_{3/2})$, if $a=2/3$. 
This applies to D3 brane adjoint scalars and D7 Wilson lines. For
adjoint D7 matter, $a=0$ and the scalar masses are comparable to
$m_{3/2}$. Of course, Standard Model matter fields are in
bifundamental representations and should correspond to D3-D7 or D7-D7
matter. In this case the only calculations for $\mc{K}$ are in the
context of toroidal orbifolds, where $0 \le a < 2/3$. In this case there
is no cancellation in (\ref{scalarmass}) and the scalars are
comparable to the gravitino mass, with positive mass squared, and heavier than the gauginos by the
small hierarchy $\ln(m_{3/2})$.

This also allows us to say something about flavour universality.
In the large volume scenario developed in this thesis, the physical picture is
of Standard Model matter suppported on
almost-vanishing small cycles within a very large internal space ($\mc{V} \sim 10^{14}$).
The physics of flavour is essentially local physics which is determined by the
geometry of the small cycles and their intersections.
 Consequently, all flavours should see the
large bulk in the same way, as the distinctive flavour physics is local
not global. Therefore the power of $a$ in
(\ref{MatterMetric}) should be flavour-universal. The function
$h(\tau_s)$, in contrast, \emph{is} sensitive to the local geometry
and so should not be flavour-universal.

In these circumstances we can both show 
universality for the soft scalar masses and also
estimate the fractional level of non-universality. In the sum (\ref{scalarmass})
the leading $m_{3/2}^2$ term and the terms involving $F^b$ are
flavour-universal and give a universal contribution of
$\mc{O}(m_{3/2}^2)$. Universality fails due to the F-terms associated
with the small moduli. We can then rewrite (\ref{scalarmass}) as
\be
\label{Universality}
m_i^2 \sim \underbrace{\Bigg( m_{3/2}^2 + F^b \bar{F}^b \partial_b
\partial_{\bar{b}} \ln \tilde{K} \Bigg)}_{\hbox{universal}} + \underbrace{\left( \sum_s F^s \bar{F}^s
\partial_s \partial_{\bar{s}} \ln \tilde{K} \right)}_{\hbox{non-universal}}.
\ee
The $F^b F^{\bar{s}} + F^{\bar{b}} F^s$ cross-terms vanish.
As $F^s \sim \frac{m_{3/2}}{\ln(m_{3/2})}$ we obtain
\bea
\label{ScalarUniversality}
m_i^2 & \sim & \, m_{3/2}^2 (1+
 \epsilon_i), \nonumber \\
\Rightarrow m_i & \sim & m_{3/2} \left( 1 + \frac{\epsilon_i}{2} \right),
\eea
where non-universality is encoded in $\epsilon_i \sim \frac{1}{\ln (m_{3/2})^2}$.
As we require $m_{3/2} \sim 1 \hbox{TeV}$, we estimate the fractional
non-universality for soft masses as $\sim 1/(\ln (10^{18}/10^3))^2
\sim 1/1000$.

It is remarkable that these general
results can be extracted despite our ignorance of the precise
dependence of the K\"ahler potential on the matter fields. This is
possible because in the above scenario flavour physics is local while
supersymmetry breaking is global, and there exists a controlled
expansion in $\frac{1}{\mc{V}}$.

We also note that the large-volume scenario naturally addresses the
$\mu$ problem. This is because the natural scale for any mass term,
susy or non-susy, is $\mu \sim \frac{M_P}{\mc{V}} \sim 1
\hbox{TeV}$. Indeed the dilaton and complex structure moduli, which
are stabilised supersymmetrically by the fluxes, do acquire masses of
this order. This arises because the scalar potential has a
prefactor $e^{\mc{K}} \sim \frac{1}{\mc{V}^2}$. A superpotential term
$\mu \phi \phi$ generates a scalar potential 
$$
V_\phi \sim e^{\mc{K}} (DW)(DW) \sim \frac{\mu^2}{\mc{V}^2} \phi^2.
$$
and so masses are naturally suppressed by $\frac{1}{\mc{V}}$ and at the
gravitino mass scale.

The phenomenology of flux compactifications has been much studied
recently. However the above suggests new well-motivated scenarios to consider.
The most obvious case is that of an 
intermediate string scale and thus a TeV scale gravitino mass, 
with squarks and sleptons heavy and comparable
to the gravitino mass, while gauginos are suppressed by a $(\ln m_{3/2})$
factor. 
As the gaugino masses are suppressed, it is necesary to include 
anomaly mediated contributions in addition to the gravity-mediation
expressions above. However as the scalar masses are heavy and
comparable to the gravitino mass
the contribution of anomaly mediation is in that case negligible
- this is just as
well given the notorious problem of tachyonic sleptons for pure
anomaly mediation. It will also be interesting to analyse the
phenomenology of the non-universality predicted in equation
(\ref{ScalarUniversality}).
A detailed investigation of these and related scenarios is in progress \cite{ACQSinprogress}.

There is one final note of caution. In attempting to build realistic
models, there are sound reasons to
suppose that not all K\"ahler moduli are stabilised by
nonperturbative effects. In particular, if all moduli were stabilised
by nonperturbative effects then their axionic parts would all also be
heavy and a QCD axion capable of solving the strong CP problem would
not exist. If the K\"ahler modulus corresponding to the QCD cycle
is partially stabilised through perturbative effects, it is possible that
gluinos may be heavier than the remaining gauginos. 
It is therefore also interesting to analyse the phenomenology of a mixed
scenario with an ordering $m_{3/2} \sim m_i \gtrsim m_{\tilde{g}} >
m_{\tilde{W}} \sim \frac{m_{3/2}}{\ln(m_{3/2})}$.

\chapter{Axions and Moduli Stabilisation}
\label{ChapterAxions}

This chapter is based on the paper \cite{hepth0602233}.

In chapter \ref{chapterSoftSusy} we analysed the structure of soft
supersymmetry-breaking terms for the large-volume minimum constructed
in chapter \ref{chapterLargeVol}. Through judicious assumptions about
the loci of Standard Model matter we were able to analyse the
magnitude of soft terms without having a detailed construction of the
Standard Model. This is as well, since obtaining the Standard Model is a
hard problem typically involving substantial amounts of algebraic
geometry. In this chapter we continue with this style of approach,
focusing on the axionic solution to the strong CP problem.

This chapter is structured as follows.
Section \ref{StrongCP} reviews both the strong CP problem and the ways axions can arise
in different string theory constructions. In section
\ref{AxionsAndModuliStabilisation} I investigate how to stabilise
moduli while keeping an axion sufficiently light to solve the
strong CP problem.
This section also contains a no-go theorem, showing that there exist no
supersymmetric minima of the F-term potential consistent with
stabilised moduli and unfixed axions.
Section \ref{AxionDecayConstantSec} addresses the axion decay
constant and in it I show how the large
volume compactifications 
described in chapter \ref{chapterLargeVol} can generate
a phenomenologically acceptable value of $f_a$
together with the relationship $f_a \sim \sqrt{M_{SUSY} M_P}$.

\section{Axions}
\label{StrongCP}

\subsection{Axions and the Strong CP Problem}

CP (charge conjugation and parity reversal) is a very good approximate
symmetry of the Standard Model. It is violated in the electroweak
sector through complex phases in the CKM matrix. CP violation was
first observed experimentally in kaon decays \cite{CPKaonDecay}. Recently CP
violation has also been observed in the decays of B mesons
\cite{hepex0207042, hepex0208025}.

Experimentally CP violation has only been observed in weak
interactions.
In principle the strong interactions could also violate CP. The QCD
Lagrangian is
\be
\label{QCDthetacoupling}
\mc{L}_{QCD} = -\frac{1}{4 g_s^2} \int F_{\mu \nu}^a F^{a,\mu \nu}
+ \frac{i \theta}{16 \pi^2} \int F_{\mu \nu}^a \tilde{F}^{a, \mu \nu} + \mc{L}_{matter}.
\ee
The coupling $F \tilde{F}$, which violates CP, is instantonic in nature. Being
topological, it vanishes in perturbation theory. However, the
amplitude of QCD instantons are sensitive to the value of $\theta$,
and if $\theta \neq 0$ this coupling will lead to observable CP
violation in strong interactions. In particular, if $\theta \neq 0$
a nonzero neutron electric dipole moment is generated.
As this is not observed the magnitude of $\theta$ is highly
constrained, with an experimental limit of $\vert \theta \vert \lesssim
10^{-10}$ \cite{NeutronEDM}. The `strong CP problem' is the fact that $\theta$, a
periodic constant whose natural range is
$- \pi < \theta < \pi$, is so
small.

The strong CP problem is of course a well-known problem with a well-known answer: a Peccei-Quinn axion
\cite{PecceiQuinn1, PecceiQuinn2}.
There do exist other resolutions. The measured value of $\theta$
involves not just the $\theta$ of equation (\ref{QCDthetacoupling})
but also a contribution from the phases in the quark mass matrix,
$$
\theta = \theta_0 + \arg \det (M_{q_i q_j}).
$$
If a quark - say the $u$ - is massless, this phase can be used to gauge
away the $\theta$-angle, which becomes unphysical. However, this
approach is disfavoured by lattice data, and $m_u = 0$ does not seem
possible \cite{heplat0407028}.
A second approach - the Nelson-Barr mechanism - is to assume that CP
is an exact symmetry of the high energy theory, which is spontaneously
broken at low energies. The smallness of $\theta$ is to be understood
as a legacy of exact CP conservation at high scales.
However in this thesis I assume the Peccei-Quinn solution to be correct.
As I review, 
this entails promoting $\theta$ to a
field whose potential is dynamically minimised for $\theta = 0$.

Promoting $\theta$ to a scalar field is natural in string theory, where all `coupling constants' are
vevs of dynamical fields. In this respect one essential feature of the Peccei-Quinn
solution is always present in string theory, as $\theta$ is the imaginary part of a
complex field which is the scalar component of a modulus superfield.
However, this then creates a modulus anti-stabilisation problem.
There are many string theory axions that may in principle solve the strong CP problem.
However, to do so an axion must remain massless
throughout the thicket of moduli stabilisation effects and down to the QCD scale.
This problem is clean and sharply posed, as effects very weak
on the Planck scale may be very large on the QCD scale.
Given the necessity of moduli stabilisation, this question is best analysed within the context of
string constructions for which all moduli have been stabilised. Once
we have
reviewed the Peccei-Quinn solution we shall analyse this problem in
section \ref{AxionsAndModuliStabilisation}.

The Peccei-Quinn approach is characterised by promoting $\theta$ to a dynamical field $\theta(x)$, with Lagrangian
\be
\label{axionLagrangian1}
\mc{L} = \mc{L}_{SM} + \frac{1}{2} f_a^2 \partial_\mu \theta(x)
\partial^\mu \theta(x) +
\frac{\theta(x)}{16 \pi^2} F^a_{\mu \nu} \tilde{F}^{a, \mu \nu}.
\ee
In (\ref{axionLagrangian1}) $f_a$ has dimensions of mass and is known as the axion decay constant.
Conventionally a scalar has mass dimension one,
and so we redefine
$a \equiv \theta f_a$. This gives
\be
\label{axionLagrangian}
\mc{L} = \mc{L}_{SM} + \frac{1}{2} \partial_\mu a \partial^\mu a +
\frac{a}{16 \pi^2 f_a} F^a_{\mu \nu} \tilde{F}^{a, \mu \nu}.
\ee
In equation (\ref{axionLagrangian}) there exists an anomalous global $U(1)$ symmetry, $a \to a + \epsilon$.
This symmetry is violated by QCD instanton effects, which break it to a discrete subgroup.
These generate a potential for $a$,
\be
\label{QCDAxionPot}
V_{\textrm{1-instanton}} \sim \Lambda_{QCD}^4 \left( 1- \cos \left( \frac{a}{f_a} \right) \right).
\ee
In the absence of other effects, this potential is minimised at
$a=0$. This sets the $\theta$-angle to zero and dynamically solves the
strong CP problem.
The mass scale for the axion $a$ is
\be
\label{axionMasses}
m_a \sim \frac{f_\pi^2 m_\pi^2}{f_a} = \left( \frac{10^{11}
  \hbox{GeV}}{f_a} \right) 10^{-4} \hbox{eV}.
\ee
The replacement of the $\Lambda_{QCD}^2$ of (\ref{QCDAxionPot}) by
$f_\pi m_\pi$, where $f_{\pi} \sim 90 \textrm{MeV}$ is the pion decay constant, comes about
through a precise calculation.

By its nature, the Peccei-Quinn symmetry is anomalous - it is violated
by QCD instanton effects that generate a potential for the axion.
It is necessary that QCD instanton effects
dominate the axion potential - the limit on $\vert \theta \vert$ is so
strong that even tiny additional contributions to the potential may
displace the minimum of the axion potential sufficiently to be in
conflict with experiment. In particular, higher-order Planck
suppressed contributions such as $a^5/M_P$ must be entirely absent.
This may seem unnatural, but is easily accomplished if the $U(1)$ Peccei-Quinn symmetry is
only broken to a discrete subgroup. This is in fact what happens in
string theory.

Even if an axion solves the strong CP problem,
there are strong constraints on the allowed value of $f_a$. $f_a$
gives the axion-matter coupling.
The smaller the value of $f_a$, the more strongly the axion
couples to matter. Assuming the axion to couple to QED as well as QCD,
the axion may be sought through direct searches in a strong magnetic
field, excluding
the regime $f_a \lesssim 10^6 \hbox{GeV}$. Astrophysical studies of
supernova cooling
further exclude the regime $10^6 \hbox{GeV} \lesssim f_a \lesssim 10^9
\hbox{GeV}$. In this limit, axions are produced in the core of a
supernova and are emitted without reabsorption. The energy carried
away by the axions causes the supernova to cool more rapidly than is observed.

These lower bounds are hard. There is also a cosmological upper bound on $f_a$.
The axion field presumably starts its cosmological evolution with
$\theta$ an arbitrary value between 0 and $2 \pi$.
From (\ref{QCDAxionPot}), there is a primoridal axion energy density
of $\rho_{\hbox{axion}} \sim (0.1 \hbox{GeV})^4$. Cosmologically, a scalar field evolves
according to
\be
\ddot{\phi} + 3 H \dot{\phi} = -\frac{\partial V}{\partial \phi}.
\ee
Once $H \sim m_a$ at time $t_0$ the axion field starts oscillating and the energy
density subsequently dilutes as matter,
\be
\rho_{\hbox{axion}}(t) \sim (0.1 \hbox{GeV})^4 \frac{a(t_0)^3}{a(t)^3}.
\ee
As $m_a \propto 1/f_a$, larger values of $f_a$ imply the axion starts
oscillating at later cosmological times. The bound comes from
requiring the current energy density in
the axion field to be below that of the dark matter density, which gives
the approximate constraint $f_a \lesssim 10^{12} \hbox{GeV}$.

This upper bound assumes a standard cosmology. In non-standard
cosmologies - for example including an intermediate inflationary stage
to dilute the axion density - the upper bound may be avoided. String
compactifications typically have long-lived moduli and so it can be
argued that in string theory early universe cosmology should be
non-standard. However, low-reheat temperatures generically encounter problems with
baryogenesis and dark matter abundance, and so 
here we restrict to a standard cosmological evolution.

In this case
detailed recent analyses of the astrophysical and cosmological
constraints on the above `invisible axion' are \cite{hepph0509198, hepph0606014}. These
give a current bound for the axion window as
$$
10^9 \hbox{GeV} \lesssim f_a \lesssim 3 \ti 10^{11} \hbox{GeV}.
$$

\subsection{Axions in String Theory}

String compactifications generically contain fields $a_i$ which have
$a_i F \tilde{F}$ couplings and
possess the anomalous global symmetry $a \to a + \epsilon$ featuring
in the axionic solution to the strong CP problem.
We first enumerate possible axions, before considering their relation to the physics of moduli
stabilisation.

\subsubsection*{Axions in the Heterotic String}

In heterotic compactifications, the axions are traditionally divided into the universal, or model-independent,
axion and the model-dependent axions. The model-independent axion is the imaginary part of the dilaton multiplet, $S =
e^{-2 \phi} \mc{V} + i a$. It is the dual of the 2-form potential $B_{2,\mu \nu}$ arising from the NS-NS 2-form
field: $da = e^{-2 \phi} *dB_{\mu \nu}$. The dilaton superfield is the tree-level gauge kinetic function for
all gauge groups,
$$
\mc{L} \sim \textrm{Re}(S) \int F^a_{\mu \nu} F^{a \mu \nu} + \textrm{Im}(S) \int F^a_{\mu \nu} \tilde{F}^{a \mu \nu}.
$$
Consequently any realistic heterotic compactification always has an $a F_{QCD} \tilde{F}_{QCD}$ coupling.

There are also the model independent axions, $b_i$, given by the
imaginary parts of the K\"ahler moduli $T_i$.
For a basis $\Sigma_i$ of 2-cycles of the Calabi-Yau,
$T_i = t_i + i b_i$, with $t_i = \int_{\Sigma_i} J$ the string frame
volume of the cycle $\Sigma_i$ and
$b_i = \int_{\Sigma_i} B_2$.
Such axions have no tree-level couplings to QCD.
However, such a coupling may be generated through
 the one loop correction to the gauge kinetic function.
For the $E_8 \ti E_8$ heterotic string, this correction is
\bea
\label{HetOneLoop}
f_1 & = & S + \beta_i T_i, \\
f_2 & = & S - \beta_i T_i,
\eea
where $1$ and $2$ refer to the first and second $E_8$ respectively.
The factors $\beta_i$ are determined by
the gauge bundles on the compactification manifold $X$.
For gauge bundles $V_1$ and $V_2$,
\be
\beta_i = \frac{1}{8 \pi^2} \int e_i \wedge  \left( c_2(V_1) -
c_2(V_2) \right),
\ee
where $e_i$ is the 2-form dual to the cycle $\Sigma_i$.
This can be derived by dimensional reduction of the Green-Schwarz term
$\int B_2 \wedge X_8(F_1, F_2, \mc{R})$.\footnote{Strictly, this only gives the correction to $\textrm{Im}(f)$. The
corresponding correction to $\textrm{Re}(f)$ is however implied by holomorphy.} The axions associated to the K\"ahler moduli
are called model-dependent as their appearance in $f$ depends on the one-loop correction, which in turn depends on the specific
properties of the compactification.

\subsubsection*{Axions in Intersecting Brane Worlds}

The discovery of D-branes \cite{hepth9510017} led to the extension of
string model building beyond the heterotic string. The type II string theories,
or more properly orientifolds thereof, can give rise to `intersecting brane worlds'. In these the
Standard Model is localised on a stack of branes while gravity propagates in the bulk. Light bifundamental matter arises
from strings located at intersection loci and stretching between brane stacks.

The bosonic action of a single Dp-brane is the sum of DBI and
Chern-Simons terms,
\be
S_p =  \frac{- 2 \pi}{(2 \pi \sqrt{\alpha'})^{p+1}} \left(
\int_{\Sigma} d^{p+1} \xi e^{-\phi} \sqrt{ \textrm{det}(g + B + 2 \pi \alpha' F)}
+ i \int_{\Sigma} e^{B + 2 \pi \alpha' F} \wedge \sum_q C_q \right).
\ee
$\Sigma$ is the cycle wrapped by the brane and the sum is a formal sum
over all
RR potentials in which only relevant terms are picked out.

The gauge kinetic term $F_{\mu \nu} F^{\mu \nu}$ comes from the DBI action
and the instanton action $F \wedge F$ from the Chern-Simons term.
The gauge coupling corresponds to the inverse volume of $\Sigma$ and the $\theta$ angle to the
component of $C_{p-3}$ along $\Sigma$. These fields pair up to become the scalar component of the chiral multiplet which is
the gauge kinetic function of the resulting gauge theory.

As IIB compactifications are our main focus, we shall be more explicit here. In principle, IIB string theory
allows, consistent with supersymmetry, space-filling D3, D5, D7 and D9-branes. However, in an orientifold setting we are restricted
to either D3/D7 or D5/D9 pairs. In this thesis we have been concerned
with the former case. The branes must wrap appropriate cycles to cancel
the negative charge and tension carried by the orientifold planes; we assume this has been done.

In such compactifications, the relevant superfields
are those of the dilaton and K\"ahler moduli multiplets. Their scalar
components are\footnote{As noted in chapter \ref{ChapterModuliAndFluxes},
technically this is only for the case that $h^{1,1}_{-} = 0$. This will not be
important for
the issues we discuss, and so we will use this
simplifying assumption.}
\bea
S & = & e^{-\phi} + i c_0, \\
T_i & = & \tau_i + i c_i.
\eea
$c_0$ is the Ramond-Ramond 0-form and $e^\phi \equiv g_s$ the string coupling. For $\Sigma_i$ a 4-cycle of the Calabi-Yau,
$$
c_i = \frac{1}{l_s^4} \int_{\Sigma_i} C_4 \qquad \textrm{ and } \qquad
\tau_i = \int_{\Sigma_i} \frac{e^{-\phi}}{2} J \wedge J,
$$
where $l_s = 2 \pi \sqrt{\alpha'}$ denotes the string length.
Indeed, $S$ is the universal gauge kinetic function for D3-branes,
whereas $T_i$ is the gauge kinetic function for the field theory on a D7-brane stack
wrapping the 4-cycle $\Sigma_i$.

The axionic couplings arise from the Chern-Simons term in the action. For D3-branes, this gives
\be
S_{F \tilde{F}} = \frac{c_0}{4 \pi} \int F \wedge F,
\ee
while for D7-branes
\be
S_
{F \tilde{F}} = \frac{c_i}{4 \pi} \int F \wedge F.
\ee
By expanding the DBI action, we obtain the field theory couplings
$$
\frac{1}{g^2}\Bigg\vert_{D3} = \frac{e^{-\phi}}{2 \pi} \qquad \textrm{
  and }
\qquad \frac{1}{g^2}\Bigg\vert_{D7} = \frac{\tau_i}{2 \pi}.
$$

There exists a similar story for IIA intersecting brane worlds, where the Standard Model must be realised
on wrapped D6-branes (a Calabi-Yau has no 1- or 5-cycles to wrap D4- or D8-branes on).
The gauge coupling now comes from the calibration form $\textrm{Re}(\Omega)$ and
the axion from the reduction of the 3-form potential $C_3$,
$$
\frac{1}{g^2}\Bigg\vert_{D6} = \int_{\Sigma_i} \textrm{Re}(\Omega) \qquad \textrm{ and } \qquad \theta \Bigg\vert_{D6} =
\int_{\Sigma_i} C_3.
$$
Our interest in this chapter is the interaction of axions with moduli stabilisation
and supersymmetry breaking, to which we now turn.

\section{Axions and Moduli Stabilisation}
\label{AxionsAndModuliStabilisation}

It is obvious from the above that potential
axions are easily found in string compactifications; indeed, they are superabundant.
For axions to solve the strong CP problem, they must also be light, with QCD instantons giving the dominant
contribution to their potential.
Light axions are not in themselves
problematic. Pure type II Calabi-Yau compactifications have
many axions, which remain exactly massless as a consequence of four-dimensional $\mc{N}=2$ supersymmetry.
However, the same $\mc{N}=2$ supersymmetry that guarantees the axions remain massless also
guarantees a non-chiral spectrum with the axions' scalar partners massless.
These are modes of the
graviton and will lead to unobserved fifth forces.

More realistic string constructions have $\mc{N}=1$ supersymmetry in four dimensions,
allowing a potential to be generated for the moduli.
To avoid bounds from fifth force experiments, the size moduli - the saxions - must at a minimum receive masses at a scale
$m_T \gtrsim (100 \mu \textrm{m})^{-1} \sim 2 \ti 10^{-3}
\hbox{eV}$. However, the expected scale is much larger
and typical constructions
give moduli masses comparable to the supersymmetry breaking scale. Much of this thesis has been concerned
with recent progress made
in moduli stabilisation,
with fluxes and non-perturbative effects being used to lift the degeneracies associated
with the geometric moduli.
In the context of strong CP, this same progress creates a modulus
anti-stabilisation problem.
There are many stringy effects that can generate
a potential for a putative QCD axion. These include worldsheet instantons, D-instantons and
gaugino condensation, the last two of which we have already made much
use of in chapters \ref{ChapterKahlerModuli} and \ref{chapterLargeVol}.
If any one of these effects is more important for a given axion than
QCD instantons, that axion does not solve the strong CP problem.

Axions do not receive a potential in perturbation theory. In heterotic
compactifications, worldsheet perturbation theory is an expansion about the
trivial (point-like) embedding of the worldsheet in
spacetime. However, from the definition of the heterotic axions, we
see these can only contribute to the action, and thus the Feynman path
integral, if the embedding of the worldsheet in spacetime is
topologically non-trivial. Such an embdedding is a worldsheet
instanton and thus a nonperturbative effect. In IIB compactifications,
the axions are components of the RR fields. As the fundamental string
is uncharged under these, string perturbation theory cannot generate a
potential for them - brane instantons, objects nonperturbative
in both $g_s$ and $\alpha'$, are necessary.

Consequently axion potentials come from nonperturbative effects whose magnitude is exponentially sensitive
to the values of the stabilised moduli.
The natural place to analyse the importance of such effects is
therefore within a context in which all moduli are stabilised. To this
end we focus on the constructions described in
chapters \ref{ChapterKahlerModuli} and \ref{chapterLargeVol}, but along the way we will prove a no-go theorem
applicable to all string compactifications and relevant to some recent
approaches to moduli stabilisation.

\subsection{The Simplest Scenarios: Why Axions are Heavy}
\label{KKLTAxionSection}

We shall examine both KKLT and large-volume scenarios.
Our first point is that the simplest versions of these
scenarios lack a QCD axion. All potential axions
receive a high scale mass and thus cannot solve the strong CP problem. For simplicity we concentrate on the
KKLT construction, but a very similar argument holds for the exponentially large volume compactifications.

We start by asking whether QCD is to be realised on D3 or D7 branes.
If we were to use D3-branes, the QCD axion must be
the imaginary component of the dilaton multiplet, $S = e^{-\phi} + i c_0$.
However, as discussed in chapter \ref{ChapterModuliAndFluxes} the dilaton is
stabilised at tree-level by the fluxes, with a mass $m_S \sim
\frac{N}{R^6} M_P$, with $R$ the radius in units of $l_s$.
We note this perturbative flux stabilisation may seem at odds with the axionic shift
symmetry $c_0 \to c_0 + 2 \pi$ and our earlier statement that axions
do not get a potential in perturbation theory. However, in this case
the axion shift
symmetry is a subgroup of the fundamental $SL(2,\mbb{Z})$ symmetry, under
which the fluxes also transform.
The fluxes appear in the low-energy theory as coupling constants through the flux superpotential.
If the fluxes have vevs, then at low energy the $SL(2,\mbb{Z})$
symmetry is spontaneously
broken and cannot prevent the axion acquiring a mass.
As $R \lesssim 5$ in KKLT, the axion obtains a mass $m_{c_0} \gtrsim 10^{15} \textrm{GeV}$ and cannot be a QCD axion.
For the exponentially large volume compactifications, $R$ is larger but the conclusion unchanged: QCD on a D3 brane stack is
inconsistent with the existence of a Peccei-Quinn axion.

This implies that QCD ought to be realised
on D7 branes. The axions are now the imaginary parts of the K\"ahler moduli,
and the instanton effects used to stabilise these moduli will also give the axions a mass.
To estimate the scale of this mass,
it is simplest just to construct the potential explicitly.

We take the superpotential
\be
W = W_0 + \sum_{i=1}^{h^{1,1}} A_i \exp (-a_i T_i),
\ee
and the K\"ahler potential
\be
\label{KahlerPotentialStandard}
\mc{K} = - 2 \log \mc{V}.
\ee
Note that $\mc{K} = \mc{K}(T_i + \bar{T}_i)$, and so the axions 
appear neither in $\mc{K}$ nor its derivatives.
The supergravity F-term potential is
\be
\label{FTermPot}
V = e^{\mc{K}} (\mc{K}^{i \bar{j}} D_i W D_{\bar{j}} \bar{W} - 3 \vert W \vert^2).
\ee
The no-scale property of the K\"ahler potential simplifies (\ref{FTermPot}) to
\be
\label{NoScaleScalarPotential}
V = e^{\mc{K}} \left( \mc{K}^{i \bar{j}} \partial_i W
\partial_{\bar{j}} \bar{W} +
\mc{K}^{i \bar{j}}\left( (\partial_i \mc{K}) W \partial_{\bar{j}}
\bar{W} + (\partial_{\bar{i}} \mc{K} ) \bar{W} \partial_j W \right) \right).
\ee
It is a property of the K\"ahler potential (\ref{KahlerPotentialStandard}) that $\mc{K}^{i \bar{j}} \partial_i \mc{K} = -2 \tau_j$.
(\ref{NoScaleScalarPotential}) becomes
\bea
V & = & e^{\mc{K}} \Bigg( \mc{K}^{i \bar{j}} a_i a_j \left( A_i
\bar{A}_j e^{-a_i T_i -a_j \bar{T}_j} + \bar{A_i} A_j e^{-a_i \bar{T}_i -a_j \bar{T}_j} \right)
  \nonumber  \\
  & & - 2 a_i \tau_i
\left( W \bar{A}_i e^{-a_i \bar{T}_i} + \bar{W} A_i e^{-a_i T_i} \right) \Bigg) .
\label{LatestPotential}
\eea
It is easy to extract the axionic dependence of the potential (\ref{LatestPotential}).
$\mc{K}$ and its derivatives are all real and phases only come
from the superpotential. The potential becomes
\bea
V & = & e^{\mc{K}} \left( \mc{K}^{i \bar{j}} \left( 2 a_i a_j \vert A_i A_j \vert e^{-a_i \tau_i - a_j \tau_j} \cos(a_i \theta_i + a_j \theta_j + \gamma_{ij}) \right) \right. \\
& & - 4 a_i \tau_i \vert W_0 A_i \vert e^{-a_i \tau_i} \cos(a_i \theta_i + \beta_i) - 4 a_i \tau_i \vert A_i A_j \vert \cos (a_i \theta_i + a_j \theta_j +
\gamma_{ij}) \Big). \nonumber
\eea
$\theta_i$ denote the axions and the
phases $\gamma_{ij}$ and $\beta_i$ come from the phases of $A_i \bar{A}_j$ and $\bar{A}_i W$ respectively.
The axionic mass matrix is
\be
M^2_{ij} = \frac{\partial^2 V}{\partial \theta_i \partial \theta_j},
\ee
and we obtain\footnote{We emphasise this does not say that the mass
  matrix is a direct product $a \otimes a$ - such a matrix is
  degenerate. The point is simply that if the coefficient of the
  exponent is large, the mass matrix may receive an enhancement.}
\be
M^2_{ij} \sim \mc{O}(a_i a_j) V_{min},
\ee
with $V_{min}$ the magnitude of the potential at the AdS minimum. In KKLT, $D_{T_i} W = 0$ for all $i$
and $V_{min} \sim - 3 \frac{\vert W_0 \vert^2}{\mc{V}^2}$. As there are
$h^{1,1}$ independent phases in the superpotential, there is no reason for $M^2_{ij}$ to be degenerate and
we expect all eigenvalues to be $\mc{O} ( a^2 V_{min})$, where $a$ is the typical magnitude of the $a_i$.

The determination of physical masses also requires the K\"ahler potential. In general there is no explicit expression for the
overall volume $\mc{V}$ in terms of 4-cycle volumes $\tau_i$. The K\"ahler metric may however be written as \cite{hepth0403067}
\be
\label{GeneralKahlerP}
\mc{K}_{i \bar{j}} = \frac{G_{i \bar{j}}^{-1}}{\mc{V}^2}, \qquad \qquad
G_{i \bar{j}} = -\frac{3}{2} \left( \frac{k_{ijk}v^k}{\mc{V}} - \frac{3}{2} \frac{k_{imn}t^m t^n k_{jpq} t^p t^q}{\mc{V}^2} \right).
\ee
If $\mc{V} \sim \textrm{ (a few)} l_s^6$, $\mc{K}_{i \bar{j}} \sim \mc{O}(1)$ and the mass matrix $M^2_{ij}$
gives a good estimate of the scale of axion masses. If $\mc{V} \gg l_s^6$,
then $\mc{K}_{i \bar{j}} \ll \mc{O}(1)$, and $M^2_{ij}$ underestimates the axion masses.
$M^2_{ij}$ could only overestimate the axion masses if $\mc{V} \ll 1$.
This realm of moduli space is not accessible in a controlled fashion and
we do not concern ourselves with it.

In units where $M_P = 1$, we therefore have
\be
m_{\tau_i} \sim m_{c_i} \sim a_i \sqrt{V_{min}} \sim \frac{a_i
  W_0}{\mc{V}} \sim a_i m_{3/2}.
\ee
The axion masses are consequently set by the value of the tree-level superpotential
$W_0$. This also determines the vacuum
energy, the gravitino mass and, implicitly, the energy scale of supersymmetry breaking
required to cancel the vacuum energy.
TeV-scale soft terms require hierarchically small $W_0$. For the
(gravity-mediated) case
studied in \cite{hepth0503216, hepph0504036, hepph0504037}, this required $W_0 \sim 10^{-13}$, with
\be
m_{\tau_i} \sim m_{a_i} \sim m_{3/2} \sim 10 \textrm{TeV}.
\ee
This scale is vastly greater than that associated with QCD instanton effects, and thus the axions
are incapable of solving the strong CP problem.
We could insist on a QCD axion, and require that $W_0$ be sufficiently small
that QCD instantons
dominate over the D-instanton effects of moduli stabilisation.
This would require $W_0 \sim 10^{-40}$. However, this scenario is entirely
excluded as the size moduli are light enough to violate fifth force experiments and the susy breaking scale would be $\mc{O}(10^{-14}
\textrm{eV})$.
Consequently, in the simplest KKLT scenario it is impossible to generate a QCD axion. The D3 axion
receives a high scale mass from fluxes, whereas the instanton effects
give the D7 axions
large masses comparable to the size moduli.

A similar argument holds for the exponentially large volume compactifications. As in KKLT, the D3 axion
receives a large flux-induced mass. The `small' cycles are stabilised by instanton effects, and these give the
corresponding axions masses of a similar scale to the size moduli, $m_{a_i} \sim m_{\tau_i} \sim m_{3/2}$.
One difference is that there is a modulus, the `large' modulus $\tau_b$, which
need not appear in the superpotential. We recall this is stabilised through the
K\"ahler potential, and while it is massive its axionic partner indeed
remains massless.
However, this cycle is exponentially large, and
any gauge group supported on this cycle is far too weakly coupled to be QCD.
The same conclusion holds: the simplest version of this scenario does not generate a QCD axion.

The above formulates the `modulus anti-stabilisation problem': naive
scenarios of moduli stabilisation are incompatible with a QCD axion.

We next examine an apparent solution to this problem with
a subtle flaw.
We require a method of stabilising moduli without stabilising the axions.
Axions correspond to phases in the superpotential and do not appear in
the K\"ahler potential. If we included a multi-exponential term
$e^{-\alpha^i T_i}$ in the
superpotential, a massless axion would certainly survive,
as at least one phase would be absent.
As the size moduli all appear in the K\"ahler potential,
by solving the F-term equations we may hope to stabilise the size
moduli while keeping some axions massless.

To illustrate this idea, let us consider a toy KKLT model on a
toroidal orientifold,
\bea
\label{ToyTorusK}
\mc{K} & = & -2 \ln (\mc{V}) = - \ln (T_1 + \bar{T}_1) - \ln (T_2 + \bar{T}_2) - \ln
(T_3 + \bar{T}_3), \\
\label{LightAxionSuperpotential}
W & = & W_0 + Ae^{-2 \pi(T_1 + T_2 + T_3)}.
\eea
The K\"ahler potential (\ref{ToyTorusK}) is that appropriate for
toroidal orbifolds, with $\mc{V} = t_1 t_2 t_3$.
To understand where the superpotential could arise from, we can hypothesise that
the cycle (1+2+3) is the smallest cycle with only two fermionic
zero modes, and that instantons wrapping (for example) the cycle (2+3)
all have more than two zero modes and do not appear in the
superpotential. However, we are not here really concerned with the
microscopic origin of the superpotential: at this level we simply regard equations
(\ref{ToyTorusK}) and (\ref{LightAxionSuperpotential}) as defining the model.

The F-term equations $D_{T_1} W = D_{T_2} W = D_{T_3} W = 0$ give
\bea
\label{F1}
- 2 \pi A e^{-2 \pi (T_1 + T_2 + T_3)} - \frac{1}{T_1 + \bar{T}_1} \left(W_0 + A
e^{-2 \pi (T_1 + T_2 + T_3)}\right) & = & 0, \\
\label{F2}
- 2 \pi A e^{-2 \pi (T_1 + T_2 + T_3)} - \frac{1}{T_2 + \bar{T}_2} \left(W_0 + A
e^{-2 \pi (T_1 + T_2 + T_3)}\right) & = & 0, \\
\label{F3}
- 2 \pi A e^{-2 \pi (T_1 + T_2 + T_3)} - \frac{1}{T_3 + \bar{T}_3}
\left(W_0 + A e^{-2 \pi (T_1 + T_2 + T_3)}\right) & = & 0.
\eea
These immediately imply
\be
\label{EqualTaus}
\tau_1 = \tau_2 = \tau_3,
\ee
with equations
(\ref{F1}) to (\ref{F3}) collapsing to
\be
\label{LeftOver}
2 \pi A e^{-6 \pi \tau_1} e^{-2 \pi i(\theta_1 + \theta_2 + \theta_3)} +
\frac{1}{2 \tau_1} \left(W_0 + A e^{-6 \pi \tau_1} e^{-2 \pi i(\theta_1 +
  \theta_2 + \theta_3)}\right) = 0.
\ee
While the sum $\theta_1 + \theta_2 + \theta_3$ is fixed, there are
clearly two axionic directions not relevant for the solution of the
F-term equations. On the other hand, the combination of
(\ref{EqualTaus}) and (\ref{LeftOver}) imply there is a
unique value for the
size moduli such that the F-term equations are solved.
Except for the massless axionic directions, the scales of the masses
are unaltered from above, and we would expect
\be
m_{\tau_i} \sim m_{\theta_1 + \theta_2 + \theta_3} \sim
\frac{W_0}{\mc{V}}, \qquad
m_{\theta_1 - \theta_2} = m_{\theta_1 - \theta_3} = 0.
\ee
As this is supergravity rather than rigid supersymmetry, there is
no contradiction in having a mass splitting for the multiplet in the
presence of unbroken supersymmetry.

While this is superficially promising, in fact the above has a
serious problem. Even though all F-term
equations can be solved, numerical investigation shows that at the
supersymmetric locus the resulting scalar potential is
tachyonic, with signature $(+,-,-)$. Although supersymmetry ensures the
moduli are Breitenlohner-Freedman stable \cite{BreitenlohnerFreedman},
this notion of AdS stability ceases to be relevant
after the (required) uplift.

We now show that these tachyons are in fact generic for any attempt to stabilise the
moduli supersymmetrically while preserving one or more unfixed axions.

\subsection{A No-Go Theorem}

We start with an arbitrary $\mc{N} = 1$ supergravity theory with moduli
fields, $\Phi_{\alpha}$, $T_{\beta} = \tau_{\beta} + i b_{\beta}$,
where the $b_{\beta}$ are the axions. We write the superpotential
and K\"ahler potential as \bea
W & = & W(\Phi_{\alpha}, T_{\beta}), \\
\label{KKaxionK}
\mc{K} & = & \mc{K}(\Phi_{\alpha}, T_{\beta} + \bar{T}_{\beta}).
\eea
The Peccei-Quinn symmetry $b_{\beta} \to b_{\beta} + \epsilon_{\beta}$ implies the form
of $(\ref{KKaxionK})$ should hold in perturbation theory.

We further suppose we have solved
\be
\label{FFEqs}
D_{\Phi_{\alpha}} W = 0 \textrm{ and } \qquad D_{T_{\beta}} W = 0
\ee
for all $\alpha$ and $\beta$, but that at least one axion $b_u = \sum_{\beta}
\lambda_{\beta} b_{\beta}$ is unfixed: the solution to (\ref{FFEqs}) is
independent of $\langle b_u \rangle$.

We redefine the basis of chiral superfields so that there exists a
superfield $T_u$ with $b_u =
\textrm{Im}(T_u)$,
\bea
T_1 & \to & T_u, \nonumber \\
T_2 & \to & T_2, \nonumber \\
T_n & \to & T_n.
\eea
This is a good redefinition as it does not affect holomorphy
properties.

As the solution to all F-term equations is independent of $b_u$,
$b_u$ is a flat direction of the potential (\ref{FTermPot}) at the supersymmetric
locus.\footnote{The requirement of \emph{flatness} is stronger than
  the requirement that the axion simply be \emph{massless}. Flatness
  is the right requirement, as if an axion is fixed in any way it does
not solve strong CP.}
The potential at the supersymmetric locus is given by
\be
V = - 3 e^{\mc{K}} \vert W \vert^2.
\ee
As $b_u$ does not appear in $\mc{K}$, it follows that if $b_u$ is a
flat direction $\vert W \vert$
must be independent of $b_u$. Up to one exception this then implies that
$W$ is independent of $b_u$.

The sole exception is if $b_u$ purely represents an overall phase,
i.e. $W = e^{-a T_u}$ with no constant term. This may arise if the
flux superpotential vanishes exactly due to a discrete symmetry
\cite{hepth0411061}, while a combination of non-perturbative effects
and K\"ahler corrections stabilise the K\"ahler moduli. While
potentially interesting, this is an exceptional case for which moduli
stabilisation is not well understood, and we do not
analyse it further.

If $W$ has no explicit dependence on $b_u = \textrm{Im}(T_u)$, it
follows by holomorphy that it also has no explicit dependence on
$\tau_u = \textrm{Re}(T_u)$ and thus no explicit dependence on
$T_u$. Therefore \be
\partial_{T_u} W \equiv 0.
\ee
However, as $D_{T_u} W = 0$, it follows that at the supersymmetric locus,
$$
\textrm{either } \qquad (\partial_{T_u} \mc{K}) = 0 \qquad \textrm{ or }
\qquad W = 0.
$$
The latter is overdetermined and non-generic, so we first focus on
$(\partial_{T_u} \mc{K}) = 0$.

Direct calculation now shows that the $\tau_u$ direction is tachyonic
at the supersymmetric locus. To see this, note that from the scalar potential
(\ref{FTermPot})
\be
\label{VDeriv}
\partial_{\tau_u} V = e^{\mc{K}} \mc{K}^{i \bar{j}} \left( \partial_{\tau_u} (D_i W) D_{\bar{j}}
\bar{W} + D_i W \partial_{\tau_u} (D_{\bar{j}} \bar{W}) \right) - 3
(\partial_{\tau_u} \mc{K}) e^{\mc{K}} W \bar{W}.
\ee
We have used $\partial_{\tau_u} W \equiv 0$
and have only kept terms that
will give non-vanishing contributions to $\partial_{\tau_u}
\partial_{\tau_u} V$ at the supersymmetric locus.
Expanding $D_i W$ and again using $\partial_{\tau_u} W \equiv 0$,  (\ref{VDeriv})
simplifies to
\be
\partial_{\tau_u} V = e^{\mc{K}} \mc{K}^{i \bar{j}} \left( \partial_{\tau_u} (\partial_i \mc{K}) W
(D_{\bar{j}} \bar{W}) + D_i W \partial_{\tau_u}(\partial_{\bar{j}} \mc{K}) \bar{W} \right) - 3
(\partial_{\tau_u} \mc{K}) e^{\mc{K}} W \bar{W}.
\ee
If we again only keep terms non-vanishing at the supersymmetric locus,
the second derivative is
\be
\partial_{\tau_u} \partial_{\tau_u} V = e^{\mc{K}} \mc{K}^{i \bar{j}} \left(
2 \partial_{\tau_u} (\partial_i \mc{K}) \partial_{\tau_u} (\partial_{\bar{j}} \mc{K}) W
\bar{W} \right) - 3 (\partial_{\tau_u} \partial_{\tau_u} \mc{K}) e^{\mc{K}} W \bar{W}.
\ee
Now, as $\tau_u = \half (T_u + \bar{T}_u)$,
$$
\partial_{\tau_u} \mc{K}(T + \bar{T}) =
2 \partial_{T_u} \mc{K}(T + \bar{T}),
$$
and we have
\bea
\partial_{\tau_u} \partial_{\tau_u} V  & = &
4 e^{\mc{K}} W \bar{W} ( 2 \mc{K}^{i \bar{j}} \mc{K}_{i u} \mc{K}_{u \bar{j}} - 3 \mc{K}_{u \bar{u}} )
\nonumber \\
\label{TachyonMass}
& = & - 4 e^{\mc{K}} W \bar{W} \mc{K}_{u \bar{u}},
\eea
where we have used $\mc{K}_{i \bar{j}} = \mc{K}_{ij}$.
As $\mc{K}_{i\bar{j}}$ is a metric, $\mc{K}_{u \bar{u}}$ is positive definite and it follows
that the $\tau_u$ direction is tachyonic.

Now consider the $W = 0$ case. As indicated above, this is
non-generic: even if $W$ originally vanishes, it is expected to
receive non-perturbative corrections which make it non-vanishing.
Even so, it follows easily that in the case $W=0$
\be
\partial_{\tau_u} \partial_{\tau_u} V = 0,
\ee
and so the $\tau_u$ size modulus is massless, leading to unobserved fifth forces.

The above gives a no-go theorem: there does not exist any supersymmetric minimum of the F-term
potential consistent with stabilised moduli and unfixed axions.

It is in the nature of no-go theorems that they admit loopholes, so
let us discuss ways around this result. One point to consider is the form of the
K\"ahler potential, as we have used in the above argument the fact that
\be
\mc{K} = \mc{K}(T + \bar{T}).
\ee
While true in perturbation theory because of the axionic shift symmetry, this equation will break down
nonperturbatively, and the argument showing that the $\tau_u$
direction is tachyonic will no longer hold. However, this same breakdown will cause
$\mc{K}$, and hence the potential $V$, to depend on the axion.
As this lifts the required axionic flat direction, the no-go theorem will cease to apply.

A second loophole is that although the F-term potential might be
tachyonic, the D-term potential might come to the rescue. For example,
a Fayet-Iliopoulos term might have exactly the right structure to render the
supersymmetric locus an actual minimum of the full potential. However
this seems implausible in the presence of many tachyonic
directions. A similar approach would be to try and set $W=0$ and then
rely entirely on D-terms to stabilise the moduli.
Another possibility (discussed recently in
\cite{hepth0602120}) is that an anomalous U(1) might remove the
tachyonic directions. While the massive gauge boson will eat the
axionic degree of freedom, an axionic direction may survive in the
phase of a scalar charged under the U(1).

A third loophole is that stability might not rely on the existence of
an actual minimum for the potential.
Equation (\ref{TachyonMass}) involves the K\"ahler metric $\mc{K}_{u
  \bar{u}}$. The kinetic term for $\tau_u$ is $\mc{K}_{u \bar{u}}
\partial_\mu \tau_u \partial^\mu \tau_u$. If we just consider the
$\tau_u$ direction, the physical mass is therefore
\be
m_{\tau_u}^2 = -2 e^{\mc{K}} W \bar{W} = -\frac{8}{9} \vert m_{BF} \vert^2,
\ee
where $m_{BF}$ is the relevant
Breitenlohner-Freedman bound. As tachyonic modes can be stable in AdS,
one could argue that it is sufficient simply to solve the F-term
equations, without worrying about whether the resulting locus is an
actual minimum of the potential.

While this point is more substantial, it does not resolve the
problem. The real world is not AdS, and for stability requires a positive definite
mass matrix. For any realistic model, the vacuum energy must be
uplifted such that it vanishes.
After this uplift,
the extra geometric advantages of AdS go away and the tachyons can no
longer be supported. As there may be many tachyons present - one for
each massless axion -
the entire problem of moduli stabilisation must necessarily be solved over again in the uplifting.
As the uplift is generally the least controlled part of the procedure,
this is a hard problem.
While the uplift \emph{may} remove the tachyons, at the present level
of understanding this is pure hypothesis.
It is
then very unclear how useful the original supersymmetric AdS saddle point is, and whether
it is a suitable locus to uplift.

The argument above suggests that supersymmetric solutions are unpromising
starting points from which to address the strong CP problem. Either
all moduli appear in the superpotential, in which case there is no
light axion, or a modulus is absent from the superpotential, in which
case the potential is tachyonic.
The fourth and most obvious loophole is to give up on the requirement
of supersymmetric minima, and search for nonsupersymmetric minima of
the potential with massless axions.

We shall consider this point in the next section. Before we do so we
first note that this result has consequences for several approaches to moduli stabilisation
considered in the literature.
For example, in the weakly
coupled heterotic string, the one-loop corrections to the gauge
kinetic function (\ref{HetOneLoop}) imply that gaugino
condensation generates a superpotential
\be
\label{HetSuperpotential}
W_{n.p.} = A e^{-\alpha S + \beta_i T^i}.
\ee
It has been proposed \cite{hepth0310159} to use the superpotential
(\ref{HetSuperpotential}), together with a constant term $W_0$, to
stabilise both dilaton and K\"ahler moduli by solving $D_S W = D_{T^i} W =
0$.
However, there is only one phase - and hence only one axion -
explicitly present in (\ref{HetSuperpotential}). The above
argument shows that the resulting scalar potential will actually be
tachyonic at the supersymmetric locus, with signature $(+,-,\ldots,-)$.

A similar problem exists in several of the IIA flux compactifications
recently studied with both 2-form and 3-form fluxes turned on 
\cite{hepth0505160, hepth0506066}.
In this context it is observed that it is possible to solve all the F-term equations $D_i W = 0$, and it has been proposed to use
this to stabilise the moduli. However, the solution of these equations is independent of many of the axions present
(the imaginary parts of the complex structure moduli, namely those associated with the $C_3$ field). Consequently as long as
$h^{2,1} \neq 0$ the supersymmetric locus is always tachyonic, with one tachyon present for every unfixed axion.
Tachyons have been recognised in particular
models \cite{hepth0602120, hepth0505160, hepth0503169}, but from the above they
would seem to be a very generic problem for such approaches.

\subsection{Non-supersymmetric Minima with Massless Axions}
\label{secnonsusyminima}

The above no-go theorem shows that supersymmetric moduli stabilisation
is not a good starting point from which to solve the strong CP problem.
This is an argument in favour of non-supersymmetric moduli stabilisation.
That there is no no-go theorem for non-supersymmetric minima with
massless axions can be shown by construction: for example, the large
volume compactifications of chapter \ref{chapterLargeVol} all
contain a massless axion associated with the large cycle controlling
the overall volume. As indicated previously, this cannot be a QCD axion, as
any brane on this cycle is very weakly coupled. However, it does serve to show that the no-go
theorem does not apply to non-supersymmetric stabilisation. If we want to try and
force this cycle into being a QCD axion, we can tune the
parameters to force the minimum of this potential to relatively
small volumes. This corresponds to lowering $W_0$ to bring the minimum in to smaller volumes -
that this occurs can be confirmed numerically.
Another possibility would be a purely perturbative stabilisation of
the volume modulus, solely using K\"ahler corrections (in which axions
do not appear). This has been discussed in \cite{hepth0508171, hepth0507131}, although
without an explicit example.

I shall not dwell on these possiblities. First, because the resulting
axion decay constant would be, as we shall see shortly, close to the Planck scale and outside the
allowed window and secondly, because at such small volumes there is no
good control parameter. I shall instead focus on a version of the large-volume compactifications of chapter \ref{chapterLargeVol},
suitably modified to include a QCD axion.
In this section I only discuss
moduli stabilisation but in
section \ref{AxionDecayConstantSec} I shall show that
these can also realise phenomenological values for $f_a$.

I will illustrate the discussion with a three-modulus toy model, in which I assume the volume may be
expressed in terms of 4-cycles as
\be
\label{VolExpression}
\mc{V} = (T_1 + \bar{T}_1)^{\frac{3}{2}} - (T_2 +
\bar{T}_2)^{\frac{3}{2}} - (T_3 + \bar{T}_3)^{\frac{3}{2}}.
\ee
The use of three moduli is because this turns out to be the minimal number
required for our purposes: clearly, this is no significant restriction
on model-building.
Expressed in terms of 2-cycles, (\ref{VolExpression}) corresponds to
\be
\label{Toy2TI}
\mc{V} = \lambda( t_1^3 - t_2^3 - t_3^3).
\ee
We note (\ref{Toy2TI})
satisfies the consistency requirement that $\frac{\partial^2 \mc{V}}{\partial t_i \partial t_j}$ have signature $(+,-,-)$.
We may perhaps think of this toy model as a $\mbb{P}^3$ with two points blown up.
Denoting the cycles by 1, 2 and 3, the K\"ahler potential including
perturbative corrections is
\be
\label{ToyK}
\mc{K} = - 2 \ln \left(  (T_1 + \bar{T}_1)^{\frac{3}{2}} - (T_2 +
\bar{T}_2)^{\frac{3}{2}} - (T_3 + \bar{T}_3)^{\frac{3}{2}} \right) -
\frac{\xi}{g_s^{3/2} \mc{V}}.
\ee
$g_s$ is fixed by the
fluxes and in (\ref{ToyK}) should be regarded as a tunable parameter.
For superpotential, we shall take
\be
W = W_0 + e^{-\frac{2 \pi}{n} \left( T_2 + T_3 \right)}.
\ee
This could arise from gaugino condensation on a stack of $n$ branes
wrapping the combined cycle 2+3.\footnote{The use of gaugino
  condensation rather
  than instanton effects is necessary to ensure that the cycle 2+3 is
  large enough to contain QCD.} QCD will be realised as a stack of branes
wrapping cycle 3.

In the limit $\mc{V} \gg 1$ with $\tau_2$ and $\tau_3$ small,
the leading functional form of the scalar potential is (omitting numerical factors)
\be
\label{ToyScalarPotential}
V = \frac{(\sqrt{\tau_2} + \sqrt{\tau_3})e^{-\frac{2 \pi}{n} 2(\tau_2 +
    \tau_3)}}{\mc{V}}
- \frac{(\tau_2 + \tau_3) e^{-\frac{2 \pi}{n} (\tau_2 +
    \tau_3)}}{\mc{V}^2}
+ \frac{\xi}{g_s^{3/2} \mc{V}^3}.
\ee
The minus sign in (\ref{ToyScalarPotential}) arises from minimising the potential
for the axion $\textrm{Im}(T_2 + T_3)$. The axions $\textrm{Im}(T_1)$ and $\textrm{Im}(T_2 -
    T_3)$ do not appear in (\ref{ToyScalarPotential}) and are unfixed.
By considering the limit $\mc{V} \to \infty$, $\frac{2 \pi (\tau_2 + \tau_3)}{n} \sim
    \ln \mc{V}$, it follows that as $\mc{V} \to \infty$ the potential
    (\ref{ToyScalarPotential}) goes to zero from below. As by
    adjusting $g_s$ we can make the third term of
    (\ref{ToyScalarPotential}) arbitrarily large, we can ensure the
    potential remains positive until arbitrarily large volumes, and
    thus any minimum will be at exponentially large volumes.

Is there a minimum? The potential is clearly symmetric under $\tau_2 \leftrightarrow \tau_3$, and the potential
restricted to the locus $\tau_2 = \tau_3$ indeed has a minimum at
exponentially large volumes. Because of the symmetry $\tau_2
\leftrightarrow \tau_3$, this `minimum' is also a
critical point of the full potential. However, it is not a minimum of
the full potential. At fixed $\tau_2 + \tau_3$ and fixed $\mc{V}$,
(\ref{ToyScalarPotential}) depends only on $\sqrt{\tau_2} +
\sqrt{\tau_3}$. For fixed $\tau_2 + \tau_3$, this is \emph{maximised}
at $\tau_2 = \tau_3$, and so the mode
$\tau_2 - \tau_3$ is tachyonic at this locus. We have not
investigated whether this tachyon satisfies the Breitenlohner-Freedman
bound for AdS stability.
This is for
the same reasons as above: once we uplift, the geometric protection of
AdS ceases to be relevant. Consequently the fields in
(\ref{ToyScalarPotential}) run away either to
$\tau_2 = 0$ or $\tau_3 = 0$,
where one of the blow-up cycles collapses.

This result shows that the
above toy model does not, by itself, have a minimum of the potential with
a massless QCD axion. We may ask whether this is a feature of the
geometric details of the model - for example,
whether a different choice of triple intersection form in
(\ref{Toy2TI}) would alter this result. We have investigated several other
toy models without finding a minimum, and while we have no proof we
suspect none exists so long as the K\"ahler potential is given by (\ref{ToyK}).

This is bad news, but it is controllable bad news.
The instability above is very
particular: there is no instability either for the overall volume or for
the sum of the blow-up volumes $\tau_1 + \tau_2$,
but only for the difference $\tau_1 - \tau_2$. The effect of the
instability is to drive one
of the blow-up cycles
to collapse. Consequently, the instability can be cured by \emph{any}
effect that becomes important at small cycle
volume and prevents
collapse.

For example, the presence of a term
\be
\label{NewTerm}
\frac{1}{\sqrt{\tau_2} \mc{V}^3} + \frac{1}{\sqrt{\tau_3} \mc{V}^3}
\ee
in (\ref{ToyScalarPotential}) would obviously stabilise the cycles
$\tau_2$ and $\tau_3$ against collapse and generate a minimum of the
potential.
As this term does not affect the
argument that in the $\mc{V} \to \infty$ limit the potential approaches
zero from below, the resulting minimum would still be at exponentially
large volume.

At this point it is useful to recall the discussion of section
\ref{PerturbativeExpansions} 
on the general form of
K\"ahler corrections that are and
are not allowed.
Here we just note
the terms of (\ref{NewTerm}) may be generated from a correction to the K\"ahler potential,
\be
\label{KahlerCorrectionNewTerm}
\mc{K} + \delta \mc{K} = - 2 \ln (\mc{V}) + \frac{\epsilon
  \sqrt{\tau_2}}{\mc{V}} + \frac{\epsilon \sqrt{\tau_3}}{\mc{V}}.
\ee
For simplicity we have kept the $\tau_2 \leftrightarrow \tau_3$
symmetry.
Such a correction is motivated by the fact that it gives corrections
to the
K\"ahler metrics $\mc{K}_{2 \bar{2}}$ and
$\mc{K}_{3 \bar{3}}$ suppressed by
factors of $g^2$ for the field theory on the relevant cycle. More specifically,
\be
\mc{K}_{2 \bar{2}} + \delta \mc{K}_{2 \bar{2}} =
\frac{3}{2\sqrt{2 \tau_2} \mc{V}}\left( 1 - \frac{\epsilon}{12 \sqrt{2} \tau_2} \right).
\ee
As $\tau_2 = \frac{1}{g^2}$ for a brane wrapping the cycle 2, the
  correction is
suppressed by $g^2$.

The inverse metric involves an infinite series of terms diverging in
the $\tau_2 \to 0$ and $\tau_3 \to 0$ limit. For example,
\be
\label{InverseKDivergence}
\mc{K}^{2 \bar{2}} = \frac{2 \sqrt{2 \tau_2} \mc{V}}{3} \left( 1 +
\frac{\epsilon}{12 \sqrt{2} \tau_2} + \frac{\epsilon^2}{288 \tau_2^2} +
\ldots \right).
\ee
This is easy to understand: at $\tau_2 = \frac{\epsilon}{12
  \sqrt{2}}$, the K\"ahler metric $\mc{K}_{2 \bar{2}}$ goes to zero
and the inverse metric diverges. This divergence can be seen by resumming
(\ref{InverseKDivergence}). In the physical potential, this divergence
will create a positive wall at finite values of $\tau_2$ and
$\tau_3$. The positivity can be seen from the fact that the divergence
in $\mc{K}^{-1}$ will only appear through the term
\be
e^{\mc{K}} \mc{K}^{i \bar{j}} D_i W D_{\bar{j}} \bar{W},
\ee
which is manifestly positive
definite. Consequently the potential will diverge positively
at finite values of $\tau_2$ and $\tau_3$, and so a stable minimum
must exist for both $\tau_2$ and $\tau_3$.

We have now outlined, in the context of the large-volume scenario of
chapter \ref{chapterLargeVol}, a way to stabilise all the size
moduli while retaining a massless QCD axion. We would like
cycle 3 to support QCD:
by adjusting $\epsilon$, we can always make the correction
(\ref{KahlerCorrectionNewTerm})
sufficiently large to ensure that $\tau_3$ is stabilised with
the correct size for QCD. For intermediate string scales, this
requires $\tau_3 \sim 10$. As the actual correction would be very hard
to calculate, at this level we just adjust $\epsilon$
phenomenologically.
Of course, the complexity of a real model is much greater than that of
(\ref{KahlerCorrectionNewTerm}).
However we note again that, even though the corrections cannot easily be calculated, our
proposal for moduli stabilisation only requires that they exist and come with the right sign to prevent collapse.

While the above has been a toy example, the above
approach may be applied to any model in which the moduli are stabilised
along the lines of \cite{hepth0502058, hepth0505076}. Keeping an axion
massless introduces an instability causing a blow-up cycle to want to
collapse. K\"ahler corrections that become important at small volume
can stabilise this cycle but will not affect the overall structure of
the potential, and in particular will not affect the stabilisation of
the volume at $\mc{V} \gg 1$.

\subsection{Higher Instanton Effects in the Axion Potential}
\label{sechigherinstanton}

We have given above a K\"ahler potential and superpotential that will
stabilise the moduli while containing a candidate QCD axion.
The nonperturbative terms in the superpotential are in general
just the leading terms in an instanton expansion. Even though the higher order terms
may be highly suppressed and irrelevant to moduli stabilisation, they could still
lift the flat direction associated with the massless axion.
Given that $\Lambda_{QCD} \ll M_P$, even highly subleading terms could dominate over QCD instantons.

Let us estimate the general magnitude of such instanton effects.
The magnitude of brane instantons depends on the volumes of the cycles they can wrap. Generally there will be many such cycles,
whose sizes depend on the stabilised moduli, but
minimally there must always exist the cycle which support the QCD stack. It follows from the DBI action
that the gauge coupling for a D7-brane stack is
$$
\frac{1}{g^2} = \frac{\textrm{Re}(T)}{2 \pi} \Rightarrow \alpha^{-1} =
\frac{4 \pi}{g^2} = 2 \textrm{Re}(T) = 2 \tau.
$$
This defines the gauge coupling at the high scale where the effective field theory becomes valid.
When $m_s \sim M_P$, this is in essence the string scale, but if $m_s \ll M_P$,
the difference between $m_{KK}$ and $m_s$ becomes significant. It is a
subtle issue whether $m_s$ or $m_{KK}$ is the appropriate high
scale. If QCD is supported on a small cycle within a large internal
space, the KK modes associated with the bulk will be uncharged under
QCD and will not contribute to the running coupling. KK modes of the
QCD cycle will contribute, but these will be at masses comparable to
the string scale. In considering
the running coupling, we therefore use $m_s$ as the high scale rather
than $m_{KK}$.
We consider a wide range of string scales and take a sampling of high scale values from
$10^8 \to 10^{16}$ GeV. Given a string scale, we determine the appropriate internal volume by $m_s
\sim \frac{M_P}{\sqrt{\mc{V}}}$.

The QCD coupling runs logarithmically with energy scale, with
$$
\alpha_{QCD}^{-1}(10^2 \textrm{GeV}) \sim 9 \qquad \textrm{ and } \qquad
\alpha_{QCD}^{-1}(10^{16} \textrm{GeV}) \sim 25.
$$
The required high-scale couplings and cycle sizes are given in table \ref{Table1},
together with the action for a D3-brane instanton wrapping the same cycle as the QCD stack.
Its magnitude is set by $\sim e^{-2 \pi T}$ and
we show in table \ref{Table1} the approximate magnitude of single- and double-instanton effects.
In addition to the QCD cycle, there may be other cycles which instantons
may wrap. We do not include these for two reasons: first,
whether such instantons would generate a potential for the QCD axion
is model-dependent\footnote{It seems odd that such instantons could
  affect the QCD axion at all. However, if QCD is on cycle 3, and the
  axion $b_2 + b_3$ is fixed by effects on cycle 2+3, an instanton
  solely on cycle 2 effectively generates a potential for the QCD axion.}, and secondly, we can always arrange the
model such that the QCD cycle is smallest and hence
dominates the instanton expansion.
We observe that the magnitude of the required cycle volume, and thus the magnitude of potential instanton effects,
varies significantly with the string scale.
\begin{table}
\caption{Cycle sizes and instanton amplitudes for various UV scales}
\label{Table1}
\centering
\vspace{3mm}
\begin{tabular}{|c|c|c|c|c|c|}
\hline
$\textrm{E}_{UV}$ & $10^8$ GeV & $10^{10}$ GeV & $10^{12}$ GeV & $10^{14}$ GeV & $10^{16}$ GeV\\
\hline
$\alpha_{QCD}^{-1}(\textrm{E}_{UV})$ & 15.8 & 18.1 & 20.4 & 22.7 & 25 \\
\hline
\textrm{Re}(T) = $\frac{\alpha^{-1}}{2}$ & 7.9 & 9.1 & 10.2 & 11.4 & 12.5 \\
\hline
$e^{-2 \pi T}$ & $2.8 \ti 10^{-22}$ & $1.5 \ti 10^{-25}$ & $1.5 \ti 10^{-28}$ & $7.8 \ti 10^{-32}$ & $7.8 \ti 10^{-35}$ \\
\hline
$e^{-4 \pi T}$ & $7.7 \ti 10^{-44}$ & $2 \ti 10^{-50}$ & $2 \ti 10^{-56}$ & $6 \ti 10^{-63}$ & $6 \ti 10^{-69}$ \\
\hline
\end{tabular}
\end{table}
If present, such D-instantons would generate a potential for the QCD axion. To compare their magnitude to that of QCD effects,
we need an estimate of their contribution to the scalar potential. In this context we only care
about terms containing a phase and so contributing to the axion potential.
To this end, the relevant term from the scalar potential (\ref{NoScaleScalarPotential}) is
\be
V_{axion} = e^{\mc{K}} \left( \mc{K}^{i \bar{j}} \left( \partial_i W (\partial_{\bar{j}} \mc{K}) \bar{W} + c.c. \right) \right).
\ee
A superpotential instanton contribution $e^{-2 \pi n  T_i}$ generates a term
\be
\label{AxionPotential}
V_{axion} = \frac{- 2 a_i \tau_i W_0}{\mc{V}^2} e^{-2 \pi n \tau_i} \cos(\theta_i).
\ee
The absolute magnitude of (\ref{AxionPotential}) depends on the value
of $n$, the internal volume $\mc{V}$
and the tree-level
superpotential $W_0$. As we are looking towards phenomenology
we also take $m_{3/2} = \frac{W_0}{\mc{V}} \sim 1 \textrm{TeV} \sim 10^{-15} M_P$,
as appropriate for gravity-mediated TeV-scale soft terms.
Note that for models built around the KKLT scenario, we always have $m_s \gtrsim M_{GUT}$ and
only the largest value of $\textrm{E}_{UV}$ is achievable.
In table \ref{Table2} we give the internal volumes required for
each UV scale, as well as the resulting absolute magnitude of 1-, 2- and 3-instanton superpotential corrections to the scalar potential.
For the same reasons as above, we only consider instantons wrapping the QCD cycle.
\begin{table}
\caption{Magnitude of Axion Potentials from Superpotential Instanton Effects}
\label{Table2}
\centering
\vspace{3mm}
\begin{tabular}{|c|c|c|c|c|c|}
\hline
$\textrm{E}_{UV}$ & $10^8$ GeV & $10^{10}$ GeV & $10^{12}$ GeV & $10^{14}$ GeV & $10^{16}$ GeV\\
\hline
$\mc{V}$ & $10^{20}$ & $10^{16}$ & $10^{12}$ & $10^{8}$ & $10^4$ \\
\hline
$V_{\textrm{1-instanton}}$ & $10^{-57} M_P^4$ & $10^{-56} M_P^4$ & $10^{-55} M_P^4$ & $10^{-54} M_P^4$ & $10^{-53} M_P^4$ \\
\hline
$V_{\textrm{2-instanton}}$ & $10^{-79} M_P^4$ & $10^{-81} M_P^4$ & $10^{-83} M_P^4$ & $10^{-85} M_P^4$ & $10^{-87} M_P^4$ \\
\hline
$V_{\textrm{3-instanton}}$ & $10^{-101} M_P^4$ & $10^{-106} M_P^4$ & $10^{-111} M_P^4$ & $10^{-116} M_P^4$ & $10^{-121} M_P^4$ \\
\hline
\end{tabular}
\end{table}

As well as superpotential effects, there are also nonperturbative corrections to the K\"ahler potential (the
perturbative corrections to $\mc{K}$ do not have an axionic dependence).
While smaller, these are easier
to generate - the instantons can have four fermionic zero modes rather than only two. A correction
\be
\mc{K} = - 2 \ln (\mc{V}) \to \mc{K} = -2 \ln (\mc{V} + e^{-2 \pi n T})
\ee
will generate effects in the scalar potential at order
$$
V_{\delta \mc{K}} \sim \frac{W_0^2}{\mc{V}^3} e^{-2 \pi n T}.
$$
and thus generate a potential for the QCD axion $\theta$ of the form
$$
V_{\delta K} \cos (\theta + \alpha).
$$
Again assuming TeV-scale (visible)
SUSY breaking, $\frac{W_0}{\mc{V}} \sim 10^{-15}$, the magnitudes of such effects are shown in table \ref{Table3}.
\begin{table}
\caption{Magnitude of Axion Potentials from K\"ahler Potential Instanton Effects}
\label{Table3}
\centering
\vspace{3mm}
\begin{tabular}{|c|c|c|c|c|c|}
\hline
$\textrm{E}_{UV}$ & $10^8$ GeV & $10^{10}$ GeV & $10^{12}$ GeV & $10^{14}$ GeV & $10^{16}$ GeV\\
\hline
$V$ & $10^{20}$ & $10^{16}$ & $10^{12}$ & $10^8$ & $10^4$ \\
\hline
$V_{\textrm{1-instanton}}$ & $10^{-77} M_P^4$ & $10^{-72} M_P^4$ & $10^{-67} M_P^4$ & $10^{-62} M_P^4$ & $10^{-57} M_P^4$ \\
\hline
$V_{\textrm{2-instanton}}$ & $10^{-99} M_P^4$ & $10^{-97} M_P^4$ & $10^{-95} M_P^4$ & $10^{-93} M_P^4$ & $10^{-91} M_P^4$ \\
\hline
$V_{\textrm{3-instanton}}$ & $10^{-121} M_P^4$ & $10^{-122} M_P^4$ & $10^{-123} M_P^4$ & $10^{-124} M_P^4$ & $10^{-125} M_P^4$ \\
\hline
\end{tabular}
\end{table}

 The axion potential originating from QCD effects and relevant to the
strong CP problem is
$$
V_{QCD} \sim \Lambda_{QCD}^4 (1 - \cos(\theta)),
$$
with $\Lambda_{QCD} \sim 2 \ti 10^{-19} M_P$ and $\Lambda_{QCD}^4 \sim 10^{-75} M_P^4$. We require QCD effects to be sufficiently
dominant to be consistent with the failure to observe CP violation in strong interactions.
Suppose we have a potential
\be
V = A (1 - \cos(\theta)) + \epsilon \cos(\theta + \gamma).
\ee
If $A \gg \epsilon$, the minimum is displaced from $\theta = 0$ by
$\delta \theta \sim \frac{ \epsilon }{A}$. Observationally, $\vert \theta \vert < 10^{-10}$,
and thus non-QCD contributions must have absolute magnitude smaller than $10^{-85} M_P^4$.
By comparison with tables $\ref{Table2}$ and $\ref{Table3}$
it follows that in order for QCD instantons to dominate the axion
potential, the single instanton corrections to both the
superpotential and the K\"ahler potential must be absent,
while the 2-instanton superpotential correction may or may not be
present
depending on factors of $2 \pi$ and the precise value of the string
scale.

To contribute to a superpotential (K\"ahler potential) an instanton must have at most 2 (4) fermionic zero modes, to generate
$\int d^4x d^2 \theta$ and $\int d^4 x d^2 \theta d^2 \bar{\theta}$ terms respectively. In the absence of flux, there is a
necessary condition on a divisor to generate a superpotential
\cite{hepth9604030}: it must have holomorphic
Euler characteristic one, $\chi_g(D) = 1$.
In the presence of flux, this condition may be relaxed. The number of zero modes on an instanton, and thus
its ability to appear in either the K\"ahler or superpotential,
may also be affected by the presence of the stack of QCD branes wrapping the would-be instanton cycle.
We shall not attempt to analyse this question for specific `real' models, but
by fiat simply assume the necessary instantons to be absent from
the potential. In this regard it is encouraging that the number of
instantons required to be suppressed is quite limited.

\section{The Axion Decay Constant}
\label{AxionDecayConstantSec}

The last section was devoted to solving the strong CP problem: keeping
an axion light while stabilising the moduli.
However, even achieving this does not resolve all phenomenological problems.
Given that a QCD axion exists, as indicated
above there are strong bounds
on the axion decay constant $f_a$ of equation (\ref{axionLagrangian}):
$10^9 \textrm{GeV} \lesssim f_a \lesssim 10^{12} \textrm{GeV}$.
The lower bound, from supernova coolings is hard. While the upper bound may
be relaxed by considering non-standard cosmologies, here we shall also
treat this as hard.
We want to estimate $f_a$ in some moduli stabilisation scenarios (the
value of $f_a$ in string compactifications is also discussed in
\cite{hepth0605206, hepth0605256}).
In IIB compactifications, the axionic coupling to QCD arises from the clean and model-independent
Chern-Simons coupling.
However, to obtain the physical value of $f_a$ the axion must be
canonically normalised.
This depends on the K\"ahler metric,
and in particular on where the moduli are stabilised.

We assume we can write the K\"ahler potential as
\be
\mc{K} = \mc{K}(T_i + \bar{T}_i),
\ee
with $\mc{K}$ real.
This is true in perturbation theory, owing to the axionic shift
symmetry,
and any nonperturbative violations are small enough to be
irrelevant for this purpose.
In this case the kinetic terms for the axionic and size moduli do not mix.
Noting that $\mc{K}_{i \bar{j}} = \mc{K}_{j \bar{i}}$, we have for any $i$ and $j$
\bea
& & \mc{K}_{i \bar{j}} ( \partial_\mu T^i \partial^\mu \bar{T}^j ) +
\mc{K}_{j \bar{i}} ( \partial_\mu T^j \partial^\mu \bar{T}^i ) \nonumber \\
& = & \mc{K}_{i \bar{j}} \left( (\partial_{\mu} \tau_i + i \partial_\mu c_i)(\partial^\mu \tau_j -i \partial^\mu c_j)
+ (\partial_{\mu} \tau_j + i \partial_\mu c_j)(\partial^\mu \tau_i -i \partial^\mu c_i) \right) \nonumber \\
& = & \mc{K}_{i \bar{j}} (2 \partial_\mu \tau_i \partial^\mu \tau_j + 2 \partial_\mu c_i \partial^\mu c_j),
\eea
and the two sets of terms decouple.

Let us first show that if both the overall volume and the individual cycle
volumes are comparable to the string scale, then as expected $f_a \gtrsim 10^{16}
\textrm{GeV}$. Suppose an axion $c_i$ is to be the QCD axion. The
Lagrangian for this axion is
\be
\mc{K}_{i \bar{i}} \partial_\mu c_i \partial^\mu c_i + \frac{c_i}{4
  \pi} \int F^a \wedge F^a.
\ee
For simplicity we have not included mixing terms: these will not
greatly affect the discussion.

The simplest toy model is that of a factorisable toroidal orientifold, with
K\"ahler potential
\bea
\mc{K} & =  & - \ln \left( (T_1 + \bar{T}_1)(T_2 + \bar{T}_2)(T_3 + \bar{T}_3) \right) \nonumber \\
& = & - \ln (T_1 + \bar{T}_1) - \ln (T_2 + \bar{T}_2) - \ln (T_3 + \bar{T}_3).
\eea
and K\"ahler metric
\be
\mc{K}_{i \bar{j}} = \left( \begin{array}{ccc} (T_1 + \bar{T}_1)^{-2} & 0 & 0 \\
0 & (T_2 + \bar{T}_2)^{-2} & 0 \\
0 & 0 & (T_3 + \bar{T}_3)^{-2} \end{array} \right).
\ee
If we denote the axions by $c_1$, $c_2$ and $c_3$, the axion kinetic terms are
\be
\frac{1}{4 \tau_1^2} \partial_\mu c_1 \partial^\mu c_1 + \frac{1}{4 \tau_2^2} \partial_\mu c_2
\partial^\mu c_2 + \frac{1}{4 \tau_3^2} \partial_\mu c_3 \partial^\mu c_3.
\ee
For definiteness, let us assume QCD is realised on cycle 1. There is no inter-axion mixing and the relevant axion Lagrangian is
\be
\frac{1}{4 \tau_1^2} \partial_\mu c_1 \partial^\mu c_1 + \frac{c_1}{4 \pi} \int F^a \wedge F^a.
\ee
If we canonically normalise $c_1' = \frac{c_1}{\sqrt{2} \tau_1}$, this becomes
\be
\frac{1}{2} \partial_\mu c_1' \partial^\mu c_1' + \frac{\sqrt{2}\tau_1}{4 \pi} c_1' \int F^a \wedge F^a.
\ee
In units where $M_P = 1$, the axion decay constant is
$$
f_a = \frac{1}{4 \pi \tau_1 \sqrt{2}} .
$$
If QCD is to be realised on this cycle, we need $\tau_1 \sim 12$, and thus $f_a \sim 10^{16} \textrm{GeV}$.
Going beyond this toy example, we recall that in general the K\"ahler metric was given by
(\ref{GeneralKahlerP}),
\be
\mc{K}_{i \bar{j}} = \frac{G_{i \bar{j}}^{-1}}{\mc{V}^2}, \qquad \qquad
G_{i \bar{j}} = -\frac{3}{2} \left( \frac{k_{ijk}t^k}{\mc{V}} - \frac{3}{2} \frac{k_{imn}t^m t^n k_{jpq} t^p t^q}{\mc{V}^2} \right).
\ee
If all cycles are string scale in magnitude, then $\mc{K}_{i \bar{j}}
\sim \mc{O}(1)$ and it is impossible to lower the axion
decay constant substantially through canonical normalisation. The same conclusion
applies: $f_a \gtrsim 10^{16} \textrm{GeV}$. This conclusion
is unsuprising: the axionic coupling to matter is a stringy
coupling, and so we expect $f_a$ to be comparable to the string scale.
If the string and Planck scales are identical, $f_a$ cannot lie within
the allowed window.

\begin{figure}[ht]
\linespread{0.2}
\begin{center}
\makebox[10cm]{ \epsfxsize=10cm \epsfysize=7cm \epsfbox{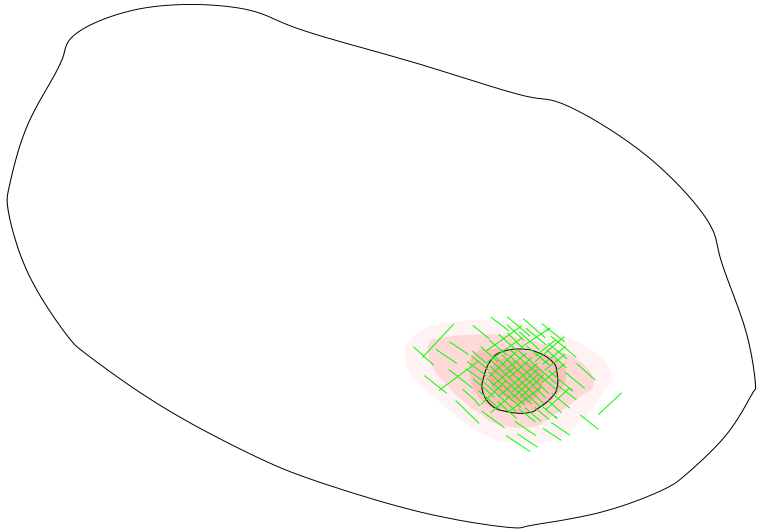}}
\end{center}
\caption{The axion and matter wavefunctions are both localised on
  small string-scale cycles. The axion decay constant measures the
  (string scale) overlap.}
\label{MatterAxionCoupling}
\end{figure}

If we lower the string scale, phenomenological values for $f_a$ can be
achieved. To analyse this, let us return to the toy model of
(\ref{VolExpression}), which had
K\"ahler potential
\be
\mc{K} =  -2 \ln \left( (T_1 + \bar{T_1})^{\frac{3}{2}} - (T_2 + \bar{T}_2)^{\frac{3}{2}} - (T_3 + \bar{T}_3)^{\frac{3}{2}} \right).
\ee
The K\"ahler metric for this model is
\be
\label{KMetricfa}
\mc{K}_{i \bar{j}} = \left( \begin{array}{ccc}
\vspace{0.2cm}
\frac{-3}{2 \sqrt{2 \tau_1}\mc{V}} + \frac{9 \tau_1}{\mc{V}^2} &
-\frac{9 \sqrt{\tau_2}}{2 \mc{V}^{5/3}} & -\frac{9 \sqrt{\tau_3}}{2
  \mc{V}^{5/3}} \\
\vspace{0.2cm}
-\frac{9 \sqrt{\tau_2}}{2 \mc{V}^{5/3}} &
\frac{3}{2 \sqrt{2 \tau_2} \mc{V}} + \frac{9 \tau_2}{\mc{V}^2} & \frac{9 \sqrt{\tau_2
    \tau_3}}{\mc{V}^2} \\
-\frac{9 \sqrt{\tau_3}}{2 \mc{V}^{5/3}} & \frac{9 \sqrt{\tau_2 \tau_3}}{\mc{V}^2}
& \frac{3}{2 \sqrt{2 \tau_3} \mc{V}} + \frac{9 \tau_3}{\mc{V}^2} \end{array} \right).
\ee
The axion kinetic terms are $\mc{K}_{i \bar{j}} \partial_\mu c_i
\partial^\mu c_j$. At small volumes there is
substantial mixing between the axions $c_1$, $c_2$ and $c_3$. However,
in the limit $\mc{V} \to \infty$ with $\tau_1 \gg \tau_2, \tau_3$,
the K\"ahler metric has the scaling behaviour
\be
\mc{K}_{i \bar{j}} \sim \left( \begin{array}{ccc} \mc{V}^{-4/3} &
\mc{V}^{-5/3} & \mc{V}^{-5/3} \\
\mc{V}^{-5/3} & \mc{V}^{-1} & \mc{V}^{-2} \\
\mc{V}^{-5/3} & \mc{V}^{-2}
& \mc{V}^{-1} \end{array} \right),
\ee
and is to a good approximation diagonal.
The requirement $\tau_1 \gg 1$ implies that QCD cannot be realised
on branes wrapping cycle 1, as the resulting field theory is far too
weakly coupled.
However, if $\tau_2 \sim \tau_3 \sim 10$ we may realise QCD by wrapping branes on one of these
cycles (for concreteness cycle 3).
The resulting axion decay constant is
\be
f_a \sim \frac{\sqrt{\mc{K}_{3 \bar{3}}}}{4 \pi} M_P \sim \frac{\mc{O}(1)}{4 \pi \sqrt{\mc{V}}} M_P.
\ee
Thus if $\mc{V} \sim 10^{14}$ and $\tau_3 \sim 10$,
the QCD gauge coupling is correct and the axion decay constant $f_a \sim 10^{10} \textrm{GeV}$
lies within the narrow phenomenological window. Up to $\mc{O}(1)$
factors, the string and Planck scales are related by
\be
m_s = \frac{g_s M_P}{\sqrt{\mc{V}}}.
\ee
Such a large volume corresponds to lowering the string scale to $m_s
\sim 10^{11} \textrm{GeV}$.
The lowered axion decay constant is easy to understand physically.
$f_a$ measures the axion-matter coupling, which is an effect localised
around the small QCD cycle.
Thus the only scale it is sensitive to is the string scale, and so up
to numerical factors $f_a \sim m_s$. This is illustrated in figure \ref{MatterAxionCoupling}.

The above is a very particular limit of moduli space, with one cycle
taken extremely large while all others are
only marginally larger than the string scale. It would thus be essentially a curiosity if it were not also
the exact regime in which the moduli are stabilised in the large-volume compactifications described in chapter
\ref{chapterLargeVol}. As the stabilised volume is exponentially sensitive to the stabilised dilaton,
\emph{a priori} the string scale can lie anywhere between the Planck and TeV scales. There is no difficulty,
and no fine-tuning, in stabilising the volume so as to achieve an intermediate string scale.

The above result on $f_a$ is independent of whether an axion remains
massless or not. As described in section \ref{KKLTAxionSection}, the simplest version of the scenarios of
\cite{hepth0502058, hepth0505076} makes all axions far too heavy
to solve the strong CP problem.
In sections \ref{secnonsusyminima} and \ref{sechigherinstanton} we have described the necessary modifications to this
scenario such that a massless QCD axion will survive to solve the
strong CP problem. Combining this with
the above, we have for the first time given a procedure to stabilise
all moduli while ensuring a
QCD axion exists with a phenomenologically allowed value for
$f_a$.

In itself this is interesting, as axions within the phenomenological
window have always been hard to achieve in string compactifications.
However, this scenario compels a further very interesting relationship between the axion decay constant
and the (visible) supersymmetry breaking scale.

We argued earlier that if a QCD axion is to be present in IIB flux
compactifications, QCD must be realised on a stack of D7-branes.
We have also described how to
stabilise the moduli such that a QCD axion can exist with a
phenomenologically allowed decay constant. As discussed in detail in chapter \ref{chapterSoftSusy}, given a procedure of
moduli stabilisation it is possible to calculate the magnitude of soft supersymmetry breaking terms.
We recall that for a gauge theory on D7 branes the scalar and gaugino masses have scaling behaviour (we neglect subleading
factors of $\ln \mc{V}$)
\be
m_{D7} \sim M_{D7} \sim m_{3/2} = e^{\mc{K}/2} W \sim \frac{M_P}{\mc{V}}.
\ee
It is an attractive feature of the large-volume compactifications that these scales are all set by the
volume $\mc{V}$. As we have
$$
f_a \sim \frac{M_P}{\sqrt{\mc{V}}} \qquad \textrm{ and } \qquad m_{soft} \sim \frac{M_P}{\mc{V}}.
$$
it follows that up to numerical factors
\be
\label{AxionSoftRelation}
f_a = \sqrt{ M_P m_{soft}}.
\ee
Thus in such models the axion decay constant is compelled to be
the geometric mean of the Planck scale and the (visible) supersymmetry breaking scale.
This is a striking result, as \emph{a priori} these two pieces of physics are
entirely unrelated. The relation (\ref{AxionSoftRelation}) is another
example of the phenomenological virtues of an intermediate string scale
\cite{hepph9810535}.

\section{Consequences}

The purpose of this chapter has been to investigate the conditions under
which a QCD axion, ideally with a phenomenological value for $f_a$,
may coexist with stabilised moduli in string compactifications. This
divides into two questions: first, how to stabilise the moduli such
that a massless axion survives, and secondly, how to obtain allowed
values of $f_a$, $10^9 \textrm{GeV} < f_a < 10^{12} \textrm{GeV}$.

In the context of the first question, we have shown that the simplest
versions of the moduli stabilisation scenarios considered in chapters
\ref{ChapterKahlerModuli} and \ref{chapterLargeVol} do not
contain any light axions. If every relevant modulus is stabilised by
nonperturbative effects, then their axionic components receive large masses
and cannot be a QCD axion.
We also proved a negative result, in that
supersymmetric moduli stabilisation is disfavoured: there exist no
supersymmetric minima of the F-term potential with flat axionic
directions.
Even if AdS stability is present due to the Breitenlohner-Freedman
bound, the tachyons must be removed by the time we are in Minkowski
space. Performing this step requires a much greater technical understanding of uplifting
AdS vacua to Minkowksi space than is currently available, and so it is
unclear how relevant the original supersymmetric AdS solutions are.

This result is pure $\mc{N}=1$ supergravity and so makes no assumptions
about the particular string model considered. This result is
parenthetical to the main thrust of this thesis, which concerns the
large volume non-supersymmetric compactifications of chapter
\ref{chapterLargeVol}. However we include it because
it applies to all
string compactifications, and in particular shows that in many of the
supersymmetric IIA flux compactifications considered recently in the
literature the complex structure moduli sector
is heavily tachyonic.

We do however view this negative result as a positive feature of the large volume
compactifications
of chapter \ref{chapterLargeVol}: as the minimum found there is
non-susy it does not suffer from the general problem identified in
this result.
In the context of these compactifications, we outlined
how to stabilise moduli while keeping axions massless.
Here we had to rely on K\"ahler corrections that will become
important as a cycle collapses to zero size. While unfortunately not
much is known about these, our main requirement was simply that they
exist.
Clearly progress in determining the form of such corrections
would be very interesting. We also specified the extent to which
subleading higher-order instantons must be absent in order for a leading-order
axion to solve the strong CP problem.

We note the result of the no-go theorem also favours gravity-mediated supersymmetry breaking. If
the moduli potential must break supersymmetry in order to solve the
strong CP problem, then this suggests that supersymmetry should be broken at
the string scale. Gravity mediation therefore always contributes to the visible
soft terms and, unless the string scale is drastically lowered, will
tend to dominate over gauge mediated effects. An
intermediate string scale may then be preferred in order to obtain
TeV-scale soft terms.

In the context of the second question, the fact that $f_a$ is
hierarchically lower than the Planck scale implies that
compactifications with $m_s \sim M_P$
are unlikely to give allowed values for $f_a$.
This problem can be avoided in the models of chapter
\ref{chapterLargeVol} in which
the string scale is hierarchically lower than the Planck
scale. In these models, $f_a \sim M_s$ and $M_{SUSY} \sim
\frac{M_s^2}{M_P}$.
An intermediate string scale therefore gives both
$10^9 \textrm{GeV} < f_a < 10^{12} \textrm{GeV}$ and visible susy
breaking at $\mc{O}(1 \textrm{TeV})$.
It is hard to find models with
phenomenological values for $f_a$, and so it is very
interesting that in the above model this also implies TeV-scale
supersymmetry breaking. 

A more general point argued in this chapter is that in the context of the
landscape the strong CP problem may serve as
an \emph{experimentum crucis}. Assuming that the solution to the strong CP
problem is a Peccei-Quinn axion and that string theory is a correct
description of nature, this is a solution that is
extremely sensitive to the physics of moduli stabilisation.
Requiring an axion to remain (essentially) massless while all
other moduli are stabilised is a technically clean problem directly
addressing the issue of vacuum selection.
Indeed, as seen above imposing this requirement directly rules
out many scenarios of moduli stabilisation.
The further condition $10^9 \textrm{GeV} < f_a < 10^{12}
\textrm{GeV}$ is even more constraining: the large-volume models of
chapter \ref{chapterLargeVol} are, as far as we know, the only models
capable of producing axions with the required decay constants.

We now turn our attention from particle physics problems to
cosmological problems, and consider a way to realise inflation in the
large-volume compactifications of chapter \ref{chapterLargeVol}.

\chapter{An Inflationary Model}
\label{InflationModel}

This chapter is based on the paper \cite{hepth0509012}.

The previous two chapters have described particle physics applications
of the moduli stabilisation scenario developed in chapter
\ref{chapterLargeVol}. As discussed there, one of the main
applications of moduli potentials is the determination of the
magnitude and pattern of supersymmetry breaking.
However, the moduli potential may also have important cosmological
applications in the theory of inflation.

Inflation is the dominant theory for the origin of structure in the
universe.
The universe is observed to be 
homogeneous on scales larger than that of galactic superclusters
($\sim 100 \hbox{Mpc}$) and
inhomogeneous below it. This behaviour can be seen primordially in the
Cosmic Microwave Background (CMB), which has a uniform temperature T
= 2.73 K with perturbations at the level of one part in
$10^5$. Simulations show that these perturbations are of the right
magnitude to generate the large-scale inhomogeneities seen today.

The term `inflation' simply refers to any time in the history of the
universe where the scale factor $a(t)$ is accelerating. If the metric of the
universe can be written
\be
ds^2 = -dt^2 + a(t)^2 (dx^2 + dy^2 + dz^2),
\ee
then inflation corresponds to any period in the history of the universe
during which 
$$
\ddot{a}(t) > 0.
$$
During a long period of inflation, the size of the universe expands
exponentially. The original advantage of inflation was that it would
dilute unwanted relics (such as gravitinos or magnetic monopoles) produced in the
very early universe and would solve the horizon and flatness problems. It
was later realised that inflation can also generate density
perturbations through the vacuum fluctuations of a scalar field. We shall not review all of the
theory behind inflation - useful references are the books
\cite{KolbTurner, LiddleLyth}.

The standard inflationary paradigm is that of \emph{slow-roll
  inflation}. It is well known that a universe dominated by a vacuum
  energy $\Lambda$ expands exponentially, with
$$
a(t) = e^{\Lambda (t-t_0)} a(t_0).
$$
Slow-roll inflation occurs as a scalar field slowly rolls down a very
  flat potential. During this time, the energy density of the universe
  is dominated by the vacuum energy of the scalar field. 
As long as the scalar potential is sufficiently flat, the universe
  undergoes $N_{e} \gg 1$ efolds of inflation during this period,
\be
\ln \frac{a(t_{end})}{a(t_{start})} = N_{e}.
\ee
The flatness of the potential can be quantified through the slow-roll
  parameters, $\epsilon$, $\eta$ and $\xi$. For single-field inflation, these
  are defined by
\bea
\epsilon & = & \frac{M_P^2}{2} \left( \frac{V'}{V} \right)^2, \\
\eta & = & M_P^2 \frac{V''}{V}, \\
\xi & = & M_P^4 \frac{V' V^{'''}}{V^2},
\eea
taking derivatives with respect to the canonically normalised
inflaton field. 
Slow-roll inflation occur if $\epsilon, \eta \ll 1$. Inflation
continues until $\eta \sim 1$ (note that typical models have $\epsilon \ll \eta$),
whereupon it ends rapidly. If these conditions
  hold, many e-folds of inflation can be expected: $N \gtrsim 60$ is
  required. The primordial density perturbations are generated by the quantum
  fluctations of the inflaton field. The task of inflationary model-building is to find
  (well-motivated) potentials realising these conditions.
Such questions are timely as observations can now provide precision tests
for inflationary models \cite{astroph0302207, astroph0603449}.

As slow-roll inflation is driven by the potential of a scalar field,
the moduli potential may in principle give rise to inflation.
There are many scalar fields in string theory, and so many candidates
for the inflaton
field. These can be classified by their origin in either
the open or closed string sector \cite{BinetruyGaillard, hepth9503114,
  hepph9812483}. The most
common open string inflaton is a brane/antibrane separation
\cite{hepth0105204, hepth0105203, hepth0111025}, whereas closed string
inflatons typically correspond to geometric moduli
\cite{hepth0406230}. As inflation is driven by the moduli potential, the
rapid recent advances in moduli stabilisation have been accompanied by
much effort devoted to inflationary model building in string theory
\cite{hepth0308055, hepth0406230, hepth0311077,
hepth0311191, hepth0312020, hepth0402047, hepth0403119, hepth0508029,
hepth0403123, hepth0403203, hepth0501130, hepth0505252, hepth0507205,
hepth0310221, hepth0404084, hepth0501184, hepth0508101}.

In phenomenological applications, the low energy
limit of string theory is generally $\mc{N} = 1$
supergravity. In this case there is a standard problem - the $\eta$ problem -
attendant on building inflationary models, whether involving
brane or modular fields as the inflaton.
This states that for the $\mc{N}=1$ F-term potential the slow-roll
$\eta$ parameter is $\mc{O}(1)$ unless a finely tuned cancellation
occurs. The F-term potential is
\be
V_F = e^{\mc{K}} \left( \mc{K}^{i \bar{j}} D_i W D_{\bar{j}} \bar{W} -
3 \vert W \vert^2 \right).
\ee
Examining the potential along the $\varphi$ direction, we then have
\be
\frac{\partial^2 V}{\partial \varphi^2} = \mc{K}^{''} V_F + \ldots
\ee
By considering $\varphi$ to be canonically normalised (i.e. $\mc{K}
\sim \phi \bar{\phi} + \ldots$), we see that
generically $\eta \sim 1$.
The $\eta$ problem is
manifest for F-term modular inflation. In brane inflation it is
not manifest, but reappears once this is embedded into a moduli stabilisation scenario
\cite{hepth0308055}.

In this chapter I present a simple inflationary scenario within
the framework of the moduli stabilisation mechanism developed in
chapter \ref{chapterLargeVol}. 
The inflaton is one of the K\"ahler moduli and inflation proceeds by
reducing the F-term energy.
The $\eta$ problem can be evaded by use of the no-scale properties of
the K\"ahler potential.
This mechanism in principle applies to a 
large class of Calabi-Yau compactifications that will be further specified
below.

\section{Keeping $\eta$ Small}
\label{seGI}

\subsection{General Idea}
\label{sseBPL}

Slow-roll inflation requires almost flat directions in
the scalar potential.
A natural source of flat directions would be a field only appearing
exponentially in the potential. Denoting this field by $\tau$, an
appropriate (and textbook \cite{LiddleLyth}) potential would be
\be
\label{TbookPot}
V(\tau) = V_0 \left( 1 - A e^{-\tau} + \ldots \right),
\ee
where the dots represent higher exponents.

As discussed above, nonperturbative effects are relevant for
stabilising many string moduli.
Examples are the K\"ahler moduli of IIB flux
compactifications and the dilaton and K\"ahler moduli in
heterotic Calabi-Yau compactifications.
In principle this discussion applies to all such fields,
but we focus here on the IIB K\"ahler moduli ($T_i = \tau_i + i c_i$) whose potential was
studied at length in chapter \ref{chapterLargeVol}.
We recall these appear nonperturbatively in the superpotential, 
\be
\label{eqNP}
W = \int G_3 \wedge \Omega + \sum_i A_i e^{-a_i T_i},
\ee
with the threshold corrections $A_i$ independent of the K\"ahler moduli.

The $\eta$ problem states that the potential from 
generic $\mc{K}$ has $\eta \sim \mc{O}(1)$.
However, the key word is `generic', and K\"ahler potentials
arising from string theory are (by definition) not generic.
A common way these potentials fail to be generic is by being no-scale,
corresponding to
\be
{\mc{K}}^{i \bar{j}} \partial_i {\mc{K}} \partial_{\bar{j}} {\mc{K}} = 3.
\ee
For a constant superpotential $W= W_0$, the no-scale scalar potential
vanishes:
\be
V_F = e^{\mc{K}} \left( \mc{K}^{i \bar{j}} D_i W D_{\bar{j}} \bar{W} - 3 \vert W
\vert^2 \right) = 0.
\ee
In contrast to the `generic' behaviour predicted by the $\eta$
problem, all directions are exactly flat. Indeed, the tree-level K\"ahler
potential for the IIB size moduli is no-scale,
\be
\label{InflatingK}
\mc{K} = - 2 \ln (\mc{V} ),
\ee
with $\mc{V}$ the internal volume.
Suppose we lift the no-scale behaviour through nonperturbative
terms in the superpotential as in (\ref{eqNP}). The scalar potential
becomes
\be
\label{eqPT}
V_F = e^{\mc{K}} \mc{K}^{i \bar{j}} \left[ a_i A_i a_j \bar{A}_j e^{-a_i T_i - a_j \bar{T}_j} -
(  (\partial_i \mc{K}) W a_j \bar{A}_j e^{-a_j \bar{T}_j} + c.c ) \right].
\ee
The $T_i$ directions are no longer exactly but exponentially flat as in (\ref{TbookPot}). It is
then natural to ask whether this flatness can drive inflation.

While the potential (\ref{eqPT}) is exponentially flat, it
also appears exponentially small. However, this is only true
so long as all $T_i$ fields are large. In the presence of several
K\"ahler moduli the variation of $V$ along the $T_i$ direction is in general
uncorrelated with the magnitude of $V$.  We can then hope to build an
inflationary model in a multi-modulus scenario.

\subsection{Embedding in IIB Flux Compactifications}
\label{ssePP}

The above gives the motivation. 
We now embed the above idea in the moduli stabilisation mechanism 
developed in chapter \ref{chapterLargeVol}.

The $\mbb{P}^4_{[1,1,1,6,9]}$ model used in chapter \ref{chapterLargeVol}
has two moduli. It will turn out that to realise inflation we need
a three-modulus model. As the determination of nonperturbative
superpotential corrections for a given Calabi-Yau is very difficult,
we use a toy model in which the existence of appropriate 
superpotential corrections is assumed.
We denote the 4-cycle volumes by $\tau_i = \textrm{Re}(T_i)$ and take
the following
simplified form for the
Calabi-Yau volume,
\bea\label{vol} \mc{V} & = &
\alpha (\tau_1^{3/2} - \sum_{i=2}^n \lambda_i
\tau_i^{3/2}) \nonumber \\
& = & \frac{\alpha}{2 \sqrt{2}} \left[ (T_1 + \bar{T_1})^{3/2} -
\sum_{i=1}^n \lambda_i (T_i + \bar{T}_i)^{3/2} \right]. \eea 
$\tau_1$ controls the overall
volume and $\tau_2, \ldots, \tau_n$ are blow-ups whose only
non-vanishing triple intersections are with themselves. 
$\alpha$ and  $\lambda_i$ are positive model-dependent constants. The
minus signs are necessary to ensure $\frac{\partial^2 \mc{V}}{\partial
\tau_i  \partial \tau_j}$ has signature $(1, h^{1,1} -1)$
\cite{CandelasDeLaOssa}. The $\mbb{P}^4_{[1,1,1,6,9]}$ case corresponds
to (\ref{vol}) with $n=1$.

The dilaton and complex structure moduli are as usual 
flux-stabilised. We take the K\"ahler moduli superpotential to
be\footnote{ More generally we could take  $W = W_0 +
\sum_{i=2}^n A_i e^{-a_{ij} T_j}$. The effect of this is to alter
the condition (\ref{eqRAT}) in a model-dependent fashion.
As long as the modified form of (\ref{eqRAT}) can be satisfied,
inflation occurs and the results for
the inflationary parameters are unaffected.
In general we expect this to be possible, although we note
that there do exist
models, such as the $\mc{F}_{11}$ model of \cite{hepth0404257}, for
which this cannot be achieved.}
 \be
\label{eqSuP}
W = W_0 + \sum_{i=2}^n A_i
e^{-a_i T_i}, \ee
where $a_i = \frac{2 \pi}{g_s N}$.
The $\alpha'$-corrected K\"ahler potential is \be \label{eqKPA}
\mc{K} = \mc{K}_{cs} - 2 \ln \left[ \alpha \left(\tau_1^{3/2} -
\sum_{i=2}^n \lambda_i \tau_i^{3/2}\right) + \frac{\xi}{2}
\right],
\ee
where $\xi = -\frac{\zeta(3) \chi (M)}{2 (2 \pi)^3}$.
As the dilaton is fixed and regarded as a constant, we can define 
the moduli using either string or Einstein-frame volumes; we use the former.
The latter would correspond to replacing $a_i \to a_i g_s$
and $\xi \to \xi g_s^{-3/2}$ in (\ref{eqSuP}) and (\ref{eqKPA}) - the physics is of
course the same.

From the results of chapter \ref{chapterLargeVol} we anticipate that
at the minimum $\tau_1 \gg \tau_i$
and $\mc{V} \gg 1$. In this limit the scalar potential is
\be
V = e^{\mc{K}} \left[ \mc{K}^{i \bar{j}} \partial_i W \partial_{\bar{j}} \bar{W}
+ \mc{K}^{i \bar{j}} \left((\partial_i K) W) \partial_{\bar{j}} \bar{W} +
c.c. \right) \right] + \frac{3 \xi W_0^2}{4 \mc{V}^3}.
\ee
As we need $\xi > 0$, we require $h^{2,1} > h^{1,1}$. For the K\"ahler
potential (\ref{InflatingK}), we have
\be
\mc{K}^{i \bar{j}} \sim \frac{8 \mc{V} \sqrt{\tau_i}}{3 \alpha \lambda_i} \delta_{ij} + \mc{O}(\tau_i \tau_j).
\ee
$\mc{K}^{i \bar{j}}$ is real and
 satisfies $\mc{K}^{i \bar{j}} \partial_{\bar{j}}
  \mc{K} = 2 \tau_i ( 1 + \mc{O}(\mc{V}^{-1}))$.
At large volume only the leading part of $\mc{K}^{i \bar{j}}$ is
relevant and the scalar potential becomes \be \label{eqSPT} V =
\sum_i \frac{8 (a_i A_i)^2 \sqrt{\tau_i}}{3 \mc{V} \lambda_i
\alpha} e^{-2 a_i \tau_i} - \sum_i 4 \frac{a_i A_i}{\mc{V}^2} W_0
\tau_i e^{-a_i \tau_i} + \frac{3 \xi W_0^2}{4 \mc{V}^3}. \ee The minus
sign in the second term arises from setting the $b_i$ axion to its
minimum. There are terms subleading in $\mc{V}$, but  
importantly these only depend on $\tau_i$ through the overall volume. 
This is crucial, as it ensures that at large
$\tau_i$ the variation of the potential with $\tau_i$ is
exponentially suppressed. We can find the global minimum by
extremising (\ref{eqSPT}) with respect to $\tau_i$. Doing this at
fixed $\mc{V}$, we obtain \be (a_i A_i) e^{-a_i \tau_i} = \frac{3
\alpha \lambda_i W_0}{2 \mc{V}} \frac{(1 - a_i \tau_i)}{(\half - 2
a_i \tau_i)} \sqrt{\tau_i}. \ee If we approximate $a_i \tau_i \gg
1$ (which is valid at large volume as $a_i \tau_i \sim \ln
(\mc{V}))$, then substituting this into the potential
(\ref{eqSPT}) generates a contribution
\be
\label{eqEffTauCont}
\frac{-3 \lambda_i W_0^2}{2 \mc{V}^3} \tau_{i,min}^{3/2} \alpha
 = \label{eqMCP} \frac{-3
\lambda_i W_0^2 \alpha}{2 \mc{V}^3 a_i^{3/2}} (\ln \mc{V} -
c_i)^{3/2}
\ee
where $c_i = \ln (\frac{3 \alpha \lambda_i W_0}{2
a_i A_i})$. At large values of $\ln \mc{V}$, the resulting
effective potential for the volume $\mc{V}$ once all $\tau_i$ fields are minimised is
\be V = \frac{-3 W_0^2}{2 \mc{V}^3} \left( \sum_{i=2}^n \left[
\frac{\lambda_i
  \alpha}{a_i^{3/2}} \right] (\ln \mc{V})^{3/2} -
\frac{\xi}{2} \right). \ee 
This effective potential is another way of understanding why the
 potential of chapter \ref{chapterLargeVol} has a minimum at exponentially large volumes.
To ensure the global minimum is at $V=0$, we must include an uplift term:
\be
\label{eqPUP}
V = \frac{-3 W_0^2}{2 \mc{V}^3} \left( \sum_{i=2}^n \left[ \frac{\lambda_i
  \alpha}{a_i^{3/2}} \right] (\ln \mc{V})^{3/2} -
\frac{\xi}{2} \right) + \frac{\gamma W_0^2}{\mc{V}^2}, \ee where
$\gamma \sim \mc{O}(\frac{1}{\mc{V}})$ parametrises 
the uplift. By tuning $\gamma$, the potential (\ref{eqPUP})
(and by extension its full form (\ref{eqSPT})) has a Minkowski or
small de Sitter minimum.

To obtain inflation we consider the potential away from the minimum.
We take one of the `small' moduli, say $\tau_n$, 
and displace it far from its minimum. At constant volume the potential is
exponentially flat along this direction, and the modulus
rolls back in an inflationary fashion. We do not need to worry about
initial conditions.
While we do not know how the moduli evolution
starts, we do know how it must end, namely with all moduli at their
minima. Given this - we have nothing new to say here on the
overshoot problem \cite{hepth9212049} - inflation occurs as the last K\"ahler modulus
rolls down to its minimum.

It is necessary that all other moduli, and in particular the volume, are
stable during inflation.
Displacing $\tau_n$ from its minimum nullifies the effective
contribution (\ref{eqEffTauCont})
made by the stabilised $\tau_n$ to the volume potential. During inflation
the effective volume potential is
\be
\label{eqMPI}
V = \frac{-3 W_0^2}{2 \mc{V}^3} \left( \sum_{i=2}^{n-1} \left[ \frac{\lambda_i
  \alpha}{a_i^{3/2}} \right] (\ln \mc{V})^{3/2} -
\frac{\xi}{2} \right) + \frac{\gamma W_0^2}{\mc{V}^2}. \ee
Provided that the ratio
 \be  \label{eqRAT}
\rho\ \equiv \ \frac{\lambda_n}{a_n^{3/2}} \, : \, \sum_{i=2}^{n}
\frac{\lambda_i}{a_i^{3/2}} \ee is sufficiently
small\footnote{This can be quantified in explicit models. For
a 3-modulus model, the condition on the ratio $\rho$ is $9.5
(\ln \mc{V} ) \rho  < 1$. 
To obtain inflation with correct density perturbations, the
  appropriate volumes are
$\mc{O}(10^5 - 10^7)$, which
can be satisfied using sensible values for
  $\lambda_i$ and $a_i$.}, there is little
difference between
  (\ref{eqPUP}) and (\ref{eqMPI}) and the volume modulus will be stable
during inflation.  
As we obviously require $\rho < 1$, it follows that at least three K\"ahler
moduli are necessary.
While (\ref{eqRAT}) can always be satisfied by an appropriate choice
  of $a_i$, the presence of the summation implies that this 
  becomes easier the more K\"ahler moduli are present.

\begin{figure}[ht]
\linespread{0.2}
\begin{center}
\makebox[16cm]{ \epsfxsize=16cm \epsfysize=10cm \epsfbox{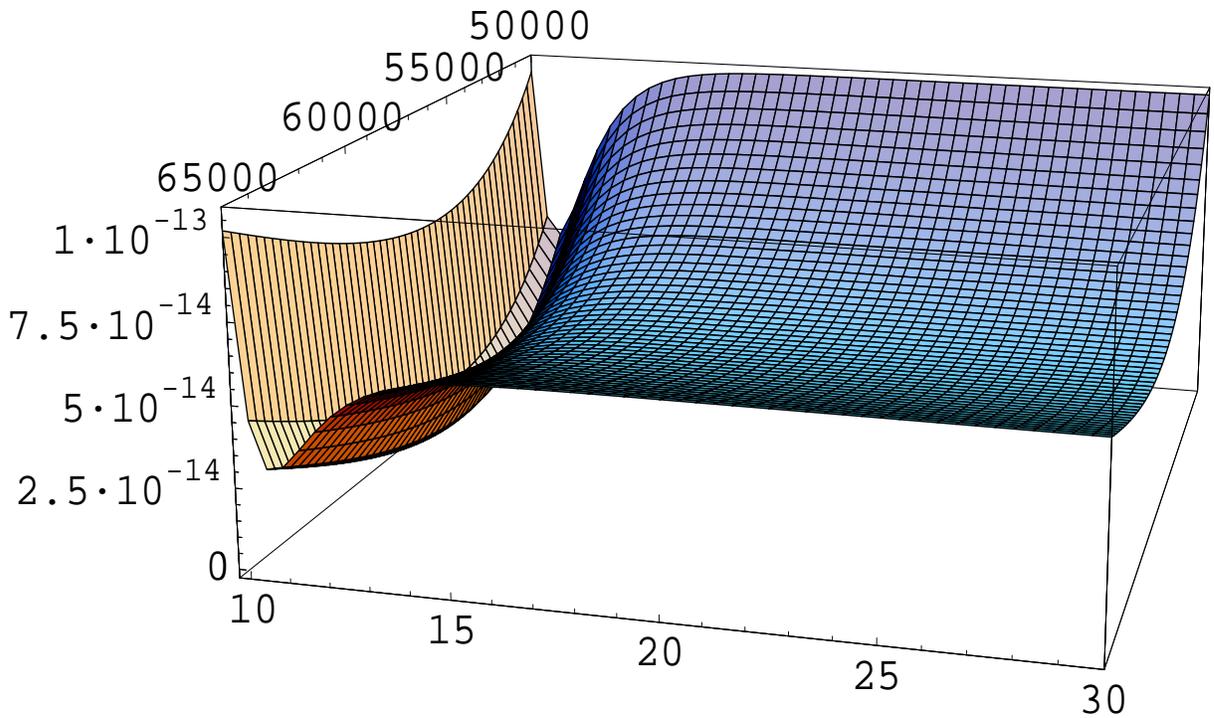}}
\end{center}
\caption{ Inflationary potential: the inflaton lies along the
  x-direction and the volume along the y-direction.}
\label{picIP}
\end{figure}

We illustrate the form of the resulting inflationary potential in figure
\ref{picIP}, showing the inflaton and volume directions.

\section{Inflationary Potential and Parameters}

Let us quantify the resulting potential and compute the
inflationary parameters. The inflationary potential is
read off from (\ref{eqSPT}) to be \be V_{inf} = V_0 - \frac{4
\tau_n W_0 a_n A_n e^{-a_n \tau_n}}{\mc{V}^2}, \ee as the double
exponential in (\ref{eqSPT}) is irrelevant during inflation.
During inflation $V_0$ is constant and can be parametrised as \be
V_0 = \frac{\beta W_0^2}{\mc{V}^3}. \ee ($\frac{1}{\mc{V}^3}$ sets the
scale of the potential during inflation). However, $\tau_n$ is not
canonically normalised, as to leading order in volume 
\be 
\mc{K}_{n \bar{n}} = \frac{3 \lambda}{8 \sqrt{\tau_n} \mc{V}}. 
\ee 
The
canonically normalised field is \be \tau_n^{c} = \sqrt{\frac{4
\lambda}{3 \mc{V}}} \tau_n^{\frac{3}{4}}. \ee
In terms of
$\tau_n^{c}$, the inflationary potential is
\be
V = V_0 - \frac{4
W_0 a_n A_n}{\mc{V}^2}
\left(\frac{3 \mc{V}}{4 \lambda} \right)^{2/3}  (\tau_n^{c})^{4/3}
\exp \left[-a_n \left(\frac{3 \mc{V}}{4
    \lambda}\right)^{2/3} (\tau_n^{c})^{4/3}\right].
\ee
This is similar, but not identical, to the textbook potential
$V = V_0 (1 - e^{-\tau})$. Note however the enormous ($\gtrsim 10^3$)
factor of $\mc{V}^{2/3}$ in the exponent.
Although $\tau_n^c$ is canonically
normalised, it has no natural geometric interpretation and for clarity
we shall express the inflationary
parameters in terms of $\tau_n$, the 4-cycle volume.

The slow-roll parameters are defined by
\bea
\epsilon & = & \frac{M_P^2}{2} \left( \frac{V'}{V} \right)^2, \\
\eta & = & M_P^2 \frac{V''}{V}, \\
\xi & = & M_P^4 \frac{V' V^{'''}}{V^2},
\eea
with the derivatives being with respect to $\tau_n^c$.
These can be evaluated to give
\bea
\epsilon & = & \frac{32 \mc{V}^3}{3 \beta^2 W_0^2} a_n^2 A_n^2
\sqrt{\tau_n} (1 - a_n \tau_n)^2 e^{-2 a_n \tau_n}, \nonumber \\
\eta & = & - \frac{4 a_n A_n \mc{V}^2}{3 \lambda \sqrt{\tau_n} \beta W_0}
\left[ (1 - 9 a_n \tau_n + 4 (a_n \tau_n)^2) e^{-a_n \tau_n} \right],
\\
\xi & = & \frac{- 32 (a_n A_n)^2 \mc{V}^4}{9 \beta^2 \lambda^2 W_0^2 \tau_n}
 (1 - a_n \tau_n) \left(1 + 11 a_n \tau_n - 30 (a_n \tau_n)^2 + 8
(a_n \tau_n)^2 \right) e^{-2 a_n \tau_n}.\nonumber
 \eea
 Then $\xi \ll
\epsilon, \eta \ll 1$ provided that $e^{-a_n
  \tau_n} \ll \frac{1}{\mc{V}^2}$.

Within the slow-roll approximation, the spectral index and its
running are given by
\bea
\label{eqSIR}
n - 1 & = & 2 \eta - 6 \epsilon + \mc{O}(\xi), \\
\frac{d n}{d \ln k} & = & 16 \epsilon \eta - 24 \epsilon^2 - 2 \xi.
\eea
The number of efoldings is given by
\be
N_e = \int_{\phi_{end}}^{\phi} \frac{V}{V'} d \phi,
\ee
which may be expressed as
\be
N_e = \frac{-3 \beta W_0 \lambda_n}{16 \mc{V}^2 a_n A_n}
\int_{\tau_n^{end}}^{\tau_n} \frac{e^{a_n \tau_n}}{\sqrt{\tau_n} (1 -
  a_n \tau_n)} d \tau_n.
\ee Matching the COBE normalisation for the density fluctuations
$\delta_H = 1.92 \times 10^{-5} $ requires \be \label{eqCNF}
\frac{V^{3/2}}{M_P^3 V'} = 5.2 \ti 10^{-4}, \ee where the LHS is
evaluated at horizon exit, $N_e = 50 - 60$ efoldings before the
end of inflation. This condition can be expressed as \be
\label{eqDP} \left( \frac{g_s^4}{8\pi} \right) \frac{3 \lambda \beta^3 W_0^2}{64 \sqrt{\tau_n} (1 -
a_n \tau_n)^2} \left( \frac{W_0}{a_n A_n} \right)^2 \frac{e^{2 a_n
\tau_n}}{
    \mc{V}^6} = 2.7 \ti 10^{-7}.
\ee
We have here included a factor of $\frac{g_s^4}{8 \pi}$ that should
properly be included as an overall normalisation in $V$. This arises
from the prefactor in $W$ and is discussed in the Appendix.
The condition (\ref{eqCNF}) determines the normalisation of the potential and
in practice we use it as a constraint on the stabilised volume.

Finally, the tensor-to-scalar ratio is
\be
\label{eqTSR}
r \sim 8 \epsilon.
\ee

\subsection{Model Footprints}

What are the predictions of the above model?
There are various undetermined
parameters arising from the detailed microphysics, such as the
threshold correction $A$ or tree-level superpotential $W_0$. In
principle, these are determined by the specific Calabi-Yau with its brane
and flux configurations, but they are prohibitively difficult to
calculate in realistic examples.
However, it turns out that the most important results are independent
of these parameters.
In particular, solving equations (\ref{eqSIR}) to (\ref{eqTSR})
numerically,
we find the robust results
\bea
\eta & \approx & -\frac{1}{N_e}, \\
\epsilon & < & 10^{-12}, \\
\xi & \approx & -\frac{2}{N_e^2}.
\eea
These results are not so surprising, given the similarity of the
potential to the textbook form $V_0 ( 1 - e^{-\tau})$.
Taking a range $N_e = 50 \to 60$, we obtain in the slow-roll approximation
\bea
0.960 & < n < & 0.967, \\
-0.0006 & < \frac{d n}{d \ln k} < & -0.0008, \\
0 & < \vert r \vert < & 10^{-10},
\eea
where the above uncertainties arise principally from the number of e-foldings. If we go beyond the
slow-roll approximation, the expression for $n$ will receive
$\mc{O}(\xi)$ corrections - these are minimal and can be neglected.

To evaluate the inflationary energy scale, it is convenient to
reformulate the COBE normalisation of density
perturbations $\delta_H
= 1.92 \times 10^{-5} $ as \be \frac{V^{1/4}}{\epsilon^{1/4}} =
6.6 \ti 10^{16} \textrm{GeV}. \ee
Unlike the predictions for the spectral index, the required internal
volume is dependent on the microscopic parameters.
For typical values of these 
this is found numerically to take a range of values \be 10^5 l_s^6
\le \mc{V} \le 10^7 l_s^6, \ee where $l_s = (2 \pi)
\sqrt{\alpha'}$. As the moduli stabilisation mechanism of
chapter \ref{chapterLargeVol} naturally generates
exponentially large volumes, these can be achieved without
difficulty. 
The range of $\epsilon$ at horizon exit
is $10^{-13} \ge \epsilon \ge 10^{-15}$, and thus the
inflationary energy scale is rather low, \be V_{inf} \sim 10^{13}
\textrm{GeV}. \ee This implies in particular that tensor
perturbations would be unobservable in this model.

There is no practical upper limit on the number of efoldings attainable. The potential is exponentially flat as the
inflaton 4-cycle increases in volume. A very large number of efoldings is
achieved by a very small variation in the original inflaton value and barring
cancellations we therefore expect $N_{e,total} \gg 60$ in these models.

In these compactifications, the lightest non-axionic modulus has a
mass \cite{hepth0505076}
\be
M \sim \frac{M_P}{\mc{V}^{3/2}}.
\ee
Thus even at the larger end of volumes $M \gg \mc{O}(10) \textrm{TeV}$ and
there is no cosmological moduli problem. Of course, this is somewhat
trivial: the cosmological moduli problem concerns a tension between
TeV-scale supersymmetry and long-lived moduli. Focusing on the
inflation scale removes the long-lived moduli, but it also removes
TeV-scale supersymmetry. It remains an open problem to build
well-motivated models containing both inflationary and weak scales.

As inflation takes place
within supergravity, 
reheating in the above model can be described within field theory. As the cycle volume is
the gauge coupling for a brane wrapped on the cycle, there
exists a coupling
$$
\tau_n \int F_{\mu \nu} F^{\mu \nu}.
$$
If $\tau_n$ is the inflaton, it can decay to radiation, rehating the
universe and recovering the Hot Big Bang.

As indicated earlier, 
we do not need to concern ourselves with initial conditions for inflation. Given that the moduli
attain their minimum, the inflaton is simply the last K\"ahler modulus
to roll down to the minimum. We do not need to worry about
interference from the evolution of the other moduli, as once they roll
down to their minimum they become heavy and rapidly decouple from
inflationary dynamics.

\subsection{Additional Corrections and Extensions}

The inflationary mechanism presented here relies on the exponential
flatness of the $\tau_n$ direction at constant volume. This is
unbroken by the tree-level K\"ahler potential, the included $\alpha'^3$
correction of \cite{hepth0204254} and the uplift term.
The uplift terms have several possible sources
\cite{hepth0301240, hepth0309187,  hepth0402135}, but all
scale inversely with the volume
\be V_{uplift} \sim \frac{1}{{\mc V}^{\alpha}}, \ee where
$\frac{4}{3} \le \alpha \le 2$. As the
modular dependence is encoded through the overall volume, rather than depending
explicitly on all the moduli,
at constant volume the $T_n$ direction is extremely flat for large
values of $T_n$.

Let us discuss possible effects that may spoil the exponential
flatness,
first considering superpotential effects. The nonrenormalisation
theorems guarantee that the K\"ahler moduli cannot appear
perturbatively in $W$.
However, the flatness could be spoiled if the functions
$A_i$ of (\ref{eqNP}) depended polynomially on the K\"ahler moduli.
From (\ref{eqSPT}), a term $A(T_j) e^{-T_i}$ in the superpotential would source an
effective polynomial term for $T_j$ once $T_i$ was stabilised.
However, as the $A_i$
must both be holomorphic in $T_i$ and respect the axion shift symmetries,
this polynomial dependence on $T_i$ cannot occur. Indeed, in the models
for which these threshold corrections have been computed
explicitly, the functions $A_i$ do not depend on the K\"ahler moduli
\cite{hepth0404087}. Combined with non-renormalisation results, this
means that the exponential flatness cannot be lifted by
superpotential effects.

The other possibility is that the exponential flatness may be lifted
by corrections to
the K\"ahler potential depending 
on $\tau_n$ such as were discussed in chapter \ref{ChapterAxions}. Both the tree-level
K\"ahler potential and the $\mc{O}(\alpha'^3)$ corrections of
\cite{hepth0204254} have the property that their contribution to the
scalar potential depends only on the overall volume. This does not
lift the exponential flatness of the $\tau_n$ direction at constant $\mc{V}$.
The possible existence of such corrections is difficult to analyse
explicitly and may depend on whether or not branes are wrapped on the
relevant cycle.
An example of such corrections would be the open string corrections computed in
\cite{hepth0508043, hepth0508171}. However this calculation does not
directly apply, as it is on an orbifold and does not
involve blow-up modes such as we have used
for the inflaton.

The upshot is that the exponential flatness of the
$\tau_n$ direction is not broken by any of the known corrections. 
In general, any correction that can be expressed in terms of the
overall volume will not alter the exponential flatness of the $\tau_n$ direction.
If
corrections existed which did break this exponential flatness, it
would be necessary to examine their form and magnitude - it is not after all
necessary that the exponential flatness survive for all values of
$\tau_n$, but merely for those relevant during the last sixty e-folds.

Finally, we have used an oversimplified form for the Calabi-Yau,
picturing it as simply a combination of a volume cycle and blow-up modes.
This is not necessary for the inflationary mechanism described
here. 
Whilst in (\ref{vol}) we assumed $h^{1,1} - 1$ moduli
to be blow-ups whose only nonvanishing triple intersection was with
themselves, 
a single such modulus would be perfectly adequate as an inflaton.
Indeed, even this is not necessary - 
the minimal requirement is simply a flat
direction, which originates from the no-scale behaviour and is broken by
nonperturbative effects. The condition necessary to ensure the volume is stable
during inflaton will then be a generalisation of (\ref{eqRAT}).

\section{Discussion}

This chapter has described a general but simple scenario of inflation in
string theory. Its advantages are that it does not require fine tuning of parameters,
applies to a very large class of compactifications and is
predictive at the level that can be ruled out within a few years.
This scenario realises large field inflation in a natural way. The
main properties of these models are the existence of flat
directions broken by non-perturbative effects. 
The flat directions
have their origins in the no-scale property of the K\"ahler
potential and are generic for IIB K\"ahler moduli, as is the
appearance of instanton-generated nonperturbative superpotentials.
The scenario is embedded in the exponentially large volume
compactifications of \cite{hepth0502058, hepth0505076} and
requires $h^{2,1} > h^{1,1}$ and $h^{1,1} > 2$. This last
requirement is necessary to ensure that the volume is stabilised
during inflation.

The main, robust numerical prediction is for a spectral index
\be
n_s = 1 - \frac{2}{N_e} = 0.96 \to 0.967.
\ee
This is in good agreement with the three-year WMAP results \cite{astroph0603449}
\be
n_s = 0.951 \pm 0.016.
\ee
Tensor perturbations are unobservable in this model, consistent with
current observations which see no evidence for them.

Notice that the volumes required to obtain inflation, while large, are not
extremely large as the string scale is only a few
orders of magnitude below the Planck scale. The necessary volumes
of $\mc{O}(10^5-10^7)$ in string units can be obtained by natural choices of
the exponential parameters $a_i$. 

Although there are many moduli, the inflationary period reduces to a
single-field case. This is because the inflaton is simply the last modulus to
roll down to its minimum, and once other moduli attain their minimum
they rapidly become heavy and decouple from inflationary dynamics.
In principle there
are at least two other fields that may have a nontrivial role during
the cosmological evolution. One is the axion partner of the
inflaton field. We have chosen this to sit at the minimum of its
oscillatory potential, at least for the last sixty efolds. This is
not a strong assumption - because the inflaton
direction is so flat, there is a lot of time for the axion to relax 
from a possibly non-zero inital value to its minimum
before the last sixty efolds start.

There is one substantial problem with the above model (which also
applies for almost all models of string inflation). The energy scale
of inflation cannot be too far removed from the GUT scale. If string moduli
are used for inflation, the same potential that gives inflation will
also generally give GUT-scale supersymmetry breaking. We saw in
chapter \ref{chapterSoftSusy} that a realistic phenomenology required
a volume $\mc{V} \sim 10^{14}$, much larger than the values $\mc{V}
\sim 10^5 \to 10^7$ 
encountered here. Thus while this inflationary 
model may be appealing by itself, it is difficult to reconcile with
supersymmetry broken at the TeV scale, and this is a problem.

This problem may be less serious in the scenario of chapter
\ref{chapterLargeVol} than in other stringy models. The
minimum of the potential is exponentially sensitive to certain
parameters (such as $g_s$). It is then possible that such parameters
change during inflation in such a way that the location of the minimum
changes from the volumes $\mc{V} \sim 10^5 \to 10^7$ suitable for
inflation to the values $\mc{V} \sim 10^{14}$ suitable for low-energy supersymmetry.
It would be interesting to build an explicit model realising this scenario.

Let us finally discuss the generality of this scenario. 
The main technical assumption we
have used is the direct expression for the volume in terms of the
K\"ahler moduli (\ref{vol}). This was overkill - the only part of the 
assumption we actually used
was that the inflaton modulus appears alone in the volume as
${\mc V}= \ldots -(T_n+\bar{T}_n)^{\frac{3}{2}}$. 
As indicated above, we can relax even this: the absolute minimal
requirement is simply the existence of a flat direction broken by
nonperturbative effects.
There may be several possible inflationary directions - in the above
model, $\tau_2, \ldots, \tau_n$ are all good candidates - with
the particular one chosen determined by which K\"ahler modulus is 
last to attain its minimum.
In each case we expect similar
physics to emerge with a robust prediction for the spectral index of density
perturbations.

\part{Conclusions and Outlook}

\chapter{Conclusions}
\label{seCON}

Let us conclude by summarising the results of this thesis and by
outlining the prospects for future work. 

The thesis has been concerned
with moduli stabilisation in IIB string theory and its phenomenological applications.
The two chapters of Part I were introductory. They motivated the use
of string theory as a framework for physics beyond the Standard Model
and reviewed the use of fluxes to stabilise moduli, from both a
four-dimensional and ten-dimensional perspective. I also described the
flux compactifications of \cite{hepth0105097} which serve as
background to most of this thesis.

Part II was concerned with developing a detailed understanding of moduli
stabilisation. In a IIB context 3-form fluxes stabilise the complex
structure moduli. The large degeneracy of flux choices suggests the
use of statistical methods to understand the loci of the stabilised
moduli. In chapter \ref{StatisticsReview} we used Monte-Carlo techniques to study this explicitly on a
particular Calabi-Yau, finding good agreement with the statistical
predictions of Douglas and collaborators.

The stabilisation of K\"ahler moduli requires the use of
nonperturbative effects. In chapter \ref{ChapterKahlerModuli} 
we reviewed the KKLT scenario in which all moduli are stabilised by
nonperturbative effects, coming from either D3-brane instantons or
gaugino condensation. Such effects are nonperturbative in both the
$g_s$ and $\alpha'$ expansions. These can only give the dominant
contribution to the scalar potential under very particular
circumstances. We gave a careful analysis of this and concluded that
in general $\alpha'$ corrections must be included to study the scalar potential.

In chapter \ref{chapterLargeVol} we gave a detailed analysis of the
scalar potential incorporating $\alpha'$ corrections. We showed that, for arbitrary values of $W_0$,
there in general exists a non-supersymmetric minimum of the scalar
potential at exponentially large volumes. This minimum requires at
least two moduli, one of which is a blow-up mode. The exponentially large
volumes come from a competition between corrections perturbative in
the overall volume and corrections nonperturbative in the 
volume of the small blow-up cycle. These compete in a logarithmic fashion leading to the
exponentially large volume. We studied this potential quantitatively
on $\mbb{P}^4_{[1,1,1,6,9]}$, explicitly finding the minimum and the
spectrum of moduli masses. The overall volume is exponentially
sensitive to the stabilised dilaton, and so in effect different flux
choices allow the string scale to be dialled arbitrarily. In
particular, the choice of an intermediate string scale naturally
produces a TeV scale gravitino mass, giving a dynamic solution of the
hierarchy problem. The small cycle volumes also have the right order
of magnitude to support Standard Model gauge couplings.

The use of some $\alpha'$ corrections invites the concern that the full (unknown)
$\alpha'$ expansion will also be needed. However at very large volumes
this is not true as the $\alpha'$ expansion is controlled. Because of
the no-scale structure, the first correction is required (as it
corrects zero) but higher corrections are not. We analysed this in
chapter \ref{chapterLargeVol} by studying the dimensional reduction of
the local terms in the ten-dimensional IIB action, showing that the
neglected terms give subleading contributions compared to those
included. We also discussed loop corrections not coming from local
ten-dimensional terms, showing how the known examples are subleading
to the $\alpha'^3$ correction we included. We also described how one
can use the existence of a classical geometric limit to constrain the 
form of corrections to the K\"ahler potential.

This large-volume scenario does not require fine-tuning, breaks
supersymmetry and stabilises all moduli in such a way as to address
the hierarchy problem. It therefore deserves a more detailed analysis
of the phenomenology, which was carried out in Part III. The
phenomenology is primarily determined by the value of the string
scale. There is here a tension between the values of the string scale
appropriate to TeV scale supersymmetry breaking and the values
giving inflationary potentials with the correct normalisation of
density perturbations.

Chapters \ref{chapterSoftSusy} and \ref{ChapterAxions} assume an
intermediate string scale $m_s \sim 10^{11} \hbox{GeV}$. In this case
the gravitino mass is $\sim 1 \hbox{TeV}$ and the phenomenology is
that of a TeV-scale MSSM - we assume a brane configuration can be
found realising the MSSM spectrum. The geometric picture is of a very
large bulk space, with approximate volume $10^{14} l_s^6$, together
with some small blow-up cycles on which branes containing Standard
Model matter live. The visible breaking of supersymmetry is
gravity-mediated.  In chapter \ref{chapterSoftSusy} we studied the
soft breaking terms in this scenario. We first calculated the overall
scale of the soft terms and then the fine structure, obtaining a
small hierarchy between scalar and gaugino masses.

In chapter \ref{ChapterAxions} we examined the strong CP problem and
the axionic solution to it. This requires an axion to remain exactly
flat down to the QCD scale, which is in some tension with most
approaches to moduli
stabilisation. We studied this in some generality for
arbitrary string compactifications and proved a no-go theorem on the
existence of supersymmetric minima of the F-term potential consistent
with unfixed axions. For a QCD axion to exist, the moduli potential
must break supersymmetry, which fits well with the large volume models
described in Part II. Lowering the string scale is one of the few ways
to ensure the axion decay constant is in a viable regime, and we
described a simple modification of the large volume models to ensure a
possible QCD axion survives.

Another possible application of moduli potentials is to
inflation. A correct normalisation of density perturbations tends to
require a relatively high string scale. In chapter
\ref{InflationModel} we described an inflationary model developed in
the context of the large volume models of part II. This requires a
string scale close to the GUT scale. The model gives a sharp
prediction of $n_s = 0.96 \to 0.967$ with negligible tensors. The
no-scale properties of the scalar potential are used to avoid the
$\eta$-problem.

The above clearly does not exhaust the possible phenomenology.
The results of chapters \ref{chapterSoftSusy} and \ref{ChapterAxions}
suggest a scenario in
which the gluino and scalars are relatively heavy while the other
gauginos are relatively light. It is interesting to investigate the
phenomenology of such a scenario \cite{ACQSinprogress} in detail and
in particular the low-energy spectrum.

The quantitative accuracy of phenomenological questions is limited by
the lack of an explicit construction of the Standard Model. If such a
construction could be found, it would enable a far more detailed study
of the phenomenology. The Standard Model gauge groups would have to be
supported on the blow-up cycles, and one can envisage the possibility
of developing models like that of \cite{hepth0508089}.

There are several cosmological questions of interest. The inflationary
scale and the MSSM scale are quite different. This leads to the
problem that models addressing
one problem generally do not address the other. 
During inflation, the stable values
of moduli fields change. In the scenario described here, the
stabilised volume is exponentially sensitive to the stabilised
dilaton. If the dilaton were to change its value during the
inflationary epoch, the scale of the potential could conceivably be
dynamically driven after inflation from the GUT to the TeV scales. It may then be
possible to merge the models of chapters \ref{chapterSoftSusy} and
\ref{ChapterAxions} with that of chapter \ref{InflationModel}. It
would be interesting to see if this possibility can be realised explicitly.

Reheating and the thermal history of the universe is another interesting question.
The potential described in chapter \ref{chapterLargeVol} has a rather
distinctive and unusual form. It would be interesting to study the
cosmological evolution of moduli in this potential: this may relate to
the problems of the paragraph above.

All the above topics I leave for work in the (near) future.

\begin{appendix}

\chapter[Appendix]{}

In the Appendix we collect together some technical results whose
derivation would interrupt the main flow of the text. Section
\ref{Kerimsbit} is based on a part of \cite{hepth0505076} primarily due to Kerim Suruliz.

\section{Dimensional Reduction of the IIB Action}
\label{appendixsec1}

In this section we derive the prefactors in the Gukov-Vafa-Witten superpotential.

The bosonic type IIB supergravity action in string frame is \cite{PolchinskiBook}
\be
S_{IIB} = \frac{1}{(2 \pi)^7 \alpha'^4} \int d^{10}x \sqrt{-g}
\left\{ e^{-2 \phi}[\mc{R} + 4(\nabla \phi)^2] - \frac{F_1^2}{2} -
  \frac{1}{2 \cdot 3!} G_3 \cdot \bar{G}_3 - \frac{\tilde{F}_5^2}{4
    \cdot 5!} \right\}.
\ee
It is convenient to work in Einstein frame. We redefine
\be
\label{Einsteinframe}
g_{MN} = e^{(\phi - \phi_0)/2} \tilde{g}_{MN},
\ee
where $\phi_0$ =  $\langle \phi \rangle$. The factor of
$e^{\frac{\phi_0}{2}}$ is to ensure that
$g$ and $\tilde{g}$ are identical in the physical vacuum.
The action is then
\be
\label{dilatonomitted}
\frac{2 \pi e^{-2 \phi_0}}{l_s^8} \int d^{10} x \sqrt{-\tilde{g}}
\left\{ \tilde{\mc{R}} - \frac{\partial_M S \partial^M \bar{S}}{2
  (\textrm{Re } S)^2}
- \frac{e^{\phi_0} G_3 \cdot \bar{G}_3}{12 \, \textrm{Re } S} -
\frac{e^{2 \phi_0} \tilde{F}_5^2}{4 \cdot 5!} \right\},
\ee
where $l_s = 2 \pi \sqrt{\alpha'}$ and $S = e^{-\phi} + i C_0$.
We neglect warping effects; as discussed in section \ref{tendpic}, these are
subleading in the large volume limit.
In the orientifold limit and in the absence of warping, $\tilde{F}_5 =
0$ and $\partial_M S = 0$.
The dimensional reduction of (\ref{dilatonomitted}) then gives
\be
S = \frac{2 \pi}{g_s^2 l_s^8} \left( \int d^{4} x \sqrt{-\tilde{g_4}} \tilde{\mc{R}}_4 \mc{V}
-  \overbrace{\int d^4 x \sqrt{-\tilde{g_4}} \left( \int d^6 x \sqrt{\tilde{g}_6}
\frac{e^{\phi_0} G_3 \cdot \bar{G}_3}{12 \, \textrm{Re } S}
\right)}^{V_{flux}} \right).
\ee
We use $g_s = e^{\phi_0}$ and define 
\be
\mc{V} \equiv \int d^6 x \sqrt{\tilde{g}_6}.
\ee
In 4d Einstein frame, the canonically normalised Einstein-Hilbert action must be
\be
S_{EH} = \frac{1}{16 \pi G} \int d^4 x \sqrt{-g_E} \mc{R}_E \equiv
\frac{M_P^2}{2} \int d^4 x \sqrt{-g_E} \mc{R}_E.
\ee
This implies $\tilde{g_4}$ and $g_E$ are related by $\tilde{g_4} = g_E \frac{{\cal V}^0_s}{\mc{V}_s}$,
where $\mc{V} \equiv \mc{V}_s l_s^6$ and ${\cal V}^0_s = \langle \mc{V}_s \rangle$.
This gives
\be
\label{UsefulRelats}
M_P^2 = \frac{4 \pi {\cal V}_s^0}{g_s^2 l_s^2} \quad \textrm{ and } \quad
m_s = \frac{g_s}{\sqrt{4 \pi {\cal V}_s^0}} M_P.
\ee
Dimensional reduction determines the K\"ahler potential to be \cite{hepth0403067}
\be
\label{FullKahlerPotential}
\frac{\mc{K}}{M_P^2} = - 2 \ln(\mc{V}_s) -
\ln( S + \bar{S}) - \ln \left( -i \int \Omega \wedge \bar{\Omega} \right).
\ee
The superpotential can be found from $V_{flux}$.
If
\be
W = \int G_3 \wedge \Omega,
\ee
we would obtain
\be
V_{flux} = \frac{4 \pi (\mc{V}_s^0)^2}{g_s l_s^8} \int d^4x
\sqrt{-g_E} e^{\mc{K}/M_P^2} \left[ \mc{K}^{a \bar{b}} D_a W D_{\bar{b}} W
- 3 W \bar{W} \right],
\ee
with $a, b$ running over all moduli.
Thus if
\be
\hat{W} = \frac{g_s^{\frac{3}{2}}M_P^3}{\sqrt{4 \pi} l_s^2} \int G_3 \wedge \Omega,
\ee
the scalar potential takes the standard $\mc{N} = 1$ form
\be
\label{scalarpotential3}
V = \int d^4 x \sqrt{-g_E} e^{\mc{K}/M_P^2} \left[ \mc{K}^{a \bar{b}}
D_a \hat{W} D_{\bar{b}} \bar{\hat{W}} - \frac{3}{M_P^2} \hat{W} \bar{\hat{W}}  \right].
\ee
In verifying this it is necessary to use the relations (\ref{UsefulRelats}).

The K\"ahler potential will receive perturbative corrections and the superpotential
non-perturbative corrections. The $\alpha'$ corrections to the K\"ahler potential
modify the 4-dimensional kinetic terms and arise from the
higher-derivative terms in the ten-dimensional IIB action.
In principle the dimensional reduction of these give the perturbative corrections to the K\"ahler
potential, although this is not the best way to determine them. We obtain \cite{hepth0204254}
\be
\frac{\mc{K}}{M_P^2} = - 2 \ln \left( \mc{V}_s + \frac{\xi
  g_s^{3/2}}{2 e^{3 \phi/2}} \right) - \ln( S + \bar{S}) -
\ln \left( -i \int \Omega \wedge \bar{\Omega} \right).
\ee
where $\xi = - \frac{ \chi(M) \zeta(3)}{2 (2 \pi)^3}$.
The superpotential can receive non-perturbative corrections causing it to depend on
the K\"ahler moduli. These can arise from D3-brane instantons or gaugino condensation.
The generic form of the superpotential is then
\be
\label{WAppendixVer}
\hat{W} = \frac{g_s^{3/2} M_P^3}{\sqrt{4 \pi}} \left( W_0 + \sum
A_i e^{-a_i T_i} \right),
\ee
with 
\be
W_0 = \frac{1}{l_s^2} \left\langle \int G_3 \wedge \Omega \right\rangle.
\ee

\section{Canonical Normalisation of K\"ahler Moduli}
\label{appendixsec2}

In this section we describe how to canonically normalise
the K\"ahler moduli for the 2-modulus $\mbb{P}^4_{[1,1,1,6,9]}$
example discussed in chapter \ref{chapterLargeVol}.
The K\"ahler potential is given by
\be
\label{KahlerPotential}
\mc{K} = \mc{K}_{cs} - 2 \ln \left(\left(T_5 + \bar{T}_5)\right)^{\frac{3}{2}} -
\left(T_4 + \bar{T}_4\right)^{\frac{3}{2}}\right) + \textrm{ constant }.
\ee
Here $T_4 = \tau_4 + i b_4$ and $T_5 = \tau_5 + i b_5$.
All terms depending on dilaton and complex structure moduli
have been absorbed into $\mc{K}_{cs}$. We
also recall that as $\tau_5 \gg 1$,
$\frac{1}{\tau_5}$ serves as a good expansion parameter.

It may be verified that
\bea
\partial_{T_5} \partial_{\bar{T}_5} K & = & \frac{3 \tau_5^{-\half}}{4(\tau_5^{\frac{3}{2}}
- \tau_4^{\frac{3}{2}})} + \frac{9 \tau_4^{\frac{3}{2}} \tau_5^{-\half}}{8 (\tau_5^{\frac{3}{2}} - \tau_4^{\frac{3}{2}})^2},
\nonumber \\
\partial_{T_4} \partial_{\bar{T_5}} K = \partial_{T_5} \partial_{\bar{T_4}} K & = &
\frac{-9 \tau_4^{\frac{1}{2}} \tau_5^{\half}}{8 (\tau_5^{\frac{3}{2}} -  \tau_4^{\frac{3}{2}})},  \nonumber \\
\partial_{T_4} \partial_{\bar{T_4}} K & = & \frac{3 \tau_4^{-\half}}{4(\tau_5^{\frac{3}{2}}
- \tau_4^{\frac{3}{2}})} + \frac{9 \tau_4}{8 (\tau_5^{\frac{3}{2}} - \tau_4^{\frac{3}{2}})^2}.
\nonumber
\eea
These results are summarised by
\be
\mc{K}_{i \bar{j}} = \left( \begin{array}{cc} \mc{K}_{4\bar{4}} & \mc{K}_{4\bar{5}} \\
\mc{K}_{5\bar{4}} & \mc{K}_{5\bar{5}} \end{array} \right) =
\left( \begin{array}{cc} \frac{3 \tau_4^{-\half}}{8\tau_5^{\frac{3}{2}}} + \mc{O}(\frac{1}{\tau_5^3}) &
\frac{-9 \tau_4^\half}{8 \tau_5^{\frac{5}{2}}} + \mc{O}(\frac{1}{\tau_5^4}) \\
\frac{-9 \tau_4^\half}{8 \tau_5^{\frac{5}{2}}} + \mc{O}(\frac{1}{\tau_5^4}) &
\frac{3}{4\tau_5^{2}} + \mc{O}(\frac{1}{\tau_5^{\frac{7}{2}}}) \end{array} \right).
\ee
We denote the values of the fields $(b_4, b_5,
\tau_4, \tau_5)$ at the minimum by $(b_4^0, b_5^0, \tau_4^0,
\tau_5^0)$ and now define
$$
\tau_5^{'} = \sqrt{\frac{3}{2}} \frac{\tau_5}{\tau_5^0}, \quad \tau_4^{'} =
\sqrt{\frac{3}{4}} \frac{\tau_4}{(\tau_5^0)^{\frac{3}{4}}(\tau_4^0)^{\frac{1}{4}}},
\quad b_5^{'} = \sqrt{\frac{3}{2}} \frac{b_5}{\tau_5^0}, \quad b_4^{'} =
\sqrt{\frac{3}{4}} \frac{b_4}{(\tau_5^0)^{\frac{3}{4}}(\tau_4^0)^{\frac{1}{4}}},
$$
$$
\textrm{ and } \quad \tau_4^{'0} =
\sqrt{\frac{3}{4}} \frac{\tau_4^0}{(\tau_5^0)^{\frac{3}{4}}(\tau_4^0)^{\frac{1}{4}}}, \quad
  \tau_5^{'0} = \sqrt{\frac{3}{2}}.
$$
Then after some manipulation, we obtain
\bea
\lefteqn{\sum_{i,j} \mc{K}_{i \bar{j}} \partial_\mu T^i \partial^\mu
  \bar{T}^j = } \nonumber \\
& & \half \Bigg[ \left( \frac{\tau_5^{'0}}{\tau_5^{'}} \right)^{\frac{3}{2}} \left( \frac{\tau_4^{'0}}{\tau_4^{'}}
\right)^{\frac{1}{2}} \partial_\mu b_4^{'} \partial^\mu b_4^{'} + \left(
\frac{\tau_5^{'0}}{\tau_5^{'}} \right)^2 \partial_\mu b_5^{'}
\partial^\mu b_5^{'} - 4\sqrt{3} \left( \frac{\tau_4^{'}}{\tau_4^{'0}}\right)^\half
\left(\frac{\tau_5^{'0}}{\tau_5^{'}}\right)^{\frac{5}{2}} \tau_4^{'0}
\partial_\mu b_4^{'} \partial^\mu b_5^{'}  \nonumber \\
& & + \left( \frac{\tau_5^{'0}}{\tau_5^{'}} \right)^{\frac{3}{2}} \left( \frac{\tau_4^{'0}}{\tau_4^{'}} \right)^{\half}
\partial_\mu \tau_4^{'} \partial^\mu \tau_4^{'} + \left(
\frac{\tau_5^{'0}}{\tau_5^{'}} \right)^2 \partial_\mu \tau_5^{'}
\partial^\mu \tau_5^{'} -4\sqrt{3} \left( \frac{\tau_4^{'}}{\tau_4^{'0}}\right)^\half
\left(\frac{\tau_5^{'0}}{\tau_5^{'}}\right)^{\frac{5}{2}} \tau_4^{'0}
\partial_\tau \tau_4^{'} \partial^\tau \tau_5^{'} \Bigg]. \nonumber
\eea
The moduli are now canonically normalised except for the crossterm,
which is suppressed by $(\tau_5^0)^{\frac{3}{4}}$ and is thus
very small. All such crossterms could be eliminated by field
redefinitions order by order in $\frac{1}{\tau_5}$; however,
negelecting the subleading corrections it is sufficient to use
$\tau_5^{'}$ and $\tau_4^{'}$ as canonically normalised fields.

\section{Soft Term Computations}
\label{Kerimsbit}

Let us introduce the notation ${\cal{V}}'=(T_5+\bar{T}_5)^{3/2}-(T_4+\bar{T}_4)^{3/2}$
(which differs by a factor of $36$ from the Calabi-Yau volume $\cal{V}$).
We write
\begin{eqnarray}
\hat{\K}_{T_5} = -3{(T_5+\bar{T}_5)^{1/2}\over{{\cal{V}}'+{\xi'/2}}},\nonumber\\
\hat{\K}_{T_4} = 3{(T_4+\bar{T}_4)^{1/2}\over{{\cal{V}}'+{\xi'/2}}}.
\end{eqnarray}
We denote $(T_4+\bar{T}_4)^{1/2} = X$ 
and  $(T_5+\bar{T}_5)^{1/2}=Y$.\footnote{We do not simply use
  $\tau_4$ and $\tau_5$ as this would not be valid if D3-branes
  were present.} 
Then ${\cal{V}}' = Y^3-X^3$ and we can calculate the metric $\K_{T_i \bar{T}_j}$:
\begin{equation}
\K_{T_i \bar{T}_j}=\left(
\begin{array}{cc}
{3\over{2Xx}}+{9X^2\over{2x^2}} &  -{{9XY}\over{2x^2}}\\
-{{9XY}\over{2x^2}} & {-3\over{2Yx}}+ {{9Y^2}\over{2x^2}}
\end{array}
\right),
\end{equation}
and the inverse metric $\K^{\bar{T}_i T_j}$
\begin{equation}
\left(
\begin{array}{cc}
-2X ({\cal{V}}'+{\xi/2})(2Y^3+X^3 -{\xi'/2})\over{3(\xi'/2-2{\cal{V}}')}
& -2X^2 Y^2{({\cal{V}}'+{\xi'/2})\over{({\xi'/2}-2{\cal{V}}')}}\\
-2X^2 Y^2{({\cal{V}}'+{\xi'/2})\over{({\xi'/2}-2{\cal{V}}')}}
& -2Y ({\cal{V}}'+{\xi'/2})(2X^3+Y^3 +{\xi'/2})\over{3(\xi'/2-2{\cal{V}}')}
\end{array}
\right).
\end{equation}
In these expressions $x = Y^3 - X^3+\xi'/2 = {\cal{V}}'+\xi'/2.$ We
can now calculate the F-terms as given by formula (\ref{fterms}).
We assume that $\partial_5 W\ll \partial_4 W$ (where $\partial_i\equiv
\partial_{T_i}$) and only include the
nonperturbative contribution corresponding to $T_4.$ The result is
\begin{eqnarray}
F^4 &=& e^{\hat{\K}/2}{{2{\cal{V}}'+\xi'}\over{2{\cal{V}}'-{\xi'/2}}} \left(
-X^2 W + {X\over3} (2Y^3+X^3 - {\xi\over2})(\partial_4 W)\right)\nonumber\\
F^5 &=& e^{\hat{\K}/2}{{2{\cal{V}}'+\xi}\over{2{\cal{V}}'-{\xi'/2}}}
\left( -Y^2 W + X^2 Y^2 (\partial_4 W))
\right)
\end{eqnarray}
After a redefinition of $c$ and $d$, the prefactor of the kinetic term for
the brane modulus $\phi$, $\tilde{\K}_i$,can be rewritten as
\begin{equation}
\tilde{\K}_i = {{cX+dY}\over{{\cal{V}}'+{\xi'/2}}}.
\end{equation}
To calculate scalar masses we need 
$$
\partial_m \partial_{\bar{n}}
\log{\tilde{\K}_i} = \partial_m \partial_{\bar{n}}\left( \log (cX+dY)
- \log \left ({\cal{V}}'+{\xi'\over2} \right) \right).
$$ 
As $ \hat{\K} = -2\log \left( {\cal{V}}'+\xi'/2 \right)+{\hbox{constant}},
$
$$ 
- \partial_m \partial_{\bar{n}} \log \left( {\cal{V}}'+{\xi'\over2} \right)
 = \half \hat{\K}_{m\bar{n}}
$$ and so
\begin{equation}
\partial_m \partial_{\bar{n}}
\log{\tilde{\K}_i} = \partial_m \partial_{\bar{n}} \left( \log (cX+dY)
\right)
+ \half \hat{\K}_{m\bar{n}}.
\end{equation}
The necessary derivatives of $\log (cX+dY)$ can be calculated using
$\partial_4 X = 1/(2X), \partial_5 Y = 1/(2Y)$, and likewise
with respect to $\bar{4}$ and $\bar{5}.$ The results are
\begin{eqnarray}
\label{logcu}
\partial_4 \partial_{\bar{4}}
\log (cX+dY) &=& -{c (2cX+dY)\over{4X^3(cX+dY)^2}},\nonumber\\
\partial_4 \partial_{\bar{5}} \log (cX+dY) &=& -{cd\over{4XY(cX+dY)^2}},
\nonumber\\
\partial_5 \partial_{\bar{5}}
\log (cX+dY) &=& -{d(2dY+cX)\over{4Y^3 (cX+dY)^2}}.
\end{eqnarray}

We now wish to find the soft masses, expressing the result as a small
deviation from the no-scale result, which is zero. To do this we
make an expansion assuming that $(\partial_4 W)$ is small (we know
it is  $\mc{O}(\mc{V}^{-1})$ at the minimum of the scalar potential) and
$\xi/{\cal{V}}$ is small. For example, we can note that $F^m \bar{F}^{\bar{n}}$ always
has a prefactor of 
$$
\frac{(2{\cal{V}}'+\xi')^2}{(2\mc{V}'- \xi'/2)^2},
$$
which may be expanded as
\begin{equation}
1+{{3\xi'}\over{2\cal{V}}'} + {\cal{O}}\left( 1\over{\cal{V}}^2 \right).
\end{equation}
From the above analysis we expect that
$F^m \bar{F}^{\bar{n}} \partial_m \partial_{\bar{n}}
\log (cX+dY)$ can be written as $e^{\hat{\K}} \left( -\half|W|^2 + \mc{O}
(\mc{V}^{-\alpha}) \right)$ for some $\alpha>0.$
Indeed, an explicit computation gives
\begin{eqnarray}
\lefteqn{F^m \bar{F}^{\bar{n}} \partial_m \partial_{\bar{n}} \log (cX+dY) =
e^{\hat{\K}}
\left( 1+{{3\xi'}\over{2\cal{V}}'}\right) \ti} \nonumber\\
& &\left(
-{1\over2}|W|^2 + {1\over{3(cX+dY)}} \left( -{{\xi c}\over2}
+ 3d X^2 Y + c X^3 + 2 c Y^3\right)W \left( \partial_4 W \right) \right). \nonumber
\end{eqnarray}
Similarly we have
\begin{equation}
F^m \bar{F}^{\bar{n}} \hat{\K}_{m\bar{n}} =
e^{\hat{\K}} {{3{\cal{V}}'\left(2{\cal{V}}'-{\xi'\over2}\right)\over{2
\left({\cal{V}}'+{\xi'\over2}\right)^2}}}
\left( |W|^2 - {W(\partial_4 W)\over{3\cal{V}}'} \left(
2X^2 {\xi'\over2} + 2X^2 {\cal{V}}' \right)\right).
\end{equation}
Expanding 
$$
\frac{\mc{V}'(2 \mc{V}' - \xi'/2)}{2(\mc{V}'+\xi'/2)^2}
$$
as $1-3\xi'/4\mc{V}'$ we get
\begin{eqnarray}
F^m \bar{F}^{\bar{n}} \partial_m \partial_{\bar{n}} \log {\tilde{\K}_i} &=&
e^{\hat{\K}} \left( 1+{{3\xi'}\over{2\cal{V}}'} \right) \Bigl[
-|W|^2  \nonumber\\
& &
+{1\over{3(cX+dY)}} \left( -{{\xi' c}\over2}+3d X^2 Y + cX^3 + 2c Y^3
\right) W(\partial_4 W) \nonumber\\
& & -W(\partial_4 W) X^2 \left( 1+ {\xi'\over{2\cal{V}}'} \right)\Bigr]
\end{eqnarray}
Further simplifying and neglecting ${\cal{O}} (1/{\cal{V}}^2)$ terms
we obtain
\begin{eqnarray}
\lefteqn{e^{\hat{\K}} \Bigg[ -|W|^2 + {{3\xi'}\over{2\cal{V}}'}|W|^2} \nonumber\\
& &+ {W\over{3(cX+dY)}} \left(
{{-\xi' c}\over 2} + 3d X^2 Y + cX^3 + 2cY^3 \right) \left( \partial_4
W \right)
- W(\partial_4 W) X^2 \Bigg]. \nonumber
\end{eqnarray}
The $-e^{\hat{\K}} |W|^2$ cancels with $m_{3/2}^2$ so the mass squared of the brane
modulus can be seen to have the form
\begin{equation}
\label{smallmass}
{{3\xi'}\over{2\cal{V}}'} + f(X,Y) (\partial_4 W),
\end{equation}
which is what we expect naively - before the no-scale structure is
broken, the mass is simply zero. There are two sources of no-scale
breaking: $\alpha'$ corrections corresponding to the first term in
(\ref{smallmass}) and nonperturbative corrections corresponding to
the second term. To estimate the size of the soft terms, note that
all the constants involved in (\ref{smallmass}) are expected to be
${\cal{O}} (1)$ (including $c, d, \xi'$) and so the first term
is ${\cal{O}}(1/\cal{V}).$ As the volume is
${\cal{V}}\sim Y^3$, we can estimate
\begin{equation}
f(u,v)\sim {Y^3\over Y} = Y^2 \sim {\cal{V}}^{2/3}.
\end{equation}
We also note that, at the minimum, $\partial_4 W \sim \frac{W_0}{\mc{V}}$.
Then the volume scaling of the second term is ${\cal{V}}^{2/3}/{\cal{V}}
={\cal{V}}^{-1/3}$, and
as $e^{\hat{\K}} \sim 1/{{\cal{V}}^2}$ the moduli masses squared
scale as ${\cal{O}} (1/{\cal{V}}^{7/3}).$
Finally, by considering the form of the superpotential
(\ref{WAppendixVer}), we see that $W (\partial_4 W) \sim W_0^2 g_s^4$.
Putting the factors together, we get
\be
m_i^2 = \mc{O}(1) \frac{g_s^4 W_0^2}{4 \pi (\mc{V}_s^0)^{7/3}} M_P^2.
\ee

\end{appendix}

\clearpage
\bibliography{thesisRefs}
\bibliographystyle{utphys}

\end{document}